\documentclass{amsart}

\usepackage{float}
\usepackage{graphicx}
\usepackage{caption}   

\usepackage{needspace}
\usepackage[section]{placeins} 

\usepackage{url,multirow}
\usepackage{amsmath, amsthm, amsfonts, enumerate}
\usepackage{comment}
\usepackage{placeins}
\usepackage{float}

\theoremstyle{definition}
\newtheorem{example}{Example}

\theoremstyle{plain}

\usepackage{multicol}
\usepackage[margin=1 in]{geometry}
\usepackage{tabularx}    
\usepackage[table]{xcolor} 
\usepackage{booktabs}    
\usepackage{colortbl}    
\usepackage{tikz, verbatim}

\usepackage[colorinlistoftodos]{todonotes}

\newcommand{\adam}{\textcolor{red}{Adam comment: }\textcolor{red}}
\newcommand{\dave}{\textcolor{blue}{Dave comment: }\textcolor{blue}}

\begin{document}

    \title[Moderating Effects of Condorcet and IRV]{Candidate Moderation under Instant Runoff and Condorcet Voting: Evidence from the Cooperative Election Survey}
    
    

\author[McCune, Jones, Schultz, et al.]{David McCune}

\author{Matthew I. Jones}

\author{Andrew Schultz}

\author{Adam Graham-Squire}

\author{Ismar Voli\'{c}}

\author{Belle See}

\author{Karen Xiao}

\author{Malavika Mukundan}

\begin{abstract}
This article extends the analysis of Atkinson, Foley, and Ganz in ``Beyond the Spoiler Effect: Can Ranked-Choice Voting Solve the Problem of Political Polarization?''. Their work uses a one-dimensional spatial model based on survey data from the Cooperative Election Survey (CES) to examine how instant-runoff voting (IRV) and Condorcet methods promote candidate moderation. Their model assumes an idealized electoral environment in which all voters possess complete information regarding candidates’ ideological positions, all voters provide complete preference rankings, etc.  Under these assumptions, their results indicate that Condorcet methods tend to yield winners who are substantially more moderate than those produced by IRV. We construct new models based on CES data which take into account more realistic voter behavior, such as the presence of partial ballots. Our general finding is that under more realistic models the differences between Condorcet methods and IRV largely disappear, implying that in real-world settings the moderating effect of Condorcet methods may not be nearly as strong as what is suggested by more theoretical models.
\end{abstract}


\maketitle
\markboth{Moderating Effects of Condorcet and IRV}{McCune, Jones, Schultz, et al.}

\section{Introduction}






\vspace{.2 in}
Instant runoff voting (IRV), also known as ranked-choice voting, is being adopted more widely and attracting increasing interest in the United States. The modern use of (single-winner) IRV in the US dates to 2004, when the city of San Francisco adopted the method for mayoral and Board of Supervisors elections.  IRV has subsequently been implemented for some municipal elections in Minneapolis, New York City, and Oakland, among many others, as well as for statewide elections in Alaska and Maine. Proponents of IRV tout its ability to mitigate the problem of vote-splitting, encourage more civil campaigning, and reduce the number of wasted votes, among other purported benefits. Whether IRV lives up to the promises of its advocates, and whether its benefits outweigh its potential downsides, is a matter of ongoing debate.

As IRV has gained traction, it has sparked renewed interest in ranked voting methods more broadly, especially among academics and voting reform advocates. The increasing use of IRV in real elections enables empirically grounded assessments of the method’s strengths and weaknesses, particularly in comparison with other ranked voting rules. Among these, Condorcet methods are frequently proposed as a theoretically superior alternative. Proponents argue that Condorcet methods yield more representative outcomes than IRV, better mitigate issues of candidate entry and electorate polarization, and are less susceptible to the spoiler effect. 

A recent notable example is  \cite{AFG}, which uses simulated elections based on Cooperative Election Study (CES) data to argue that Condorcet methods tend to elect more moderate candidates, making such methods better suited for polarized electorates. The analysis employs one-dimensional, single-peaked models of voter preferences under ``theoretically ideal'' assumptions. For example, the authors assume that each voter casts a ballot which ranks all of the candidates, and that all voters know the exact ideological position of each candidate. 

We extend the framework in \cite{AFG} by incorporating more realistic features of voter behavior, such as truncated ballots, into CES-based simulations. Our findings show that once such factors are included, Condorcet methods not only select less moderate candidates on average, but the moderating advantage of Condorcet methods over IRV 
is significantly reduced. We also find other ranked choice voting methods whose moderating effects are (on average) stronger than either Condorcet methods or IRV. 

The work in \cite{AFG}, and hence also our work in this article, connects to two inter-connected strands of the literature. The first examines which voting methods most effectively translate voter preferences into representative outcomes. An important metric in this tradition is a method’s ability to select winners ideologically close to the median voter, a criterion closely tied to mitigating polarization. Modern scholarship in this vein dates to the foundational work of Duncan Black \cite{B48} and Anthony Downs \cite{D57}, both of whom formalized how electoral competition under one-dimensional ideological preferences tends to converge toward the median voter. Black’s Median Voter Theorem established that when preferences are single-peaked along an ideological axis, a Condorcet method will elect the candidate preferred by the median voter. Downs extended this reasoning to a broader theory of party competition, arguing that ideological distributions determine whether electoral equilibria promote moderation or extremism. More recent work along these lines includes \cite{F25, FM25, RT24}, which argue for the use of Condorcet methods based on median voter or candidate ``centrist strategy'' arguments. The Median Voter Theorem has also been extended in various ways. For example, Buechel \cite{B13} shows that if a Condorcet winner exists then it coincides with the appropriately defined median voter for single-peaked preferences on median spaces.

While the first strand of literature focuses on theoretical characterizations of median-voter outcomes, the second turns to spatial modeling as a practical framework for studying how well different voting methods behave with respect to various metrics. Atkinson et al. \cite{AFG} adopt this approach, using one-dimensional spatial models to generate ballot data where voter preferences are single-peaked. The notion of single-peakedness also dates to the work of Black \cite{B48}, and has since received much attention in the social choice literature (see the classical work by Moulin \cite{M80,M84}, for example). Spatial models of various dimensions \cite{EH84, EH90} have been widely used to examine a variety of features of voting methods. Merrill \cite{M84} uses spatial models to study which voting methods maximize the social utility of the selected winner, work which was recently revisited by Holliday and Pacuit \cite{HP24}. One-dimensional spatial models in particular have been used to study the effects of ballot truncation in ranked-choice elections \cite{KGF20}, the Condorcet efficiency of various voting rules \cite{KGF25}, or the susceptibility of voting rules to various paradoxes \cite{KMT23, MW}. Such models are used so widely that modern voting software packages such as Preferential Voting Tools \cite{HP25} and VoteKit \cite{DDGGHMW25} include spatial-model ballot generators. In our specific context, spatial models have been used to examine the moderating effect of various voting methods. Recent examples are \cite{T1,T2} by Tomlinson et al., which use one-dimensional models to examine the moderating effect of IRV.

We extend the work in \cite{AFG} by building one-dimensional spatial models generating voter preferences which are approximately single-peaked and which allow for truncated ballots (almost all prior work assumes voters provide complete preferences). Our work makes three primary contributions. First, we find that under our models the moderating effect of Condorcet methods is not nearly as pronounced as in previous literature. For example, we find that the average distance from the IRV winner to the median voter is much closer to the average distance from the Condorcet winner to the median voter than was found in \cite{AFG}. We also find that when we incorporate more realistic voter behaviors into our models, other voting rules such as the Borda Count or Bucklin voting are sometimes more moderating than Condorcet methods. Second, our models incorporate truncated ballots in a natural way. We are aware of very little work concerning the incorporation of truncated ballots into models of voter preferences (\cite{HKRW} is one such example), and thus our work can be seen as contributing to this new and developing project. Finally, we examine how the underlying distribution of voters influences the electoral process and interacts with different kinds of voter behavior. 

We do not intend these results to be an argument for or against any particular ranked choice voting method over another. Instead, we highlight the effects that more nuanced assumptions about voter behavior can have on predictive models like those found in \cite{AFG}.

The paper is structured as follows. In Section \ref{section:description} we provide definitions of the voting methods we study, accompanied by examples. Section \ref{section:Atkinson_overiew} provides an overview of the work in \cite{AFG}, and Section \ref{section:our_models} describes our extension of their models. Sections \ref{section:results}-\ref{section:discussion} provide our results and discussion, while Section \ref{section:conclusion} contains some concluding remarks. There are a number of appendices at the end of the paper which give supplemental data for the interested reader.

\section{Descriptions of Voting Methods}\label{section:description}

In elections which use IRV or a Condorcet method to choose the winner, voters cast preference ballots with a (possibly partial) linear ranking of the candidates. After an election, the ballots are aggregated into a \emph{preference profile}, which shows the number of each type of ballot cast. Table \ref{profile_ex} shows an example preference profile involving the three candidates $A$, $B$, and $C$. The number 20 at the top of the first column denotes that 20 voters ranked $A$ first, $B$ second, and $C$ third. In real-world American IRV elections voters are not required to provide a complete ranking. Table \ref{profile_ex} shows that 50 voters ranked only a single candidate; such ballots are called \emph{bullet votes}. We refer to any ballot which does not provide a complete ranking as a \emph{truncated ballot}. In a 4-candidate election a ballot which ranks only two candidates is truncated but is not a bullet vote, for example. Any ranked-choice voting method takes a preference profile as input and uses the preferences to output a winner (or perhaps a set of winners, in the case of tied outcomes). 

Our focus in this article is the relationship between Condorcet methods and IRV, but we include three other voting methods in our analysis to provide context for our results. The five methods we analyze are defined below. 

\begin{itemize}
\item \textbf{Plurality}: The candidate with the most first-place votes wins.
\item \textbf{Instant Runoff Voting (IRV)}: Eliminate the candidate with the fewest first-place votes and transfer their votes to the next-ranked candidate on the ballot. Continue the process of elimination and transfer until a candidate has the majority of the remaining first-place votes; this candidate is declared the winner.
\item \textbf{Condorcet voting}: The candidate who wins all of their head-to-head match-ups is declared the winner. If no such candidate exists then Condorcet methods differ in the method for selecting a winner. In this case, we use the method of \emph{minimax}, which elects the candidate for whom the maximum pairwise score for another candidate against them  is the minimum such pairwise score among all candidates. (That is, the minimax winner is the candidate with the ``smallest worst head-to-head loss.'')

Because more than 99\% of our simulated elections contain a Condorcet winner, the specific choice of Condorcet method does not make a difference for our results. Our tables and figures all reference ``Condorcet'' as the voting method, but the underlying method is minimax.
\item \textbf{Bucklin}: The method unfolds in rounds, where in each round a candidate is given a \emph{Bucklin score}. In round $i$, each candidate receives a Bucklin score equal to the number of ballots ranking them at or above position $i$. If any candidate's score reaches or exceeds a majority of ballots cast then the process terminates and the candidate with the highest score is declared the winner.
\item \textbf{Borda}: The Borda method assigns points to each candidate according to their position on a ballot. In a $k$ candidate election, the $i$th ranked candidate on a ballot earns $k-i$ points for each ballot, and then the candidate with the most points is the winner. When truncated ballots are present we use the pessimistic model \cite{BFLR}, where candidates earn no points from a ballot in which they are not ranked. For example, in a 4-candidate election a bullet vote for candidate $A$ would give $A$ three points and give all other candidates zero points.

\end{itemize}

We include plurality in our analysis as a baseline: it serves as a “worst-case” method, one that tends to elect more ideologically extreme candidates than other ranked-choice methods. Bucklin and Borda, on the other hand, provide a useful contrast--they often select winners closer to the median voter than Condorcet or IRV under realistic voting conditions, a property we explore below.

We illustrate each of these methods using the example preference profile in Table \ref{profile_ex}.

\begin{table}
\begin{tabular}{ccccccccc}
20 &  130 & 30 & 40 & 120 & 10 & 50 & 70 & 10\\
\hline
$A$ & $A$ & $A$& $B$&$B$&$B$&$C$&$C$&$C$\\
$B$&$C$ & & $A$ & $C$ & & $A$ & $B$ & \\
$C$ & $B$ & & $C$ & $A$ & & $B$ & $A$ &\\

\end{tabular}
\caption{An example of a preference profile with three candidates.}
\label{profile_ex}
\end{table}

\begin{example}\ 
\begin{itemize}
\item From Table \ref{profile_ex} we see that $A$ receives 180 first-place votes, $B$ receives 170, and $C$ receives 130. Thus, $A$ is the plurality winner. 
\item To find the IRV winner we eliminate $C$ and redistribute their votes, causing 50 votes to transfer to $A$ and 70 to $B$ (the remaining 10, which rank only $C$, are transferred to no one). After the vote transfers $B$ has 240 votes while $A$ has 230, and thus $B$ is the IRV winner.
\item For the Bucklin winner, note that no candidate earns a majority of first-place votes and thus we add each candidate's number of second-place rankings to their first place rankings. As a result, after the second round the Bucklin scores are 270, 300, and 380 for $A$, $B$, and $C$, respectively. Each of these scores surpasses the majority threshold of 236 votes; since $C$ has the highest score, they are declared the Bucklin winner.

\item To identify the Condorcet winner, note that 250 voters prefer $C$ to $A$ while 220 prefer $A$ to $C$, and 260 voters prefer $C$ to $B$ while 230 prefer $B$ to $C$. Since $C$ wins both of their head-to-head match-ups, they are the Condorcet winner and will be declared the winner under any Condorcet method.

\item For Borda, $A$ is ranked first on 180 ballots (2 points each), second on 90 ballots (1 point each), and last or not ranked on 190 ballots (0 points), for a total of 450 points. $B$ earns 430 points, and $C$ earns 510 points, and thus $C$ is the Borda winner.
\end{itemize}
\end{example}

In this example the IRV winner $B$ is not the Condorcet winner $C$, which many social choice theorists argue is a sub-optimal outcome for IRV. Since $C$ defeats $B$ head-to-head (and also defeats $A$ head-to-head), arguably $C$ is the ``most deserving'' candidate and should be declared the winner from a majority-rule standpoint. Candidate $A$ seems to be a polarizing candidate, since they receive the most first-place but also the most last-place votes. Plurality selects this candidate despite their possible extreme placement on some sort of ideological spectrum.

\section{Overview of the Work of Atkinson et al.}\label{section:Atkinson_overiew}

In this section we briefly describe the model constructed by Atkinson et al., as well as the conclusions they draw.

The model in \cite{AFG} uses data from the \textit{Cooperative Election Survey} (CES), the largest academic survey focused on American political attitudes \cite{SAS}. The CES includes over 50{,}000 respondents who answer a wide range of questions about their political views. For our modeling purposes, the most relevant questions ask voters to place themselves along an ideological spectrum. The model in \cite{AFG} is built from questions which examine how strongly a voter identifies with the Democratic or Republican party.

Atkinson et al.\ build a one-dimensional spatial model as follows. In each state, voters are given a location in the line segment $-0.5$ to 0.5, which is divided into five bins: $(-0.5, -0.3)$, $(-0.3, -0.1)$, $(-0.1, 0.1)$, $(0.1, 0.3)$, and $(0.3, 0.5)$. The leftmost bin is the ``extreme left,'' the next bin is the ``moderate left,'' and so on. The CES  asks voters a series of questions about partisanship, including  ``\emph{Would you call yourself a strong Democrat (Republican) or a not so strong Democrat (Republican)?}''. Voters could also answer Neither, in which case they were asked whether they lean towards one party. If a voter identified as a strong Democrat then the voter's weight\footnote{The CES  assigns each voter a weight to ensure the samples from each state are representative; see the documentation in \cite{SAS} for more details.} was assigned to the bin $(-0.5, -0.3)$, if a voter identified as a not so strong Democrat or lean Democrat then their weight was assigned to the bin $(-0.3, -0.1)$; similarly for the two right bins and Republican voters. Voters who identified with neither party had their weight assigned to the bin $(-0.1, 0.1)$. In this way, each of the five bins is given a total weight for each state. 

For each state, a distribution of 100,000 voters is then created by placing the voters along the line segment using a bin's relative weight as the probability that a voter is placed in that bin. Once a voter is assigned to a bin, the voter's exact location within the bin was chosen uniformly at random.

\begin{figure}
\centering
\includegraphics[width=120mm]{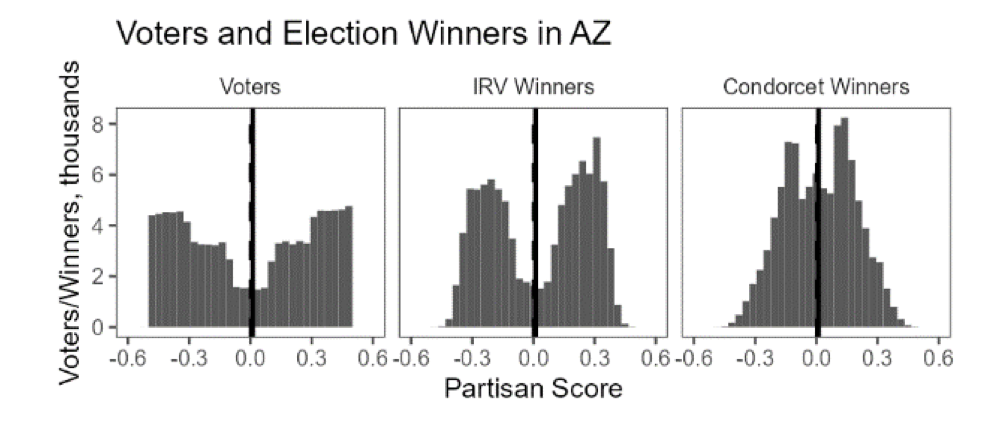}
\caption{A figure taken from \cite{AFG}, showing the voter distribution (left), distribution of IRV winner positions (center), and distribution of Condorcet winner positions (right) in Arizona. Figure reproduced with permission of the authors.}
\label{figure:Atkinson_paper_image}
    
\end{figure}

This methodology produces voter distributions that are typically bimodal, with much of the probability mass concentrated in the extreme bins. That is, most of the distributions studied by Atkinson et al.~are highly polarized. The left of Figure \ref{figure:Atkinson_paper_image} shows their voter distribution in Arizona, for example. States which do not demonstrate this bimodal behavior tend to be the firmly Democratic or firmly Republican states, such as Massachusetts or North Dakota.

With the voter distributions in place, 100,000 simulated preference profiles are generated per state using the ``citizen-candidate model'' \cite{OS96}, where four voters are chosen at random to be four candidates\footnote{The number four is chosen because of Alaska's top-4 IRV system}, and voter preferences are determined using Euclidean distance. That is, a voter's top choice is the candidate closest to them, their second choice is the candidate next closest to them, and so on. The preferences are then aggregated into a profile, and the process is repeated. It is assumed that voters provide complete preferences, so that all cast ballots have length four. As mentioned in the introduction, this methodology for generating ballot data is very much in line with previous work in the social choice literature. As far as we are aware, Atkinson et al.~are the first to use the CES data to build spatial models for generating ballot data, which is part of their contribution to the literature.

For each simulated profile, the IRV and Condorcet winners are calculated and their locations along the line segment recorded. This allows for the construction of histograms such as those in the middle and right of Figure \ref{figure:Atkinson_paper_image}, which shows the Condorcet and IRV winner distributions for Arizona from their simulations. The location of the median voter is denoted with the solid black line; note that IRV winners tend to be much farther from the median voter on average than the Condorcet winners. Similar results hold for most states across their simulations. Thus, their results reinforce theoretical findings dating to Black's Median Voter Theorem, which Atkinson et al.~interpret as evidence that Condorcet methods are better for addressing polarization than IRV. They summarize as follows: 

\begin{quote}
\emph{While $[$IRV$]$ has been offered as a solution to polarization, our results show that
IRV cannot be expected to effectively lead to representative outcomes relative to
other election systems. Reformers concerned with polarization should look to
other ranked-choice methods. As shown in our simulations, a Condorcet electoral
method will tend to elect candidates much closer to the state’s median\dots voter, especially for highly polarized states with bimodal electorates.}
\end{quote}

\section{Description of Methodology and Models}\label{section:our_models}

The question motivating our work is: How are the conclusions of Atkinson et al.~affected when we incorporate more realistic voting behavior, such as truncated ballots, into CES-based simulations? For example, do Condorcet methods still tend to elect candidates much closer to the median voter when a sizable portion of voters rank only one candidate? Furthermore, are the differences between IRV and Condorcet uniform across states, or do they vary depending on the underlying voter distribution? To address these questions, we construct multiple models that incorporate more realistic voter behavior, including the casting of bullet votes. In this section we describe how our models are built using CES data.

First, we describe our voter distributions. If we are interested in voter placement along a left-to-right spectrum, arguably the  CES question used by Atkinson et al.~is not the most useful. In particular, it is not clear that strong party identification is equivalent to an extreme placement along an ideological line. For example, extreme left voters might not identify as a strong Democrat because they perceive themselves as much farther to the left than the average Democratic policymaker; such voters might be more likely to select ``not so strong Democrat'' or ``neither.'' Similarly, a voter who selects ``neither'' might not be a centrist; such a voter might identify as a right-wing libertarian, for example. While  such points do not undermine the work of Atkinson et al., we choose to use the CES question which asks voters to identify themselves as Very Liberal, Liberal, Somewhat Liberal, Middle of the Road, Somewhat Conservative, Conservative, Very Conservative, or None/Don't Know. This question more explicitly allows voters to place themselves on an ideological spectrum.

\begin{figure}

\centering

\begin{tabular}{cc}
\includegraphics[width=63mm]{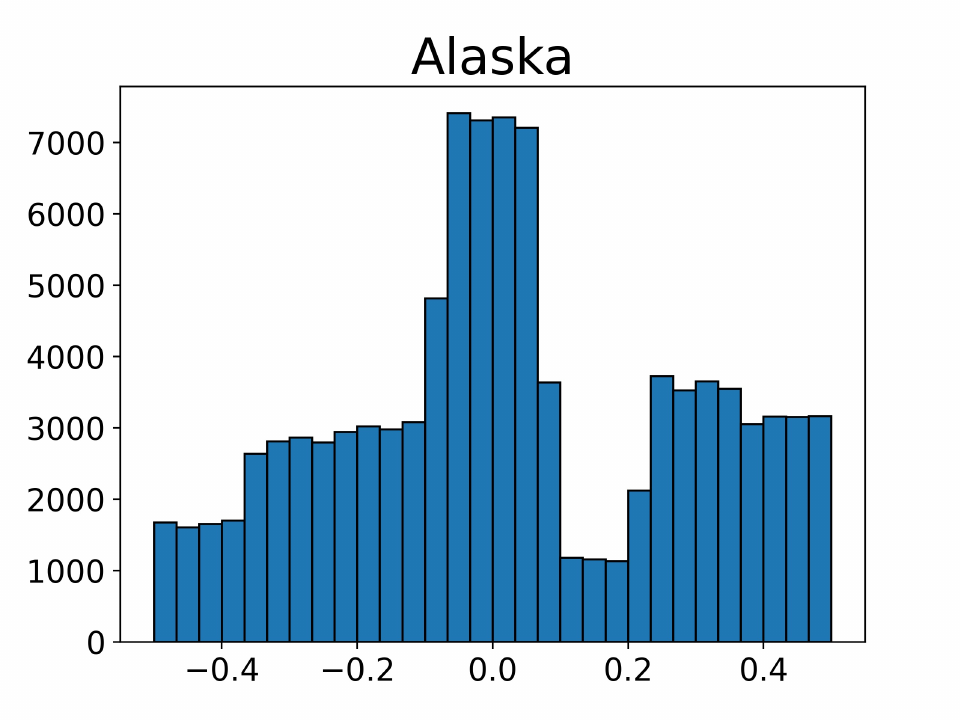}&\includegraphics[width=63mm]{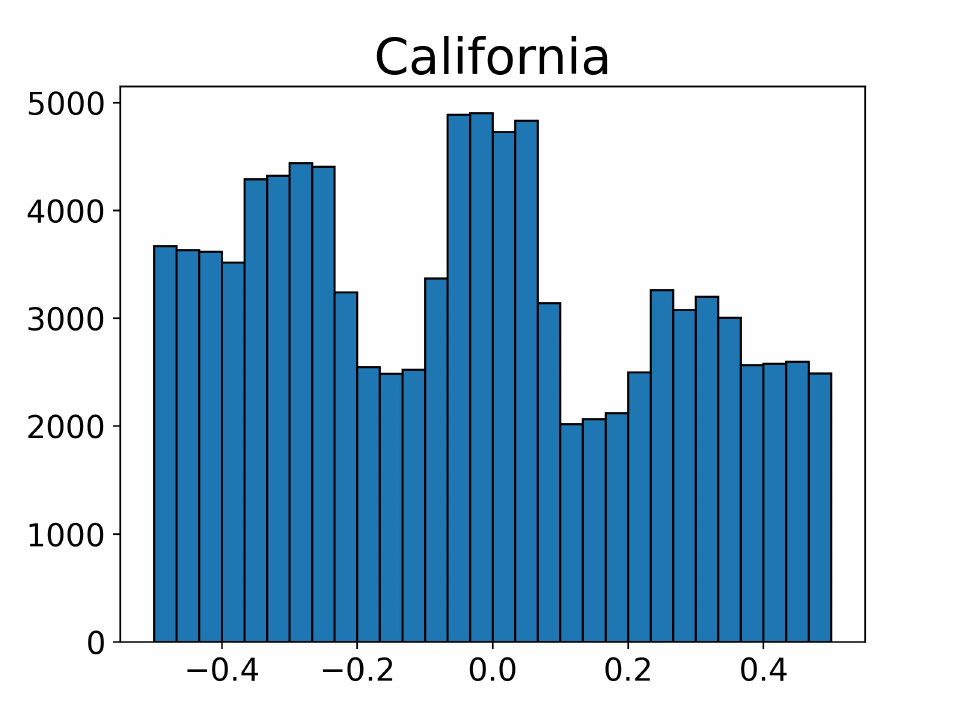}\\
\includegraphics[width=63mm]{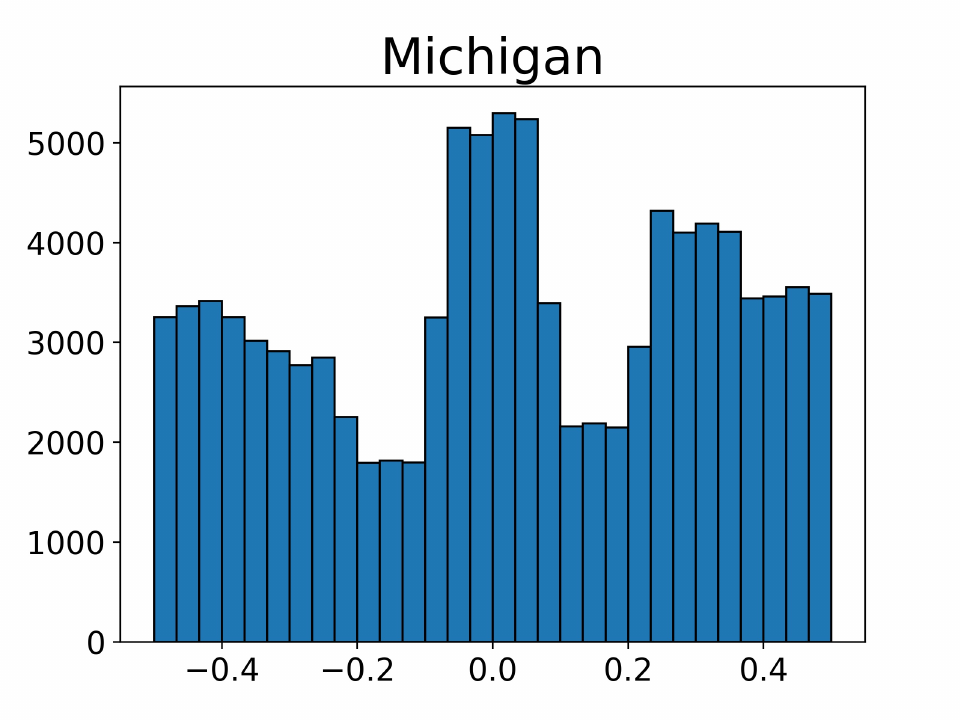}&\includegraphics[width=63mm]{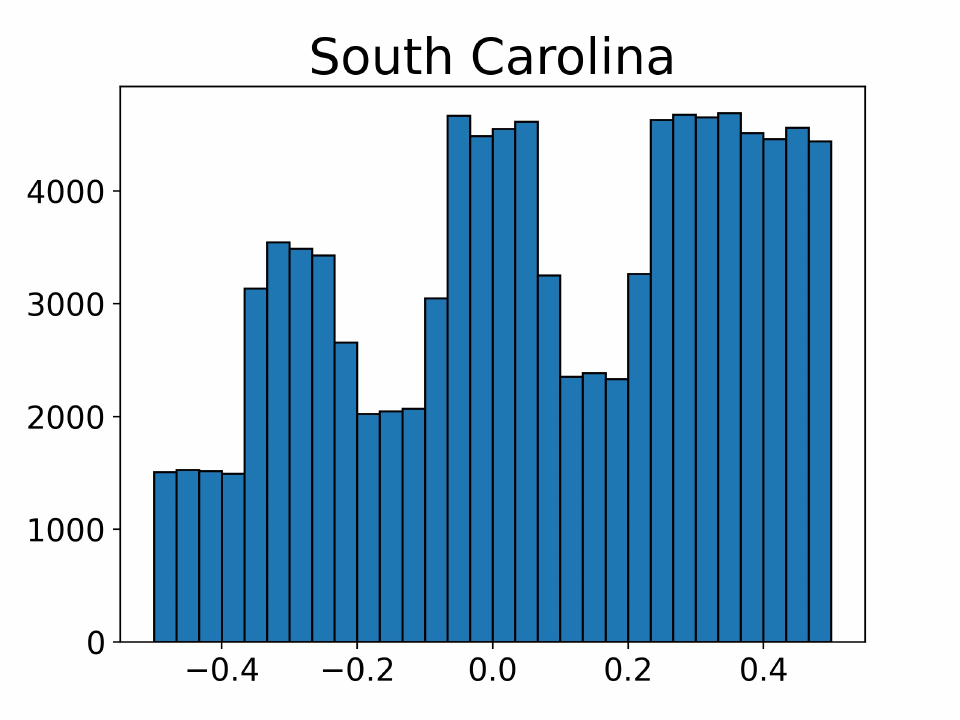}\\
\multicolumn{2}{c}{\includegraphics[width=63mm]{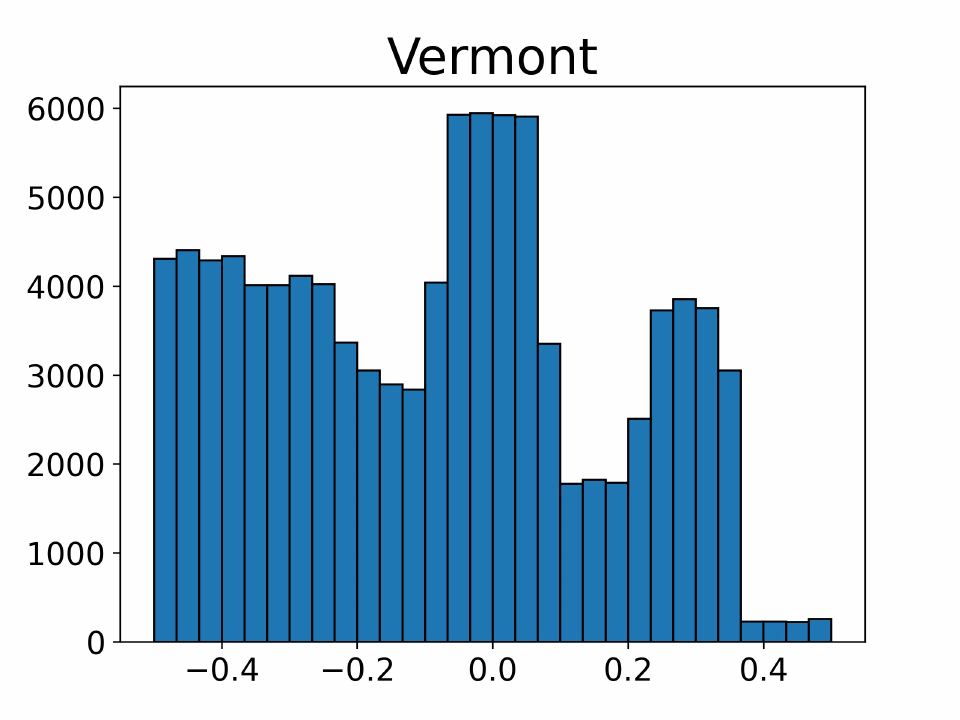}}
\end{tabular}

\caption{Trimodal voter distributions for five sample states.}
\label{figure:trimodal_voters}
\end{figure}

For each state, we use the same bin methodology as Atkinson et al.~to construct a voter distribution with 100,001 voters. In our case we use seven bins to subdivide the line segment from $-0.5$ to 0.5, corresponding to each of the responses to our question besides None/Don't Know\footnote{Very few voters selected this option, and thus we ignore such voters rather than attempt to place them on the spectrum.}. Examples of the resulting distributions are displayed for five states in Figure  \ref{figure:trimodal_voters}. As seen in the figure these distributions tend to be trimodal, since voters seem reluctant to describe themselves as ``somewhat'' liberal or conservative. Because voters tend to cluster into three distinct camps (Liberal, Middle of the Road, and Conservative) we refer to these as our trimodal distributions.

\begin{figure}

\centering

\begin{tabular}{cc}
\includegraphics[width=63mm]{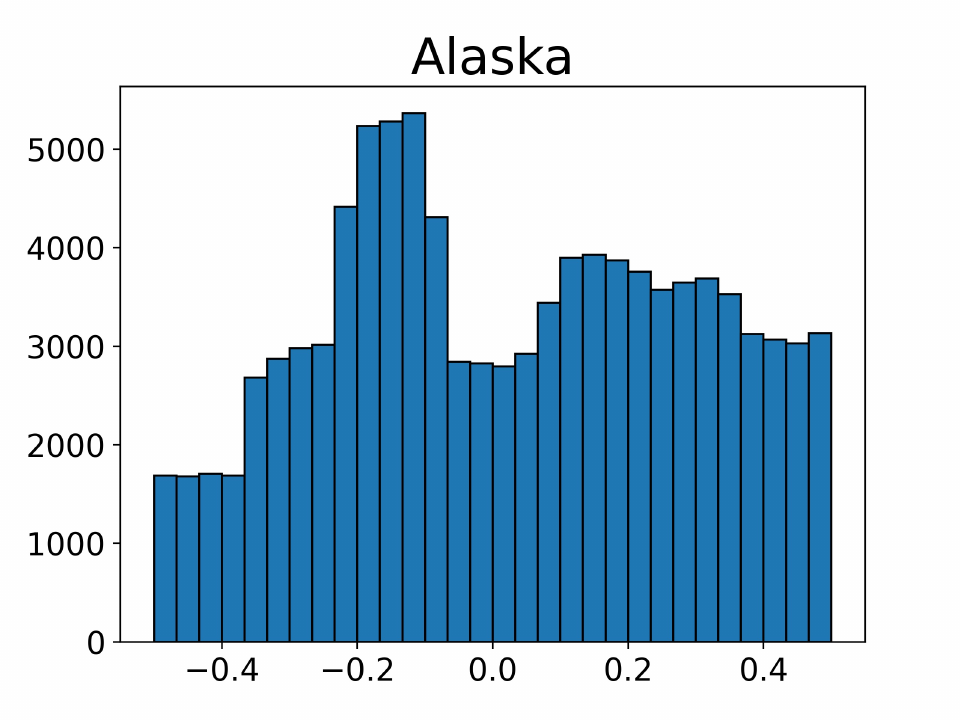}&\includegraphics[width=63mm]{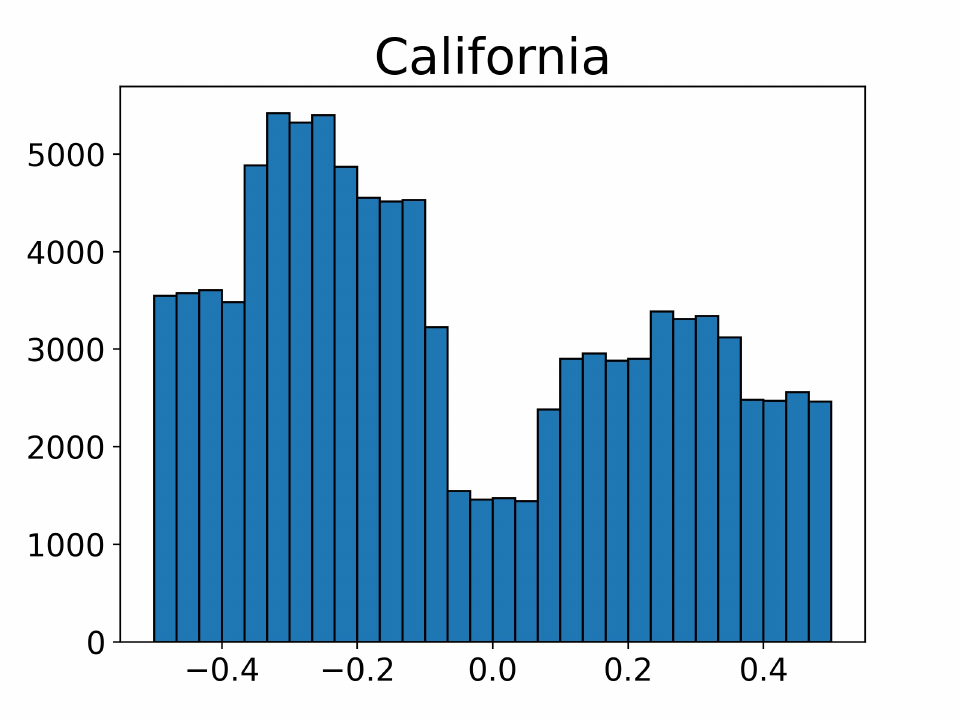}\\
\includegraphics[width=63mm]{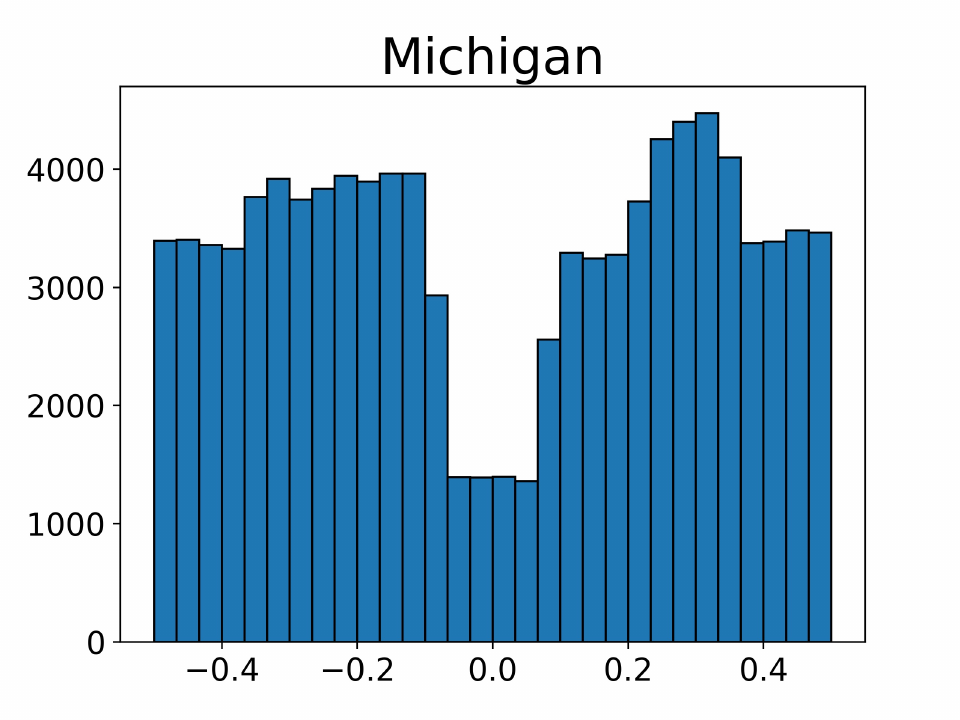}&\includegraphics[width=63mm]{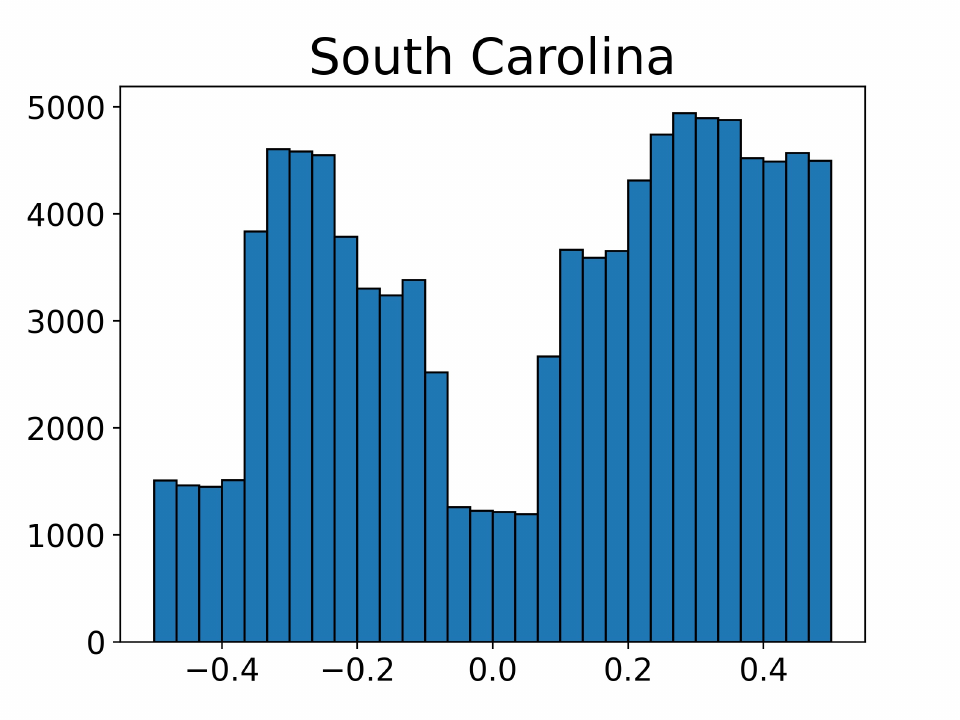}\\
\multicolumn{2}{c}{\includegraphics[width=63mm]{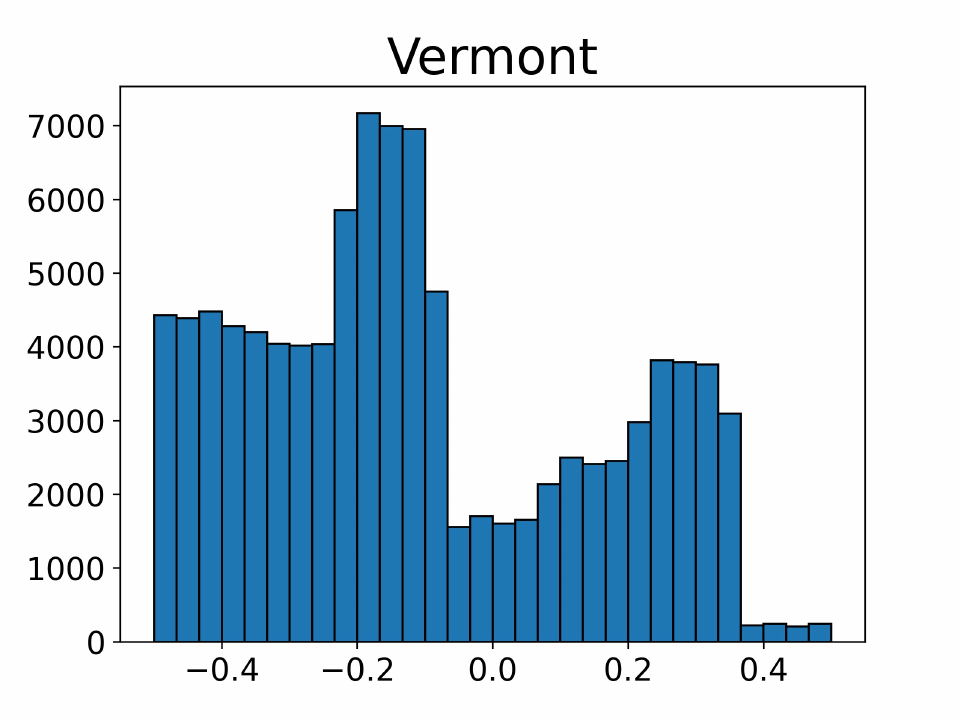}}
\end{tabular}

\caption{Bimodal voter distributions for five sample states.}
\label{bimodal_voters}
\end{figure}

To extend the polarization analysis in \cite{AFG}, we also created bimodal distributions derived from the trimodal ones. For each voter who selected Middle of the Road, if the voter selected ``strong Democrat'' or ``lean Democrat'' for the question used by Atkinson et al., we put the voter's weight in the Liberal bin. If the voter selected ``not so strong Democrat'' then we put them in the Somewhat Liberal bin. We performed similar movements for voters who expressed an identification with the Republican party. The resulting distributions for five sample states are displayed in Figure \ref{bimodal_voters}.

We emphasize that we do not claim our voter distributions accurately reflect reality; we do not know, for example, if the Michigan gubernatorial electorate follows the one-dimensional bimodal distribution shown in Figure \ref{bimodal_voters}. Rather, we claim these distributions have been created from the CES data in a reasonable fashion, and can shed light on potential electoral outcomes if voter preferences are approximately one-dimensional.

Finally, we describe our models for generating voter preferences. As outlined above, the theoretical model used in \cite{AFG} does not take into account voter behavior which is often observed in practice, such as the casting of truncated ballots. We incorporate three ``more realistic'' features of actual elections into our analysis: truncated ballots, voter abstention, and voter uncertainty about candidate placement. That is, in real elections some voters do not provide a complete ranking of the candidates, some eligible voters decide not to cast a ballot, and some voters are uncertain about a candidate's exact placement in ideological space; we incorporate these features into our one-dimensional models in natural ways. We view each feature as a binary switch which can be turned on or off, and each such choice defines a particular model. 

Each model follows the ``citizen-candidate'' framework adopted in \cite{AFG}: for each generated election, $k$ voters, $k \in \{3,4\}$, are chosen to be the candidates.\footnote{We do not analyze elections with $k\ge 5$ since we are unaware of any single-winner real-world IRV elections which contain five or more viable candidates. In practice, in 5-candidate elections a very weak fifth candidate is eliminated first, meaning that 5-candidate elections have similar dynamics to 3- or 4-candidate elections.} Voter preferences are built using Euclidean distance from the candidates.  For each choice of state, voter distribution, and number of candidates $k \in \{3,4\}$, we generate 100,000 simulated elections for each of the six models defined below, where the electorate for each election contains 100,001 voters.

\begin{enumerate}
\item \textbf{Theoretical Ideal.} All voters provide complete preferences, all voters cast a ballot, and every voter knows the exact placement of every candidate. This is the model used in \cite{AFG}, although, as explained earlier, Atkinson et al.~use slightly different voter distributions since they use a different CES question to generate voter placements.
\item \textbf{Ideological Truncation.} All voters cast a ballot (i.e., there is complete turnout) and know the exact ideological placement of the candidates, but some voters cast truncated ballots. In this model, a voter chooses not to rank candidates based on ideological distance. If there are three candidates then a voter ranks a candidate on their ballot if the distance from the voter to the candidates is 0.37 or less; this value falls to 0.26 for four-candidate elections. 

These parameters are chosen so that across our simulations, the median rate of bullet votes in simulated elections in each state is approximately in line with what we see in real ranked-choice elections. As explored in \cite{Schwab_report}, for three-candidate (respectively four-candidate) ranked-choice elections with no strong majority candidate in the US, the median bullet vote rate is 0.35 (respectively 0.34). Our parameters are chosen so that across all states we see approximately this level of bullet vote casting in generated elections. Tables \ref{tab:state_stats_three} and \ref{tab:state_stats_four} in Appendix \ref{sec:appendix_state_distribution_info} provide the median bullet vote rates by state across our simulations. For example, Table \ref{tab:state_stats_three} shows that of the 100,000 generated elections in Alabama for three-candidate elections under a bimodal electorate, the median proportion of bullet votes is 0.33. 

\item \textbf{Random Truncation.} All voters  cast a ballot and know the exact ideological placement of the candidates, but some voters cast a truncated ballot where the decision to truncate is made at random. In each generated election, if $k=3$ then each voter has a probability of 0.35 of casting a bullet vote. If $k=4$ then the probability a voter casts a bullet vote is 0.34 and the probability of casting a ballot length $2$ is $0.20$. These parameters are chosen based on American ranked-choice political elections analyzed in \cite{Schwab_report}.

\item \textbf{Abstention.} All voters who participate cast a complete ballot and know the exact ideological placement of the candidates, but some voters abstain from casting a ballot. A voter casts a ballot only if there exists a candidate whose distance from the voter is 0.14 (representing one bin length for the trimodal distribution) or less. This parameter is chosen so that three-candidate elections generally have turnout rates in the 60-70\% range for generated elections, which matches the kind of turnout observed in the 2024 US Presidential election in each state. If $k=4$ then the turnout rates rise to just over 70\% in most states. Tables \ref{tab:state_stats_three} and \ref{tab:state_stats_four} in Appendix \ref{sec:appendix_state_distribution_info} give the median abstention rates across the 100,000 simulated elections for each choice of state, number of candidates, and voter distribution. For example, in three-candidate elections in Alabama with a bimodal electorate, the median abstention rate is 0.35 (i.e., the median turnout rate is 65\%).

\item \textbf{Noise.} All voters cast a complete ballot and no voter abstains, but voters are unclear about the exact ideological location of each candidate. To model this phenomenon we introduce some random noise into voter preferences. For each voter, we perturb candidate locations by up to 0.14 in either direction uniformly at random. For example, if a  candidate's true position is 0.2 then a given voter could perceive this candidate to be at any position in the interval $[0.06,0.34]$ where this choice is made at random. Once the candidate positions are perturbed in this manner, we construct the voter's preferences as before. Note we perform this perturbation for each voter, so that different voters will perceive a candidate as being at different positions. 

 Unlike the previous parameters, there is no empirical justification for the choice of 0.14. The basic idea behind this number is that a voter could mistake a somewhat conservative candidate for a centrist or a conservative, but would not mistake this candidate for a liberal or an extreme conservative. While it is impossible to set this parameter to any empirically-guided value, 0.14 is enough noise to change a significant fraction of voter ballots: roughly 35\% in the case of $3$ candidate elections and 55\% in the case of $4$ candidate elections. This is certainly enough to potentially affect the outcome of an election. Figure \ref{fig:noise_tuning} in Appendix \ref{sec:appendix_state_distribution_info} gives more specific information on the number of changed ballots.

\item \textbf{Most Realistic.} This model incorporates ideologically truncated ballots, voter abstention, and noise as described above. We use the term \emph{most realistic} as a short-hand to indicate that all three features are included, but we do not know in practice if this model is ``more realistic'' than a model like Ideological Truncation, for example. 
\end{enumerate}

Our models allow us to examine the effects of ballot truncation driven by ideological considerations or by randomness. We are unaware of studies which thoroughly examine why voters choose to cast short ballots in real elections, and thus we do not know which of these ballot truncation models is more realistic. Of course, there could be other reasons for voters to cast a bullet vote which are beyond the scope of the article. For example, it is possible that voters cast bullet votes because their favorite candidate encouraged short ballots. Future work could investigate how our results change under other models of ballot truncation.

We also created models which incorporated random abstention, where voters were given a fixed probability of not casting a ballot. These models did not provide interesting results beyond the six models listed above, and thus we omit them.

As with any models of this nature, we must make a series of choices about their construction and a reader can make reasonable objections to any of our choices. For example, it is not clear that using a distance cutoff is the best way to model truncated ballots. For example, if a voter is located at $-0.5$ and the candidates $A$, $B$, and $C$ are positioned at $-0.4$, $0.2$, and 0.4, respectively, then it is not clear this voter will rank only $A$ on their ballot. The voter might dislike $C$ strongly enough that they will cast a complete ballot, ensuring that $C$ is ranked third. Nevertheless, we believe our modeling approach offers an empirically grounded way to examine how partial ballots and turnout interact with the performance of Condorcet methods and IRV. We view these models as an initial framework which could be refined in future work.

\section{Results}\label{section:results}

In this section we present our results, first focusing on how the different models of voter behavior affect moderating outcomes, and next analyzing the difference in moderating effects based on the underlying voter distribution.

\subsection{Moderating effects of IRV and Condorcet under the six models}\label{sec:results_part1}

First, we examine how the distributions of ideological placement of the Condorcet and IRV winners compare under our six models for each of the two  voter distributions. Following \cite{AFG}, we focus on how far a method's winner is from the median voter, on average. For context, we also include some results for the methods of Borda, Bucklin, and plurality.

We sometimes provide complete results for all states, but this is not always feasible. To keep the narrative streamlined, for some results we provide complete details for the five states in Figure \ref{figure:trimodal_voters}: Alaska, California, Michigan, South Carolina, and Vermont. We choose these five states for the following reasons. Alaska uses a top-4 IRV system for federal elections. California is the most populous state in the US and has an unbalanced bimodal structure reflecting a left-leaning population but with a strong right-wing presence. Michigan is a swing state, as evidenced by its balanced bimodal distribution. South Carolina is somewhat a partisan reflection of California where the majority of voters are on the right of the spectrum but there is still a sizable left-wing population. Vermont is an outlier in the CES data, exhibiting bimodal and trimodal structures dissimilar from the other states. Most  states have distributions similar to California, Michigan, or South Carolina, depending on the state's partisan lean.

\begin{figure}
\begin{tabular}{cc}
\includegraphics[width=66mm]{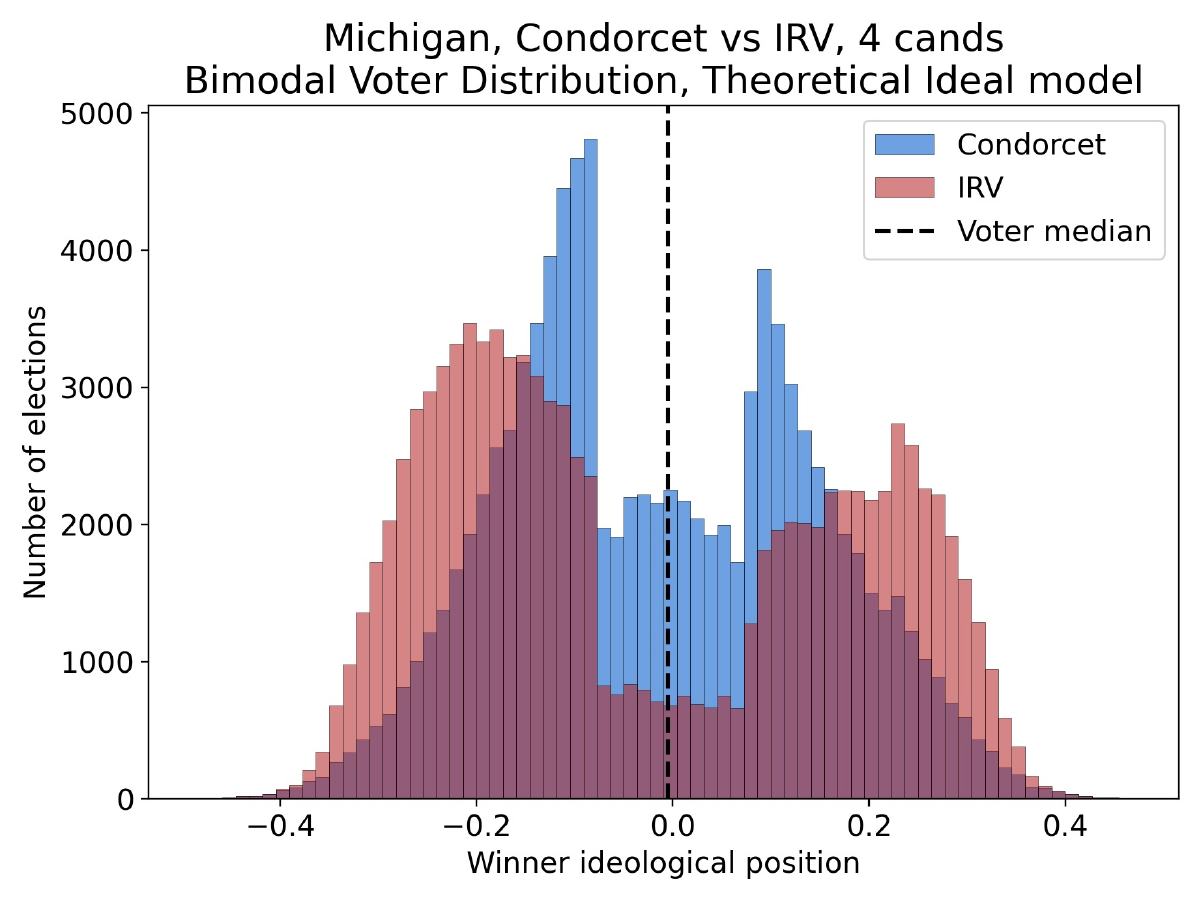} & \includegraphics[width=66mm]{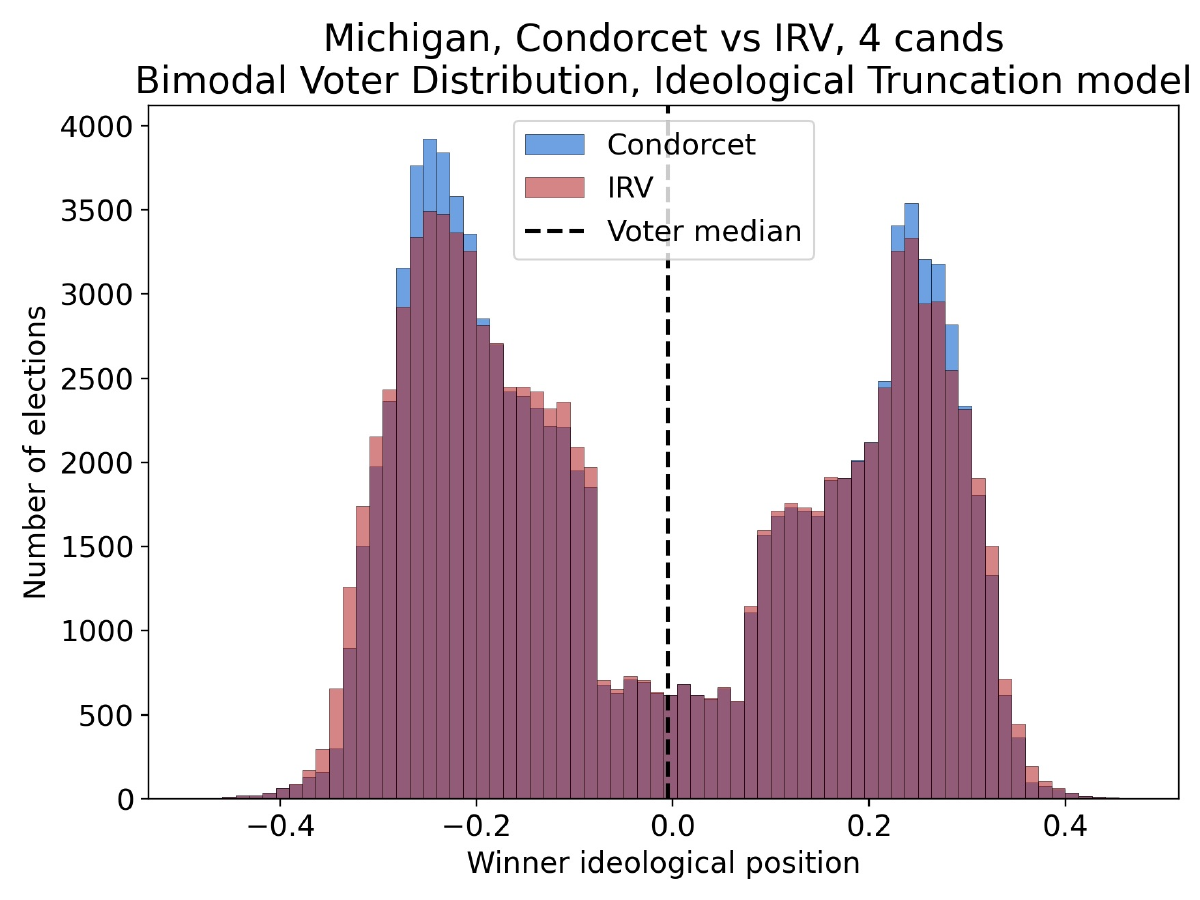}\\
\includegraphics[width=66mm]{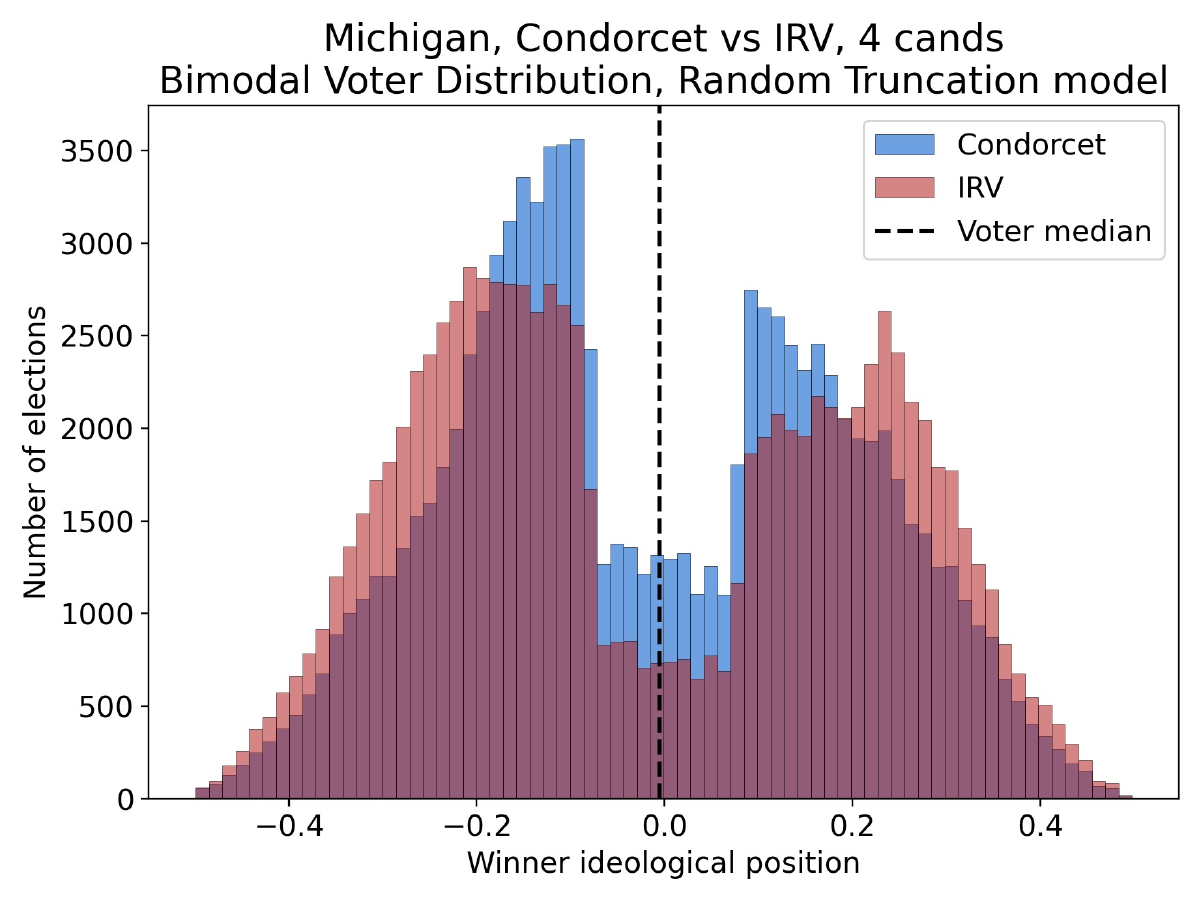}
 & \includegraphics[width=66mm]{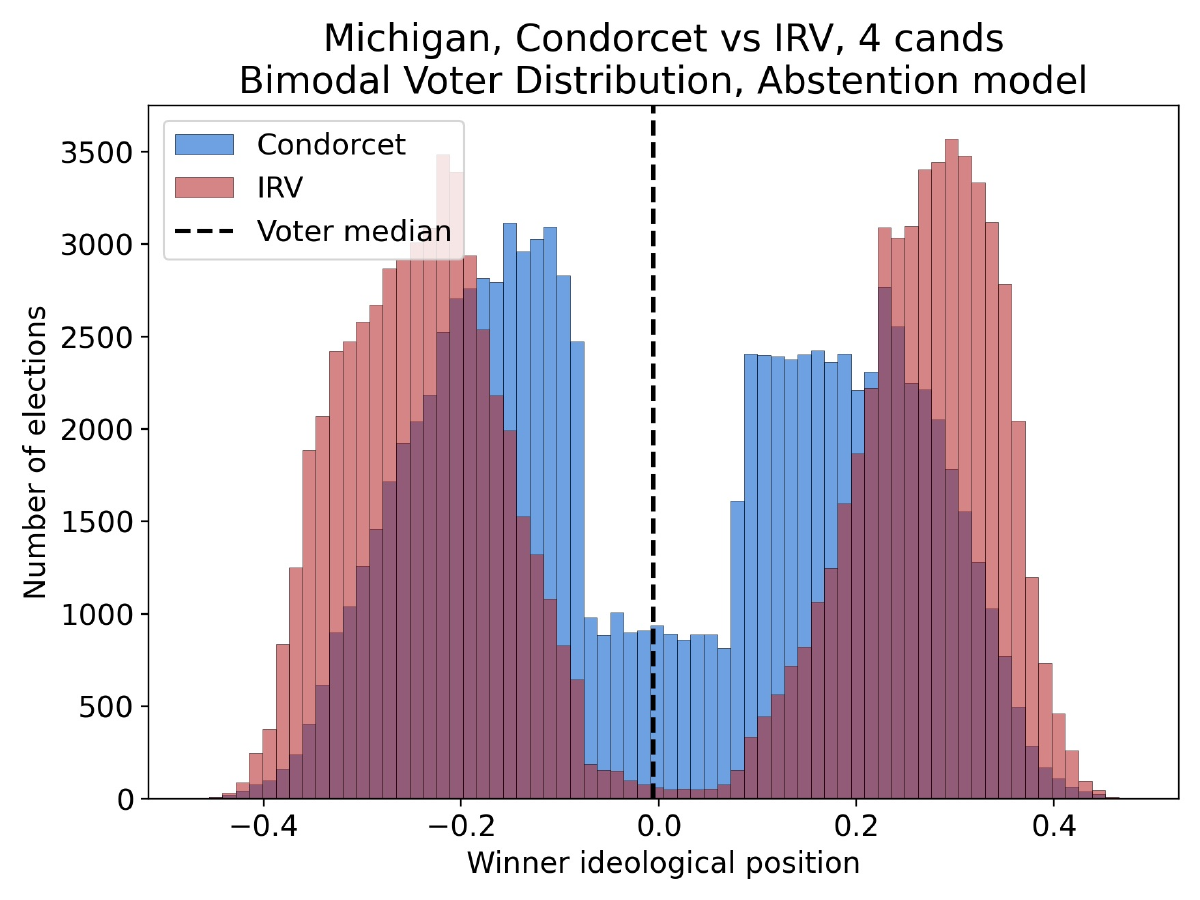}\\
\includegraphics[width=66mm]{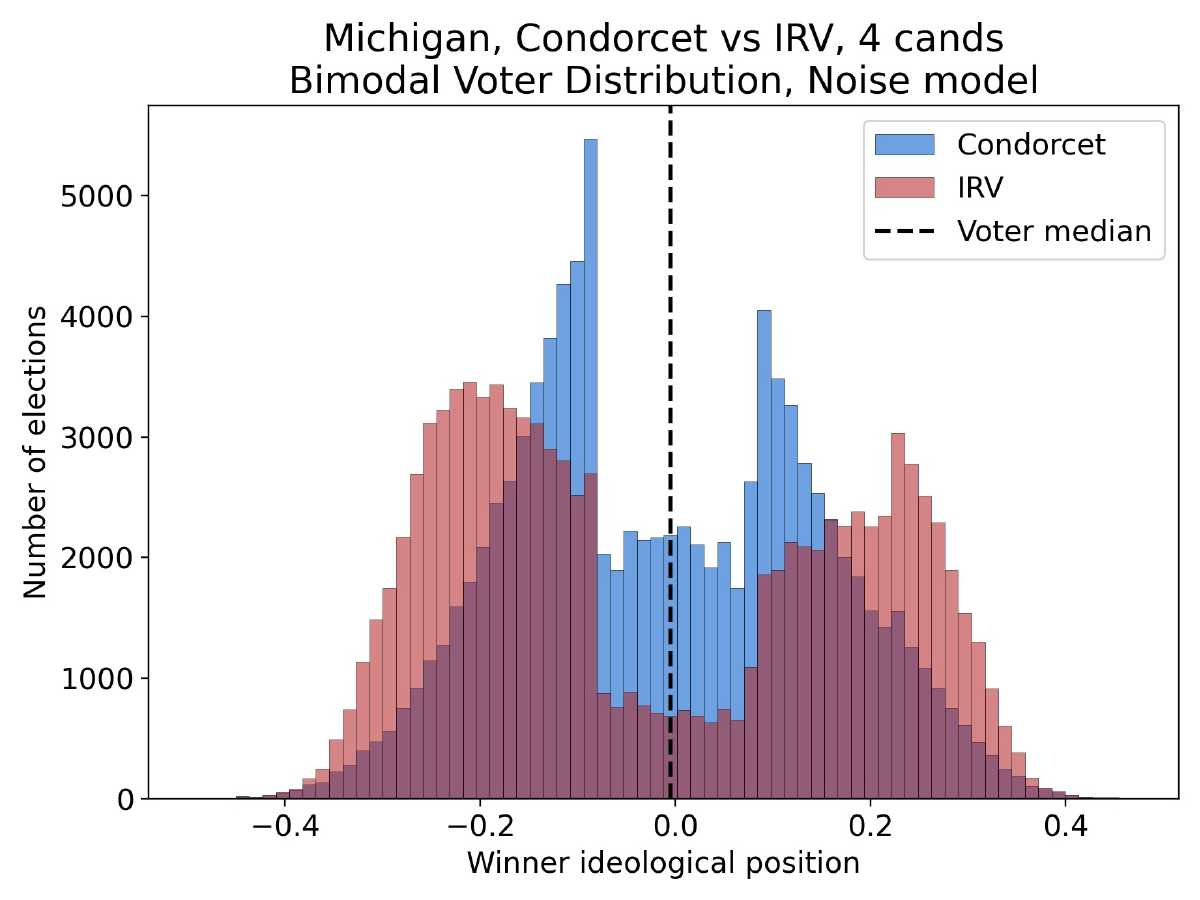} & \includegraphics[width=66mm]{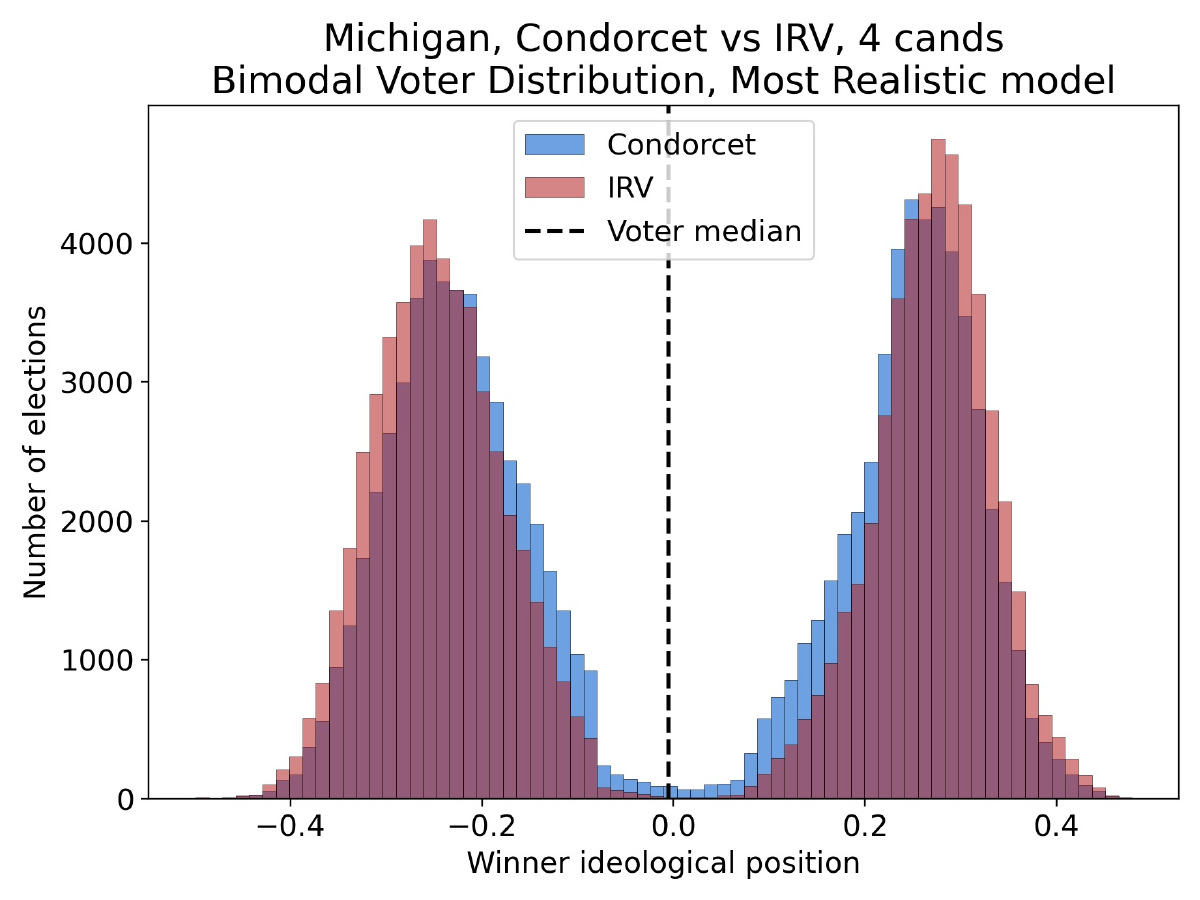}\\

\end{tabular}
\caption{Histograms showing the ideological positions of Condorcet and IRV winners across 100,000 simulations in Michigan, using 4-candidate elections with the bimodal voter distribution.}
\label{fig:MI_results_bimodal}
\end{figure}

We present our results in three forms. First, for our five sample states we provide histograms showing the ideological placements of the Condorcet and IRV winners across the simulations, similar to the histograms provided in \cite{AFG} (see Figure \ref{figure:Atkinson_paper_image}). Figure \ref{fig:MI_results_bimodal} shows these histograms for 4-candidate elections in Michigan using the bimodal voter distribution. The top left image shows how Condorcet and IRV compare under the Theoretical Ideal model; as in \cite{AFG}, the Condorcet winners cluster much more tightly around the median voter than the IRV winners, suggesting Condorcet is much more moderating than IRV. The top right and middle left images shows how the winners compare under the two truncation models. Interestingly, ideological truncation causes the histograms to become almost identical, while random truncation does not (although the histograms for random truncation are much more similar than those for Theoretical Ideal model). In Michigan, the Abstention model (middle right) does not result in similar histograms, although in other states abstention does cause substantial convergence. The bottom left image is essentially the same as the top left, showing that just adding uncertainty about candidates' ideological position makes little difference to the results in Michigan. The bottom right image shows how the two voting rules compare under the Most Realistic model. In this case, the histograms are roughly identical when voters can truncate ballots, abstain, and are influenced by noise.


Appendix \ref{sec:winner_distribution_histograms} contains the rest of the possible histograms comparing Condorcet and IRV in Michigan; see Figures \ref{fig:MI_results_bimodal_3cands}, \ref{fig:MI_results_trimodal_3cands}, and \ref{fig:MI_results_trimodal_4cands}. Appendix \ref{sec:winner_distribution_histograms} also contains similar histograms for Alaska, California, South Carolina, and Vermont (see Figures \ref{fig:AK_results_bimodal}-\ref{fig:VT_results_bimodal}), although for the sake of space we restrict to 4-candidate elections and the bimodal voter distribution, which is the focus of \cite{AFG}. For most states, the dynamic observed in California or South Carolina is typical: under the Theoretical Ideal model the IRV winners tend to be markedly ``more extreme'' than Condorcet winners on average, while this difference in extremity mostly disappears under the Most Realistic and Ideological Truncation models. Furthermore, when we incorporate ballot truncation into a model, it is much more difficult for a candidate near the median voter to win under a Condorcet method.

Second, we provide summary statistics regarding the average distance from the median voter. Tables \ref{table:avg_distance_stats_bimodal_3cands}, \ref{table:avg_distance_stats_bimodal_4cands}, \ref{table:avg_distance_stats_trimodal_3cands}, and \ref{table:avg_distance_stats_trimodal_4cands} in Appendix \ref{sec:sum_stats_tables} show the average distance from the median voter for various voting methods across the simulations in our 5 sample states. For example, for a bimodal electorate in Alaska, Table \ref{table:avg_distance_stats_bimodal_3cands} shows that in 3-candidate elections the average distance of the Condorcet winner from the median voter is 0.118 under the Theoretical Ideal model, which increases to 0.151 under the Most Realistic model. The bottom of the table shows the average across all of the states. We ran 100,000 simulations per state, and thus each entry in the ``Average Across All States'' portion of the table represents an average across 5.1 million simulated elections.

\begin{figure}
    \centering
    \includegraphics[width=\textwidth]{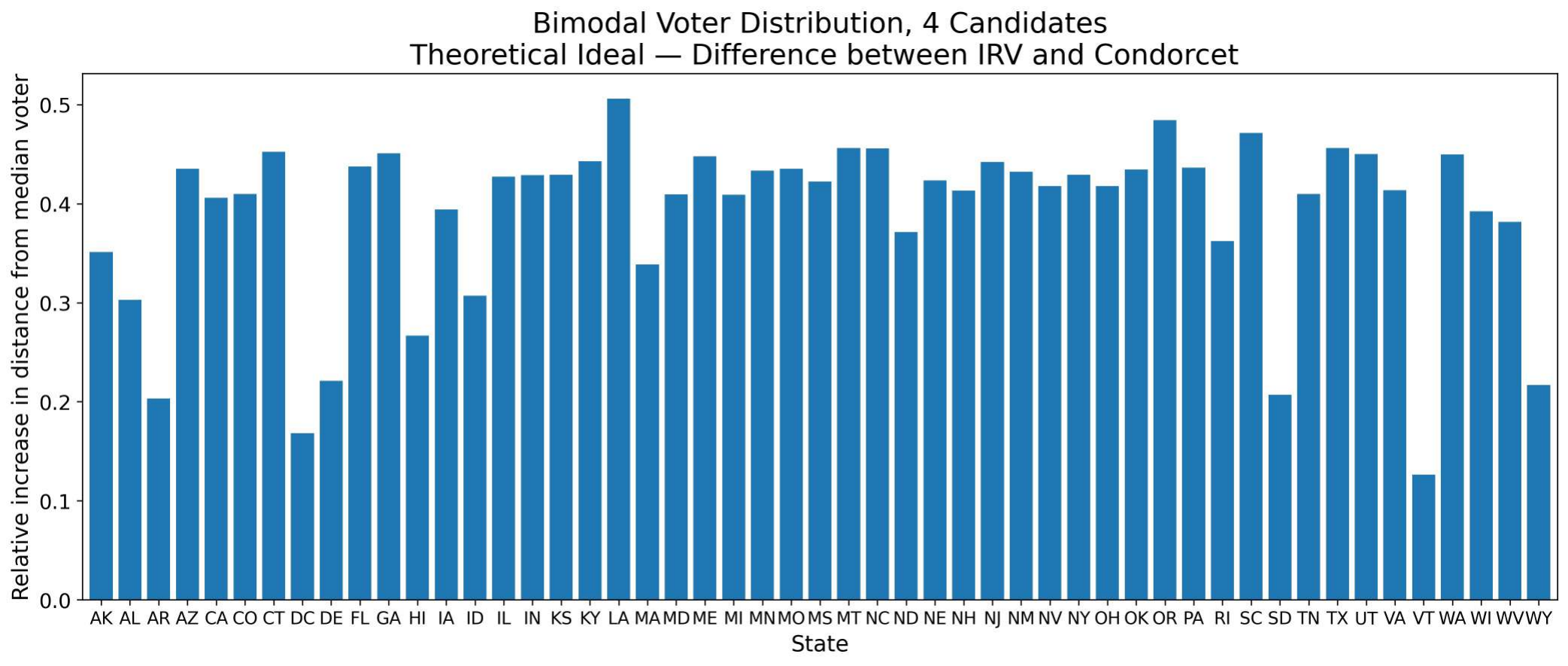}
    \caption{The relative difference of the average distance of the IRV winner and the average distance of the Condorcet winner to the median voter.}
    \label{fig:avg_dist_bar_graph_TI_bimodal_4cands}
\end{figure}

Third, this data has been compressed into figures like Figure \ref{fig:avg_dist_bar_graph_TI_bimodal_4cands}, which reports the \textbf{relative difference in the average distance from the median voter between Condorcet and IRV}. To be precise, the height of a bar shows how much farther on average the IRV winner is to the median voter than the Condorcet winner, assuming a bimodal voter distribution and $k=4$.  For example, Table \ref{table:avg_distance_stats_bimodal_4cands} shows that in Alaska the average distance of the IRV winner to the median voter is 0.132 under the Theoretical Ideal model, a 36.1\% increase over the Condorcet value of 0.097. Thus, the height of the bar corresponding to Alaska in Figure \ref{fig:avg_dist_bar_graph_TI_bimodal_4cands} is 0.361. We calculate these statistics following Atkinson et al.~\cite{AFG}, who interpret the 36.1\% increase as implying that the IRV winner tends to be 36.1\% ``more extreme'' than the Condorcet winner in these elections. The percentages for the bimodal distribution when $k=4$ are largely in line with those reported in \cite{AFG} under their model (see Figure 2 of \cite{AFG}).

Appendix \ref{sec:avg_dist_images} contains bar graphs showing these relative differences for IRV over Condorcet for the models Most Realistic (Figures \ref{fig:avg_dist_bimodal_3cands_mostrealistic}-\ref{fig:avg_dist_trimodal_4cands_mostrealistic}), Ideological Truncation (Figures \ref{fig:avg_dist_bimodal_3cands_ideotrunc}-\ref{fig:avg_dist_trimodal_4cands_ideotrunc}), Random Truncation (Figures \ref{fig:avg_dist_bimodal_3cands_randomtrunc}-\ref{fig:avg_dist_trimodal_4cands_randomtrunc}), Abstention (Figures \ref{fig:avg_dist_bimodal_3cands_abstention}-\ref{fig:avg_dist_trimodal_4cands_abstention}), and Noise (Figures \ref{fig:avg_dist_bimodal_3cands_noise}-\ref{fig:avg_dist_trimodal_4cands_noise}). The bar graphs for these models are displayed in orange and overlaid on the Theoretical Ideal graph, in blue. This allows for immediate visual comparison between candidate moderation results for a given model and the Theoretical Ideal model. Note that under the Most Realistic model, the difference between IRV and Condorcet tends to become much smaller than the difference under the Theoretical Ideal model. Indeed, for some combinations of voter distribution, state, and number of candidates, the IRV winner is closer to the median voter on average than the Condorcet winner (such outcomes occur when the orange bar goes beneath the horizontal axis).  Of the three realistic features we incorporate into our models, ballot truncation causes the largest overall deviation from the Theoretical Ideal results across the states, while ideological uncertainty causes the least deviation (note the orange bar graphs for Noise are very similar to the Theoretical Ideal graphs). That is, when IRV and Condorcet produce similar average distances from the median voter, the main driver seems to be ballot truncation.

\begin{figure}
\includegraphics[width=\textwidth]{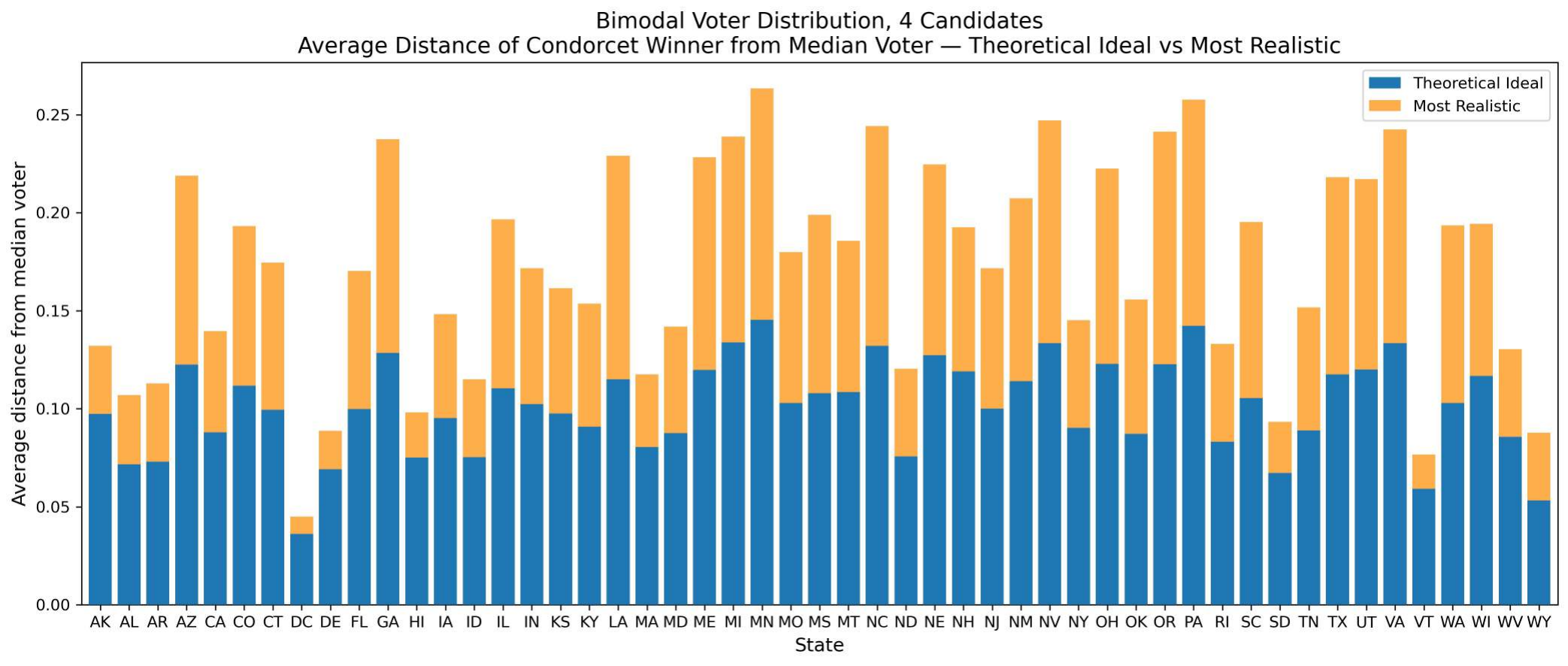}
\caption{For each state, the height of the blue bar (respectively orange bar) indicates the average distance of the Condorcet winner to the median voter under the Theoretical Ideal (respectively Most Realistic) model.}
\label{fig:Cond_vs_Cond}
\end{figure}

The figures in Appendix \ref{sec:avg_dist_images} compare two different voting methods for the same model, but we can construct similar images comparing the same voting method across two different models. Figure \ref{fig:Cond_vs_Cond} shows the average distance of the Condorcet winner from the median voter in 4-candidate elections using the bimodal voter distribution, under the Theoretical Ideal and Most Realistic models. Note this figure gives the average distance, not a relative increase in average distance as shown in the figures in Appendix \ref{sec:avg_dist_images}. In some states, the average distance of the Condorcet winner to the median voter almost doubles when moving from the theoretical ideal to more realistic voter behavior, indicating that centrist candidates win much less frequently under Condorcet methods in more realistic electoral conditions.  This point reinforces an argument made in \cite{PH25}, where the authors state that a Condorcet method may not elect the ``true'' Condorcet winner in real elections when voters deviate from theoretically ideal behavior. 

\subsection{Variations in moderating effects by voter distributions}\label{sec:distributions}

In Section \ref{sec:results_part1} we established that Condorcet methods are significantly more moderating than IRV under the Theoretical Ideal model, and this advantage of Condorcet over IRV mostly  disappears under the Most Realistic and Ideological Truncation models. However, these effects are not uniform across all states. For example, in 4-candidate elections and a bimodal voter distribution (see Table \ref{table:avg_distance_stats_bimodal_4cands}), in Michigan the relative difference in the average distance from the median voter is 41\% (0.189 versus 0.134), while in Vermont this difference is only 14\% (0.067 versus 0.059).  Similarly, in both Alaska and Michigan, the relative differences between IRV and Condorcet are close in the Theoretical Ideal model, but much different under the Ideological Truncation model (see Figure \ref{fig:avg_dist_bimodal_4cands_mostrealistic} in Appendix \ref{sec:avg_dist_images}). In this section, we analyze how the shape of a state's voter distribution influences the relative difference between IRV and Condorcet, as well the change in behavior of IRV relative to Condorcet as we move from the Theoretical Ideal model to a more realistic model.

\begin{figure}
    \centering

    \includegraphics[width=0.45\linewidth]{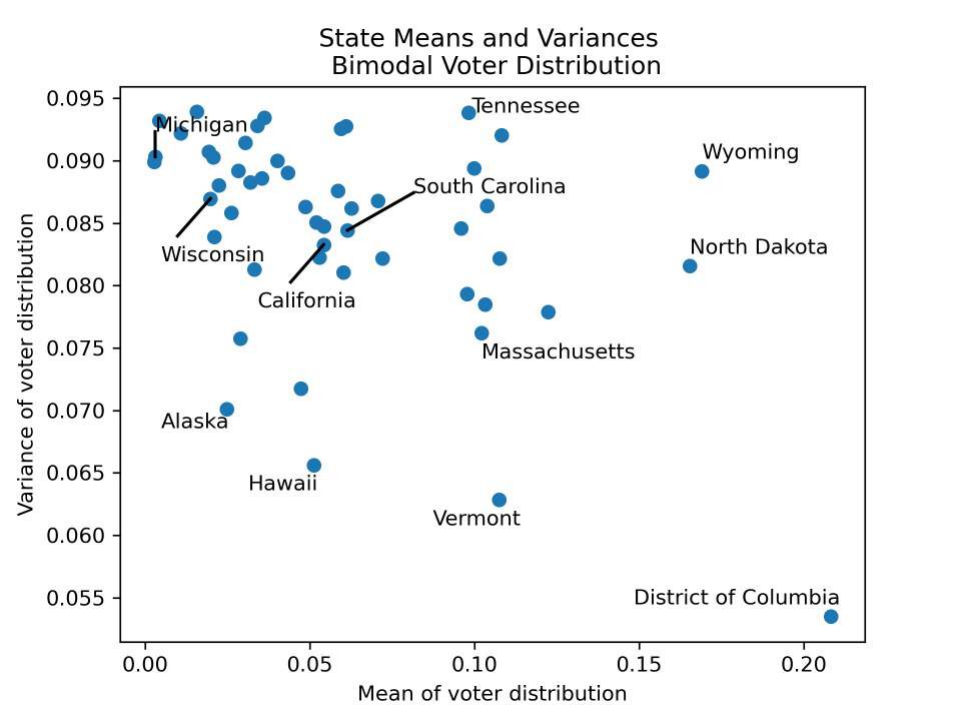}
    \includegraphics[width=0.45\linewidth]{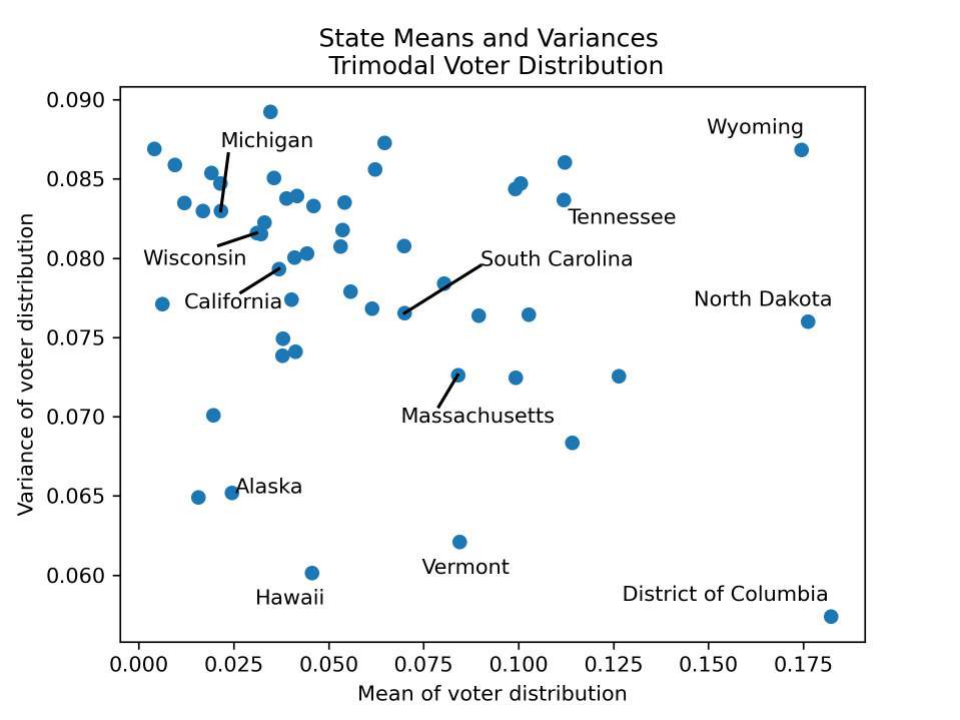}
    \caption{The positions of all 50 states with the names of all states mentioned directly in the paper. 
    }
    \label{fig:state_names}
\end{figure}

To understand why different states behave differently, we consider two basic statistical properties of the voter distributions for each state: the mean and the variance. Because extremely right-skewed and left-skewed states should behave similarly (by a simple reflection around 0), we use the absolute value of the mean. Thus, a state's positive mean is a measure of how ``un-centered'' or ``partisan'' a state's voter distribution is. The variance is a measure of how spread out or polarized a state is. The mean and variance allow us to embed the states in two-dimensional plots in a way that shows some of the structure of the voter distributions. Each state's mean and variance of voter distributions can be found in the ``Distribution Mean'' and ``Distribution Variance'' columns of Table \ref{tab:state_stats_three} in Appendix \ref{sec:appendix_state_distribution_info}. For example, when the electorate is bimodal the mean of the voter distribution is 0.10 in Alabama and the variance is 0.08, allowing us to plot Alabama at (0.10, 0.08). In California the mean of the bimodal voter distribution is $-0.05$ and the variance is 0.08, and thus we  plot this state at (0.05, 0.08), taking the absolute value of the mean. Figure \ref{fig:state_names} shows the 2D placements of the states for bimodal and trimodal voter distributions.

We quantify the change from model to model using \emph{relative change}, defined as the percent decrease in the relative difference from the median voter. For example, in 4-candidate elections using the bimodal distribution in Michigan, the relative difference of IRV and Condorcet is 0.322 in the Theoretical model but is 0.0 in the Ideological Truncation model, and thus the relative change when moving from the former model to the latter is $-1$, since the difference between IRV and Condorcet has collapsed to 0. When moving from the Theoretical Model to the Most Realistic model, the relative difference falls from 0.322 to 0.079, for a relative change of $-0.755$, or a 75.5\% decrease. In Alaska, on the other hand, the relative change when moving from the Theoretical Ideal model to the Most Realistic model is $-34.9\%$, a decrease not nearly as large as what we observe in Michigan. We examine if the mean and variance of the voter distribution can help explain the variability in relative change across states.

\begin{figure}[!htbp]
\centering

\begin{tabular}{@{}p{0.5\textwidth}p{0.5\textwidth}@{}}

\includegraphics[width=\linewidth]{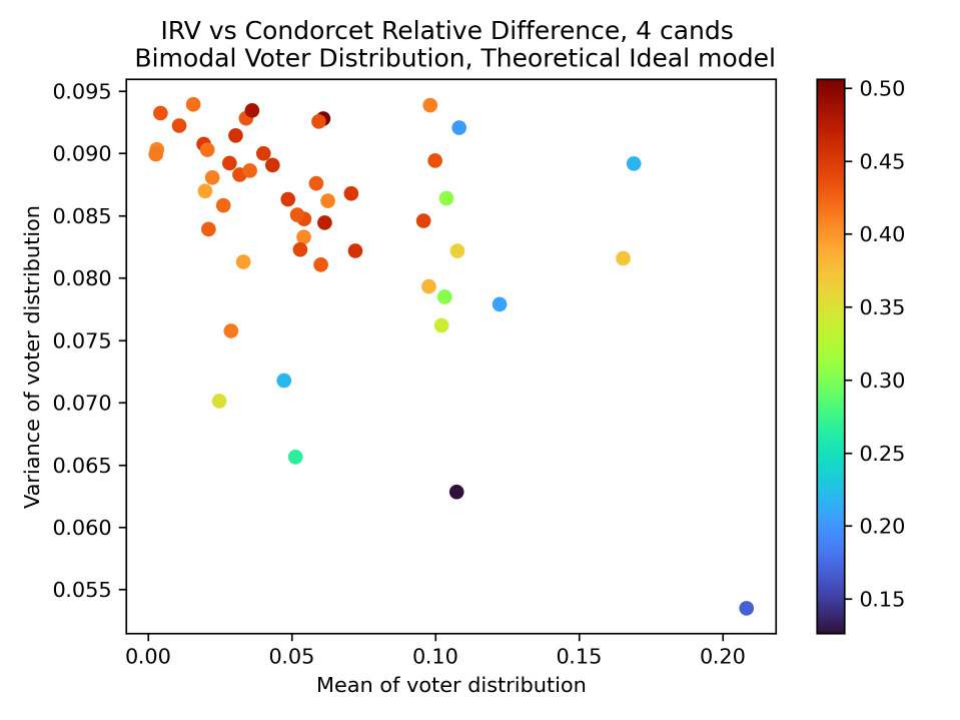}
&

\includegraphics[width=\linewidth]{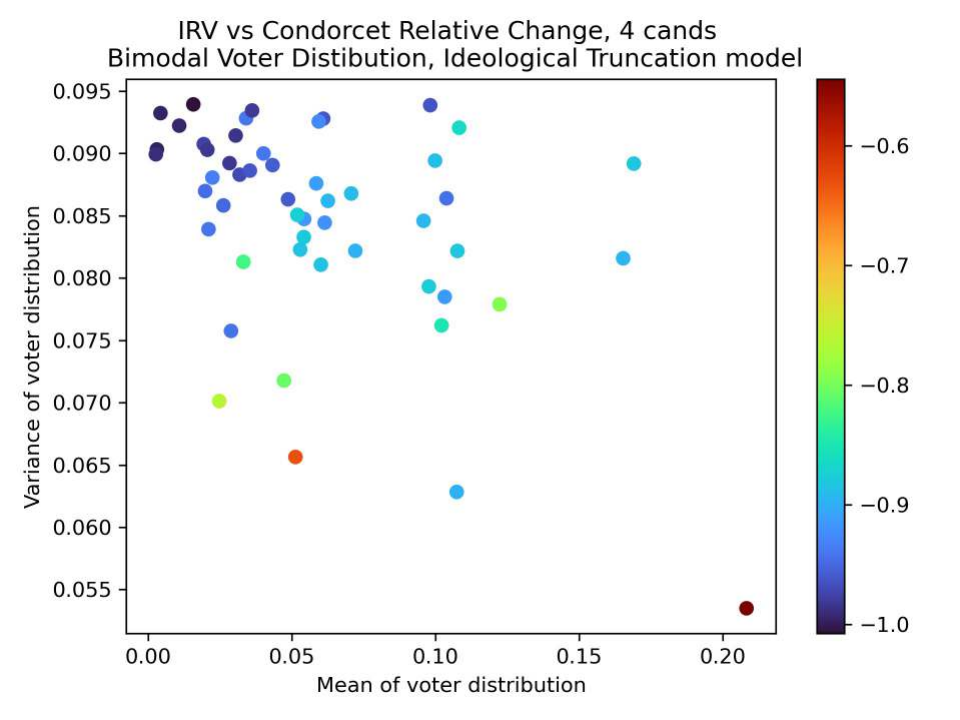}
\\[6pt]

\includegraphics[width=\linewidth]{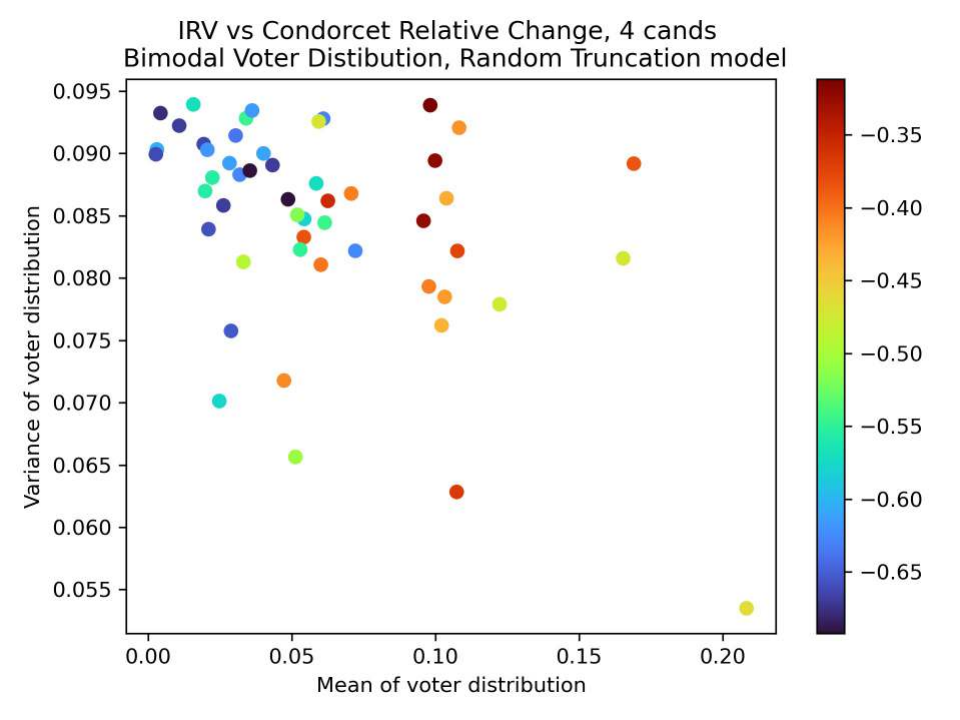}
&

\includegraphics[width=\linewidth]{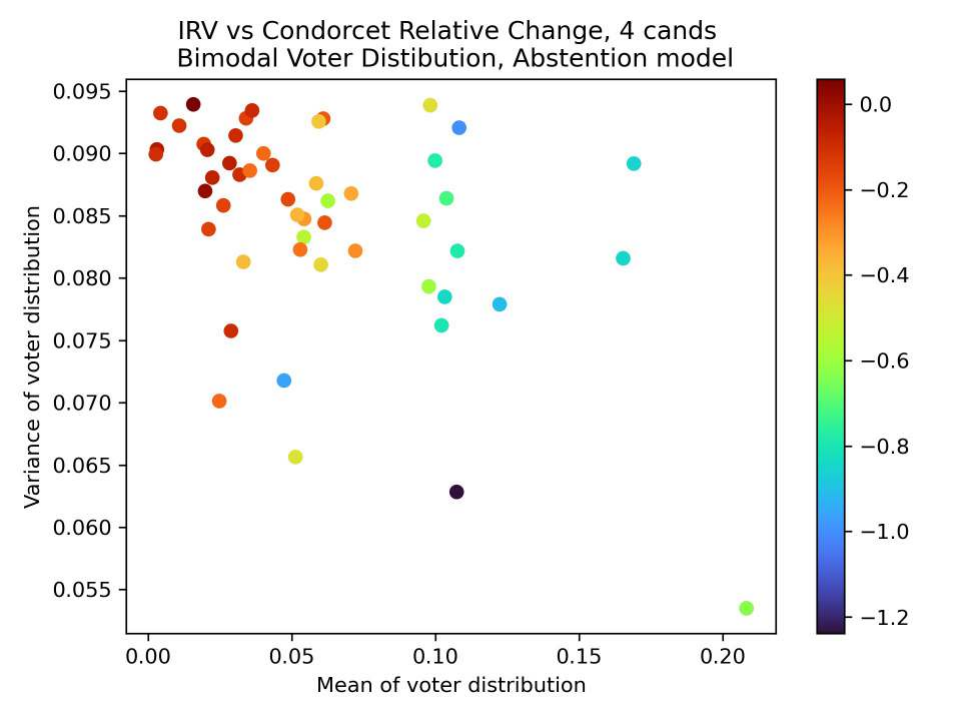}
\\[6pt]

\includegraphics[width=\linewidth]{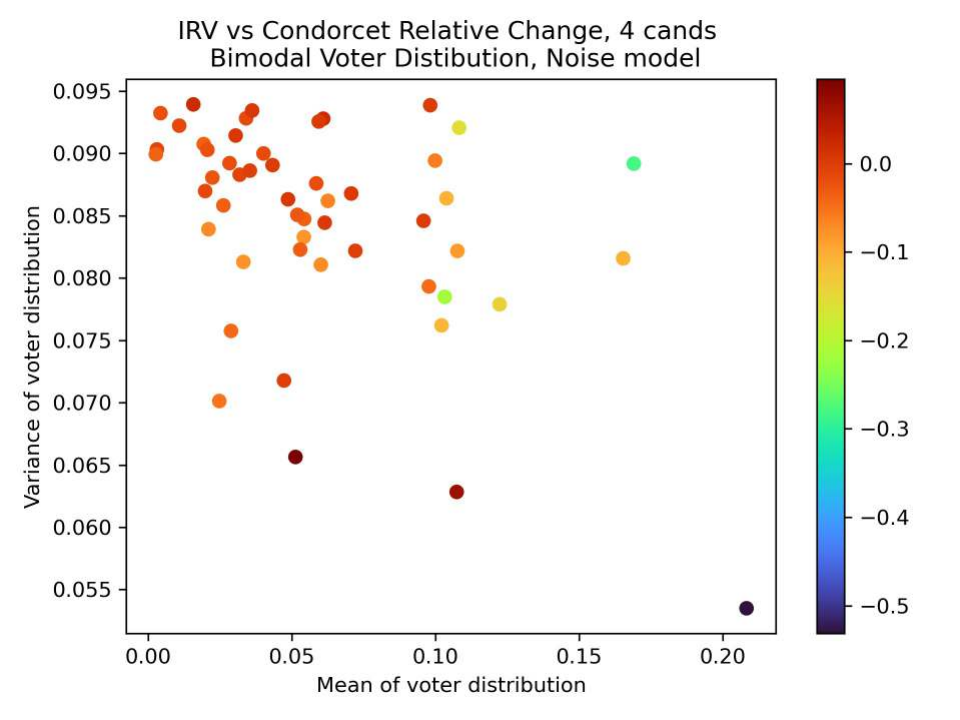}
&

\includegraphics[width=\linewidth]{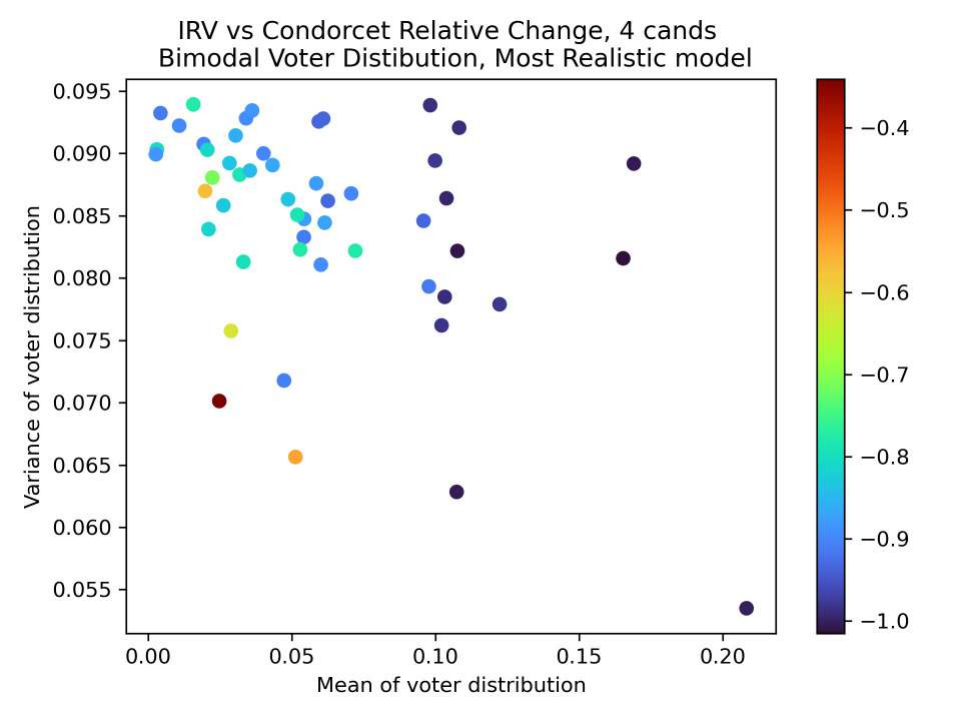}
\\

\end{tabular}

\caption{All 50 states and the District of Columbia plotted according to their (symmetrized) means and variances from the binomial distributions. Color is used to represent the relative difference between IRV and Condorcet in the top left, and the relative change when adjusting the voter model in the other five images.}
\label{fig:bimodal_4cands}

\end{figure}

In the top left of Figure \ref{fig:bimodal_4cands}, each state is colored according the relative difference between IRV and Condorcet in the Theoretical Ideal. Large values, where IRV and Condorcet are very different, are limited to states with a low mean but high variance, suggesting the differences are largest for states with dispersed, centered populations.

In the top right of Figure \ref{fig:bimodal_4cands}, which shows the relative change when moving from the Theoretical Ideal to Ideological Truncation for 4-candidate elections and bimodal distributions,  most points are deep blue for a value near $-1$, indicating that the addition of ideological truncation almost completely eliminates the difference between IRV and Condorcet. However, states with low variance, particularly Hawaii and the District of Columbia, have smaller values Even after adding ideological truncation, a substantial difference remains between IRV and Condorcet in these states.

By comparing the top right and middle left of Figure \ref{fig:bimodal_4cands}, we see that random truncation  behaves qualitatively differently than ideological truncation. Under the Random Truncation model, the mean seems to be the primary factor determining if truncation collapses IRV and Condorcet. States with high means see smaller effects from truncation than states whose means are close to zero. A state like Tennessee, located in the top right of the figure, sees a large relative change from ideological truncation but not from random truncation. 

The effect of abstention is shown in the middle right of Figure \ref{fig:bimodal_4cands}. Abstention  seems to have the largest effect in states with high means or low variances. Interestingly, it is quite similar to the top left of Figure \ref{fig:bimodal_4cands}, suggesting that ideological abstention will collapse IRV and Condorcet, but only in the states where the differences between the two methods were not very large in the Theoretical Ideal model.

We might expect that adding random noise to all voters would have no effect, because everything would ``cancel out.'' But the bottom left of Figure \ref{fig:bimodal_4cands} shows this is not always the case. Noise can slightly collapse IRV and Condorcet, but mainly in states with high means. The effect is particularly strong in the District of Columbia.

The relative change for the Most Realistic model is shown in the bottom right of Figure \ref{fig:bimodal_4cands}. We see that IRV and Condorcet collapse in most states, but to a lesser extent in states with low means, particularly Alaska, Hawaii, and Wisconsin.

The results in Figure \ref{fig:bimodal_4cands} are only for bimodal distributions with four candidates. We also examine the trimodal distributions and three candidates. Many results are the same and all the relevant plots can be found in Appendix \ref{sec:all_mean_vars_plots}. We did find some differences for certain distributions and number of candidates, which we summarize below.

\begin{figure}
    \centering

    \begin{tabular}{@{}p{0.5\textwidth}p{0.5\textwidth}@{}}

    \includegraphics[width=\linewidth]{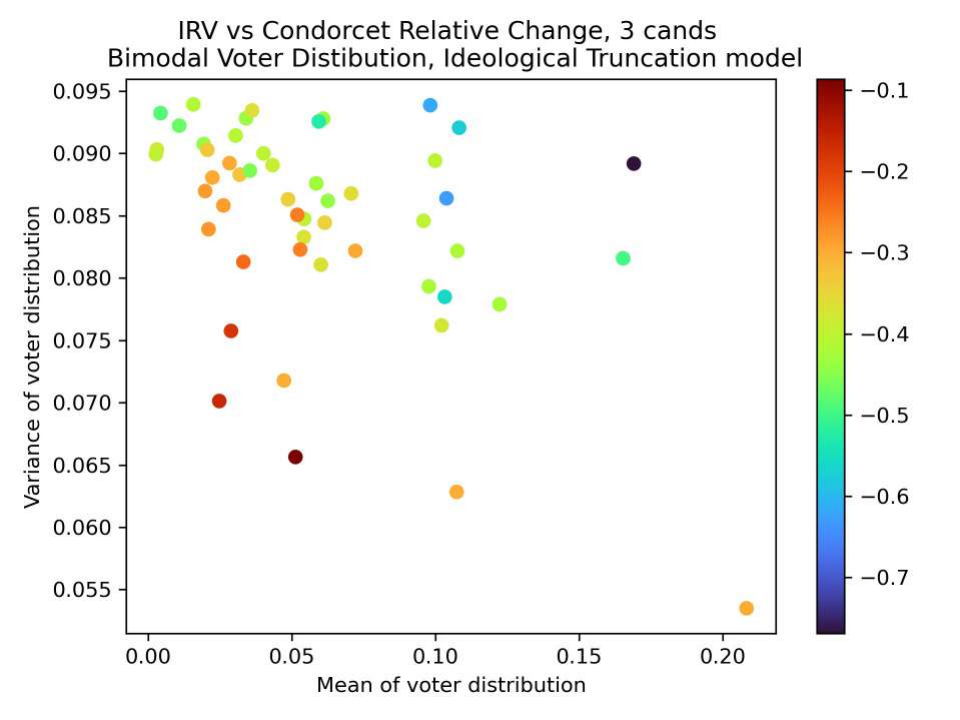}
    &
    
    \includegraphics[width=\linewidth]{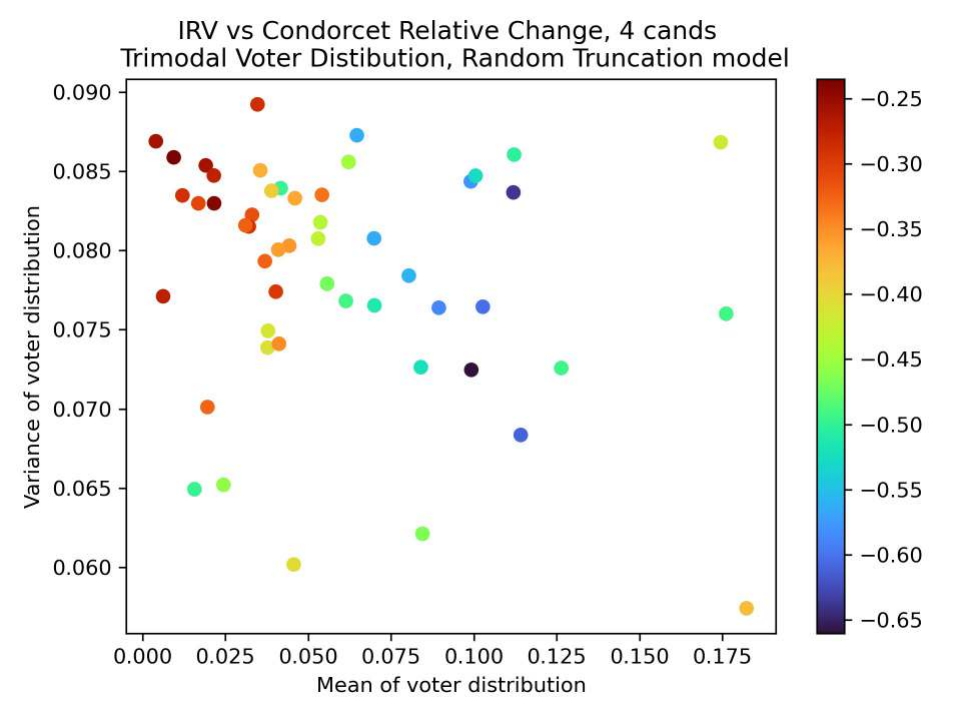}
    \\[6pt]

    \includegraphics[width=\linewidth]{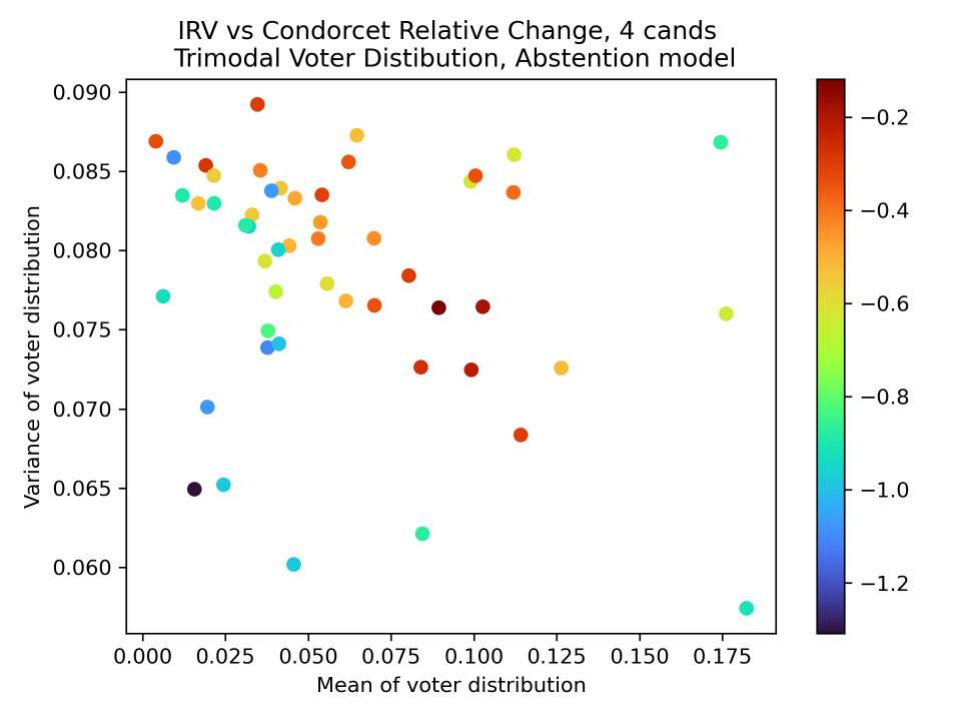}
    &
    
    \includegraphics[width=\linewidth]{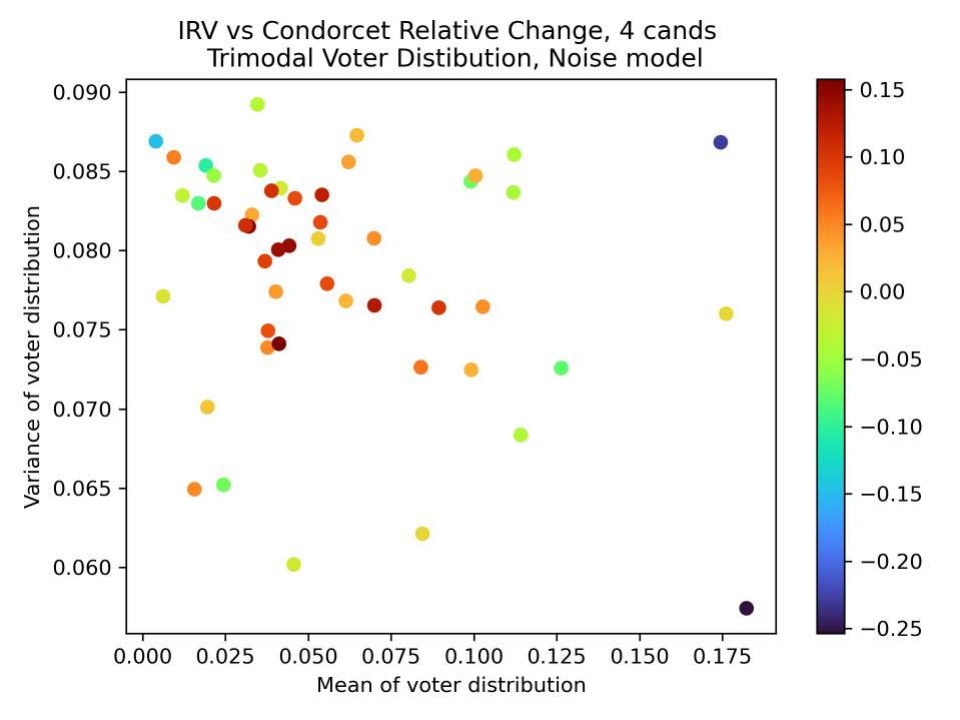}
    \\[6pt]
    
    \end{tabular}
    \caption{Plots highlighting the differences for three and four candidates, and for the bimodal and trimodal distributions. The top left image shows the relative change for bimodal distributions with three candidates. The other images show the effects of random truncation, ideological abstention, and noise, for trimodal distributions. }
    \label{fig:3cands_trimodal}
\end{figure}

First, in general, effects are stronger for four candidates than for three. This is unsurprising because it is easier for two election methods to disagree on four candidates than on three. The relative difference between the Condorcet and IRV winners is smaller for three candidates than for four, but many of the relative changes by deviating from the Theoretical Ideal model are similar in size. The only model that seems to behave qualitatively differently between three and four candidates is Ideological Truncation. The top left of Figure \ref{fig:3cands_trimodal} shows the relative change from Ideological Truncation. Comparing it to the top right of Figure \ref{fig:bimodal_4cands}, we see a drastic difference. For four candidates, the states with the largest relative change have high standard deviations and low means, but with three candidates, it is the states with high variance and high means, particularly Wyoming, that see the largest relative changes. 

The shift from bimodal to trimodal distributions yields even more dramatic results. The plots for Random Truncation and Abstention flip along the mean axis. For Random Truncation, low means have large relative changes for bimodal distributions in the middle left of Figure \ref{fig:bimodal_4cands} but small relative changes for trimodal distributions in the top right of Figure \ref{fig:3cands_trimodal}. Abstention is reversed, with high means having large effects for bimodal distributions (middle right of Figure \ref{fig:bimodal_4cands}) and small effects for trimodal distributions (bottom left of Figure \ref{fig:3cands_trimodal}). One final interesting observation is that for trimodal distributions, the smallest (or largest) effect sizes sometimes happen for moderate mean or variance values. The best example of this with Noise, shown in the bottom right of Figure \ref{fig:3cands_trimodal}, where we see that the smallest relative changes from noise are states with variance around 0.085 and means 0.05. This contrasts sharply with Figure \ref{fig:bimodal_4cands} where the extreme values are typically in the corners of the subplots. 

The cause of the differences across plots is not clear, and warrants future investigation.

\section{Discussion}\label{section:discussion}

Our results provide strong evidence that incorporating more realistic voter behavior significantly changes the relationship between Condorcet and IRV outcomes. Tables \ref{table:avg_distance_stats_bimodal_3cands}-\ref{table:avg_distance_stats_trimodal_4cands} in Appendix \ref{sec:sum_stats_tables} show that when we make only one change to a ``theoretically ideal'' model by introducing truncated ballots in the Ideological Truncation model, the average distance from the median voter tends to become much closer for the two methods for both bimodal and trimodal electorates. The most dramatic change occurs for bimodal electorates in 4-candidate elections, where the relative difference across all states drops from $(0.142-0.101)/0.101 = 40.6\%$ in the Theoretical Ideal model to $(0.148-0.144)/0.144=2.8\%$ in the Ideological Truncation model (see Table \ref{table:avg_distance_stats_bimodal_4cands} in Appendix \ref{sec:sum_stats_tables} or Figure \ref{fig:avg_dist_bimodal_4cands_ideotrunc} in Appendix \ref{sec:avg_dist_images}). Similarly, the relative difference across all states drops from  40.6\% in the Theoretical Ideal model to $(0.183-0.173)/0.173 = 5.8\%$ in the Most Realistic model. 

We note that the relative difference between IRV and Condorcet can be very small even when the two methods choose different winners at a high frequency. To see how this can occur, consider the top right image of Figure \ref{fig:MI_results_trimodal_4cands} in Appendix \ref{sec:winner_distribution_histograms}, which shows the winner position histograms in 4-candidate Michigan elections for a trimodal electorate under the Most Realistic model. The average distance to the median voter is similar for both IRV and Condorcet, yet the methods choose different winners in approximately 35\% of the generated elections. The reason is that IRV tends to elect more winners who are Somewhat Liberal or Middle of the Road, while Condorcet tends to elect more winners who are Somewhat Conservative; in the aggregate, the IRV winner tends to be slightly closer to the median voter on average than the Condorcet winner. Thus, in this example the choice between the two methods is not about whether winners tend to be moderate, but is instead about which type of moderate candidate each voting method favors. There is a small red peak just to the right of the rightmost blue peak, and thus IRV does give far-right candidates a slightly better chance of winning than Condorcet.

\begin{figure}
\centering
\begin{tabular}{cc}
\includegraphics[width=68mm]{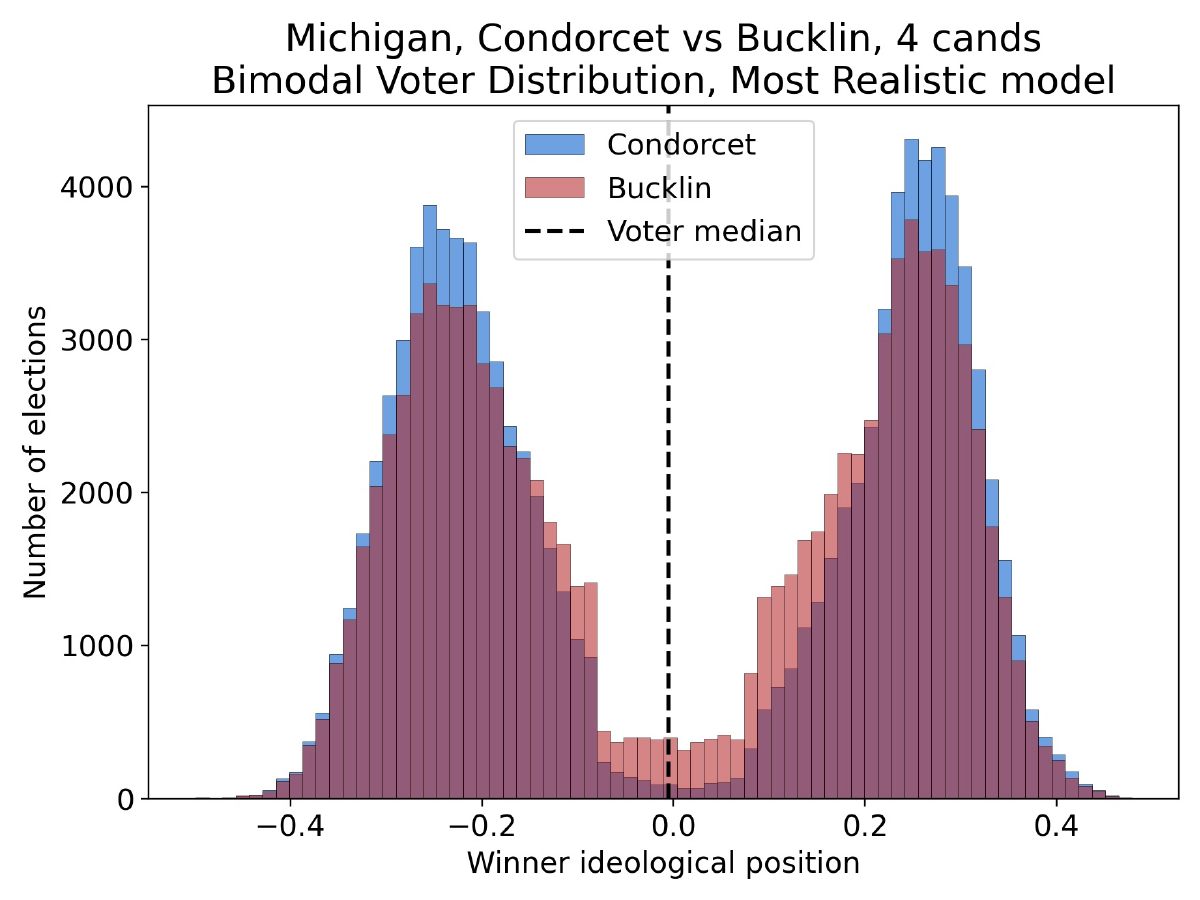}&\includegraphics[width=68mm]{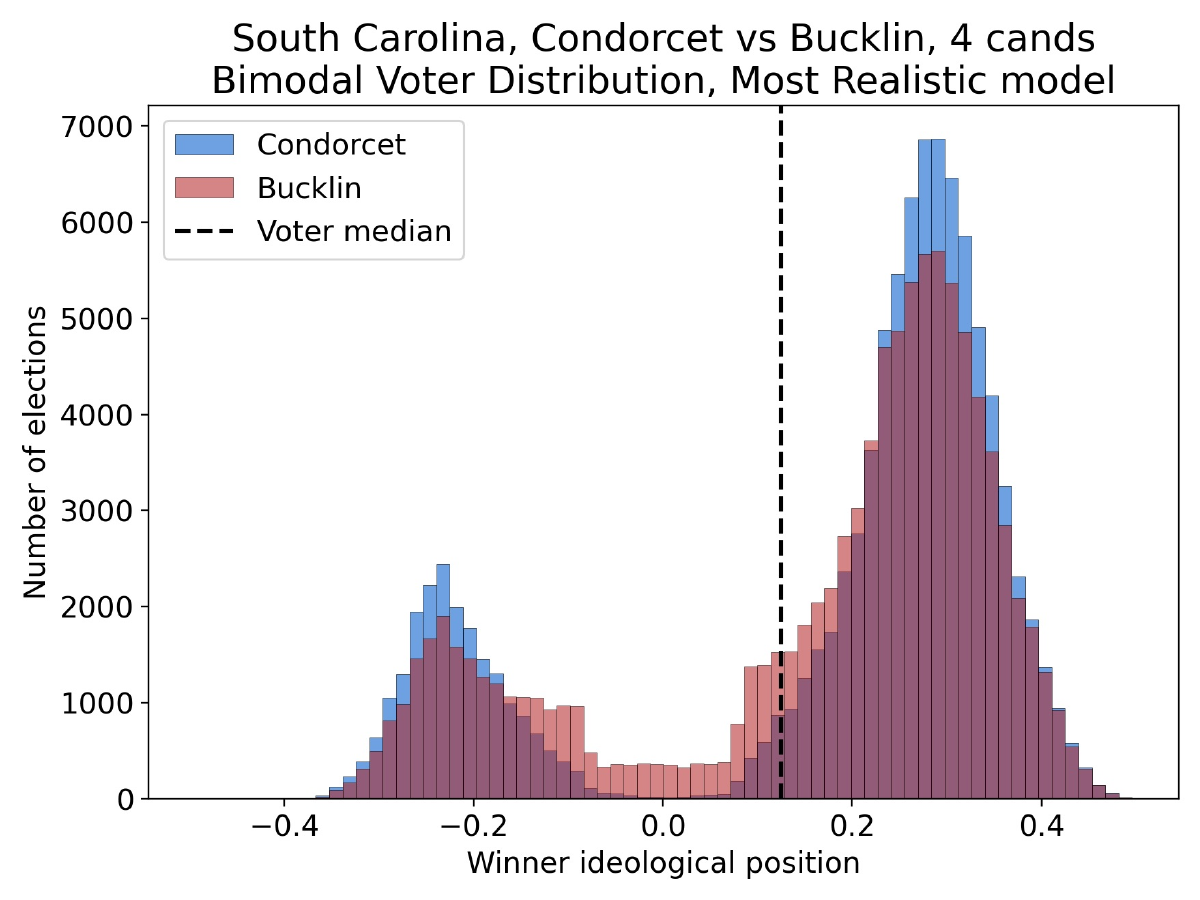}
\end{tabular}
\caption{Bucklin winner position distribution (red) versus Condorcet winner position distribution (blue) in Michigan and South Carolina under the Most Realistic model and bimodal voter distribution in 4-candidate elections.}
\label{figure:Bucklin}
\end{figure}

How well do other voting methods perform under models with more realistic features? Plurality, unsurprisingly, still tends to perform worse than other methods, while Borda and Bucklin  often perform better than Condorcet or IRV, particularly for bimodal electorates (see Tables \ref{table:avg_distance_stats_bimodal_3cands} and \ref{table:avg_distance_stats_bimodal_4cands} in Appendix \ref{sec:sum_stats_tables}). In fact, under the Most Realistic Model, in bimodal electorates with four candidates the average distance from the median voter is not minimized by Condorcet methods in any state; this minimum is achieved by Borda in 21 states, Bucklin in 29 states, and IRV in one state (Vermont). A sample of Bucklin winner distributions versus Condorcet winner distributions under the Most Realistic model is given in Figure \ref{figure:Bucklin}, where  the Bucklin winners tend to cluster more closely around the median voter than the Condorcet winners. Figures \ref{fig:avg_dist_Borda_bimodal_4cands} and \ref{fig:avg_dist_Bucklin_bimodal_4cands} in Appendix \ref{sec:avg_dist_images} show the relative difference for Borda and Bucklin over Condorcet in 4-candidate elections and a bimodal electorate. The figures show that in most states, the Borda or Bucklin winner tends to be ``less extreme'' on average than the Condorcet winner in most states.

\begin{figure}[!htbp]
\centering

\begin{tabular}{@{}p{0.5\textwidth}p{0.5\textwidth}@{}}

\includegraphics[width=\linewidth]{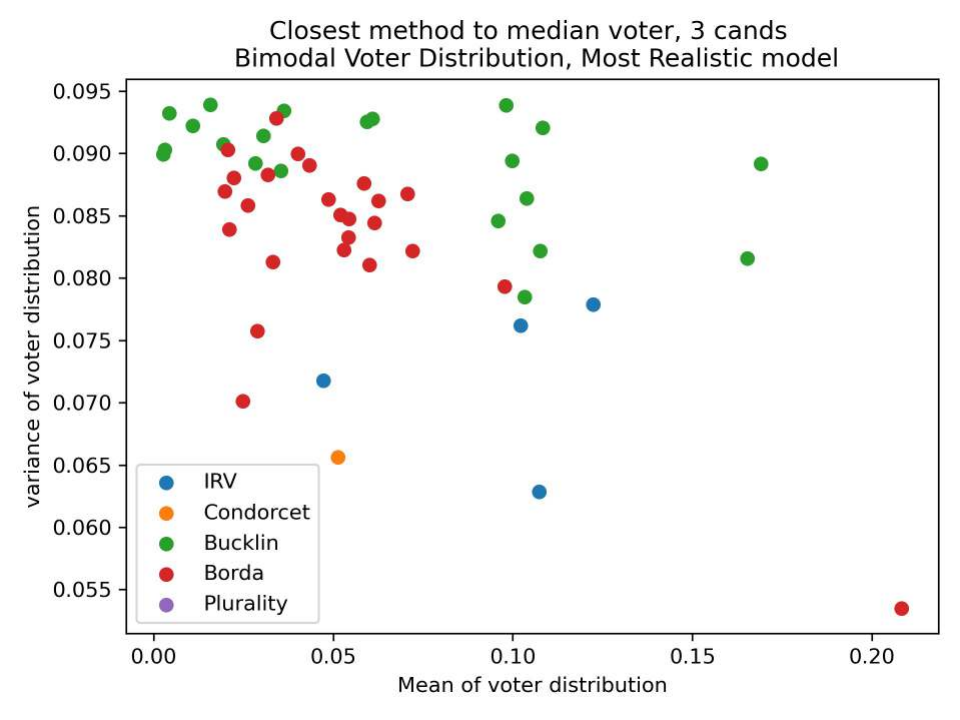}
&
\includegraphics[width=\linewidth]{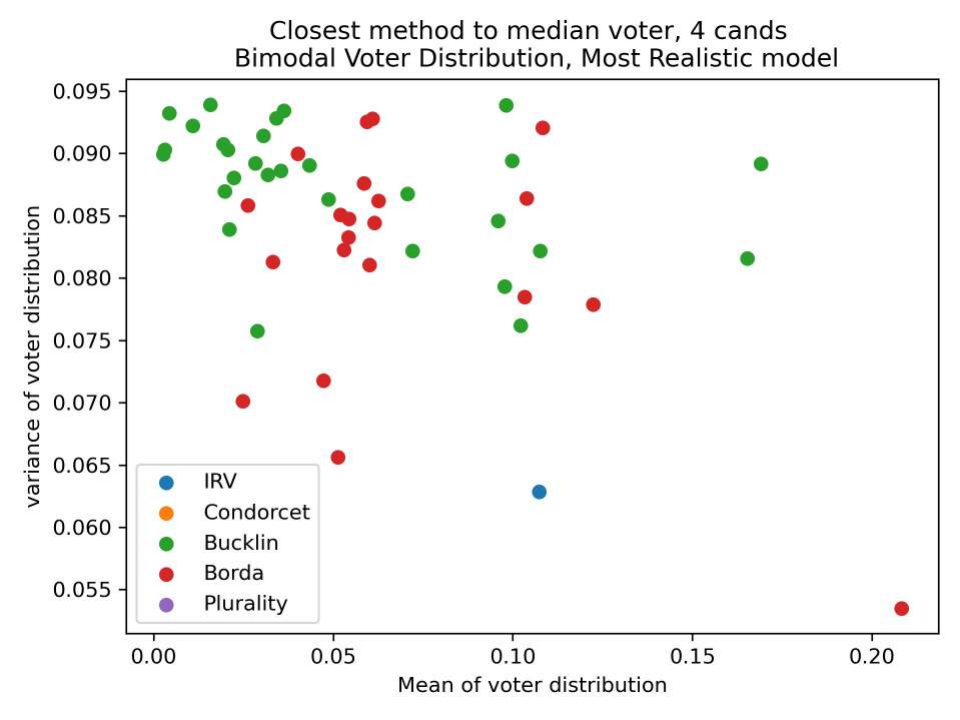}
\\[6pt]

\includegraphics[width=\linewidth]{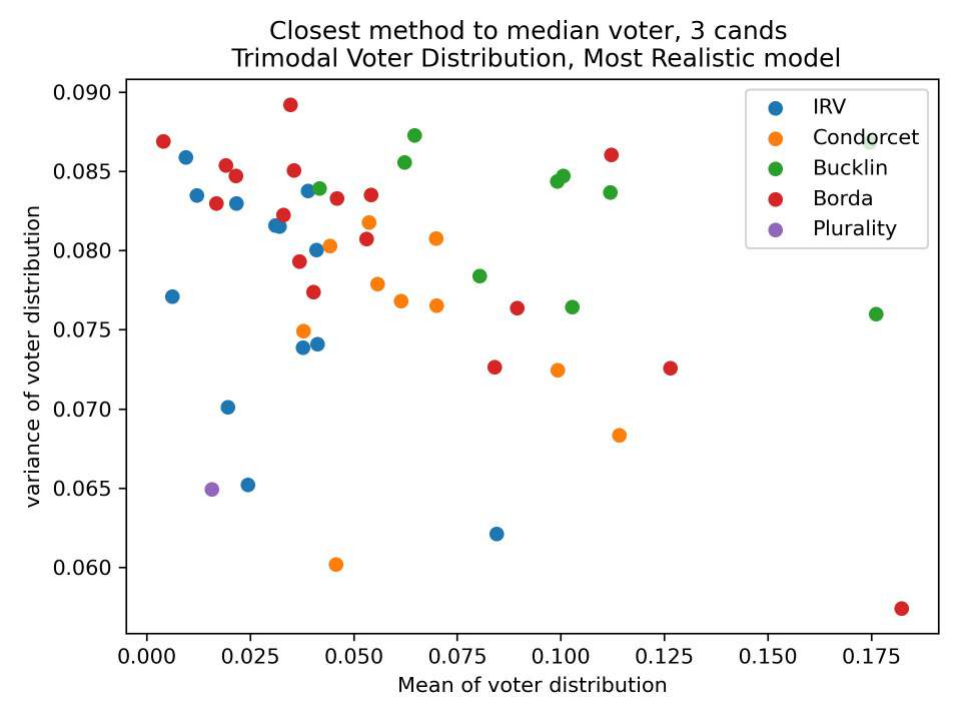}
&
\includegraphics[width=\linewidth]{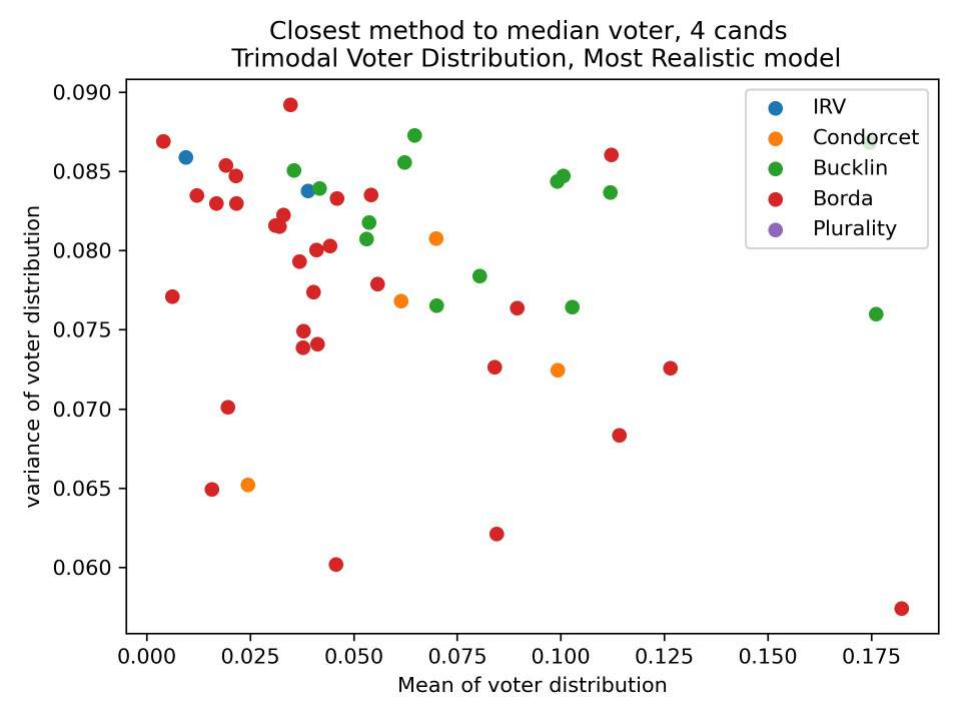}
\\

\end{tabular}

\caption{All 50 states and the District of Columbia plotted according to their (symmetrized) means and variances from the bimodal and trimodal distributions. Each dot is colored by which election method elects, on average, the candidate closest to the median voter.}
\label{fig:moderating_election_methods}

\end{figure}

Figure \ref{fig:moderating_election_methods} shows the voting method which achieves the minimum average distance from the median voter for each state, voter distribution, and number of candidates, under the Most Realistic model. Overall, Bucklin and Borda seem to be the ``most moderating'' methods, particularly with bimodal electorates. Interestingly, all four sub-figures in Figure \ref{fig:moderating_election_methods} seem to have quite different patterns. It is not clear, for example, why trimodal distributions with three candidates have the most diversity in terms of which method is most moderating. Surprisingly, in Delaware, plurality is the most moderating method.

Overall, our results suggest that if one's priority is to use a method which incentivizes candidate moderation (interpreted as a tendency to elect candidates near the median voter) then it is not clear Condorcet methods are the best tool for combating polarization under realistic conditions.  With that said, our results are limited in two important ways. First, it is possible there are ``realistic'' models under which the Condorcet and IRV winner distributions are very different. That is, perhaps there exist realistic models which reinforce the primary conclusion of \cite{AFG}, that Condorcet methods are much more ``moderating'' than IRV. We use distance-based and random approaches to generate elections with bullet vote rates similar to what is observed in real-world IRV elections, but other approaches are also sensible (and perhaps better in some sense). Second, it is  possible that if a Condorcet method were used then voters would respond by casting complete ballots, in which case the models used in \cite{AFG} would arguably be more realistic than ours. We do not have strong reasons to think a Condorcet method would elicit more complete ballots: in practice there is no real incentive for voters to cast short ballots in an IRV election, and yet voters often do so anyway \cite{Schwab_report}. It is not clear why using a Condorcet method would significantly change voter behavior in this respect.


If anything, it is conceivable that if a Condorcet method were used then we might see more truncated ballots. There are numerous ways in which this could occur. If candidates respond to the Condorcet moderation incentive and all converge to the center, then a voter might cast a bullet vote simply because they cannot distinguish between candidates beyond their favorite. It is also possible that candidates would encourage their voters to cast bullet votes, an action which candidates have no real incentive to take under IRV. To see why, 
consider the 2022 Special election for US House in Alaska, which used IRV to elect the winner\footnote{For in-depth analysis of this election see \cite{AFG,C26,GSM24}.}. The election contained the three candidates Mary Peltola, Nick Begich, and Sarah Palin, arranged ideologically from left to right. Begich was the Condorcet winner but received the fewest first-place votes and was thus eliminated in the first round, and leftmost candidate Peltola won against Palin in the final round. If a Condorcet method had been used and it was clear from pre-election polling that Begich was the Condorcet winner then Palin and Peltola would mostly likely strategize about how to win under such a method.  It seems they have three main options: (i) they could move to Begich's ideological position, (ii) they could encourage their voters to ``bury'' Begich on the ballot, insincerely ranking him last, or (iii) they could encourage their voters to cast bullet votes to raise the probability Begich loses his status as the Condorcet winner. In other words, a candidate like Peltola might reason that if she wants to remain left-leaning then she has essentially no chance of winning if voter cast complete ballots, but if she and Palin encourage their supporters to bullet vote, both converging on strategy (iii), then Peltola has a non-zero probability of winning, either through the creation of a Condorcet cycle or a change in the Condorcet winner. In this example, if candidates encouraged more bullet voting then Peltola would have been the Condorcet winner if about 8500 additional Palin supporters cast ballots of length one. This does not benefit Palin, but she might reason that if there are enough bullet votes cast then she at least has a chance to win against Peltola head-to-head. Of course, it is possible that voters would not respond when encouraged to bullet vote\footnote{Palin actually did encourage her supporters to bullet vote, but only 36\% did so.} (or bury), but until a Condorcet method is used in practice we cannot know how strongly candidates are incentivized to encourage bullet voting and the extent to which voters will listen. 

Finally, the comparison between Condorcet methods and IRV raises broader normative considerations that go beyond questions of candidate moderation. One concern is that Condorcet winners may face a legitimacy problem: the public might resist an outcome where a candidate with the fewest first-place votes is declared the winner.  If $k=3$ then the Condorcet winner is the candidate with the fewest first-place vote in every election in which the Condorcet winner is not the IRV winner. The public might interpret such outcomes as declaring the ``candidate with the fewest votes'' to be the winner. While such skepticism might diminish as voters become more familiar with a Condorcet method, early elections could face backlash over outcomes perceived as unfair.

A second consideration is whether incentivizing candidates to converge toward the median voter is always normatively desirable. If candidates respond to these incentives by placing themselves at approximately the same ideological position, such convergence could produce a field of nearly indistinguishable candidates, increasing the likelihood that no Condorcet winner exists. Moreover, if the dominant strategy is to occupy the same ideological position, a third or fourth candidate may see little reason to enter the race,  potentially collapsing competition to a de facto two-candidate system, which few would regard as ideal.

Together, these considerations suggest that while Condorcet methods may promote moderation in theory, their practical adoption could involve complex trade-offs between perceived legitimacy, an increased number of short ballots, and candidate convergence. Furthermore, if the encouragement of candidate moderation is an important feature of a voting method, our results suggest that Condorcet methods may not be the best choice, since they are often outperformed by Bucklin or Borda in the more realistic models.


\section{Conclusion}\label{section:conclusion}

We do not claim our models accurately capture real-world voter preferences; such preferences may not even be approximately single-peaked, for example. We do claim that our models are more realistic than that used in \cite{AFG}, if only because of our incorporation of truncated ballots. When we adapt the model of \cite{AFG} in this way, the differences between Condorcet and IRV largely disappear, and the incentive for candidates to move to the center under Condorcet methods is much less strong. We do not wish our results to be interpreted as an argument against the use of Condorcet methods; to the contrary, we would be interested to see a jurisdiction adopt a Condorcet method so we can better evaluate how such methods perform  in practice. We also do not wish our results to be taken as an endorsement of a particular voting method.  Instead, our results imply that findings obtained from purely theoretical models, while certainly valuable, may not accurately predict what would occur in practice. In particular, the strong performance of Condorcet methods under theoretical models may not translate to real-world elections.

Future work could focus on the construction of other empirically grounded spatial models from the CES data. It would be interesting to find ``realistic'' models which make the same predictions as those of Atkinson et al. In particular, perhaps there are realistic voter behaviors unrelated to the casting of partial ballots which would cause the Condorcet and IRV winner distributions to pull farther apart, thereby supporting the argument that Condorcet methods are significantly more ``moderating'' than IRV. Furthermore, we use the bin methodology of Atkinson et al. so that we can examine the effects of adding realistic behaviors to their models, but other ``smoother'' models could be constructed which don't rely on discrete bins. Ultimately, we probably cannot fully anticipate how candidates, parties, voters, etc., will respond to the use of a Condorcet method until such a method is actually implemented.

\clearpage
\appendix

\section{Winner Ideological Position Histograms}\label{sec:winner_distribution_histograms}

\begin{figure}[h]
\begin{tabular}{cc}
\includegraphics[width=66mm]{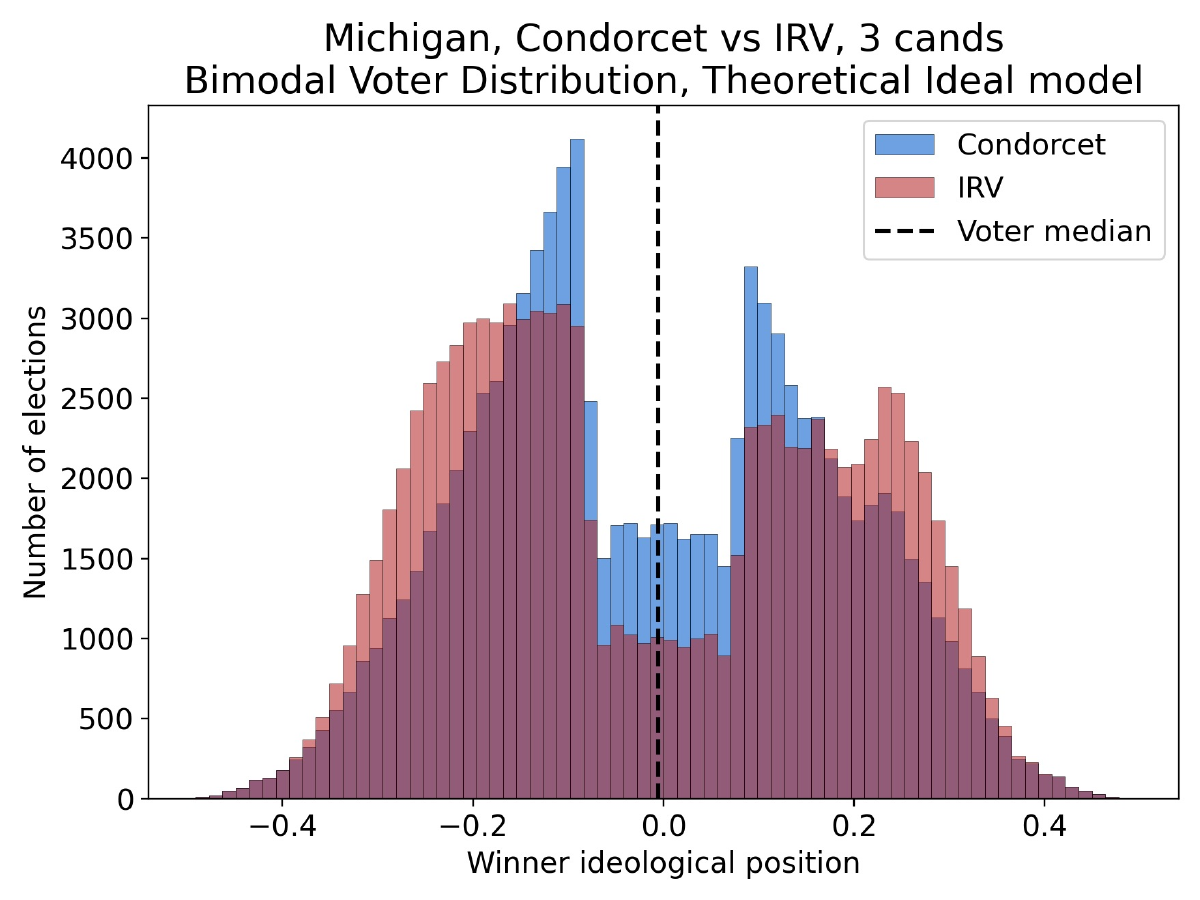} 
& 

\includegraphics[width=66mm]{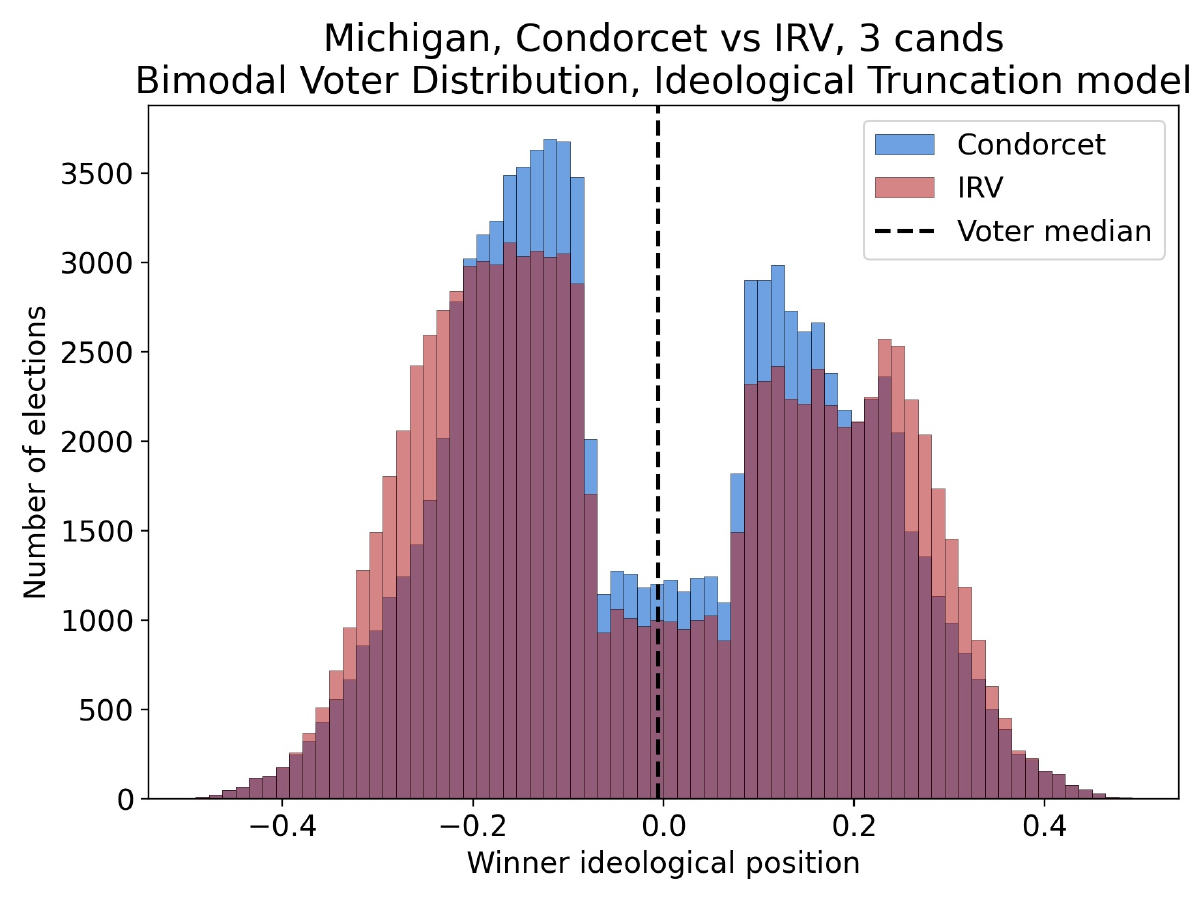} 

\\
\includegraphics[width=66mm]{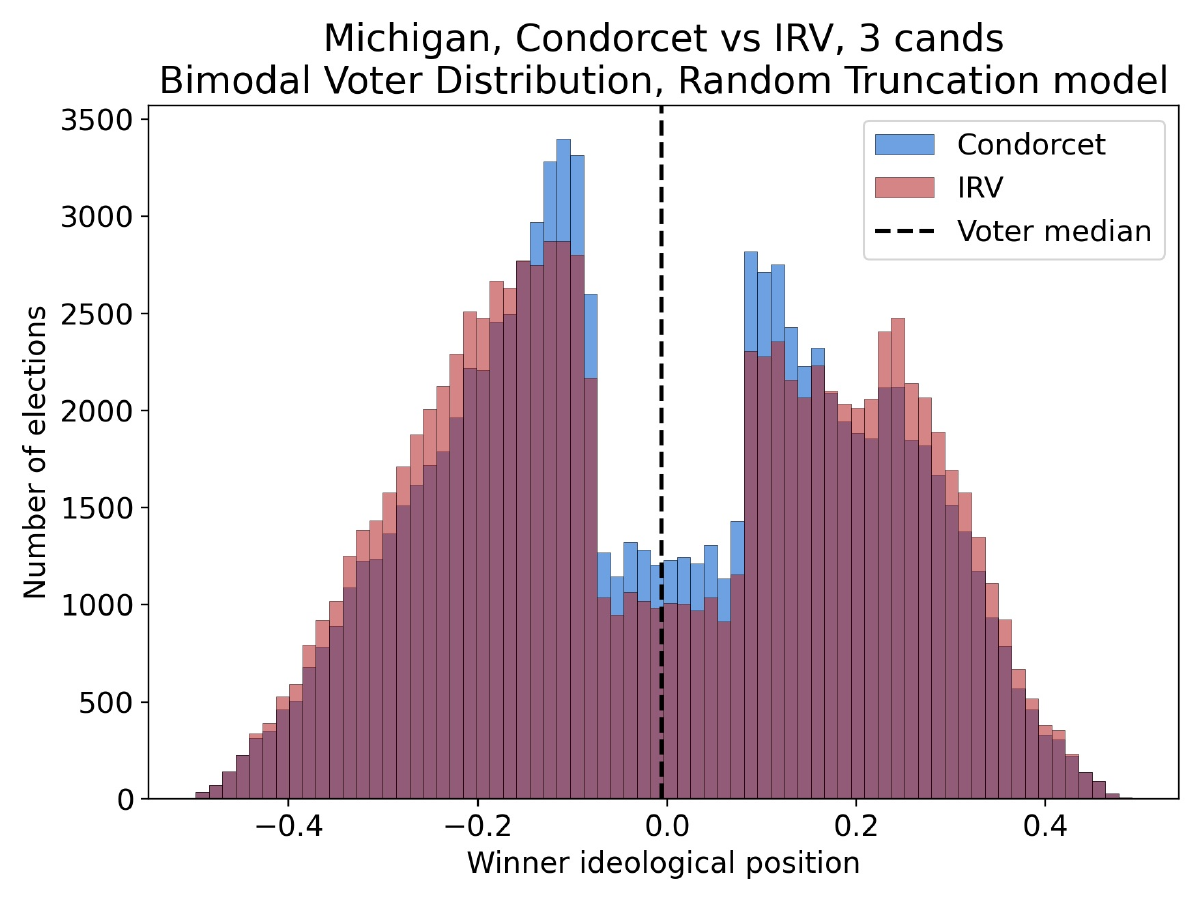}
& 

\includegraphics[width=66mm]{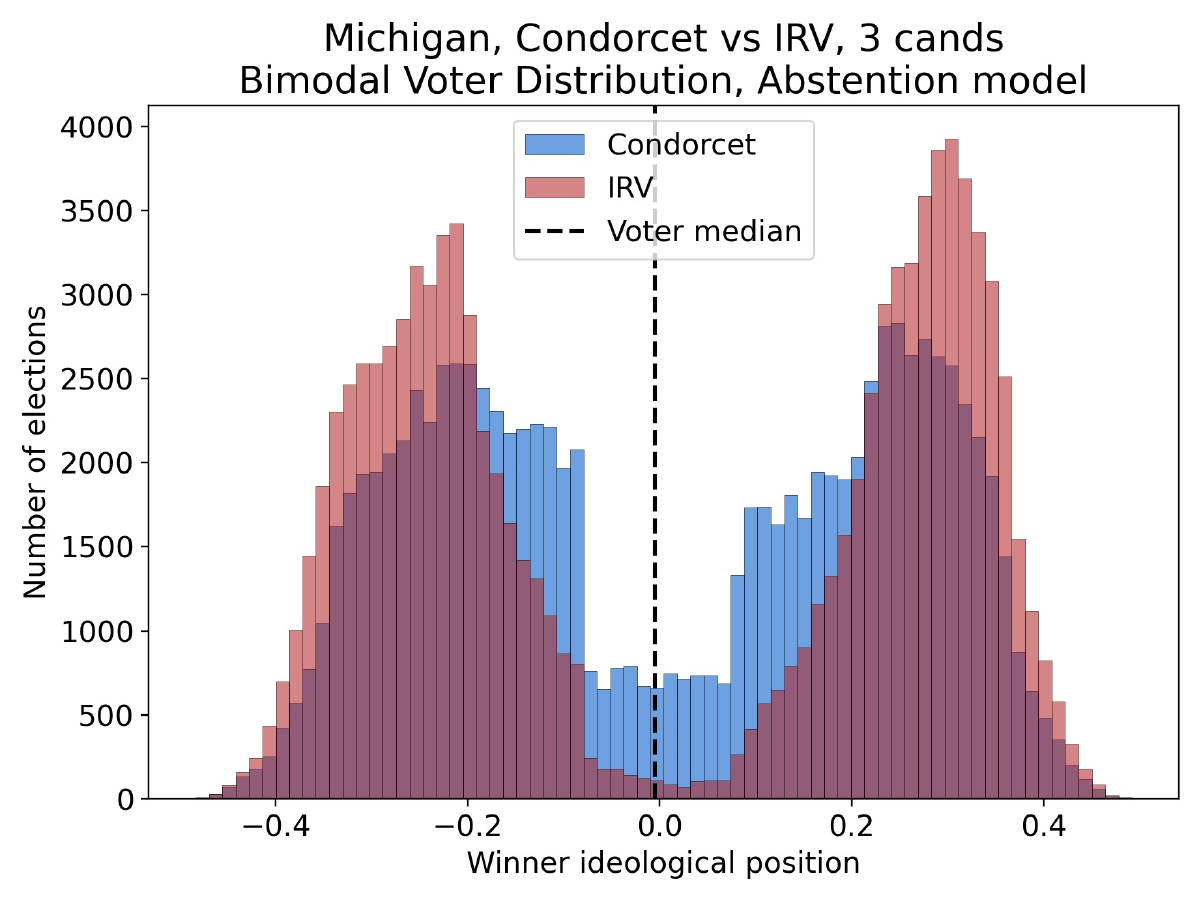} 
\\

\includegraphics[width=66mm]{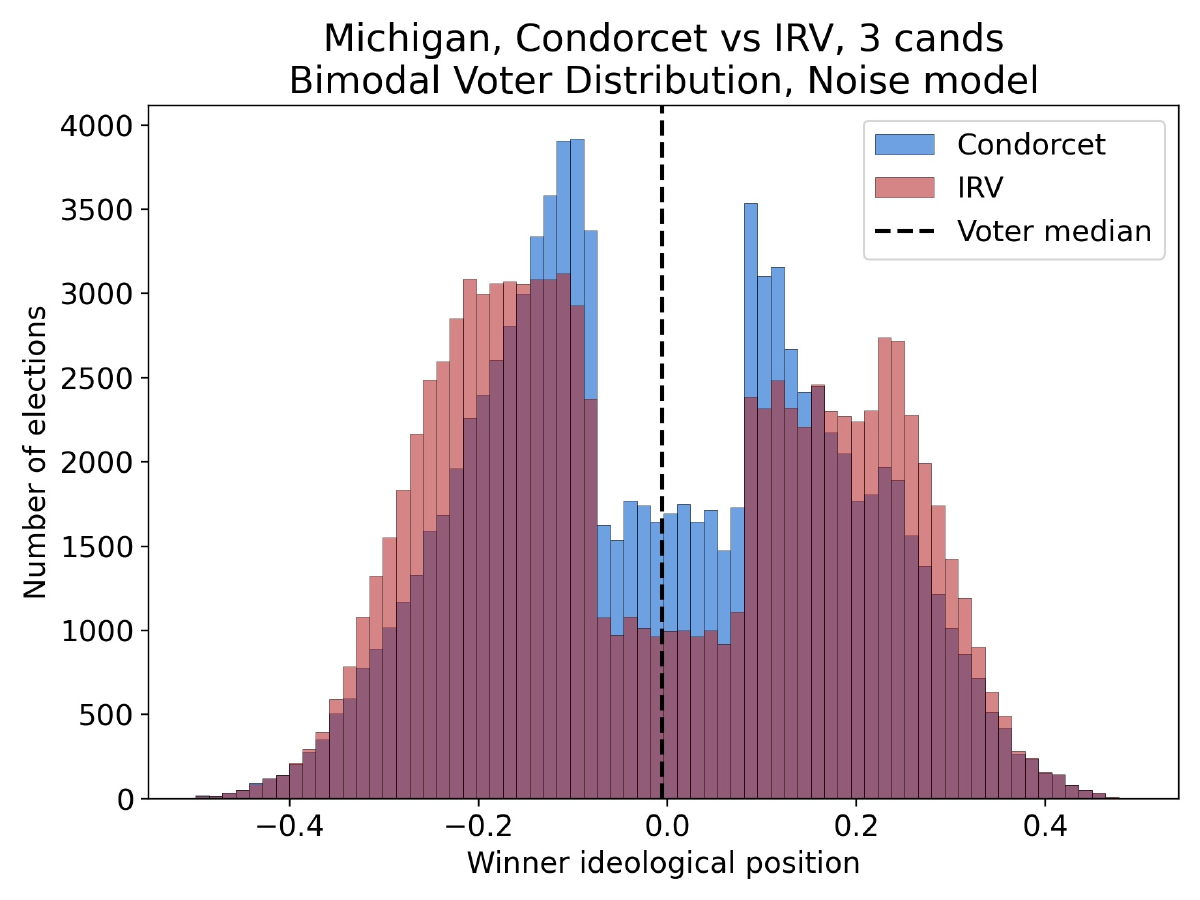}
& 

\includegraphics[width=66mm]{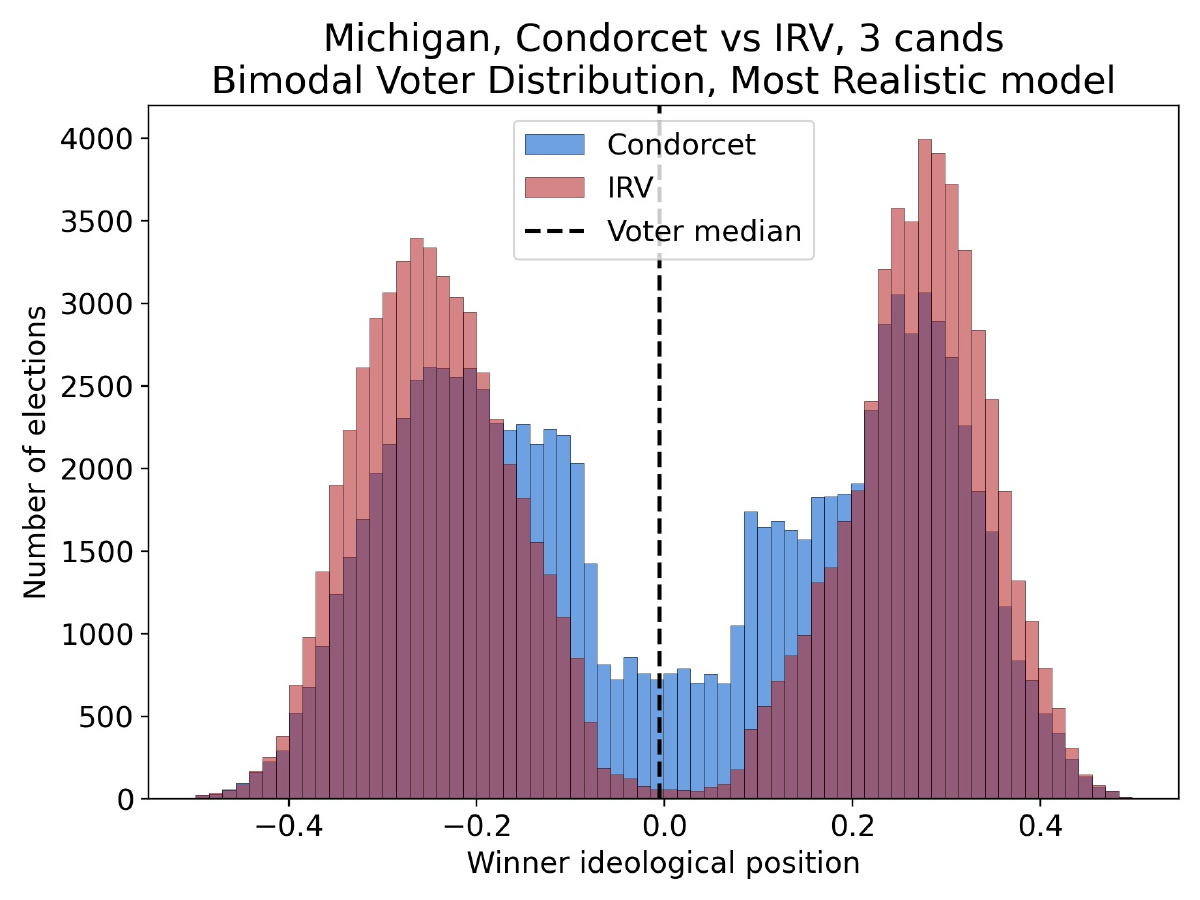}
\\
\end{tabular}
\caption{Histograms showing the ideological positions of Condorcet and IRV winners across 100,000 simulations in Michigan, using 3-candidate elections with the bimodal voter distribution.}
\label{fig:MI_results_bimodal_3cands}
\end{figure}

\begin{figure}[ht]
\begin{tabular}{cc}
\includegraphics[width=66mm]{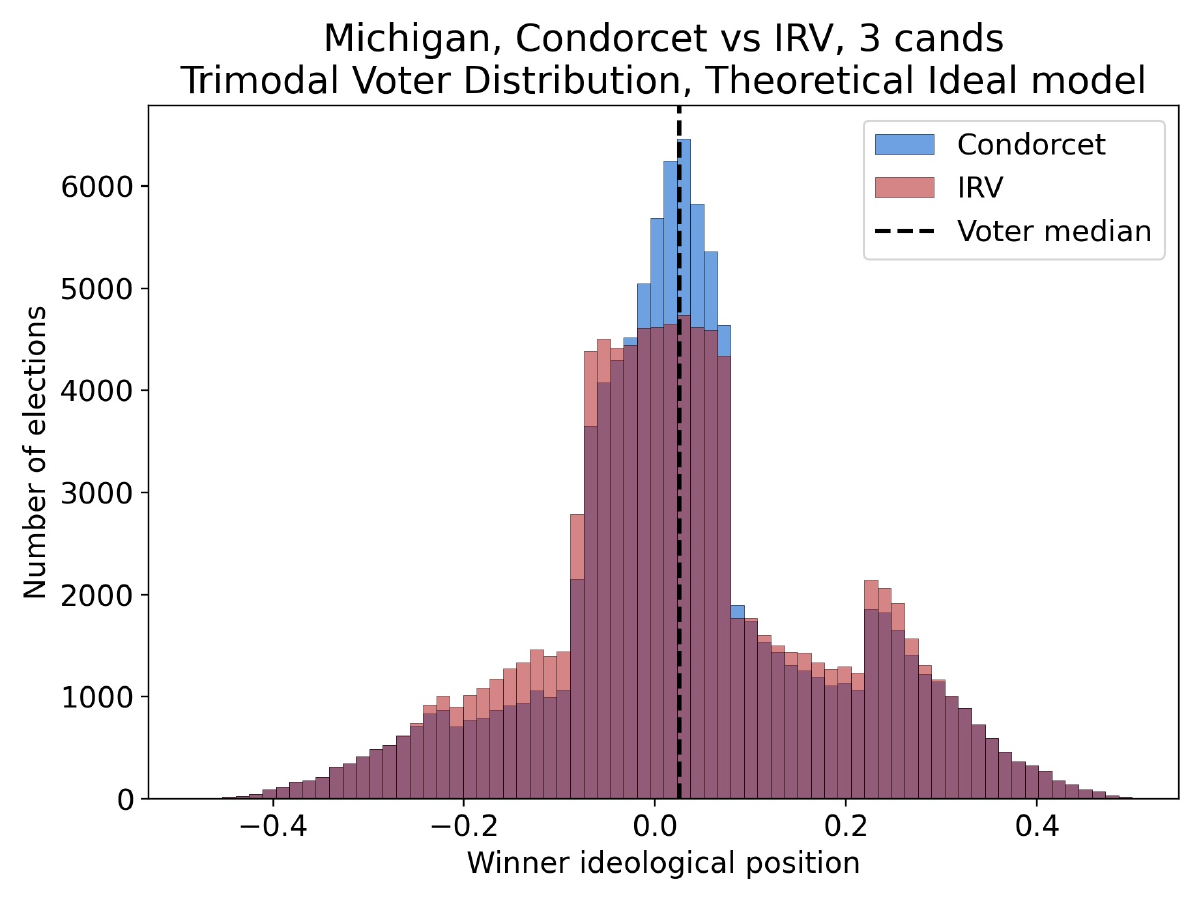} & 
\includegraphics[width=66mm]{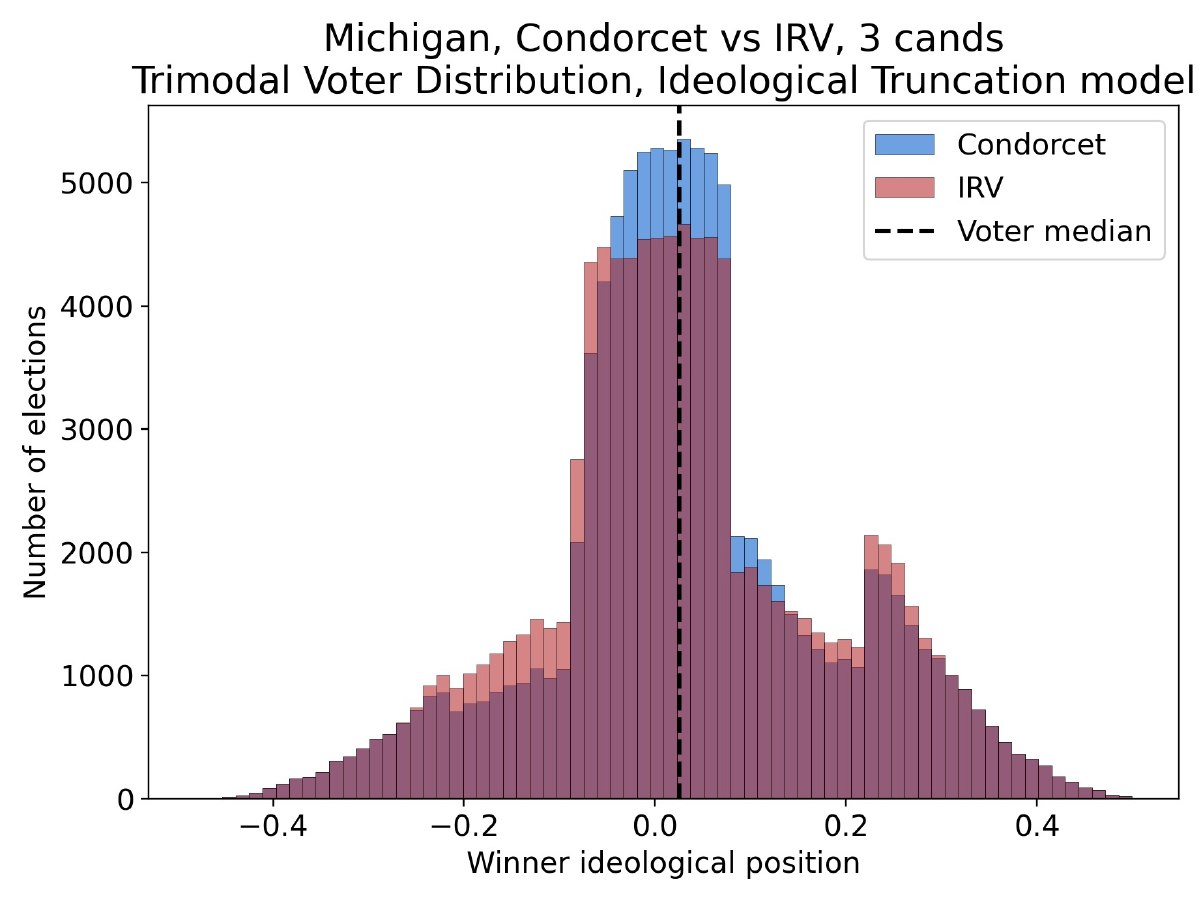} 
\\

\includegraphics[width=66mm]{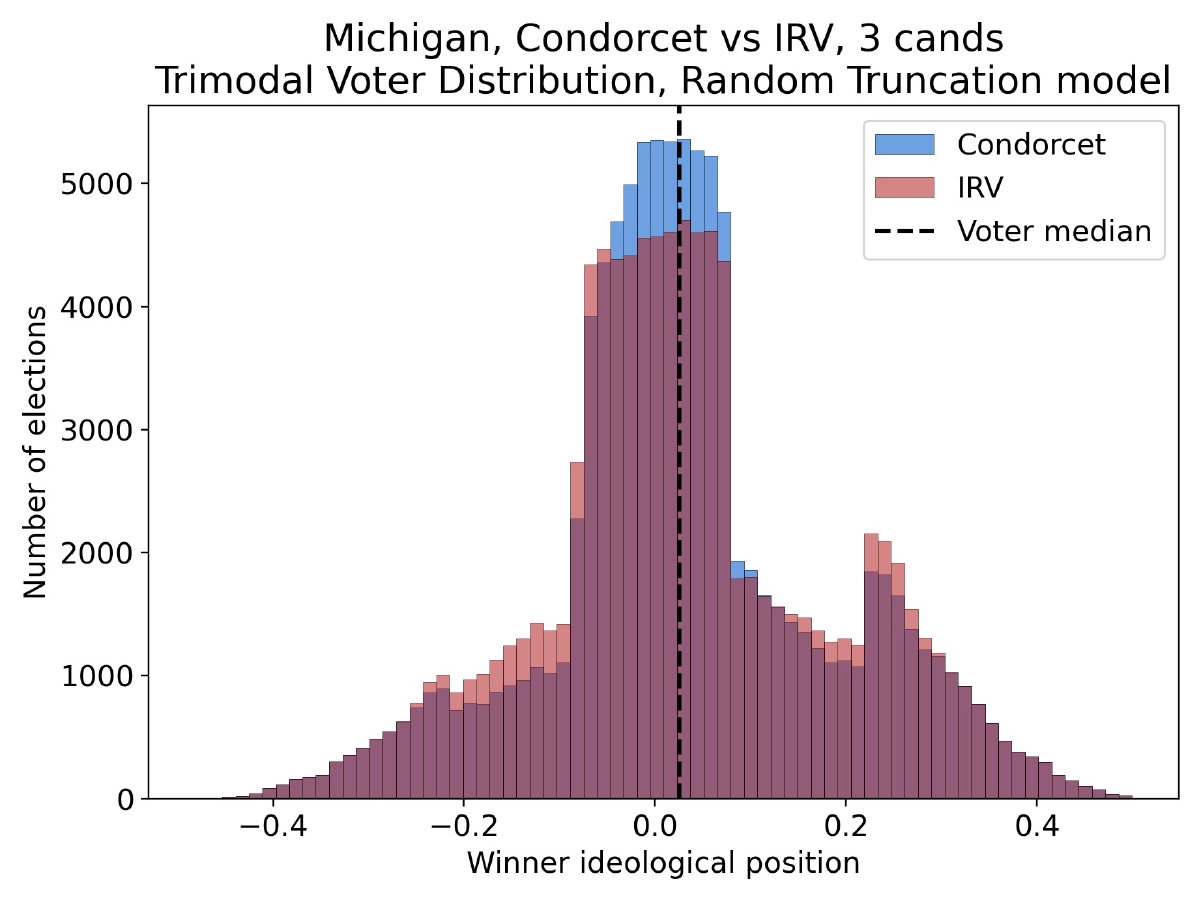}& 

\includegraphics[width=66mm]{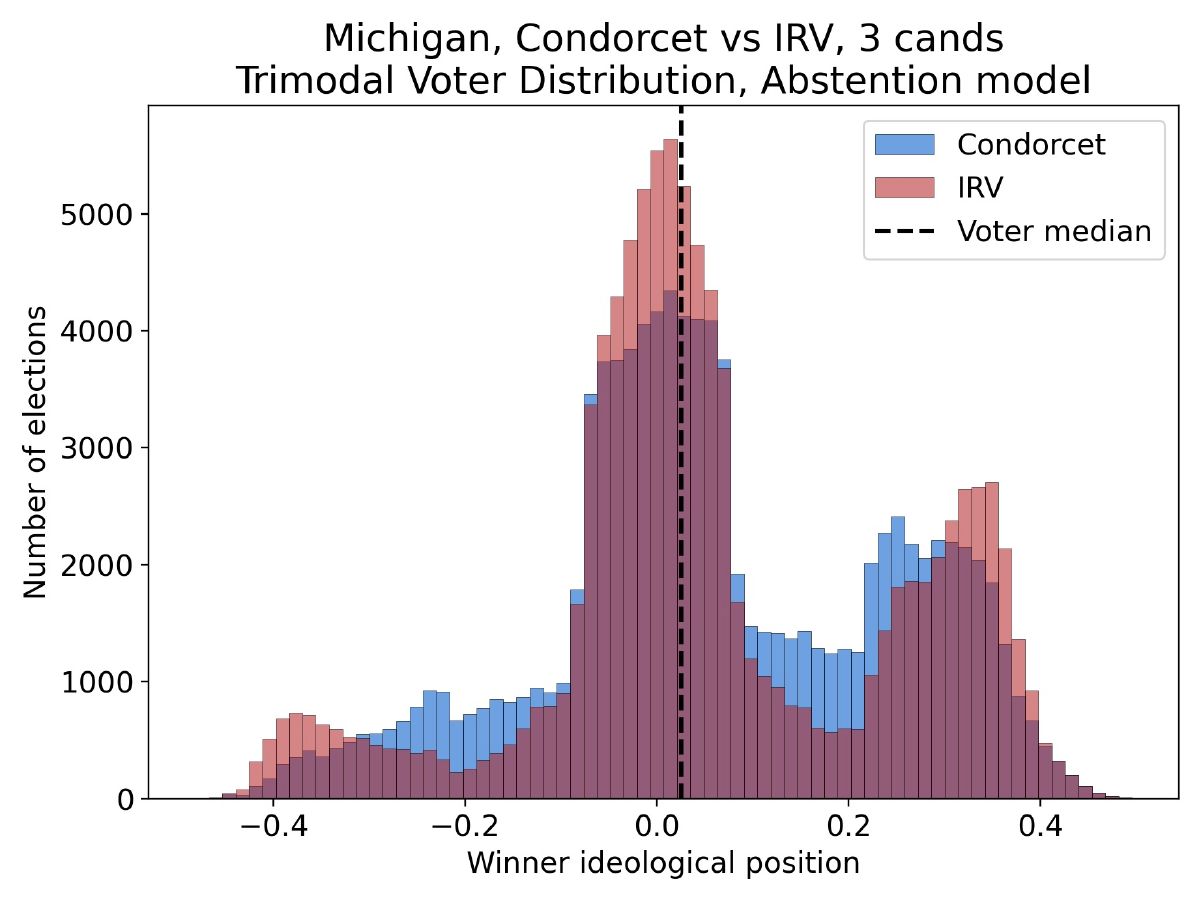}\\

\includegraphics[width=66mm]{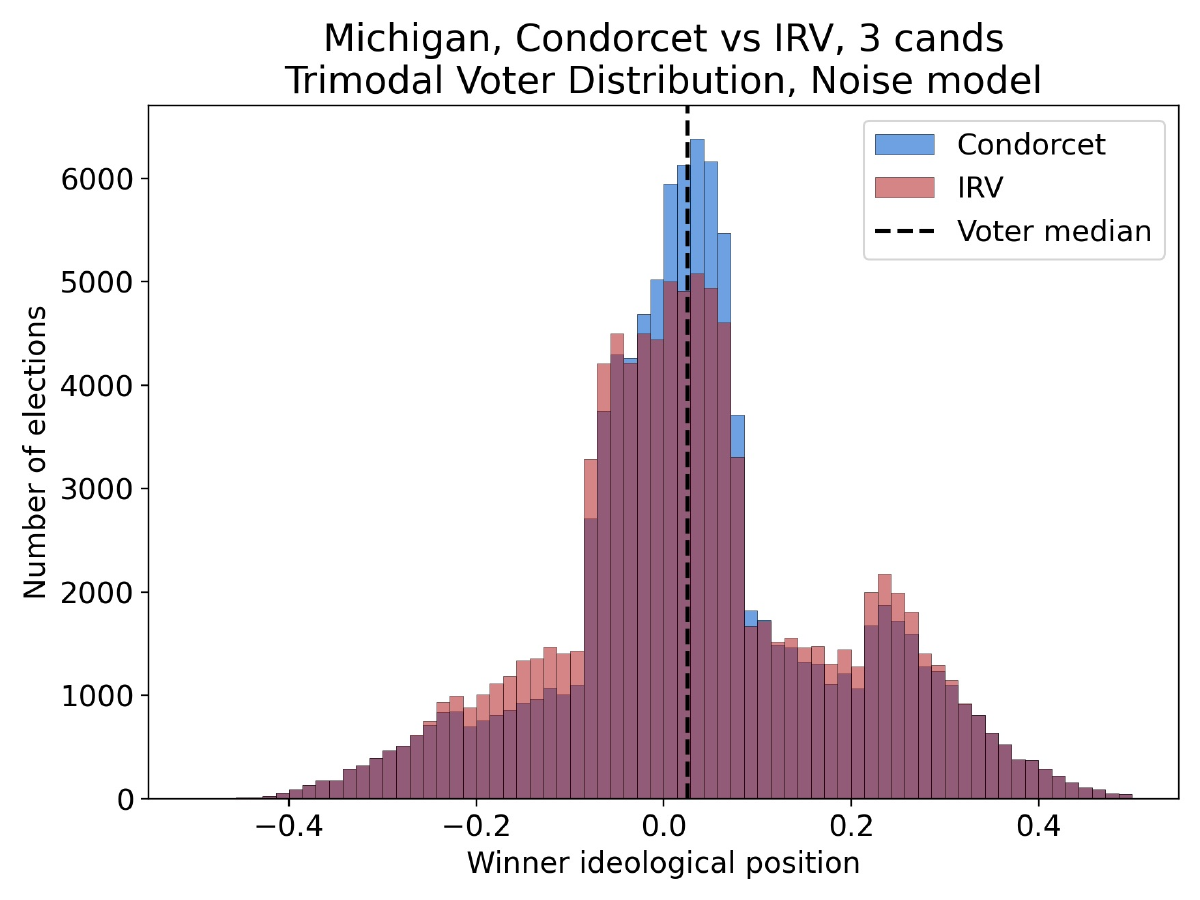}& 

\includegraphics[width=66mm]{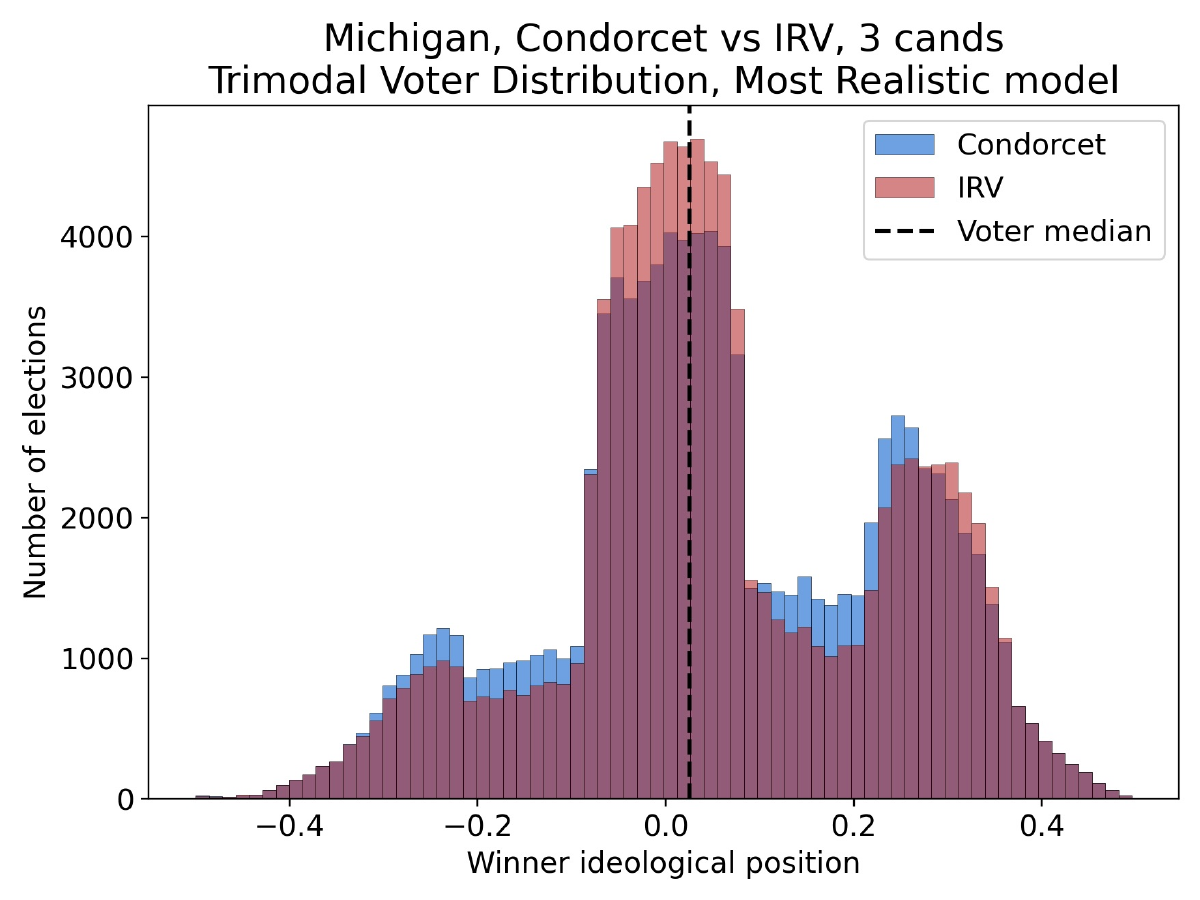}\\

\end{tabular}
\caption{Histograms showing the ideological positions of Condorcet and IRV winners across 100,000 simulations in Michigan, using 3-candidate elections with the trimodal voter distribution.}
\label{fig:MI_results_trimodal_3cands}
\end{figure}

\begin{figure}[ht]
\begin{tabular}{cc}
\includegraphics[width=66mm]{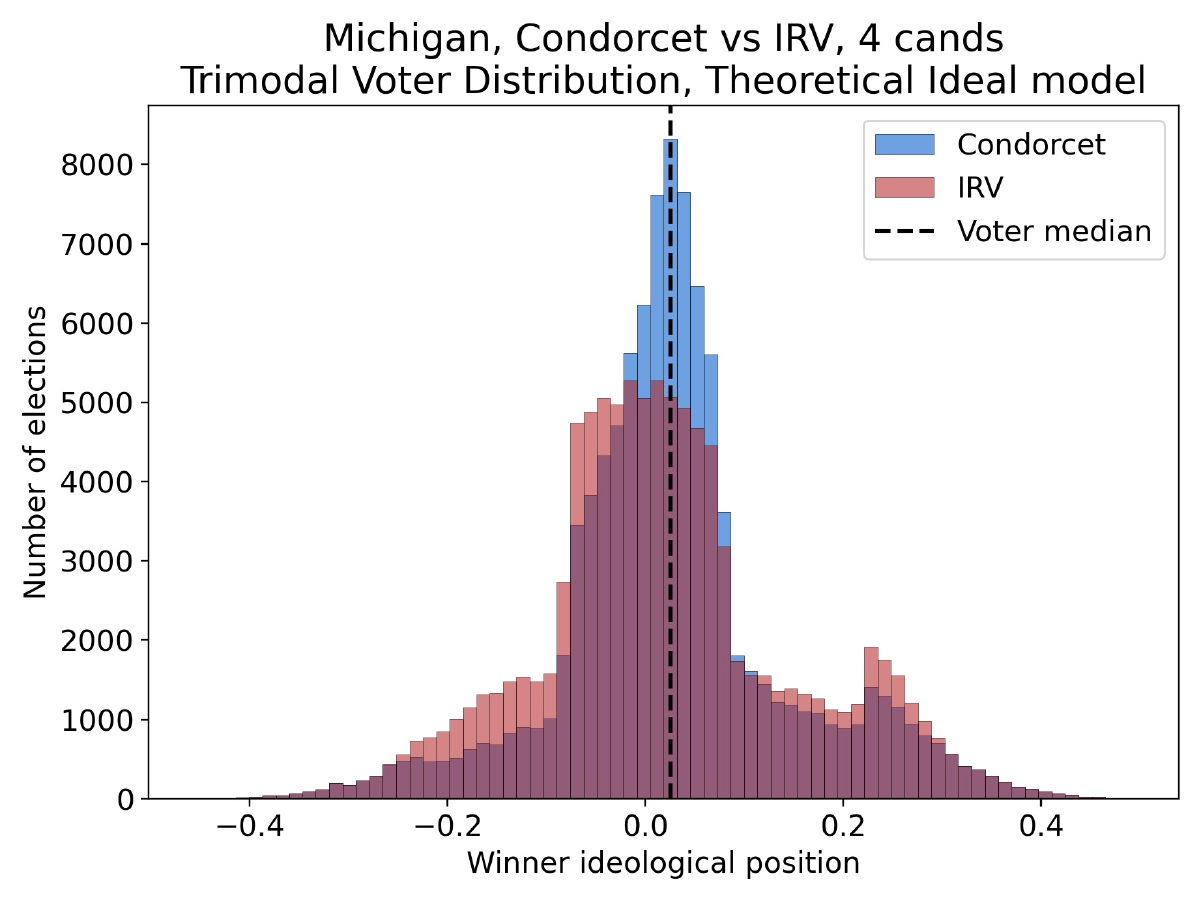} &\includegraphics[width=66mm]{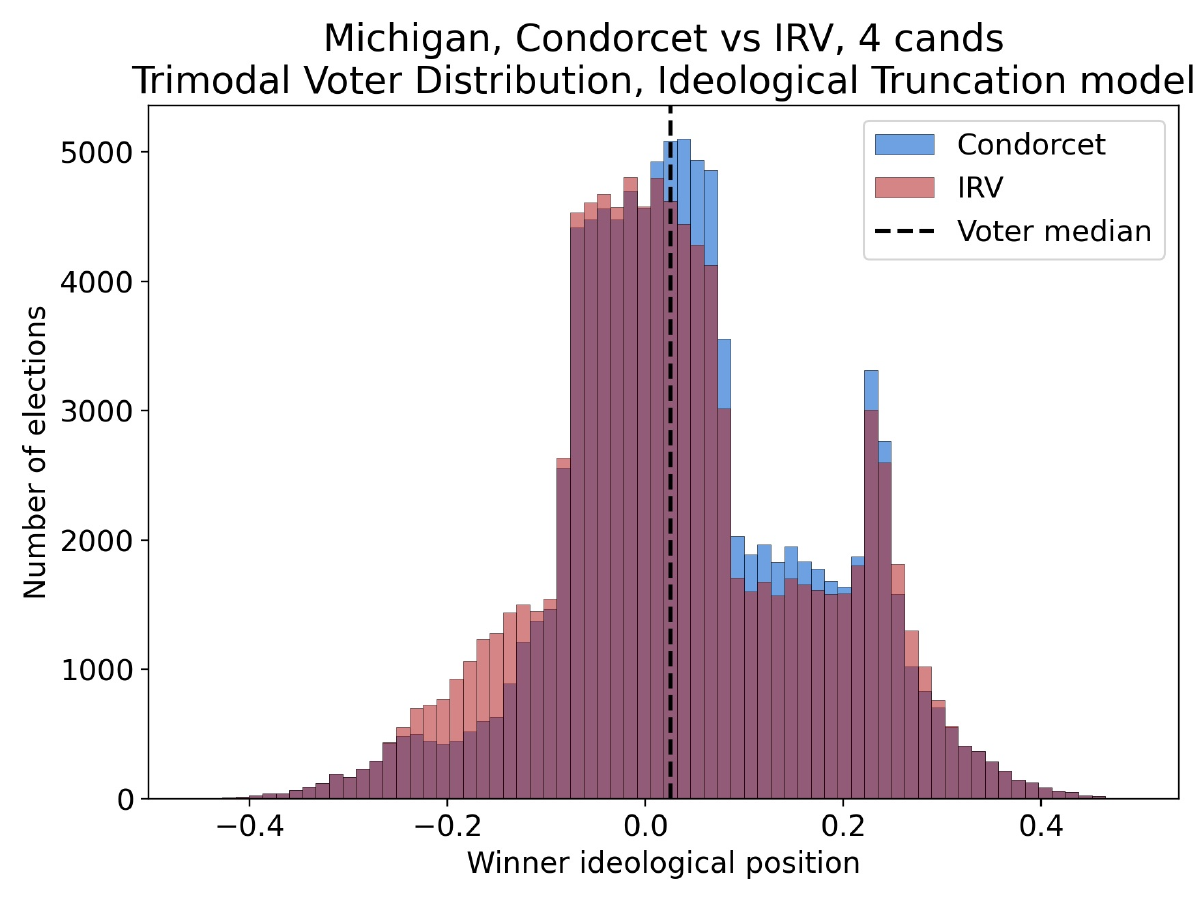}  \\

\includegraphics[width=66mm]{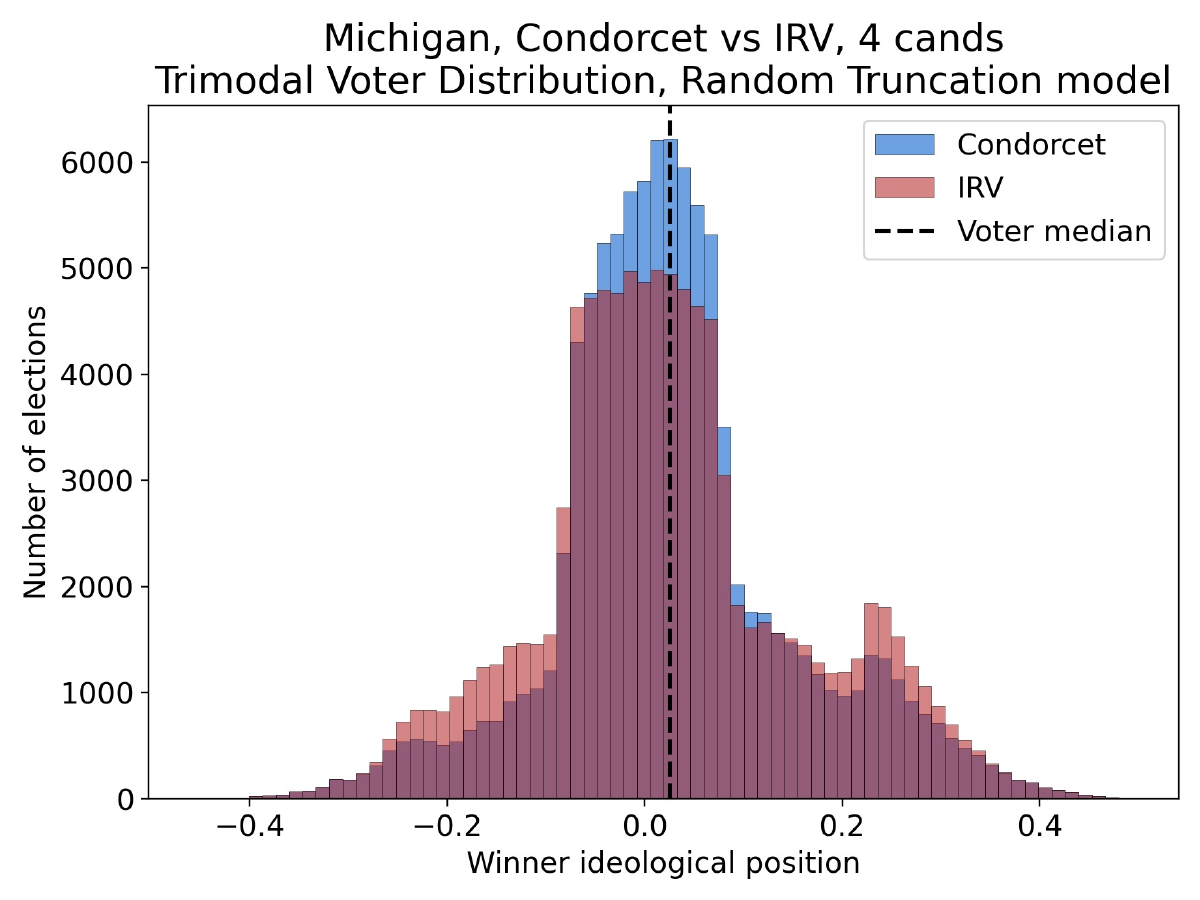}& 

\includegraphics[width=66mm]{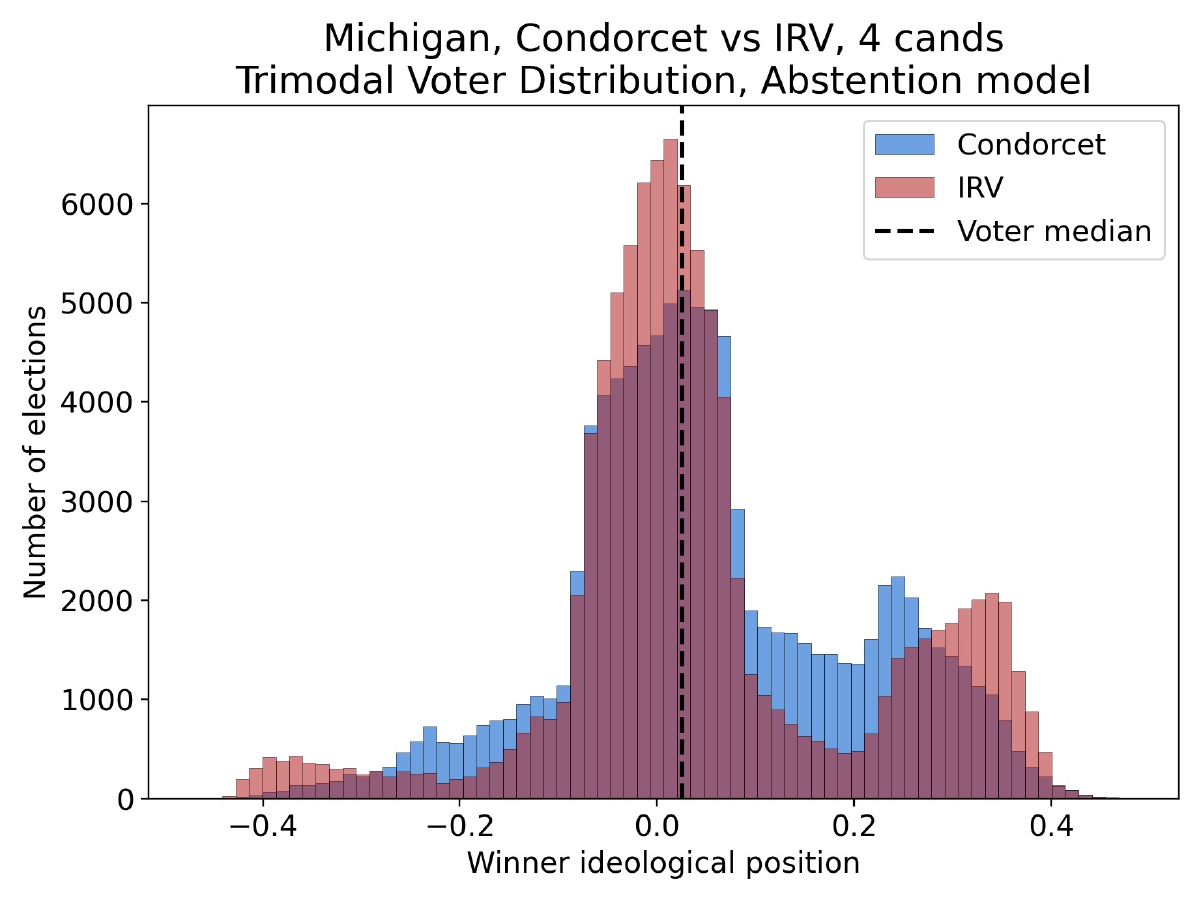}\\

\includegraphics[width=66mm]{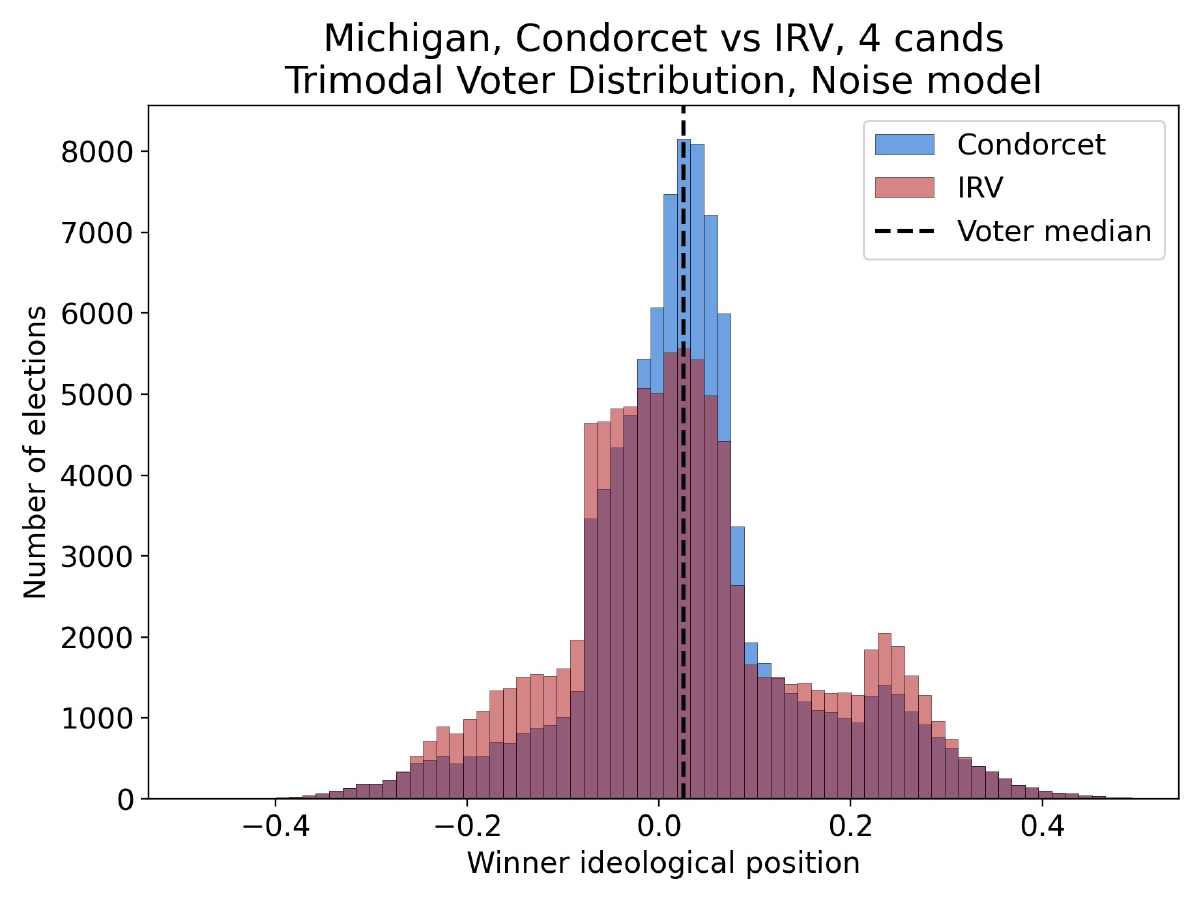}& 
\includegraphics[width=66mm]{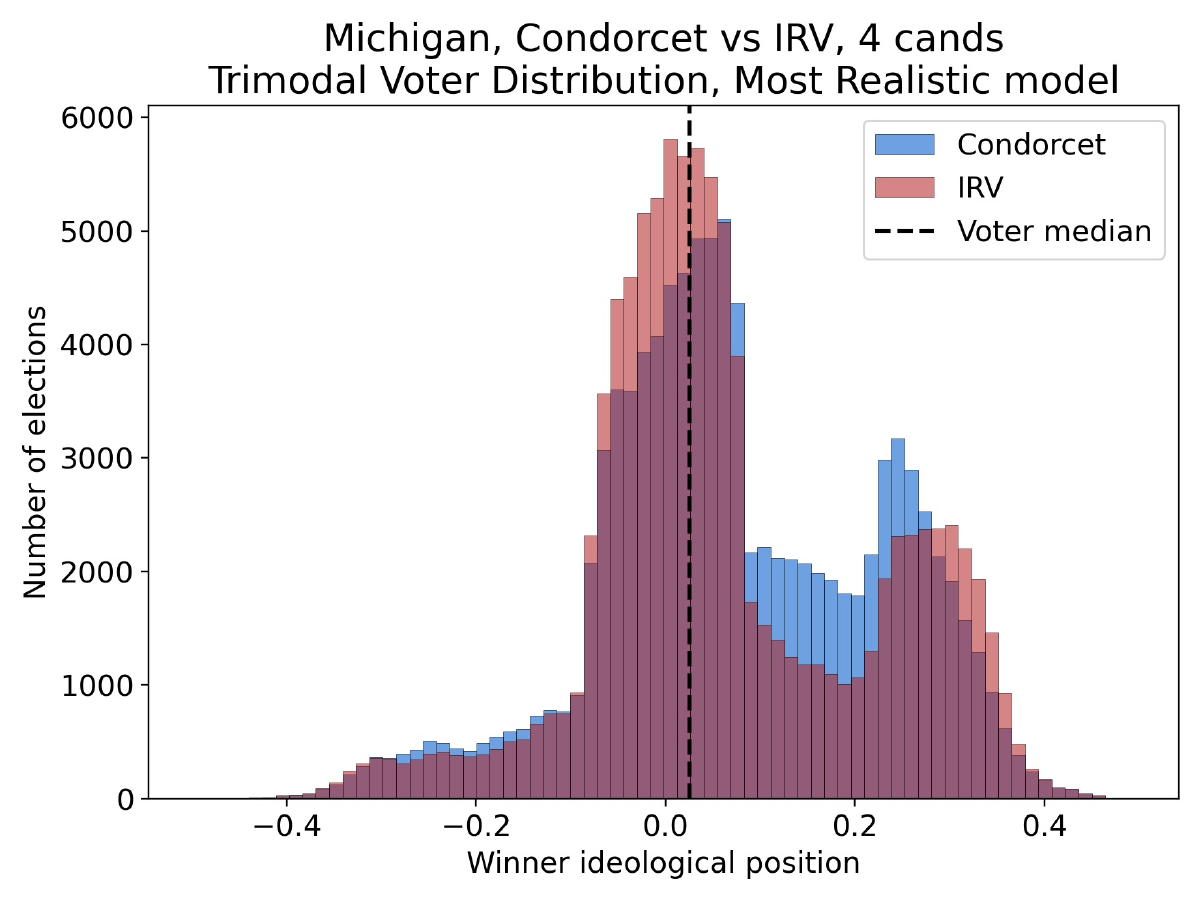}\\

\end{tabular}
\caption{Histograms showing the ideological positions of Condorcet and IRV winners across 100,000 simulations in Michigan, using 4-candidate elections with the trimodal voter distribution.}
\label{fig:MI_results_trimodal_4cands}
\end{figure}

\begin{figure}
\begin{tabular}{cc}
\includegraphics[width=66mm]{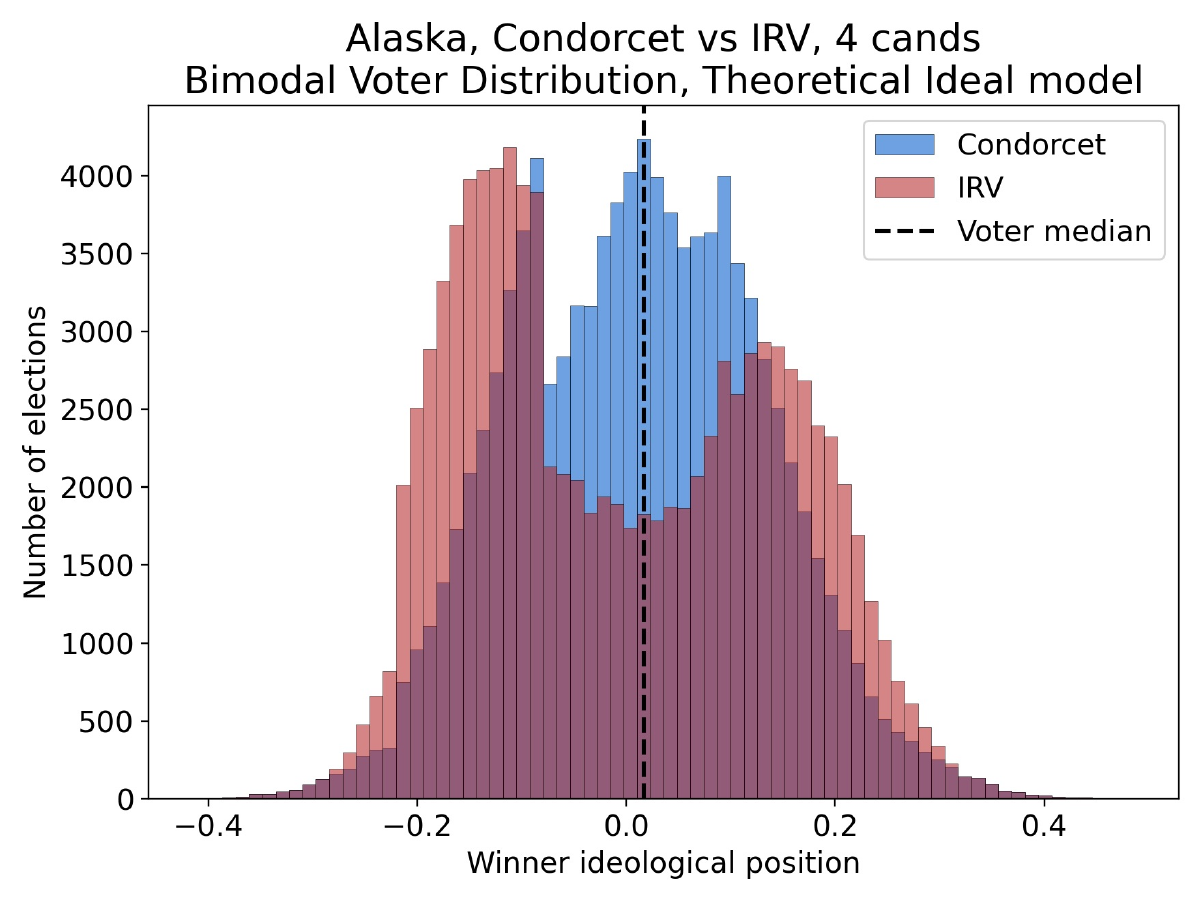} & \includegraphics[width=66mm]{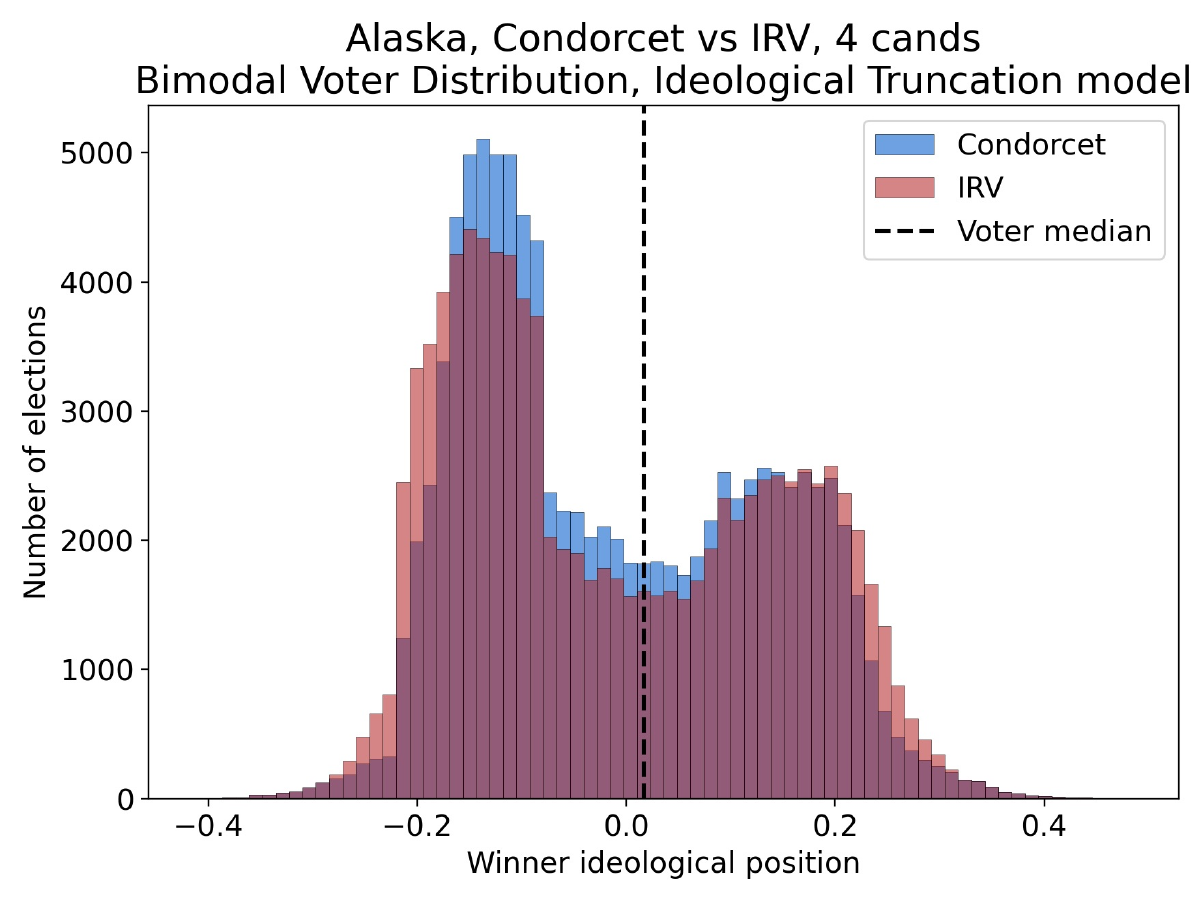} \\

\includegraphics[width=66mm]{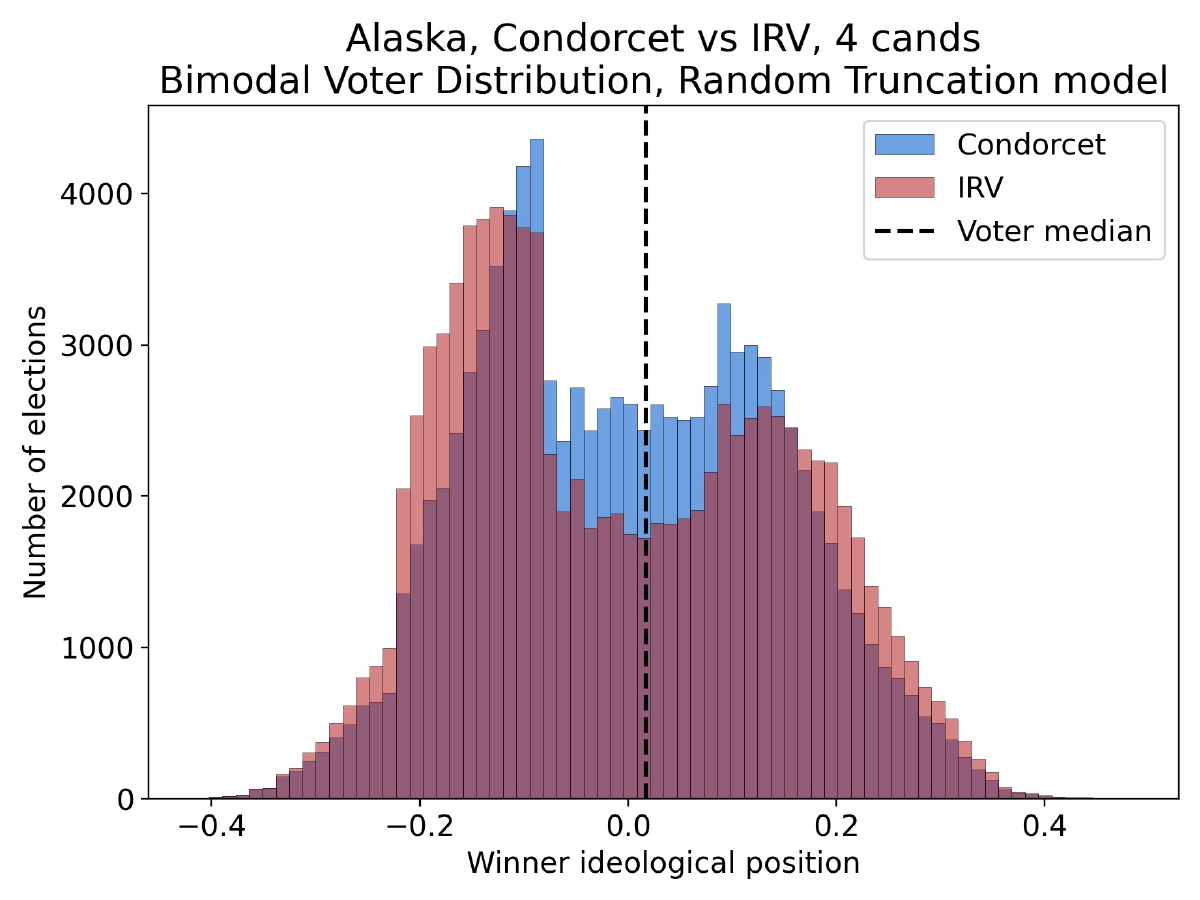}& 
\includegraphics[width=66mm]{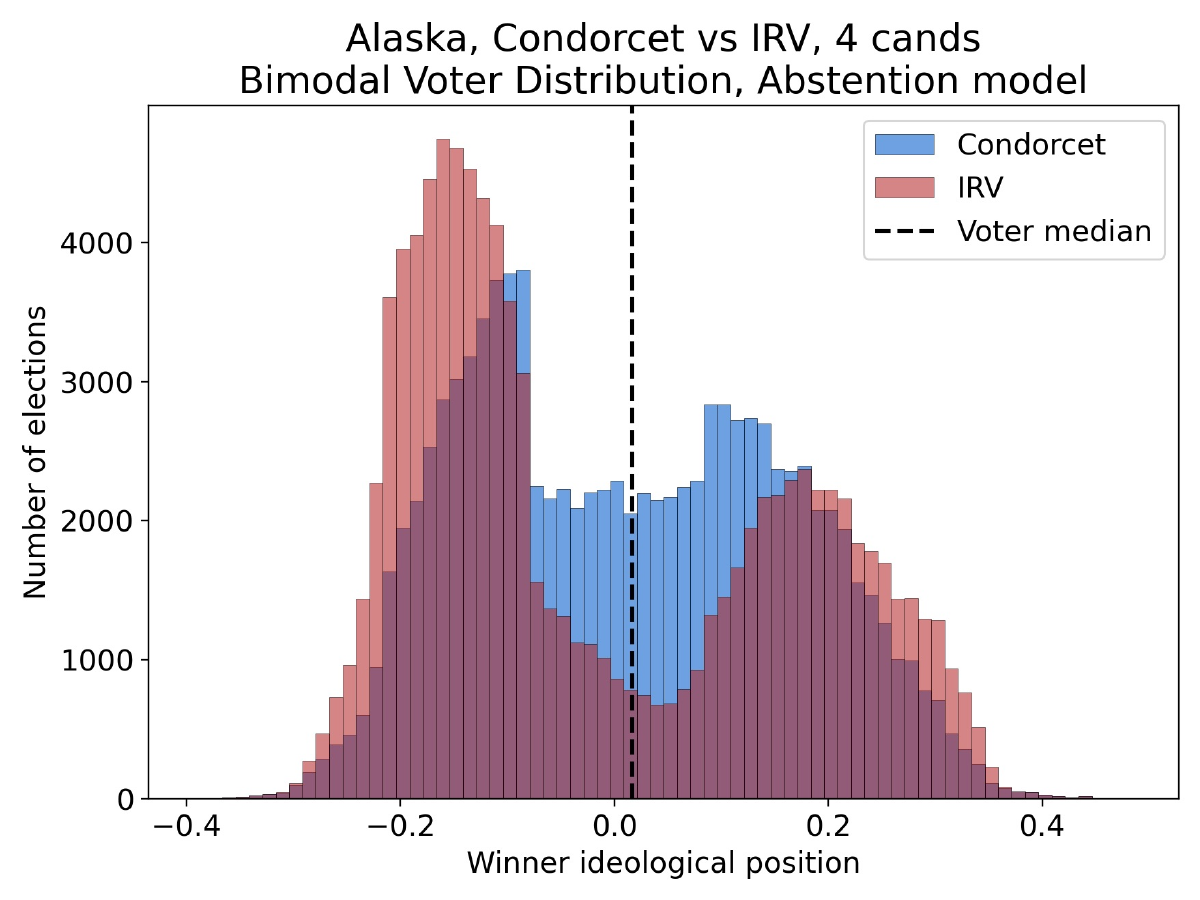}\\
\includegraphics[width=66mm]{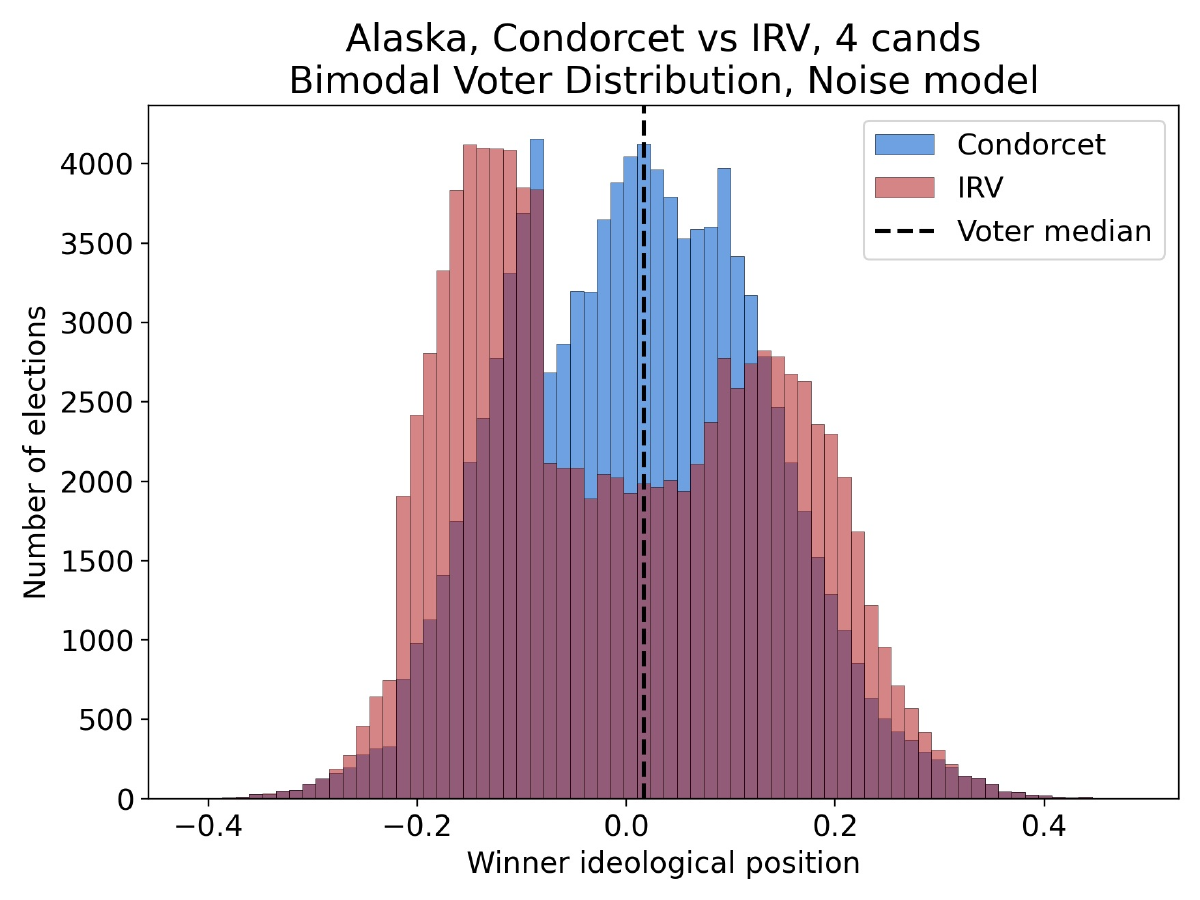}& 

\includegraphics[width=66mm]{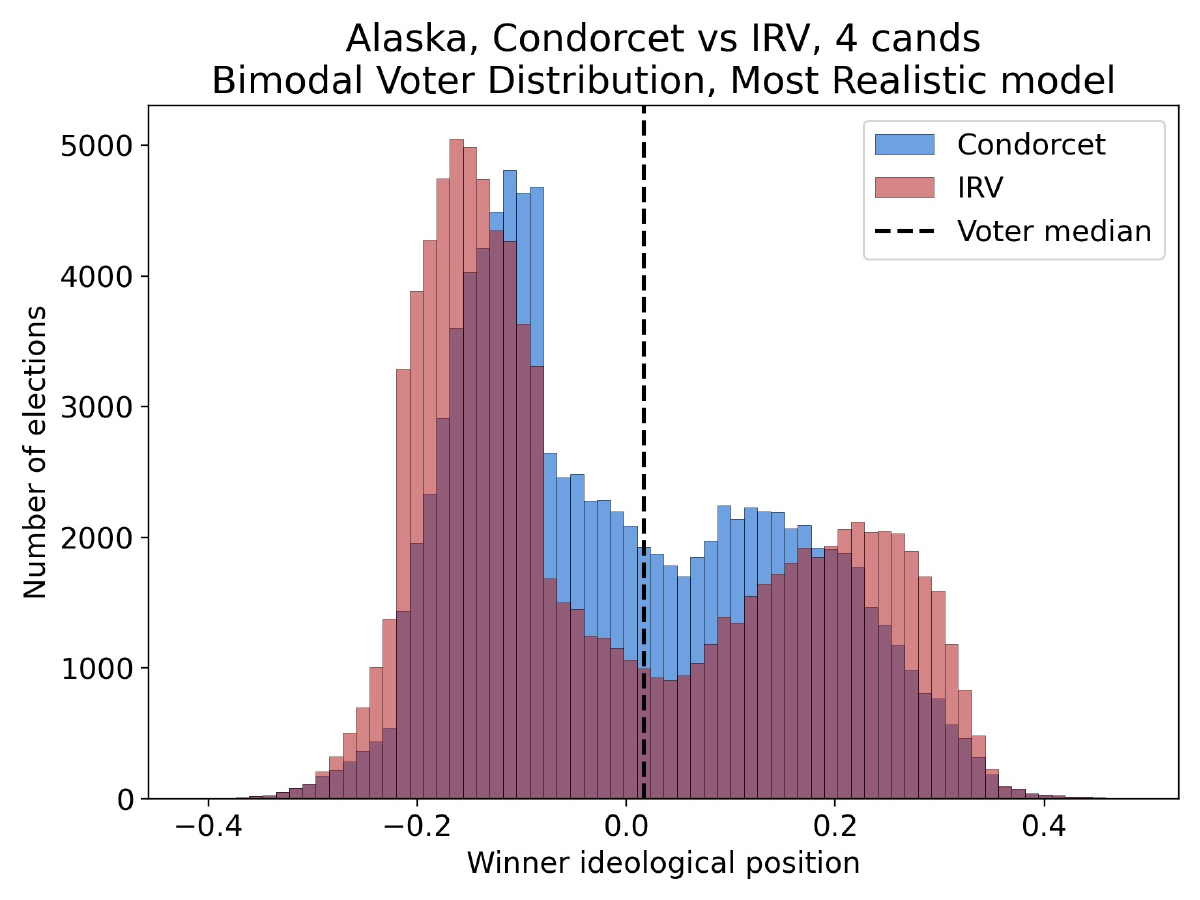}\\

\end{tabular}
\caption{Histograms showing the ideological positions of Condorcet and IRV winners across 100,000 simulations in Alaska, using 4-candidate elections with the bimodal voter distribution.}
\label{fig:AK_results_bimodal}
\end{figure}

\begin{figure}
\begin{tabular}{cc}
\includegraphics[width=66mm]{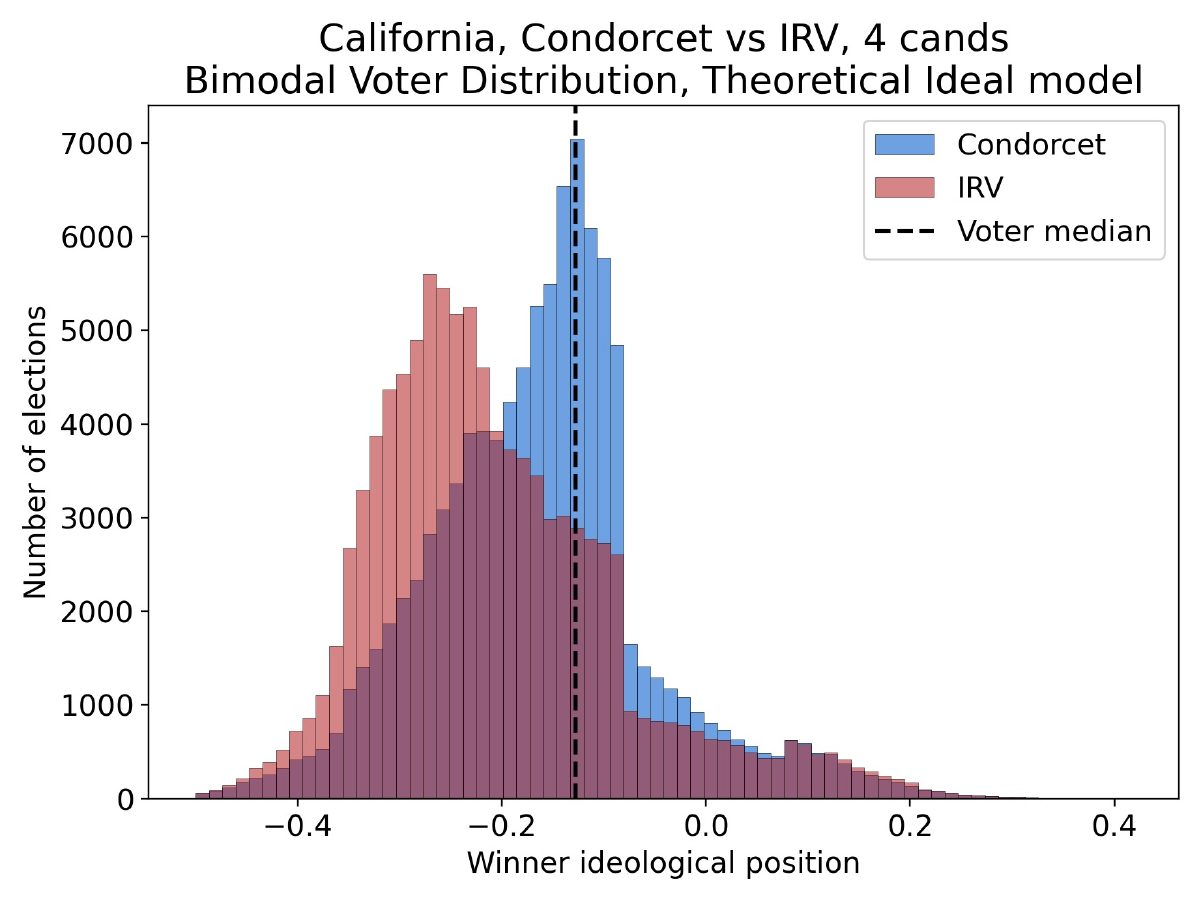} & \includegraphics[width=66mm]{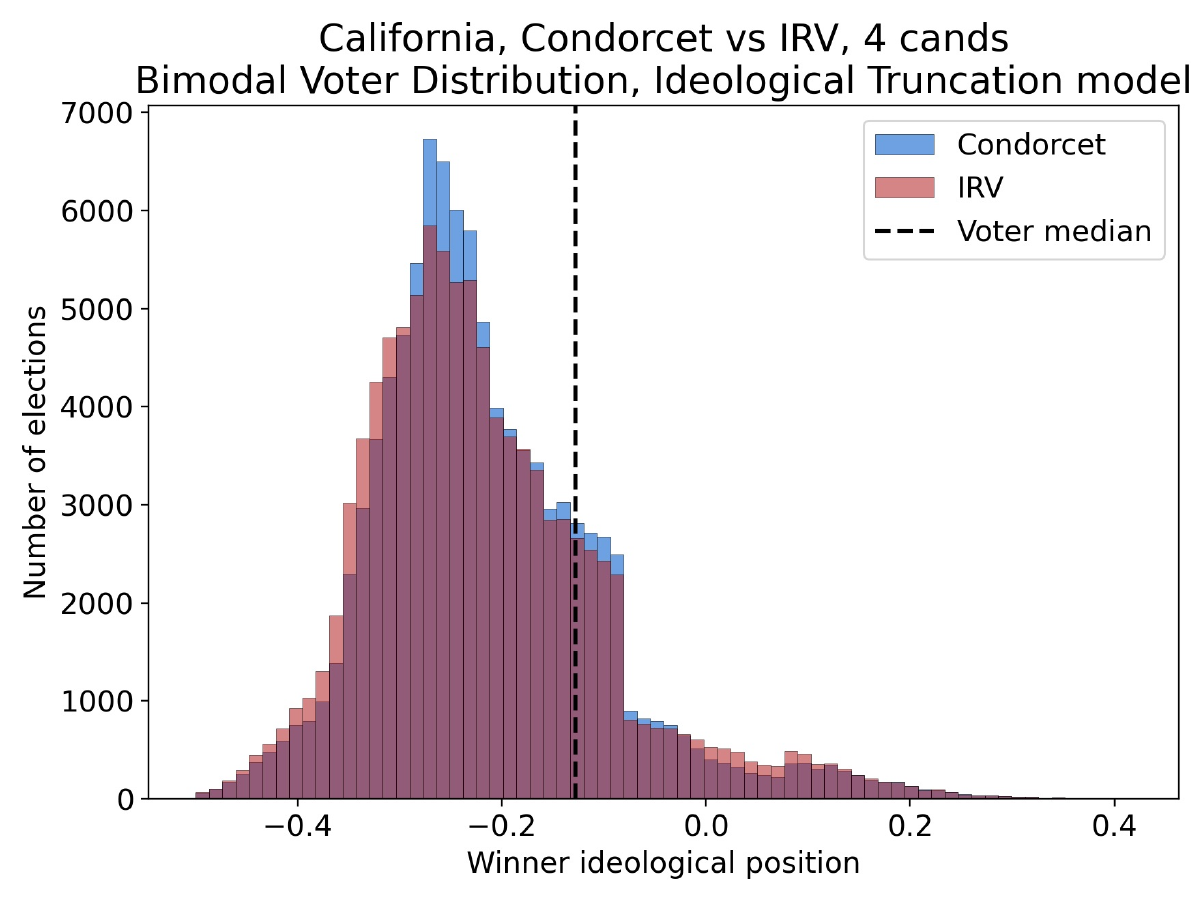}\\

\includegraphics[width=66mm]{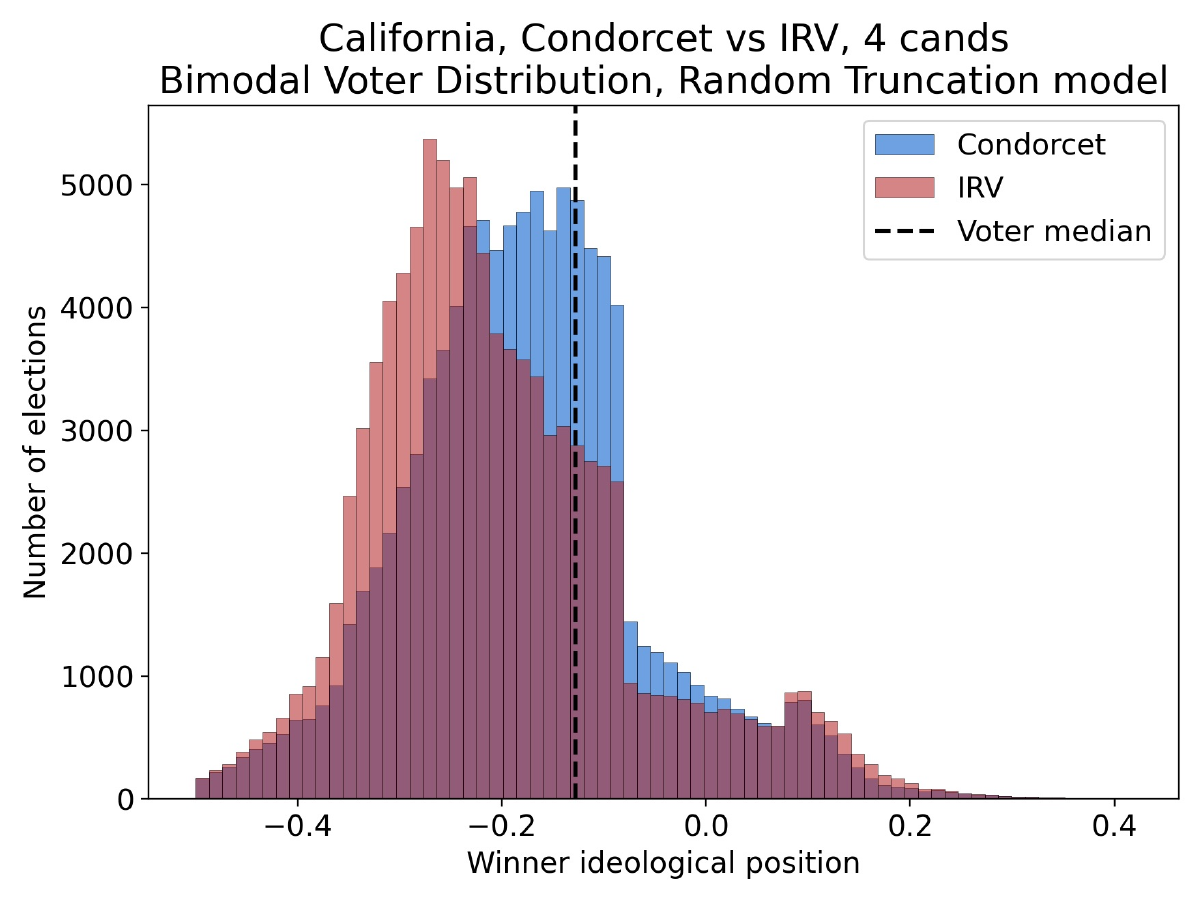}& 

\includegraphics[width=66mm]{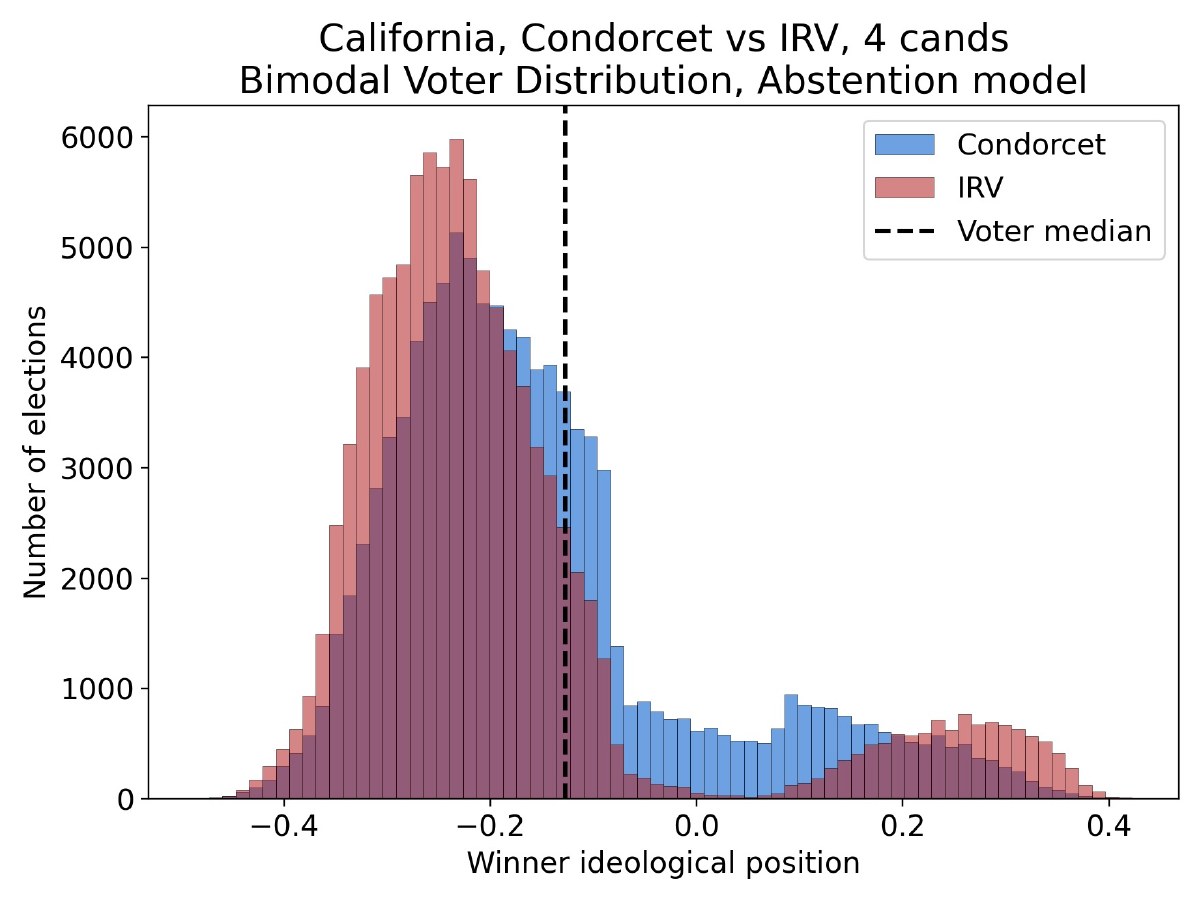}\\
 \includegraphics[width=66mm]{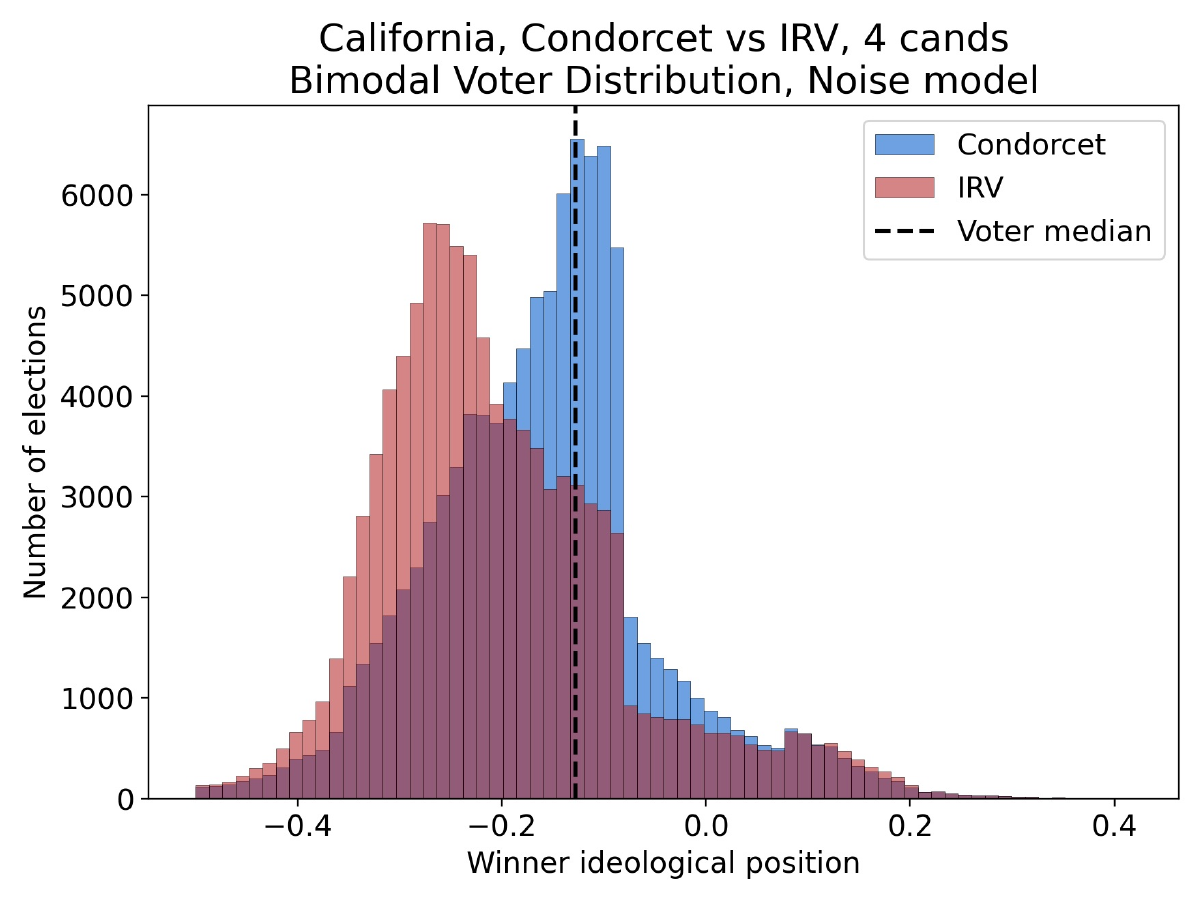}& 

\includegraphics[width=66mm]{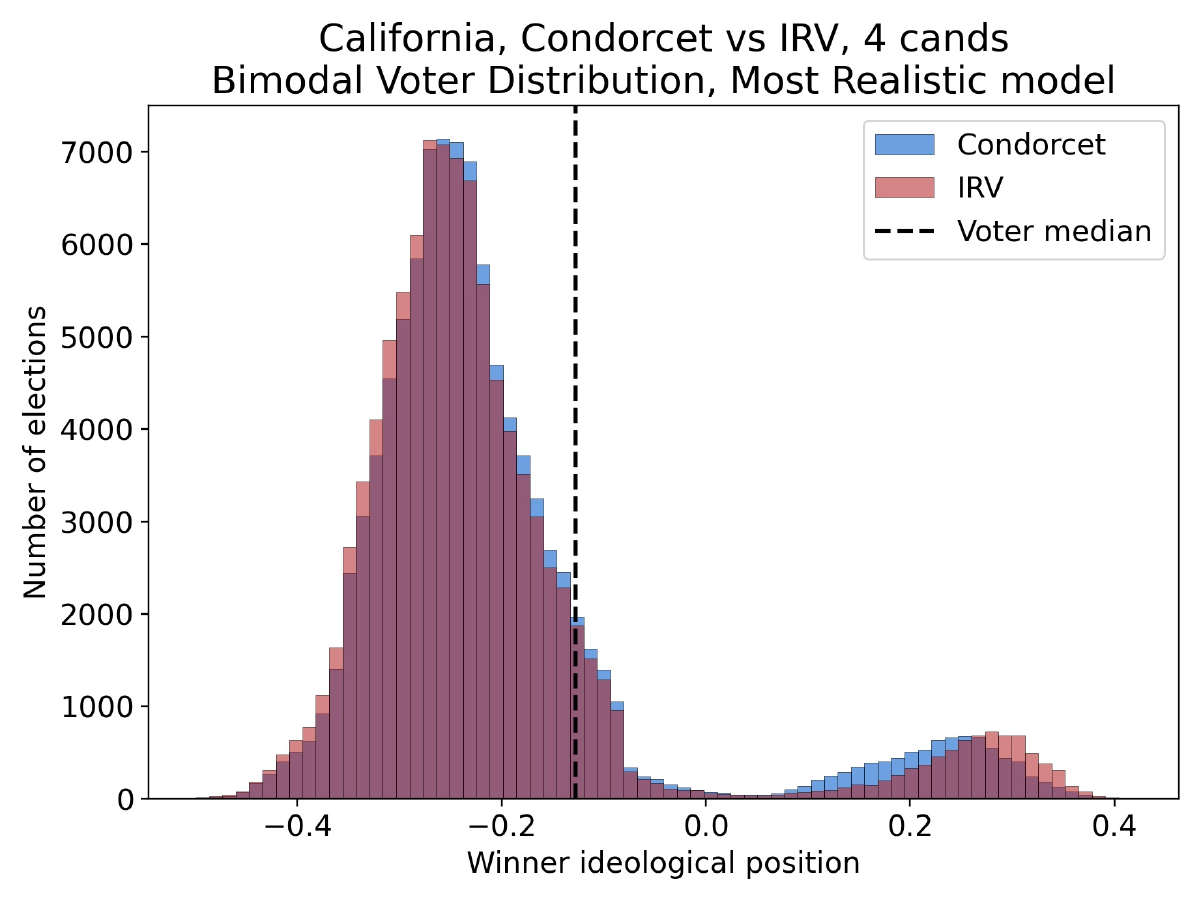}\\

\end{tabular}
\caption{Histograms showing the ideological positions of Condorcet and IRV winners across 100,000 simulations in California, using 4-candidate elections with the bimodal voter distribution.}
\label{fig:CA_results_bimodal}
\end{figure}

\begin{figure}
\begin{tabular}{cc}
\includegraphics[width=66mm]{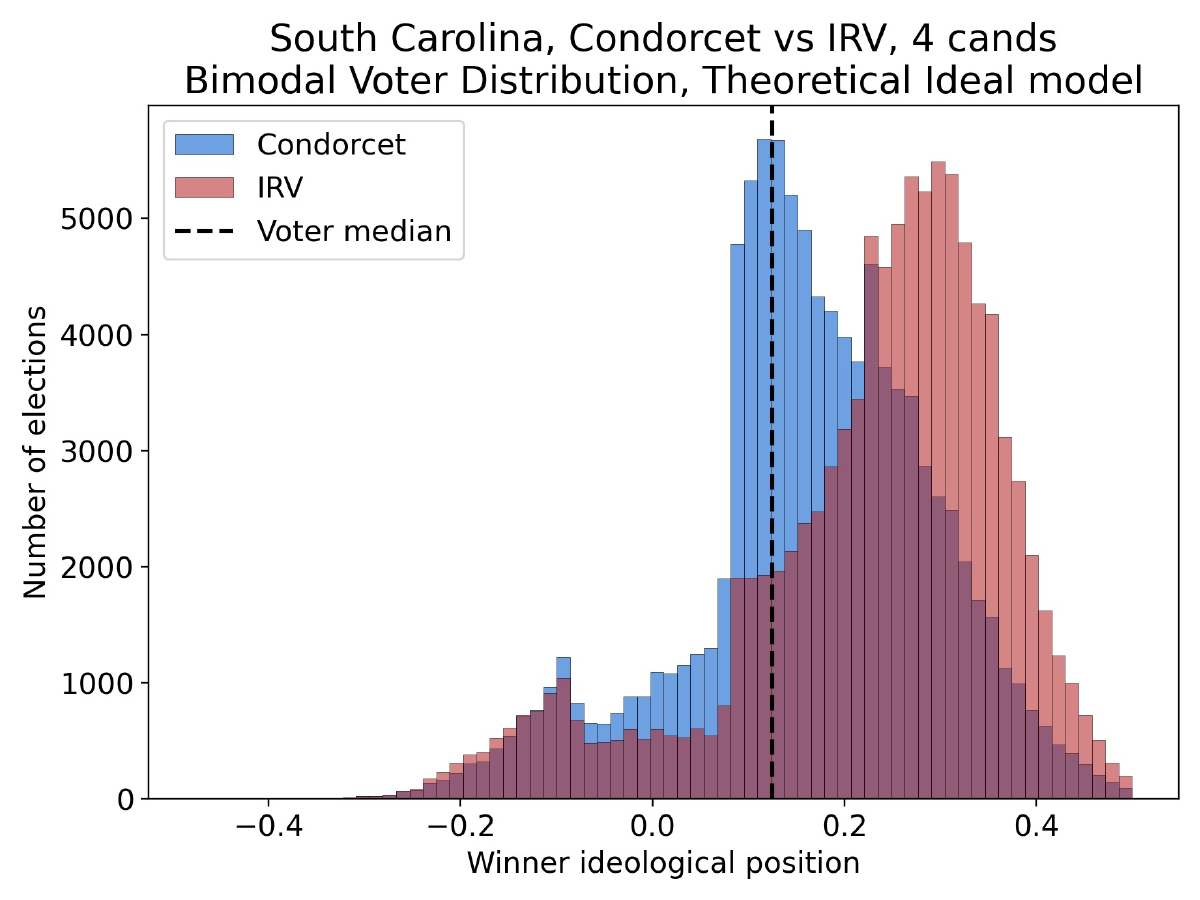} & \includegraphics[width=66mm]{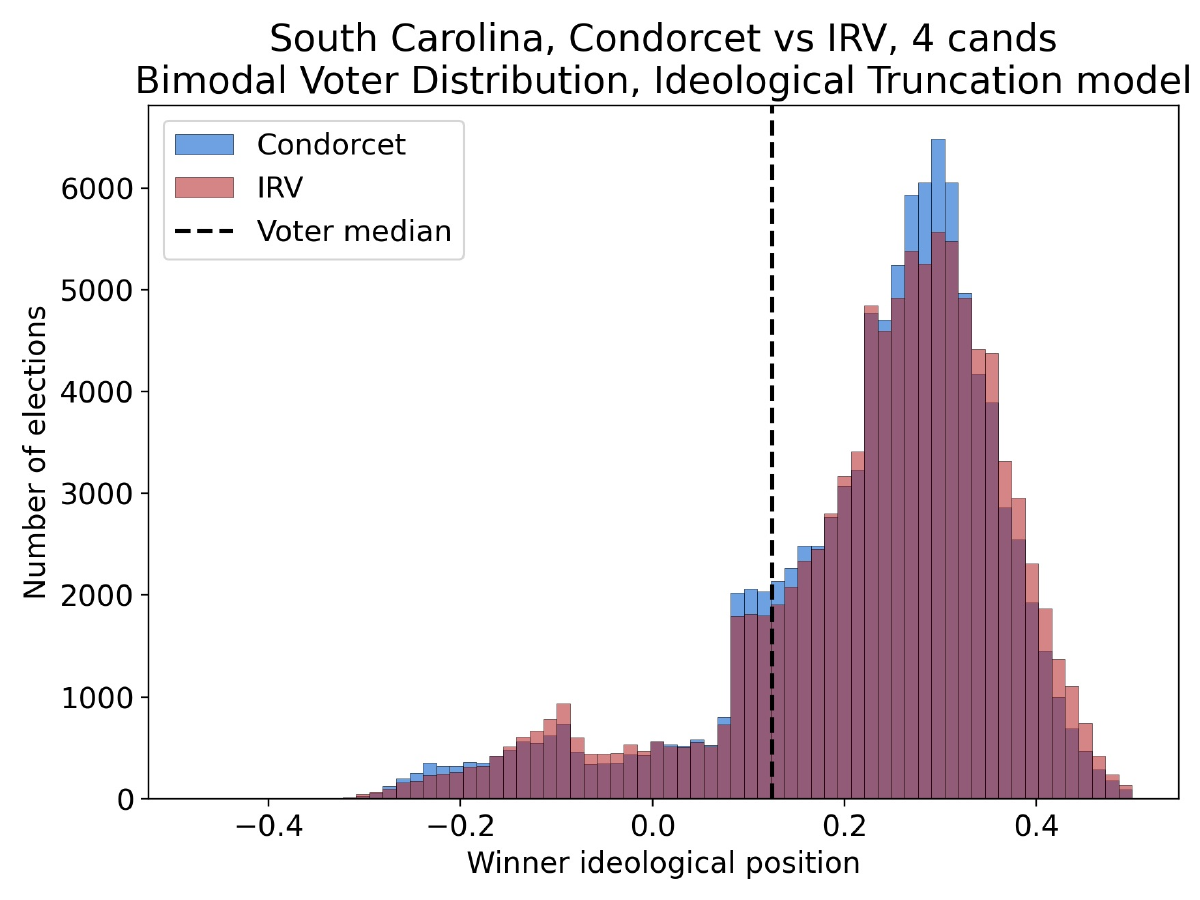} \\
\includegraphics[width=66mm]{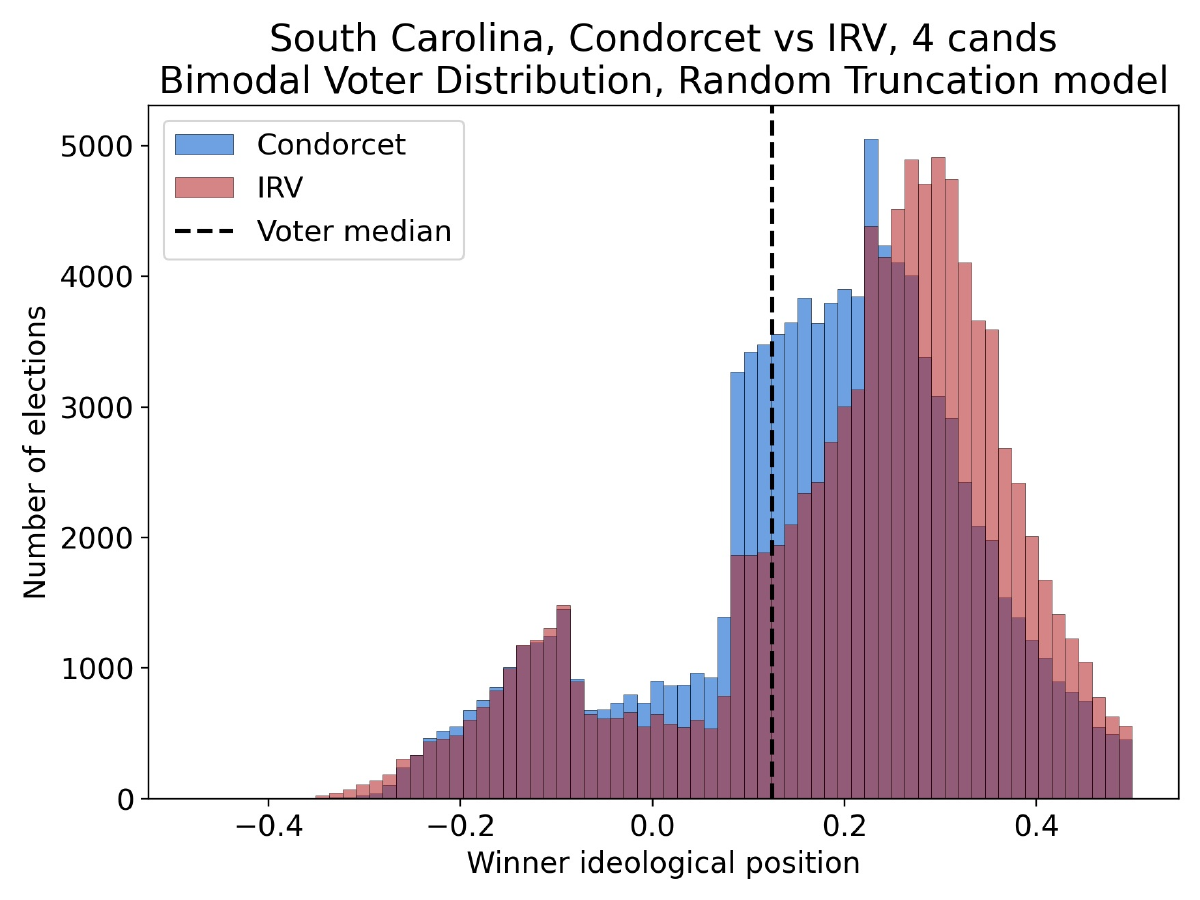}& 

\includegraphics[width=66mm]{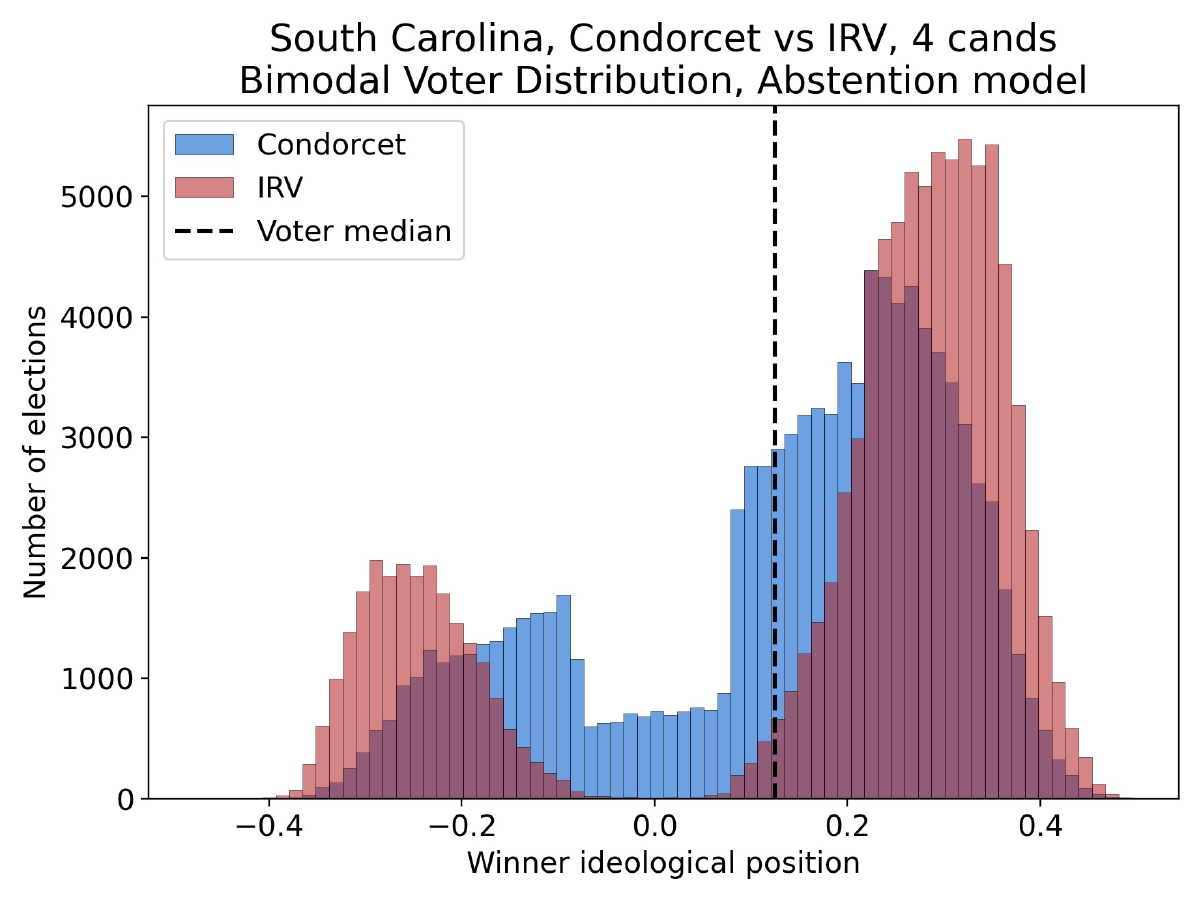}\\

\includegraphics[width=66mm]{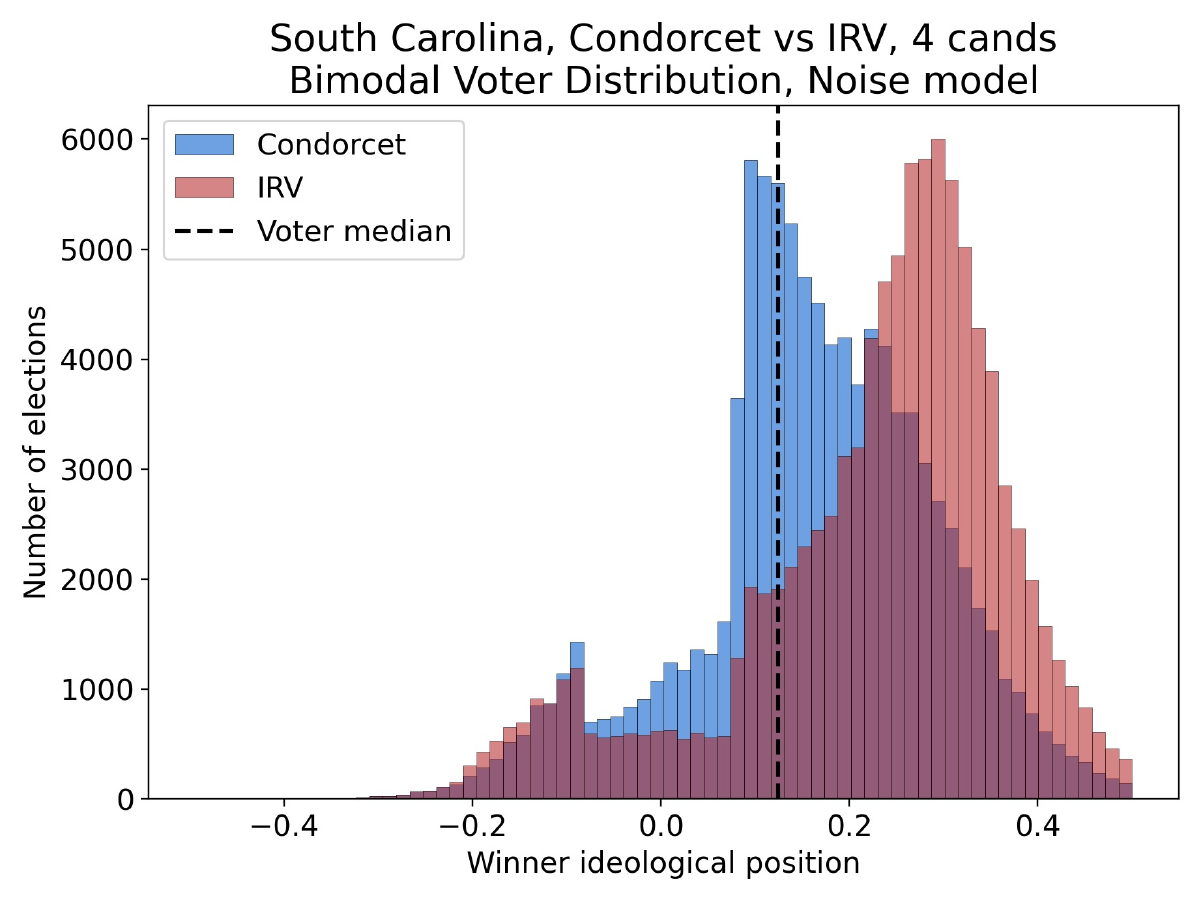}& 

\includegraphics[width=66mm]{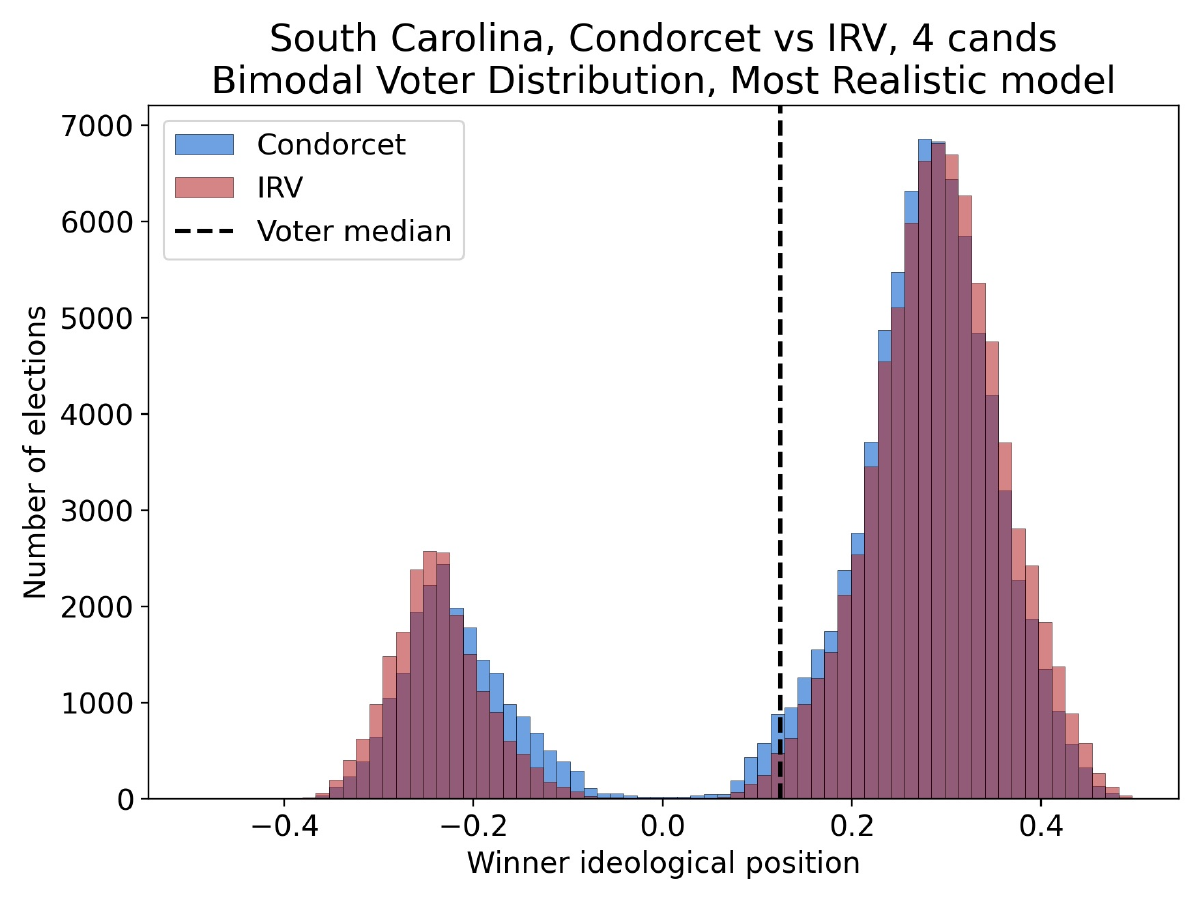}\\

\end{tabular}
\caption{Histograms showing the ideological positions of Condorcet and IRV winners across 100,000 simulations in South Carolina, using 4-candidate elections with the bimodal voter distribution.}
\label{fig:SC_results_bimodal}
\end{figure}

\begin{figure}
\begin{tabular}{cc}
\includegraphics[width=66mm]{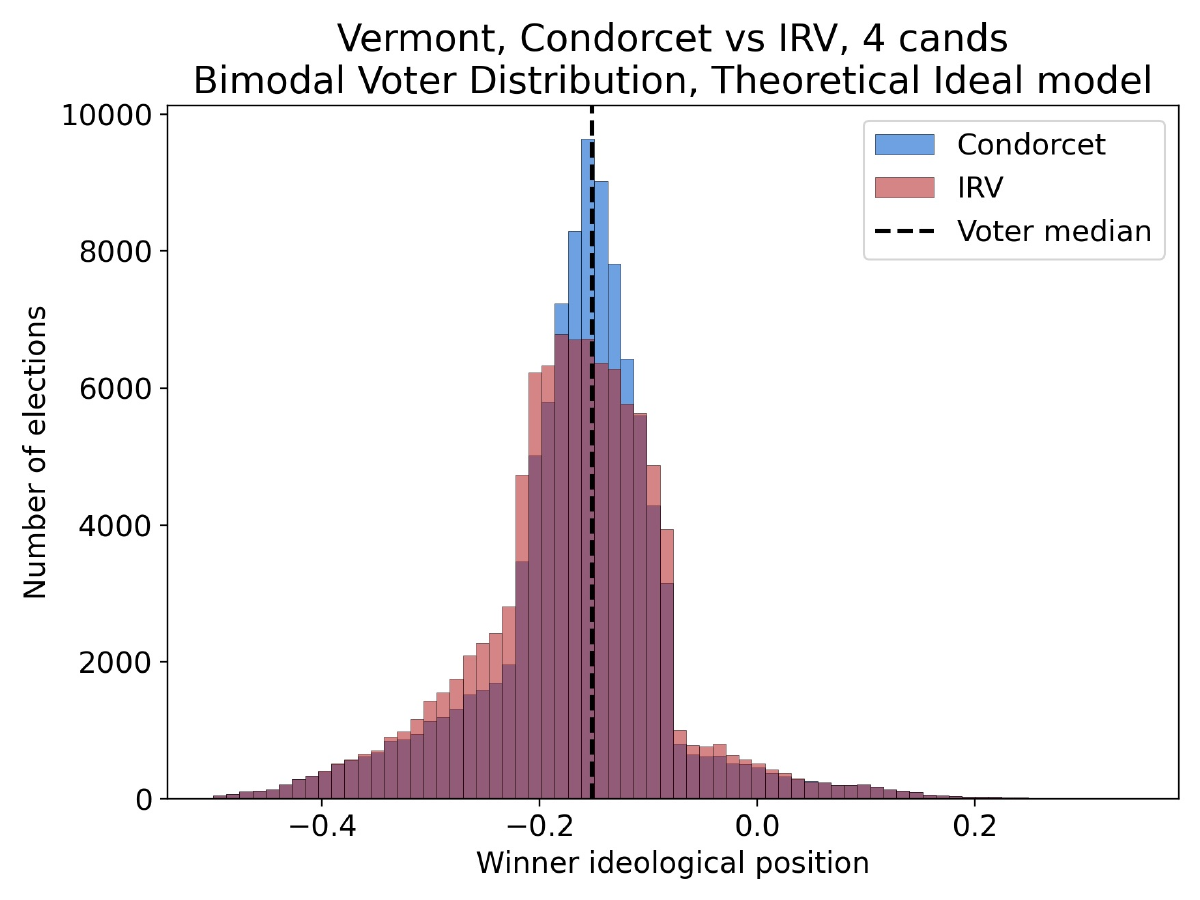} 
& \includegraphics[width=66mm]{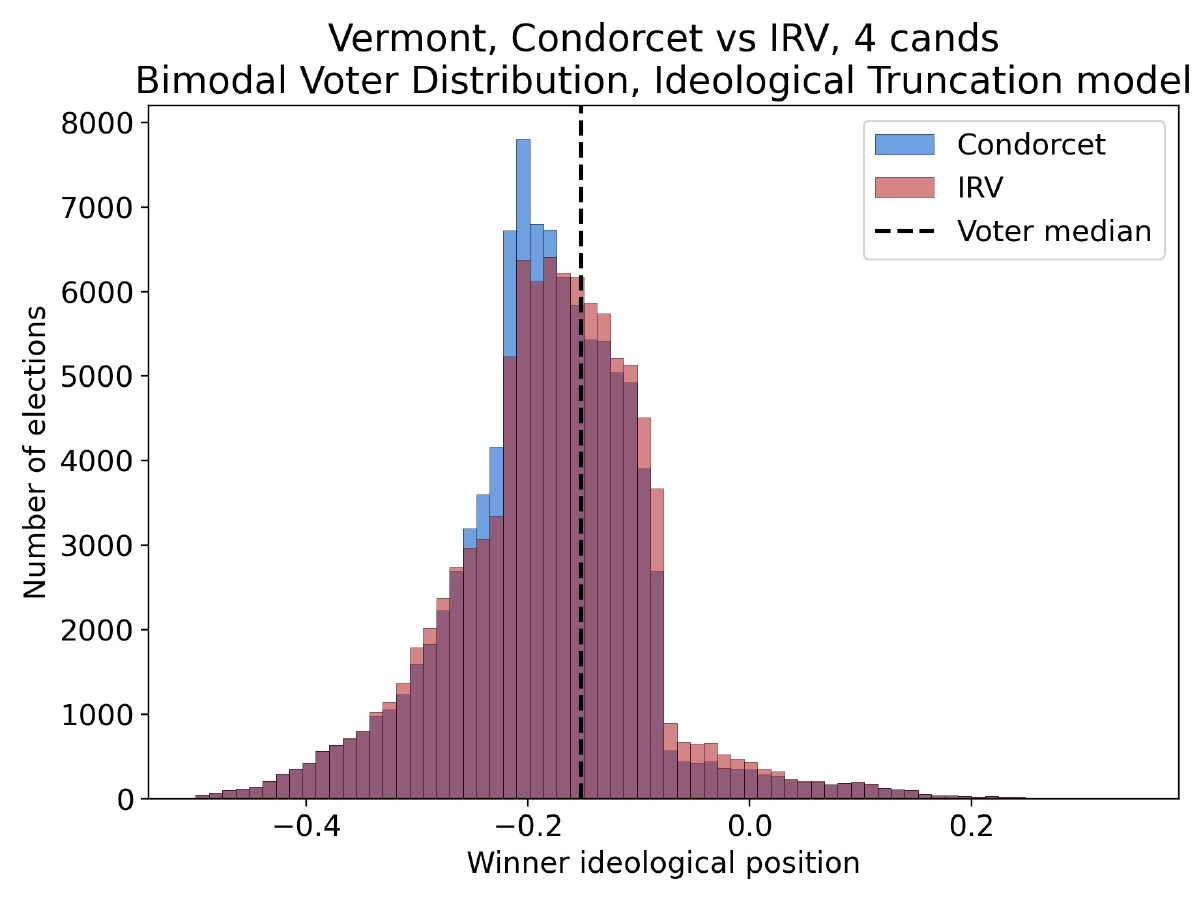}\\

\includegraphics[width=66mm]{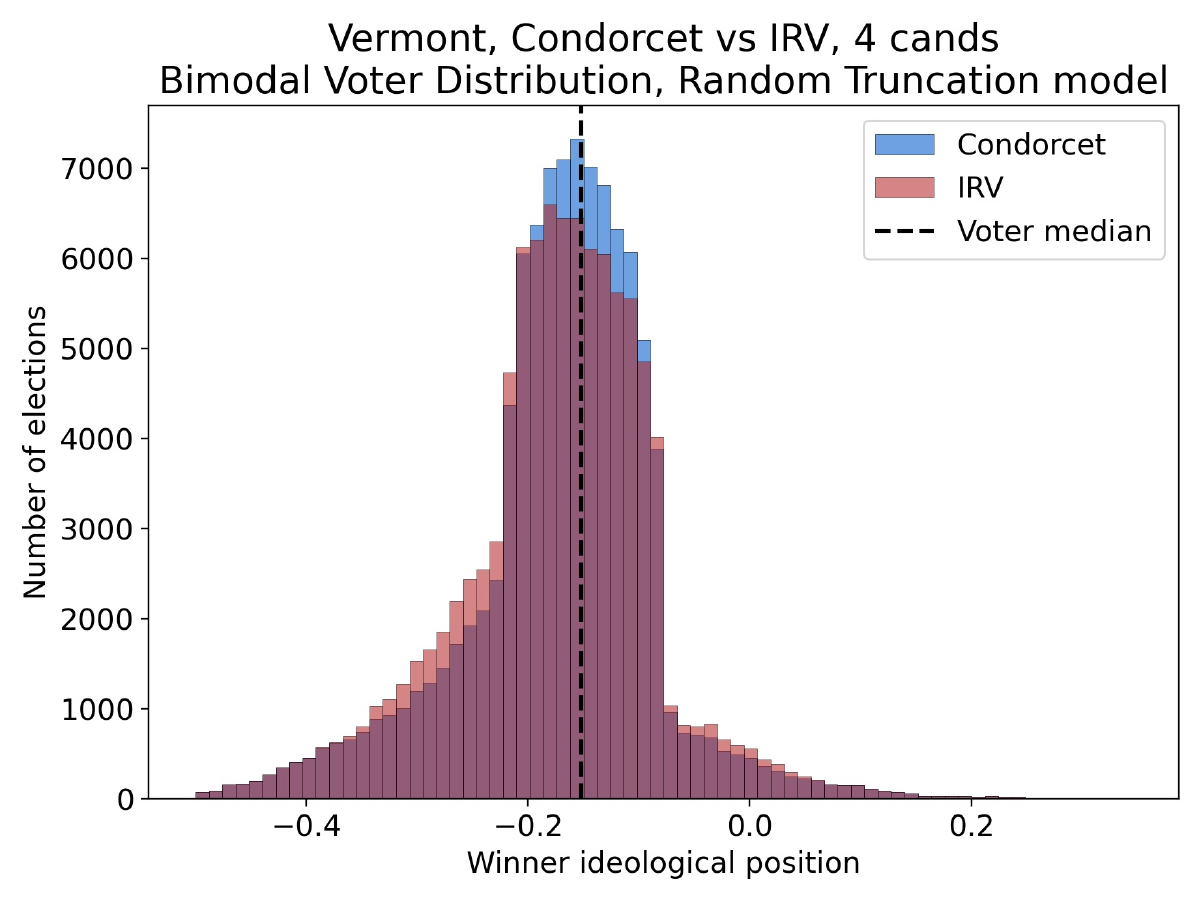}& 

\includegraphics[width=66mm]{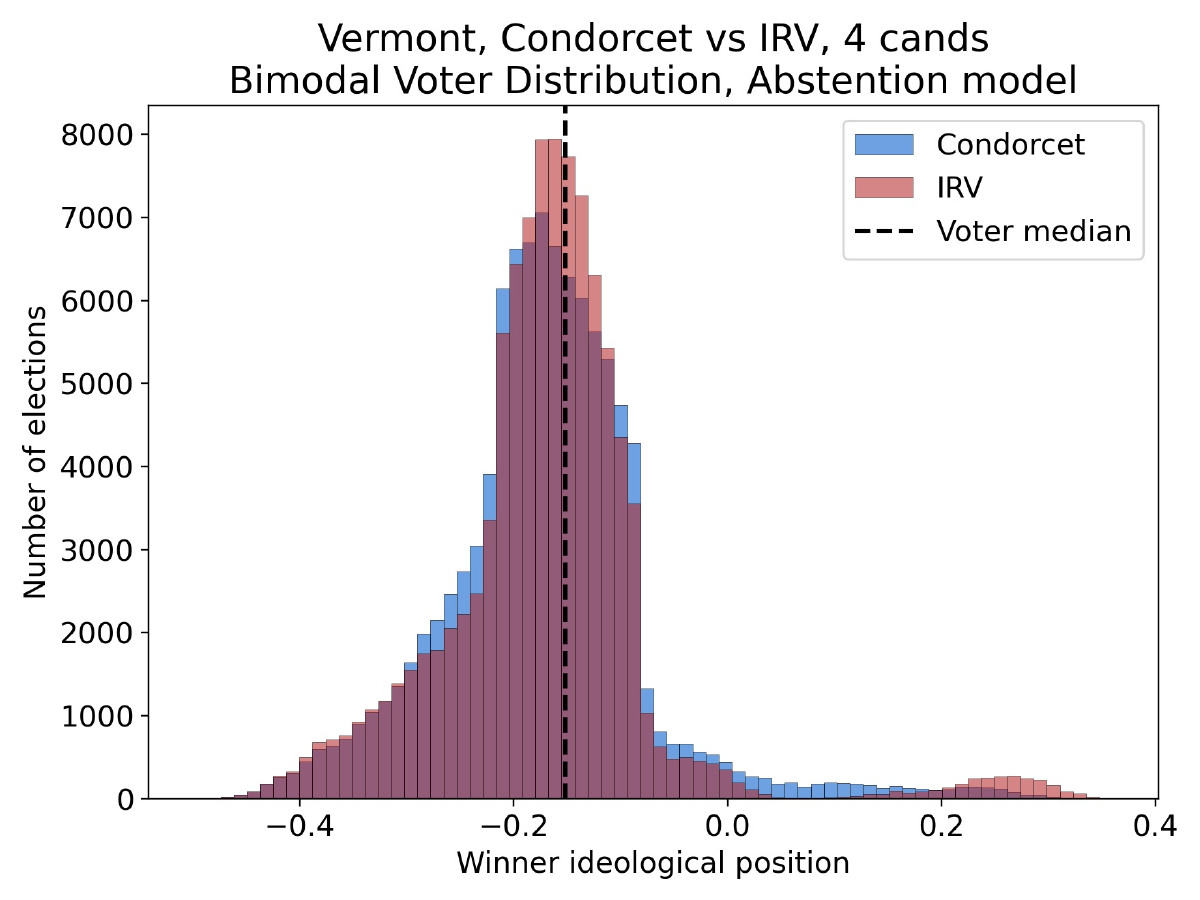} \\

\includegraphics[width=66mm]{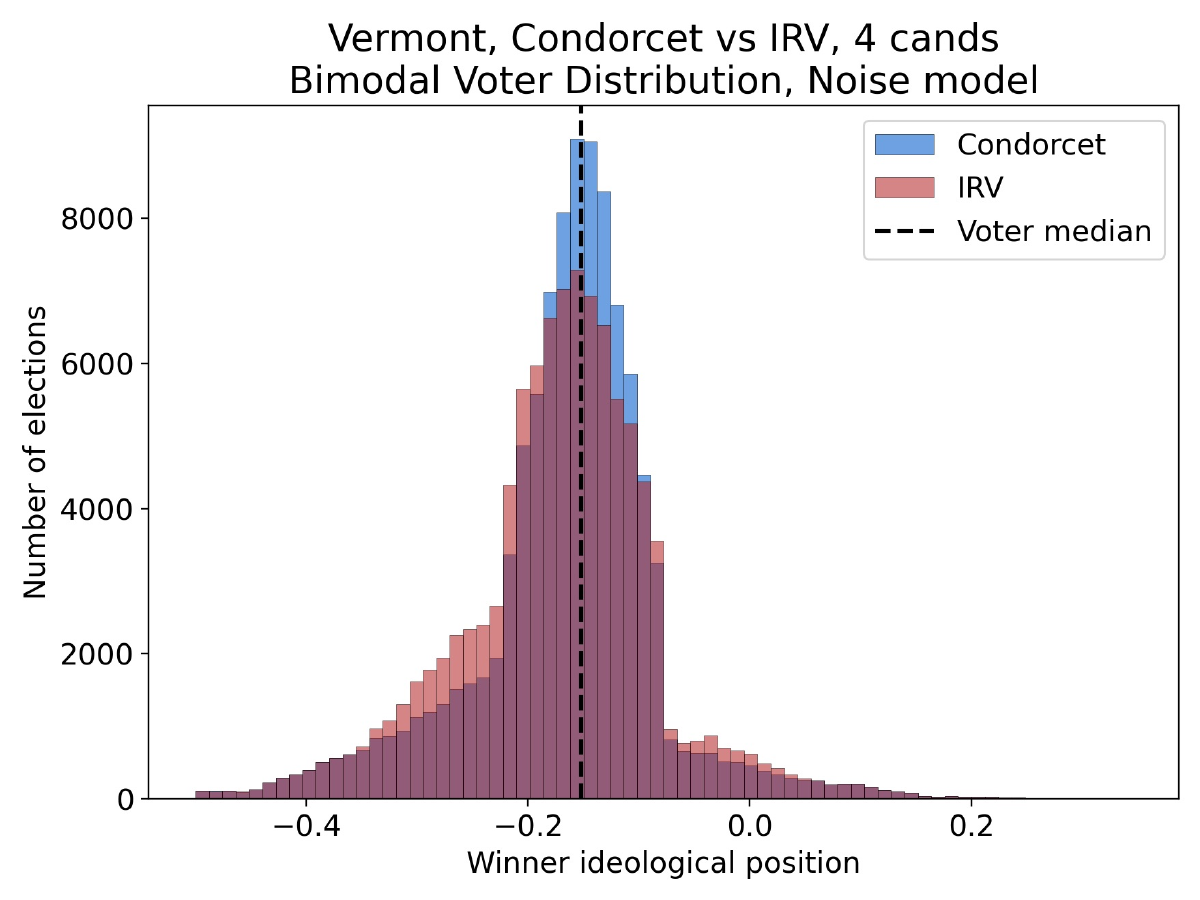}& 

\includegraphics[width=66mm]{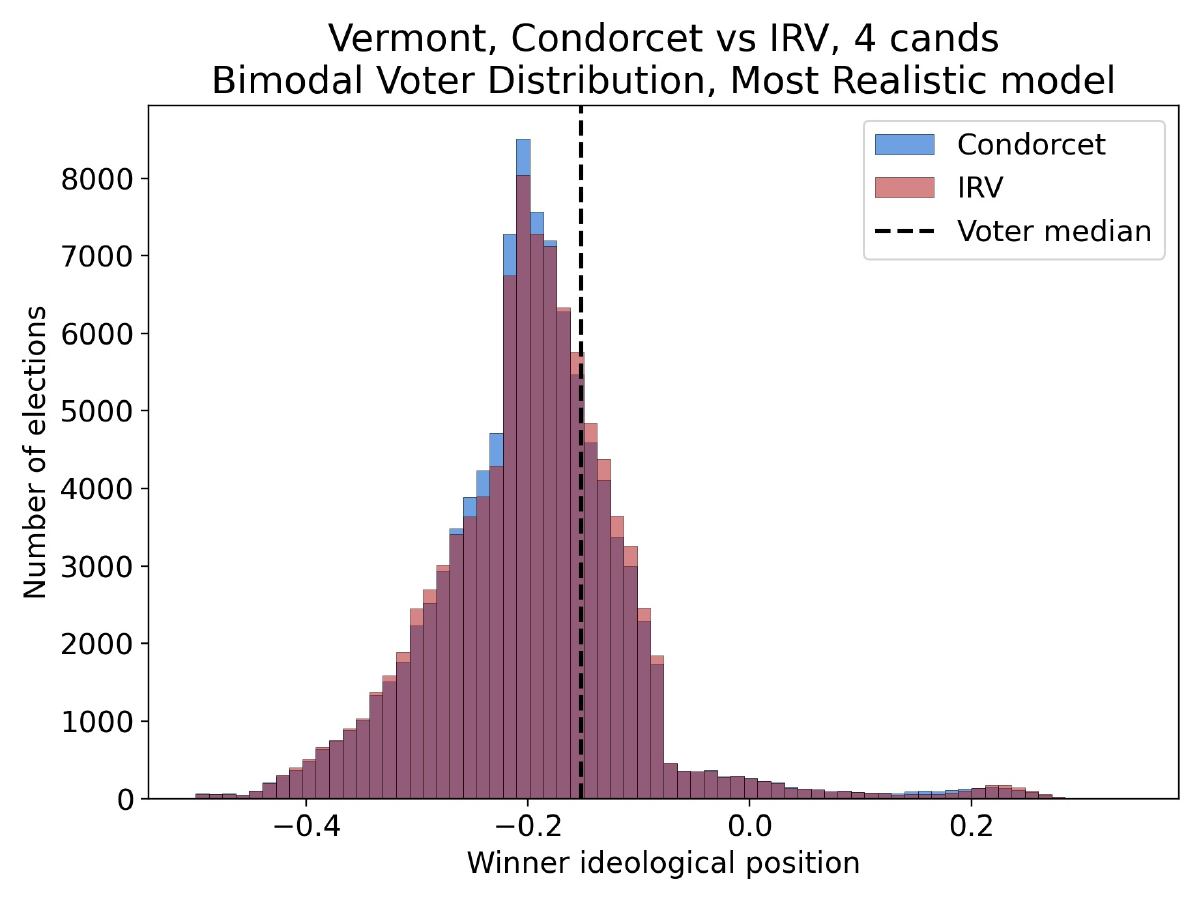}\\

\end{tabular}
\caption{Histograms showing the ideological positions of Condorcet and IRV winners across 100,000 simulations in Vermont, using 4-candidate elections with the bimodal voter distribution.}
\label{fig:VT_results_bimodal}
\end{figure}

\clearpage

\section{Average Distance from Median Voter Sample Summary Statistics}\label{sec:sum_stats_tables}

\begin{table}[H]
\begin{tabular}{l|cccccc}
\multicolumn{7}{c}{\LARGE\textbf{Bimodal Voter Distribution, 3 Candidates}}\\
\hline
\hline
\multicolumn{7}{c}{\textbf{Alaska}}\\
\hline
 & Theo. Ideal & Most Realistic & Ideo. Trunc. & Random Trunc. & Abstention & Noise\\
\hline
Borda & 0.136 & 0.148 & 0.124 & 0.128 & 0.166 & 0.124\\
Bucklin & 0.118 & 0.156 & 0.118 & 0.118 & 0.153 & 0.123\\
Condorcet & 0.118 & 0.151 & 0.121 & 0.129 & 0.153 & 0.119\\
IRV & 0.134 & 0.165 & 0.134 & 0.137 & 0.170 & 0.132\\
Plurality & 0.160 & 0.167 & 0.160 & 0.160 & 0.183 & 0.165\\
\hline
\hline
\multicolumn{7}{c}{\textbf{California}}\\
\hline
 & Theo. Ideal & Most Realistic & Ideo. Trunc. & Random Trunc. & Abstention & Noise\\
\hline
Borda & 0.133 & 0.146 & 0.119 & 0.117 & 0.178 & 0.119\\
Bucklin & 0.112 & 0.150 & 0.112 & 0.112 & 0.151 & 0.116\\
Condorcet & 0.112 & 0.151 & 0.118 & 0.116 & 0.151 & 0.112\\
IRV & 0.129 & 0.155 & 0.130 & 0.130 & 0.161 & 0.128\\
Plurality & 0.162 & 0.161 & 0.162 & 0.162 & 0.174 & 0.177\\
\hline
\hline
\multicolumn{7}{c}{\textbf{Michigan}}\\
\hline
 & Theo. Ideal & Most Realistic & Ideo. Trunc. & Random Trunc. & Abstention & Noise\\
\hline
Borda & 0.185 & 0.214 & 0.189 & 0.179 & 0.215 & 0.169\\
Bucklin & 0.158 & 0.210 & 0.158 & 0.158 & 0.215 & 0.162\\
Condorcet & 0.158 & 0.217 & 0.167 & 0.184 & 0.215 & 0.158\\
IRV & 0.184 & 0.258 & 0.184 & 0.197 & 0.260 & 0.183\\
Plurality & 0.227 & 0.266 & 0.227 & 0.227 & 0.268 & 0.235\\
\hline
\hline
\multicolumn{7}{c}{\textbf{South Carolina}}\\
\hline
 & Theo. Ideal & Most Realistic & Ideo. Trunc. & Random Trunc. & Abstention & Noise\\
\hline
Borda & 0.165 & 0.183 & 0.144 & 0.142 & 0.203 & 0.143\\
Bucklin & 0.132 & 0.189 & 0.132 & 0.132 & 0.191 & 0.136\\
Condorcet & 0.132 & 0.195 & 0.139 & 0.138 & 0.191 & 0.132\\
IRV & 0.156 & 0.219 & 0.156 & 0.158 & 0.224 & 0.156\\
Plurality & 0.205 & 0.228 & 0.205 & 0.205 & 0.238 & 0.220\\
\hline
\hline
\multicolumn{7}{c}{\textbf{Vermont}}\\
\hline
 & Theo. Ideal & Most Realistic & Ideo. Trunc. & Random Trunc. & Abstention & Noise\\
\hline
Borda & 0.082 & 0.093 & 0.082 & 0.083 & 0.116 & 0.081\\
Bucklin & 0.080 & 0.099 & 0.080 & 0.080 & 0.092 & 0.085\\
Condorcet & 0.080 & 0.092 & 0.082 & 0.082 & 0.092 & 0.080\\
IRV & 0.084 & 0.092 & 0.085 & 0.085 & 0.091 & 0.084\\
Plurality & 0.091 & 0.101 & 0.091 & 0.091 & 0.104 & 0.098\\
\hline
\hline
\multicolumn{7}{c}{\textbf{Average Across All States}}\\
\hline
 & Theo. Ideal & Most Realistic & Ideo. Trunc. & Random Trunc. & Abstention & Noise\\
\hline
Borda & 0.149 & 0.169 & 0.140 & 0.136 & 0.185 & 0.134\\
Bucklin & 0.124 & 0.169 & 0.124 & 0.124 & 0.171 & 0.129\\
Condorcet & 0.124 & 0.175 & 0.132 & 0.136 & 0.171 & 0.125\\
IRV & 0.144 & 0.192 & 0.145 & 0.149 & 0.194 & 0.144\\
Plurality & 0.180 & 0.198 & 0.180 & 0.180 & 0.205 & 0.192\\
\hline
\hline
\end{tabular}
\caption{The average distance from the median voter across 100,000 simulations in 5 states, and the average across all states, in 3-candidate elections and the bimodal voter distribution.}
\label{table:avg_distance_stats_bimodal_3cands}
\end{table}

\begin{table}[H]

\begin{tabular}{l|cccccc}
\multicolumn{7}{c}{\LARGE\textbf{Bimodal Voter Distribution, 4 Candidates}}\\
\hline
\hline
\multicolumn{7}{c}{\textbf{Alaska}}\\
\hline
 & Theo. Ideal & Most Realistic & Ideo. Trunc. & Random Trunc. & Abstention & Noise\\
\hline
Borda & 0.110 & 0.124 & 0.140 & 0.119 & 0.138 & 0.104\\
Bucklin & 0.102 & 0.133 & 0.121 & 0.102 & 0.135 & 0.101\\
Condorcet & 0.097 & 0.132 & 0.128 & 0.122 & 0.129 & 0.097\\
IRV & 0.132 & 0.163 & 0.139 & 0.140 & 0.164 & 0.130\\
Plurality & 0.168 & 0.175 & 0.168 & 0.168 & 0.182 & 0.177\\
\hline
\hline
\multicolumn{7}{c}{\textbf{California}}\\
\hline
 & Theo. Ideal & Most Realistic & Ideo. Trunc. & Random Trunc. & Abstention & Noise\\
\hline
Borda & 0.110 & 0.135 & 0.129 & 0.102 & 0.141 & 0.097\\
Bucklin & 0.091 & 0.137 & 0.121 & 0.094 & 0.137 & 0.090\\
Condorcet & 0.088 & 0.140 & 0.122 & 0.101 & 0.122 & 0.088\\
IRV & 0.124 & 0.144 & 0.128 & 0.126 & 0.144 & 0.121\\
Plurality & 0.182 & 0.177 & 0.182 & 0.182 & 0.171 & 0.205\\
\hline
\hline
\multicolumn{7}{c}{\textbf{Michigan}}\\
\hline
 & Theo. Ideal & Most Realistic & Ideo. Trunc. & Random Trunc. & Abstention & Noise\\
\hline
Borda & 0.159 & 0.244 & 0.221 & 0.177 & 0.198 & 0.146\\
Bucklin & 0.137 & 0.223 & 0.193 & 0.153 & 0.197 & 0.137\\
Condorcet & 0.134 & 0.239 & 0.201 & 0.179 & 0.183 & 0.134\\
IRV & 0.189 & 0.258 & 0.201 & 0.208 & 0.254 & 0.188\\
Plurality & 0.242 & 0.268 & 0.242 & 0.242 & 0.270 & 0.243\\
\hline
\hline
\multicolumn{7}{c}{\textbf{South Carolina}}\\
\hline
 & Theo. Ideal & Most Realistic & Ideo. Trunc. & Random Trunc. & Abstention & Noise\\
\hline
Borda & 0.131 & 0.179 & 0.167 & 0.130 & 0.170 & 0.118\\
Bucklin & 0.108 & 0.180 & 0.143 & 0.113 & 0.168 & 0.108\\
Condorcet & 0.105 & 0.195 & 0.152 & 0.135 & 0.153 & 0.106\\
IRV & 0.155 & 0.207 & 0.157 & 0.164 & 0.211 & 0.156\\
Plurality & 0.221 & 0.234 & 0.221 & 0.221 & 0.240 & 0.230\\
\hline
\hline
\multicolumn{7}{c}{\textbf{Vermont}}\\
\hline
 & Theo. Ideal & Most Realistic & Ideo. Trunc. & Random Trunc. & Abstention & Noise\\
\hline
Borda & 0.064 & 0.079 & 0.079 & 0.064 & 0.087 & 0.061\\
Bucklin & 0.063 & 0.085 & 0.087 & 0.062 & 0.084 & 0.062\\
Condorcet & 0.059 & 0.077 & 0.069 & 0.064 & 0.070 & 0.059\\
IRV & 0.067 & 0.077 & 0.070 & 0.069 & 0.068 & 0.067\\
Plurality & 0.095 & 0.102 & 0.095 & 0.095 & 0.096 & 0.122\\
\hline
\hline
\multicolumn{7}{c}{\textbf{Average Across All States}}\\
\hline
 & Theo. Ideal & Most Realistic & Ideo. Trunc. & Random Trunc. & Abstention & Noise\\
\hline
Borda & 0.122 & 0.166 & 0.157 & 0.127 & 0.158 & 0.112\\
Bucklin & 0.105 & 0.163 & 0.137 & 0.110 & 0.155 & 0.104\\
Condorcet & 0.101 & 0.173 & 0.144 & 0.129 & 0.140 & 0.101\\
IRV & 0.142 & 0.183 & 0.148 & 0.152 & 0.182 & 0.141\\
Plurality & 0.197 & 0.203 & 0.197 & 0.197 & 0.203 & 0.212\\
\hline
\hline
\end{tabular}
\caption{The average distance from the median voter across 100,000 simulations in 5 states, and the average across all states, in 4-candidate elections and the bimodal voter distribution.}
\label{table:avg_distance_stats_bimodal_4cands}
\end{table}

\begin{table}[H]
\begin{tabular}{l|cccccc}
\multicolumn{7}{c}{\LARGE\textbf{Trimodal Voter Distribution, 3 Candidates}}\\
\hline
\hline
\multicolumn{7}{c}{\textbf{Alaska}}\\
\hline
 & Theo. Ideal & Most Realistic & Ideo. Trunc. & Random Trunc. & Abstention & Noise\\
\hline
Borda & 0.080 & 0.089 & 0.082 & 0.081 & 0.109 & 0.080\\
Bucklin & 0.079 & 0.093 & 0.079 & 0.079 & 0.088 & 0.083\\
Condorcet & 0.079 & 0.087 & 0.080 & 0.081 & 0.088 & 0.079\\
IRV & 0.082 & 0.086 & 0.082 & 0.083 & 0.088 & 0.082\\
Plurality & 0.088 & 0.087 & 0.088 & 0.088 & 0.093 & 0.093\\
\hline
\hline
\multicolumn{7}{c}{\textbf{California}}\\
\hline
 & Theo. Ideal & Most Realistic & Ideo. Trunc. & Random Trunc. & Abstention & Noise\\
\hline
Borda & 0.122 & 0.146 & 0.119 & 0.116 & 0.172 & 0.115\\
Bucklin & 0.112 & 0.154 & 0.112 & 0.112 & 0.148 & 0.116\\
Condorcet & 0.112 & 0.146 & 0.113 & 0.115 & 0.148 & 0.112\\
IRV & 0.123 & 0.147 & 0.123 & 0.124 & 0.153 & 0.123\\
Plurality & 0.150 & 0.171 & 0.150 & 0.150 & 0.183 & 0.156\\
\hline
\hline
\multicolumn{7}{c}{\textbf{Michigan}}\\
\hline
 & Theo. Ideal & Most Realistic & Ideo. Trunc. & Random Trunc. & Abstention & Noise\\
\hline
Borda & 0.116 & 0.145 & 0.117 & 0.113 & 0.174 & 0.112\\
Bucklin & 0.109 & 0.153 & 0.110 & 0.109 & 0.144 & 0.114\\
Condorcet & 0.109 & 0.146 & 0.111 & 0.112 & 0.144 & 0.110\\
IRV & 0.120 & 0.138 & 0.120 & 0.120 & 0.142 & 0.120\\
Plurality & 0.142 & 0.165 & 0.142 & 0.142 & 0.174 & 0.149\\
\hline
\hline
\multicolumn{7}{c}{\textbf{South Carolina}}\\
\hline
 & Theo. Ideal & Most Realistic & Ideo. Trunc. & Random Trunc. & Abstention & Noise\\
\hline
Borda & 0.136 & 0.157 & 0.126 & 0.124 & 0.176 & 0.122\\
Bucklin & 0.117 & 0.161 & 0.117 & 0.117 & 0.162 & 0.122\\
Condorcet & 0.117 & 0.156 & 0.120 & 0.123 & 0.162 & 0.118\\
IRV & 0.130 & 0.168 & 0.131 & 0.132 & 0.175 & 0.131\\
Plurality & 0.168 & 0.193 & 0.168 & 0.168 & 0.202 & 0.173\\
\hline
\hline
\multicolumn{7}{c}{\textbf{Vermont}}\\
\hline
 & Theo. Ideal & Most Realistic & Ideo. Trunc. & Random Trunc. & Abstention & Noise\\
\hline
Borda & 0.107 & 0.122 & 0.098 & 0.100 & 0.142 & 0.099\\
Bucklin & 0.096 & 0.129 & 0.096 & 0.096 & 0.121 & 0.101\\
Condorcet & 0.096 & 0.123 & 0.096 & 0.099 & 0.121 & 0.096\\
IRV & 0.104 & 0.120 & 0.104 & 0.105 & 0.119 & 0.104\\
Plurality & 0.127 & 0.136 & 0.127 & 0.127 & 0.141 & 0.133\\
\hline
\hline
\multicolumn{7}{c}{\textbf{Average Across All States}}\\
\hline
 & Theo. Ideal & Most Realistic & Ideo. Trunc. & Random Trunc. & Abstention & Noise\\
\hline
Borda & 0.122 & 0.145 & 0.118 & 0.116 & 0.163 & 0.114\\
Bucklin & 0.110 & 0.149 & 0.110 & 0.110 & 0.147 & 0.115\\
Condorcet & 0.110 & 0.146 & 0.113 & 0.115 & 0.147 & 0.110\\
IRV & 0.122 & 0.151 & 0.124 & 0.124 & 0.155 & 0.122\\
Plurality & 0.148 & 0.170 & 0.148 & 0.148 & 0.177 & 0.154\\
\hline
\hline
\end{tabular}
\caption{The average distance from the median voter across 100,000 simulations in 5 states, and the average across all states, in 3-candidate elections and the trimodal voter distribution.}
\label{table:avg_distance_stats_trimodal_3cands}
\end{table}

\begin{table}[H]
\begin{tabular}{l|cccccc}
\multicolumn{7}{c}{\LARGE\textbf{Trimodal Voter Distribution, 4 Candidates}}\\
\hline
\hline
\multicolumn{7}{c}{\textbf{Alaska}}\\
\hline
 & Theo. Ideal & Most Realistic & Ideo. Trunc. & Random Trunc. & Abstention & Noise\\
\hline
Borda & 0.061 & 0.064 & 0.073 & 0.063 & 0.081 & 0.059\\
Bucklin & 0.062 & 0.073 & 0.084 & 0.060 & 0.074 & 0.060\\
Condorcet & 0.057 & 0.064 & 0.064 & 0.062 & 0.066 & 0.058\\
IRV & 0.064 & 0.065 & 0.066 & 0.066 & 0.066 & 0.064\\
Plurality & 0.090 & 0.093 & 0.090 & 0.090 & 0.085 & 0.114\\
\hline
\hline
\multicolumn{7}{c}{\textbf{California}}\\
\hline
 & Theo. Ideal & Most Realistic & Ideo. Trunc. & Random Trunc. & Abstention & Noise\\
\hline
Borda & 0.098 & 0.124 & 0.131 & 0.095 & 0.130 & 0.091\\
Bucklin & 0.094 & 0.134 & 0.125 & 0.090 & 0.135 & 0.091\\
Condorcet & 0.086 & 0.130 & 0.115 & 0.096 & 0.119 & 0.087\\
IRV & 0.108 & 0.134 & 0.118 & 0.113 & 0.130 & 0.111\\
Plurality & 0.158 & 0.178 & 0.158 & 0.158 & 0.178 & 0.174\\
\hline
\hline
\multicolumn{7}{c}{\textbf{Michigan}}\\
\hline
 & Theo. Ideal & Most Realistic & Ideo. Trunc. & Random Trunc. & Abstention & Noise\\
\hline
Borda & 0.093 & 0.115 & 0.126 & 0.090 & 0.123 & 0.087\\
Bucklin & 0.091 & 0.128 & 0.125 & 0.086 & 0.131 & 0.087\\
Condorcet & 0.083 & 0.122 & 0.101 & 0.090 & 0.113 & 0.083\\
IRV & 0.101 & 0.116 & 0.108 & 0.105 & 0.116 & 0.103\\
Plurality & 0.152 & 0.177 & 0.152 & 0.152 & 0.172 & 0.172\\
\hline
\hline
\multicolumn{7}{c}{\textbf{South Carolina}}\\
\hline
 & Theo. Ideal & Most Realistic & Ideo. Trunc. & Random Trunc. & Abstention & Noise\\
\hline
Borda & 0.108 & 0.153 & 0.148 & 0.109 & 0.146 & 0.100\\
Bucklin & 0.102 & 0.152 & 0.130 & 0.099 & 0.151 & 0.099\\
Condorcet & 0.093 & 0.152 & 0.137 & 0.113 & 0.134 & 0.094\\
IRV & 0.117 & 0.171 & 0.141 & 0.127 & 0.157 & 0.121\\
Plurality & 0.173 & 0.198 & 0.173 & 0.173 & 0.201 & 0.184\\
\hline
\hline
\multicolumn{7}{c}{\textbf{Vermont}}\\
\hline
 & Theo. Ideal & Most Realistic & Ideo. Trunc. & Random Trunc. & Abstention & Noise\\
\hline
Borda & 0.086 & 0.093 & 0.106 & 0.084 & 0.106 & 0.079\\
Bucklin & 0.082 & 0.107 & 0.105 & 0.078 & 0.107 & 0.080\\
Condorcet & 0.075 & 0.097 & 0.087 & 0.085 & 0.100 & 0.076\\
IRV & 0.091 & 0.103 & 0.096 & 0.095 & 0.102 & 0.092\\
Plurality & 0.132 & 0.148 & 0.132 & 0.132 & 0.143 & 0.148\\
\hline
\hline
\multicolumn{7}{c}{\textbf{Average Across All States}}\\
\hline
 & Theo. Ideal & Most Realistic & Ideo. Trunc. & Random Trunc. & Abstention & Noise\\
\hline
Borda & 0.099 & 0.130 & 0.130 & 0.099 & 0.131 & 0.091\\
Bucklin & 0.094 & 0.134 & 0.122 & 0.091 & 0.135 & 0.091\\
Condorcet & 0.086 & 0.133 & 0.116 & 0.100 & 0.120 & 0.087\\
IRV & 0.112 & 0.144 & 0.123 & 0.117 & 0.137 & 0.112\\
Plurality & 0.156 & 0.175 & 0.156 & 0.156 & 0.176 & 0.172\\
\hline
\hline
\end{tabular}
\caption{The average distance from the median voter across 100,000 simulations in 5 states, and the average across all states, in 4-candidate elections and the trimodal voter distribution.}
\label{table:avg_distance_stats_trimodal_4cands}
\end{table}

\clearpage

\section{Relative Increase in Distance from Median Voter Images}\label{sec:avg_dist_images}

\begin{figure}[!h]
\includegraphics[width=\linewidth]{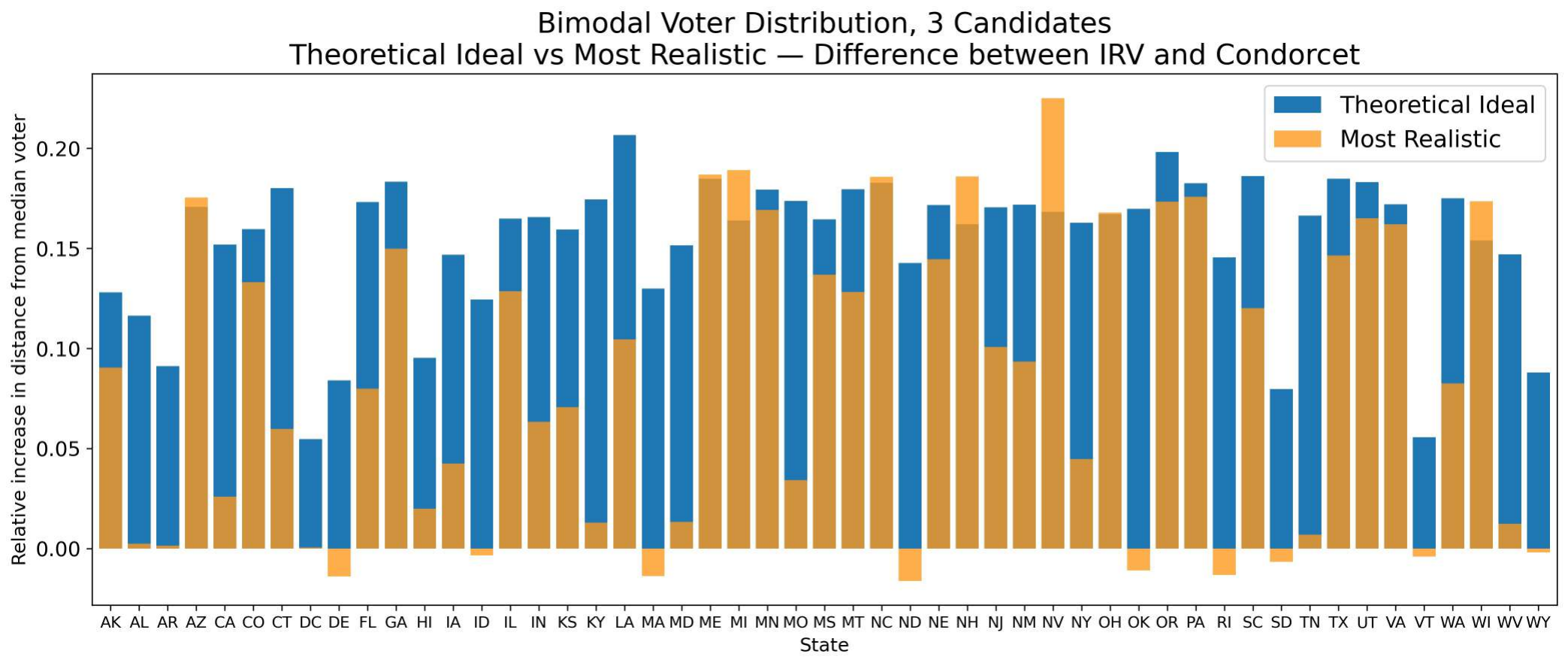}
\caption{The relative difference of the average distance of the IRV winner and the average distance of the Condorcet winner to the median voter, under the Theoretical Ideal and Most Realistic models. Bimodal Distribution, 3-Candidate Elections.}
\label{fig:avg_dist_bimodal_3cands_mostrealistic}
\end{figure}

\begin{figure}[!h]
\includegraphics[width=\linewidth]{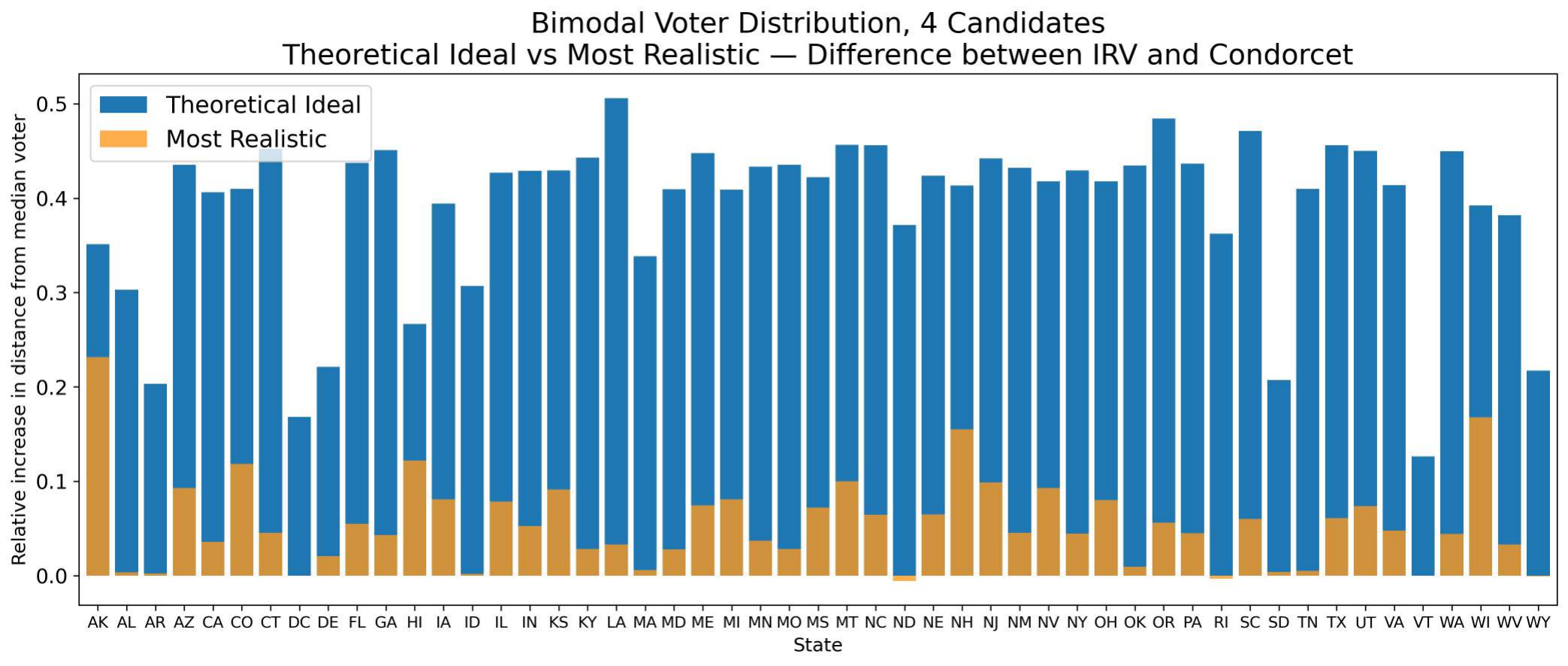}
\caption{The relative difference of the average distance of the IRV winner and the average distance of the Condorcet winner to the median voter, under the Theoretical Ideal and Most Realistic models. Bimodal Distribution, 4-Candidate Elections.}
\label{fig:avg_dist_bimodal_4cands_mostrealistic}
\end{figure}

\begin{figure}[!h]
\includegraphics[width=\linewidth]{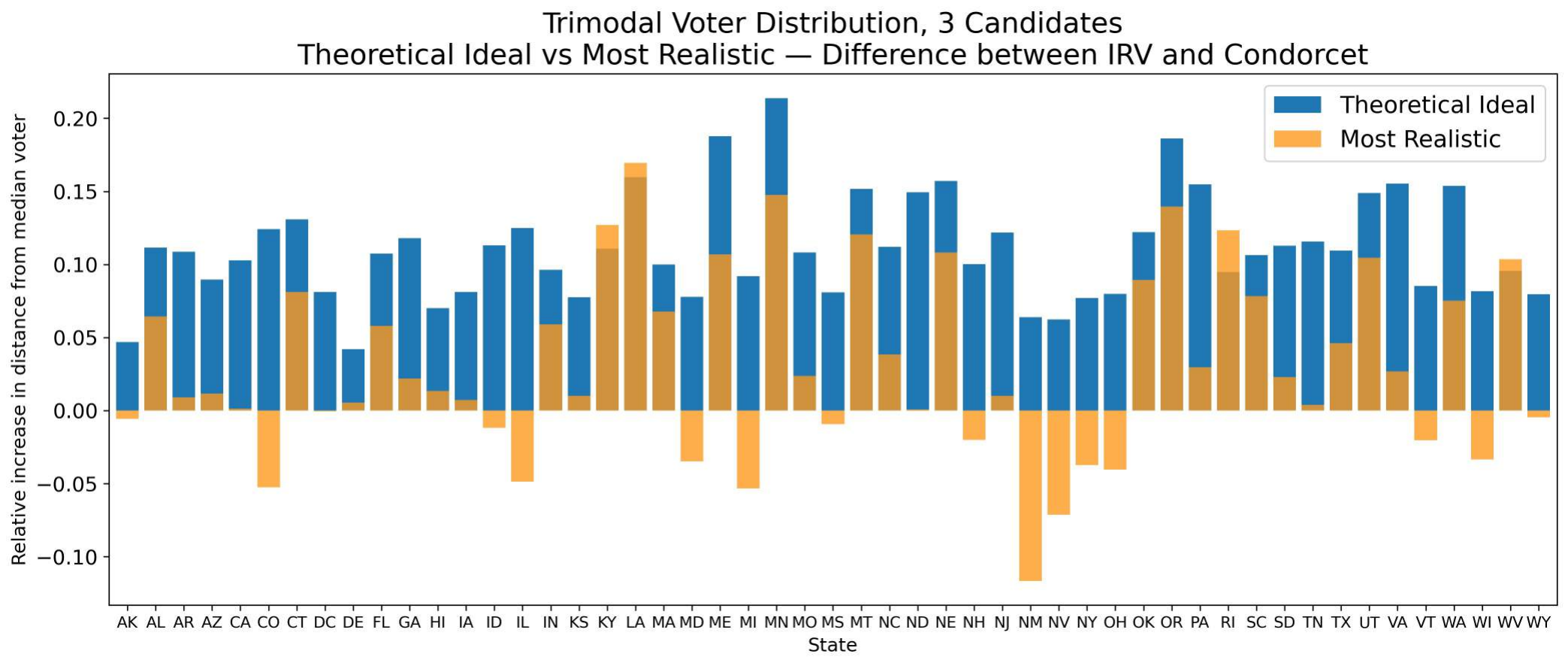}
\caption{The relative difference of the average distance of the IRV winner and the average distance of the Condorcet winner to the median voter, under the Theoretical Ideal and Most Realistic models. Trimodal Distribution, 3-Candidate Elections.}
\label{fig:avg_dist_trimodal_3cands_mostrealistic}
\end{figure}

\begin{figure}[!h]
\includegraphics[width=\linewidth]{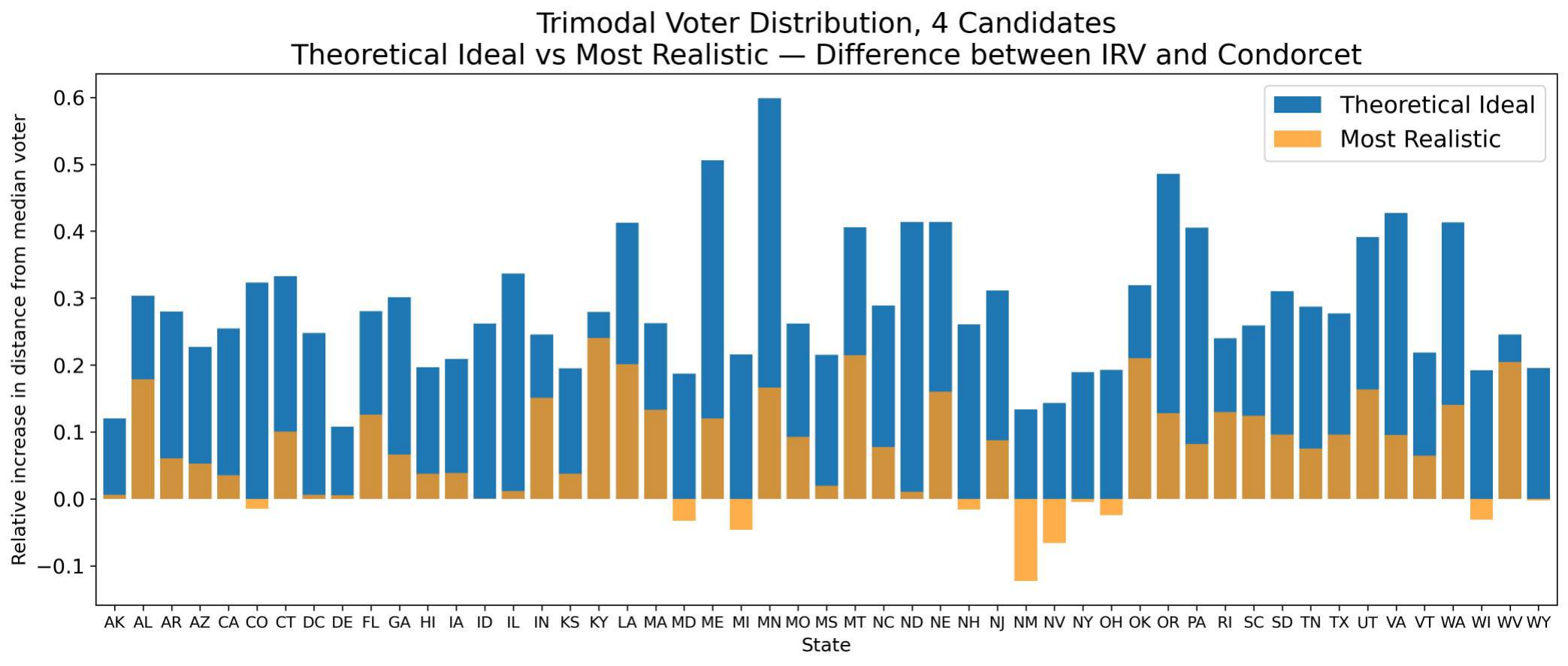}
\caption{The relative difference of the average distance of the IRV winner and the average distance of the Condorcet winner to the median voter, under the Theoretical Ideal and Most Realistic models. Trimodal Distribution, 4-Candidate Elections.}
\label{fig:avg_dist_trimodal_4cands_mostrealistic}
\end{figure}

\begin{figure}[!h]
\includegraphics[width=\linewidth]{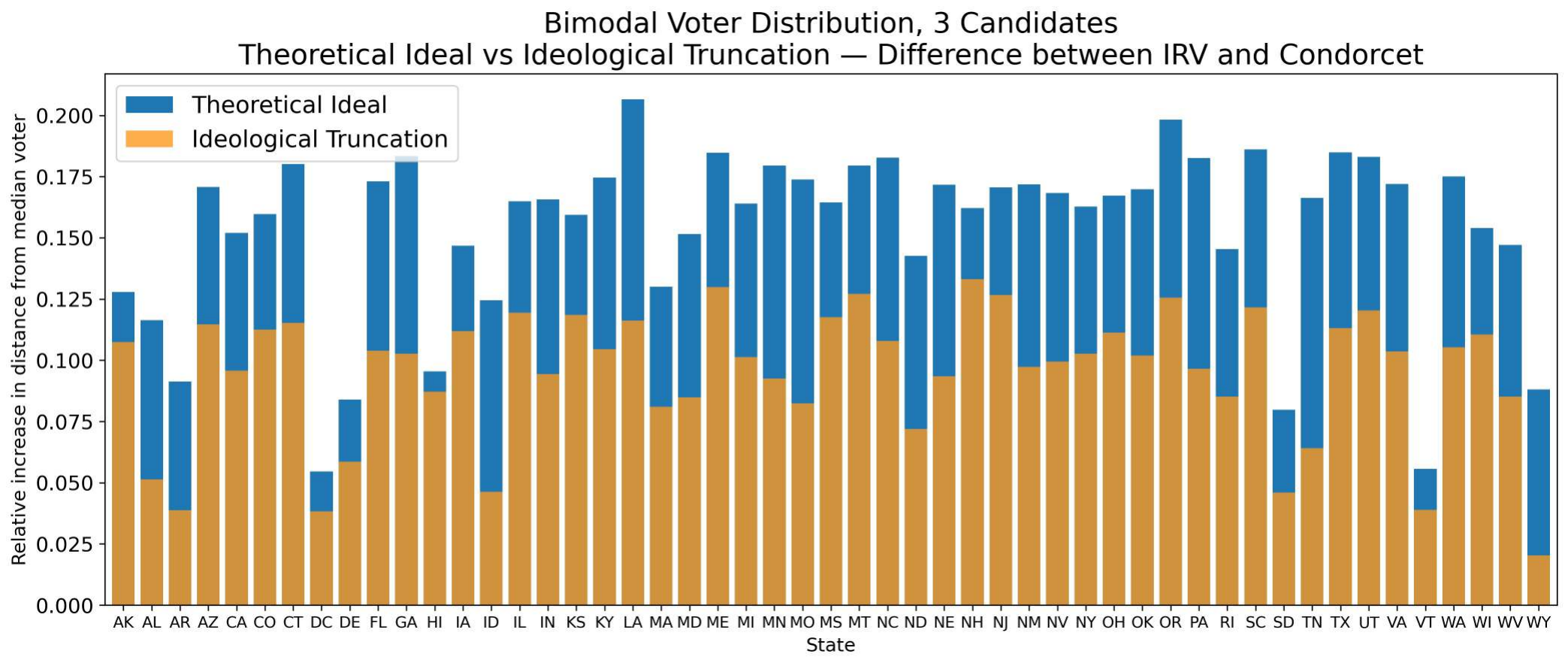}
\caption{The relative difference of the average distance of the IRV winner and the average distance of the Condorcet winner to the median voter, under the Theoretical Ideal and Ideological Truncation models. Bimodal Distribution, 3-Candidate Elections.}
\label{fig:avg_dist_bimodal_3cands_ideotrunc}
\end{figure}

\begin{figure}[!h]
\includegraphics[width=\linewidth]{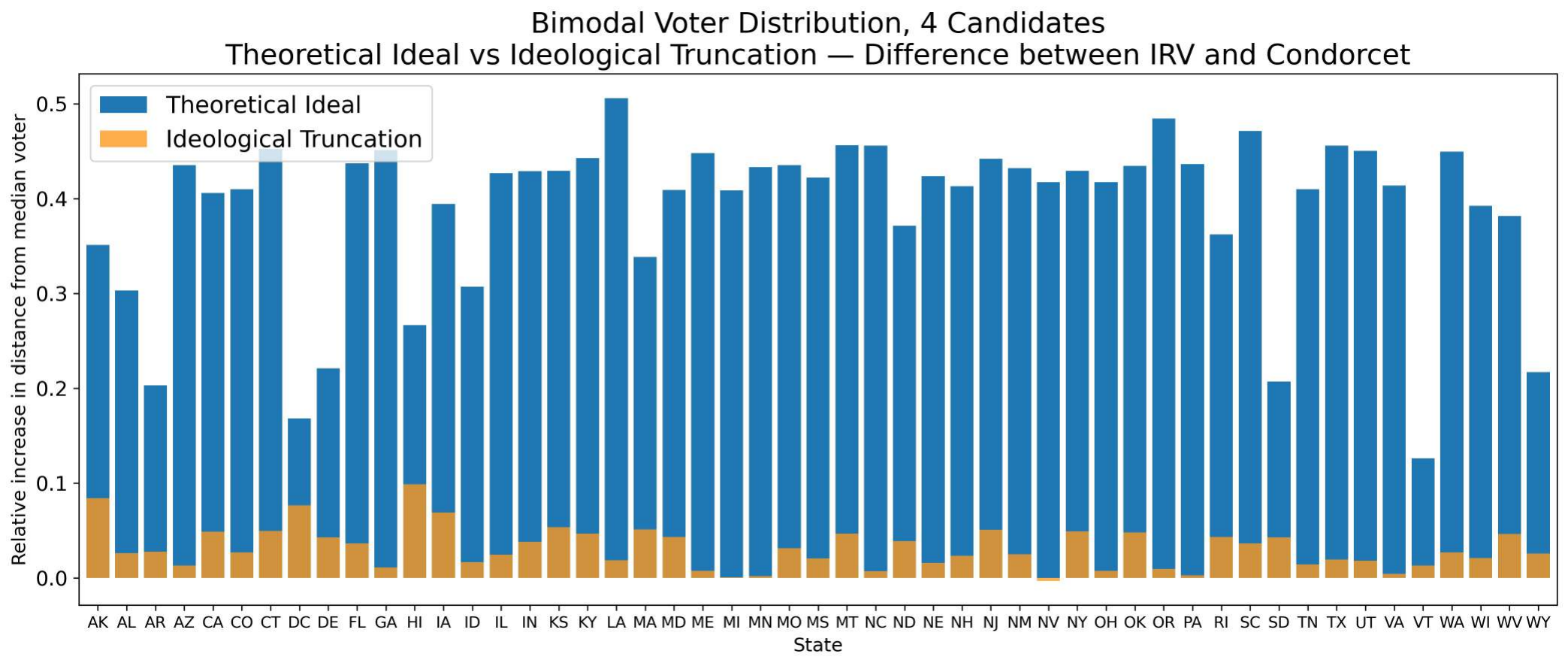}
\caption{The relative difference of the average distance of the IRV winner and the average distance of the Condorcet winner to the median voter, under the Theoretical Ideal and Ideological Truncation models. Bimodal Distribution, 4-Candidate Elections.}
\label{fig:avg_dist_bimodal_4cands_ideotrunc}
\end{figure}

\begin{figure}[!h]
\includegraphics[width=\linewidth]{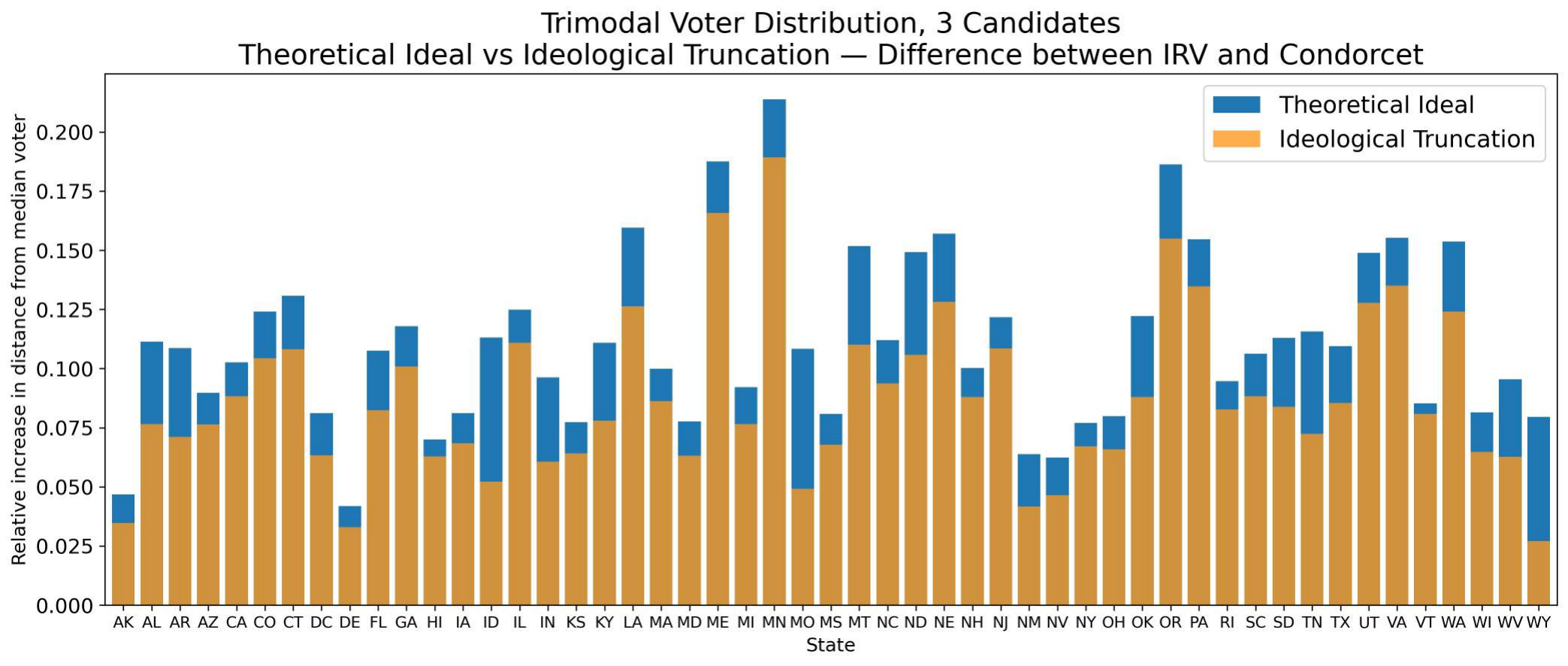}
\caption{The relative difference of the average distance of the IRV winner and the average distance of the Condorcet winner to the median voter, under the Theoretical Ideal and Ideological Truncation models. Trimodal Distribution, 3-Candidate Elections.}
\label{fig:avg_dist_trimodal_3cands_ideotrunc}
\end{figure}

\begin{figure}[!h]
\includegraphics[width=\linewidth]{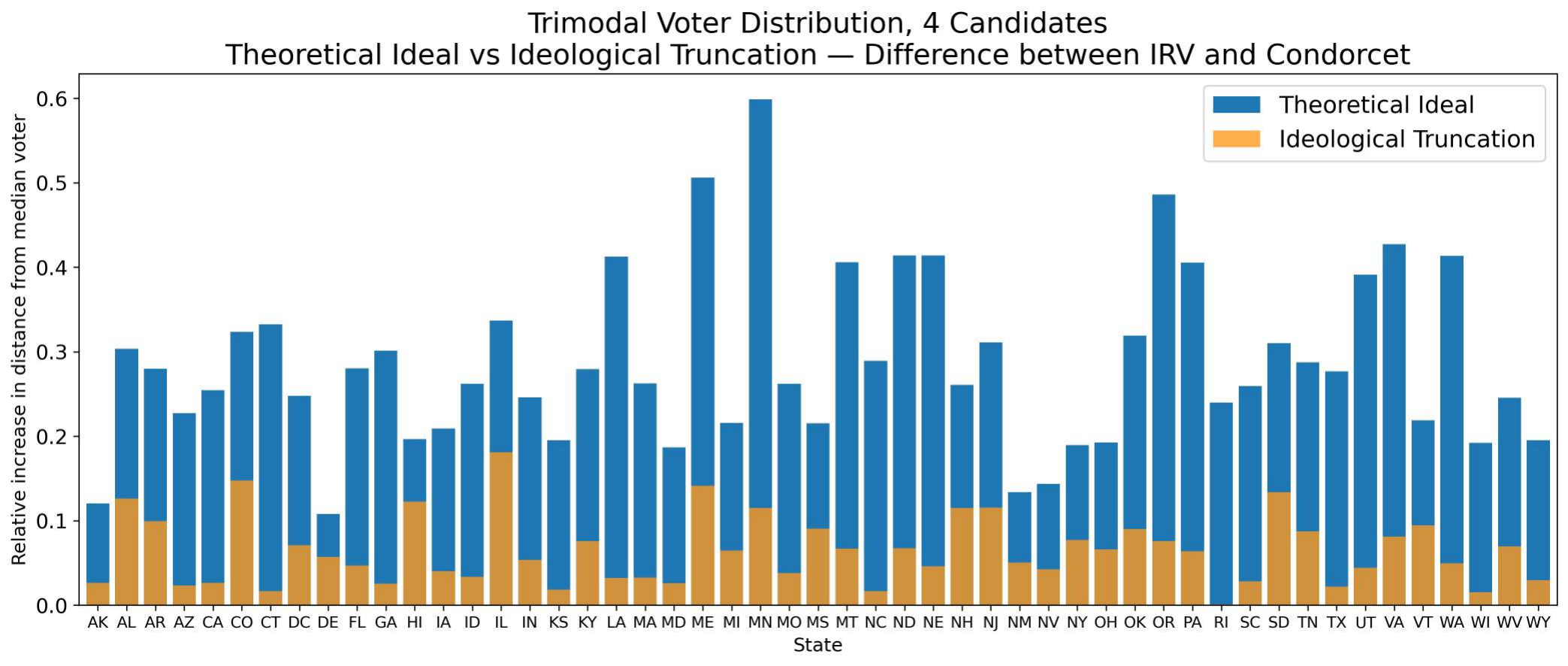}
\caption{The relative difference of the average distance of the IRV winner and the average distance of the Condorcet winner to the median voter, under the Theoretical Ideal and Ideological Truncation models. Trimodal Distribution, 4-Candidate Elections.}
\label{fig:avg_dist_trimodal_4cands_ideotrunc}
\end{figure}

\begin{figure}[!h]
\includegraphics[width=\linewidth]{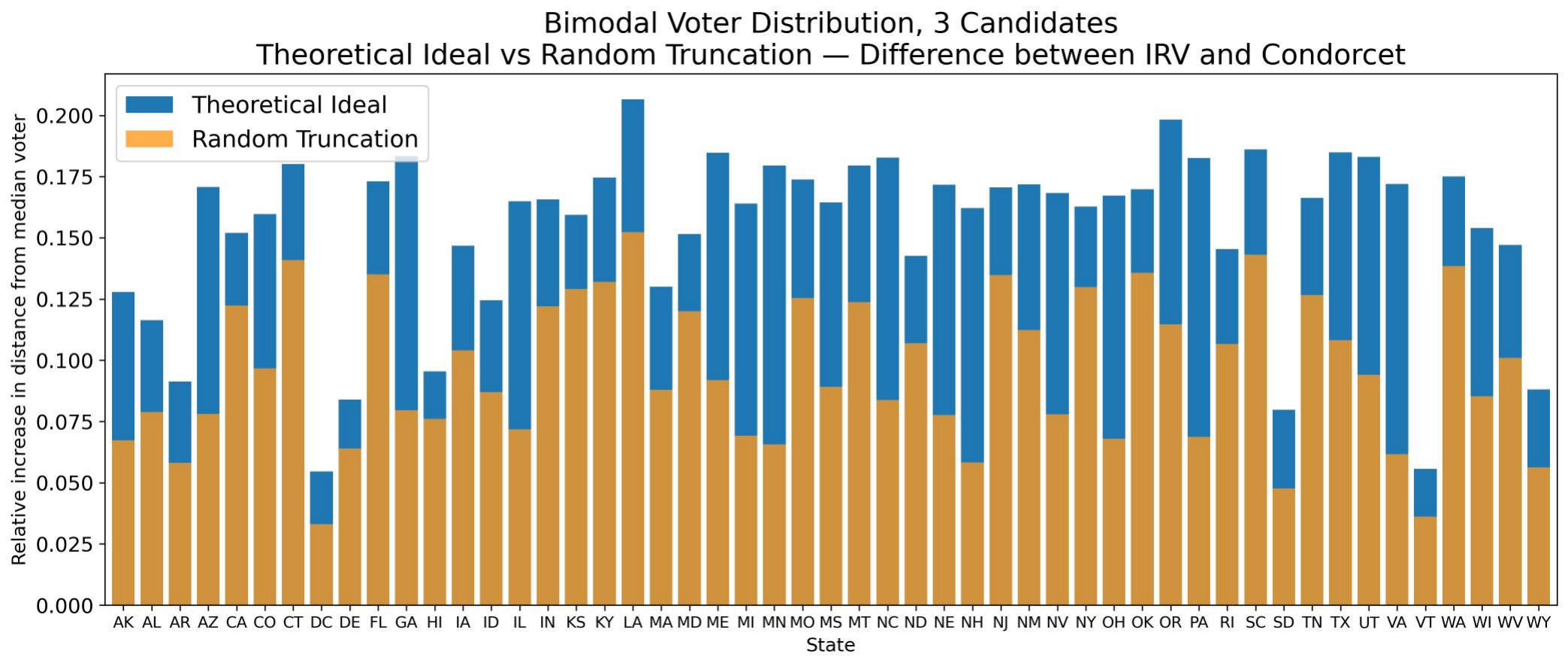}
\caption{The relative difference of the average distance of the IRV winner and the average distance of the Condorcet winner to the median voter, under the Theoretical Ideal and Random Truncation models. Bimodal Distribution, 3-Candidate Elections.}
\label{fig:avg_dist_bimodal_3cands_randomtrunc}
\end{figure}

\begin{figure}[!h]
\includegraphics[width=\linewidth]{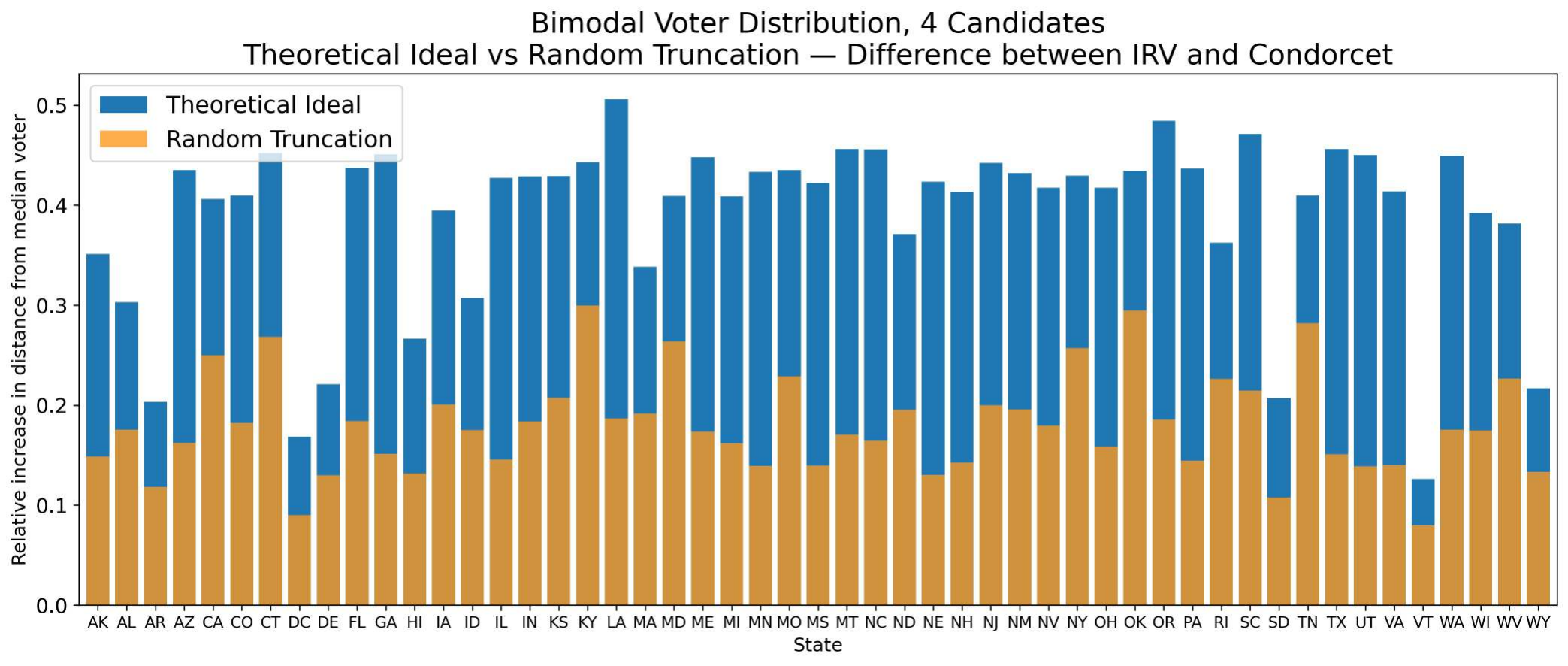}
\caption{The relative difference of the average distance of the IRV winner and the average distance of the Condorcet winner to the median voter, under the Theoretical Ideal and Random Truncation models. Bimodal Distribution, 4-Candidate Elections.}
\label{fig:avg_dist_bimodal_4cands_randomtrunc}
\end{figure}

\begin{figure}[!h]
\includegraphics[width=\linewidth]{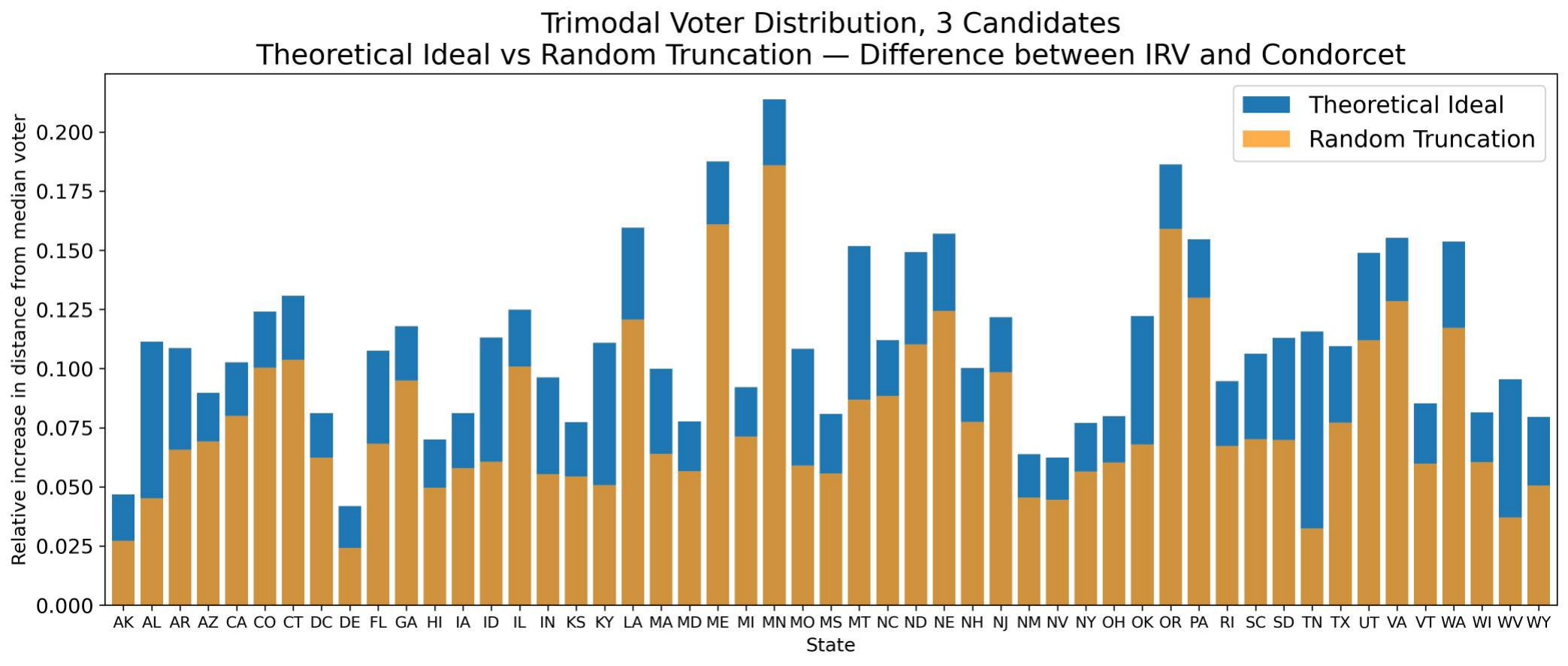}
\caption{The relative difference of the average distance of the IRV winner and the average distance of the Condorcet winner to the median voter, under the Theoretical Ideal and Random Truncation models. Trimodal Distribution, 3-Candidate Elections.}
\label{fig:avg_dist_trimodal_3cands_randomtrunc}
\end{figure}

\begin{figure}[!h]
\includegraphics[width=\linewidth]{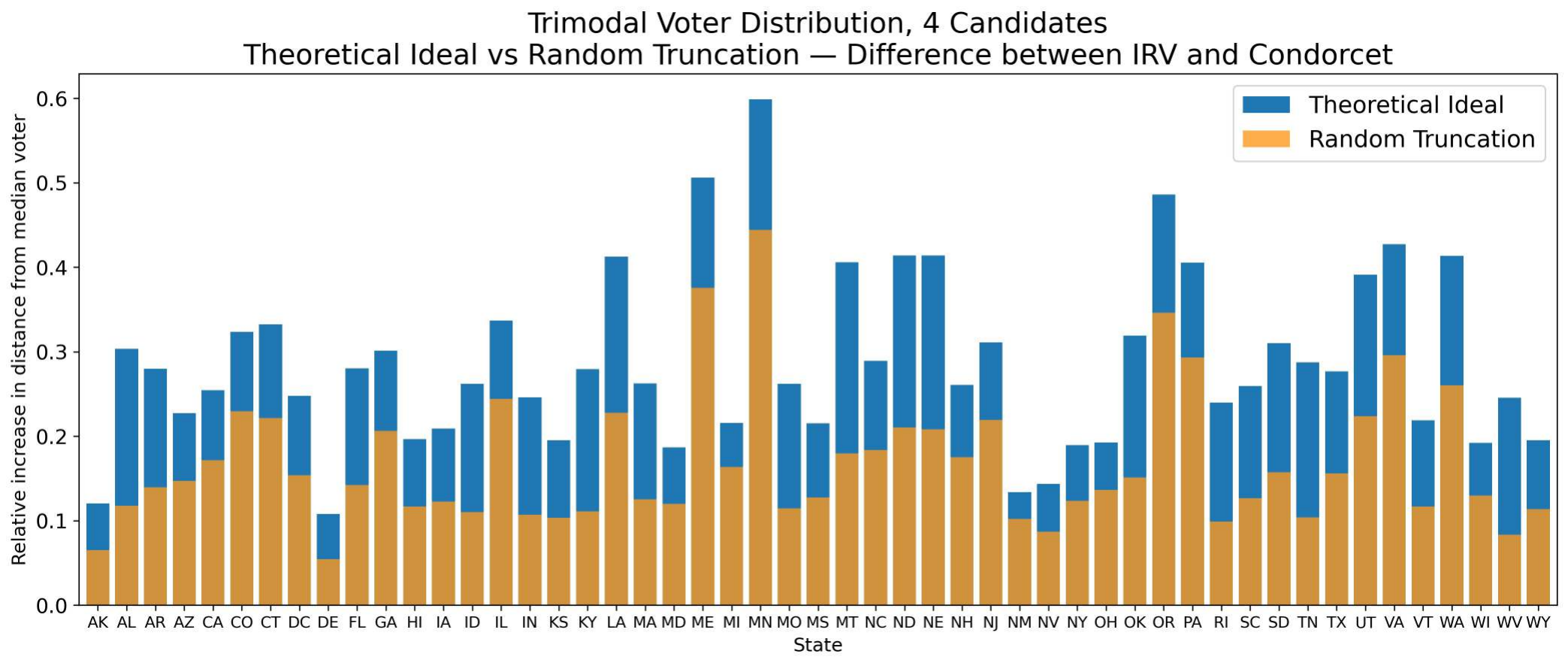}
\caption{The relative difference of the average distance of the IRV winner and the average distance of the Condorcet winner to the median voter, under the Theoretical Ideal and Random Truncation models. Trimodal Distribution, 4-Candidate Elections.}
\label{fig:avg_dist_trimodal_4cands_randomtrunc}
\end{figure}


\begin{figure}[!h]
\includegraphics[width=\linewidth]{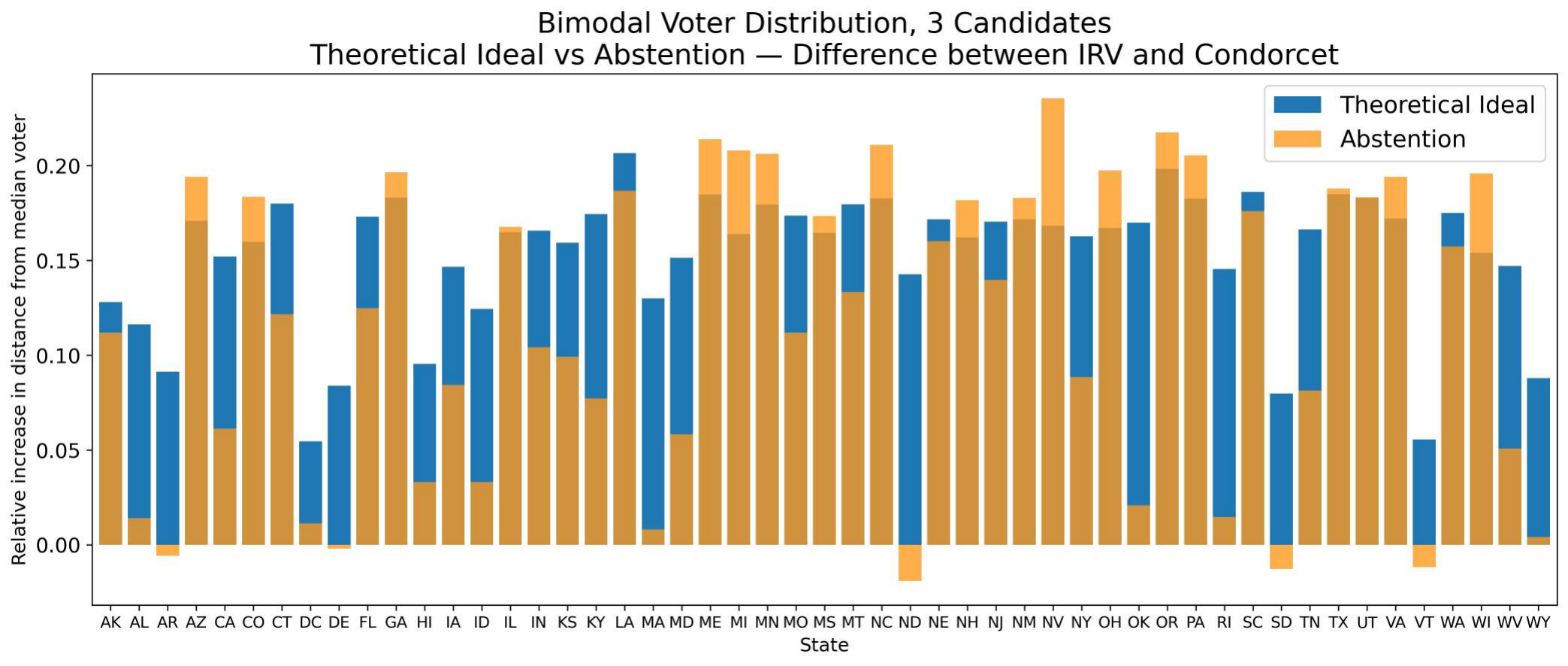}
\caption{The relative difference of the average distance of the IRV winner and the average distance of the Condorcet winner to the median voter, under the Theoretical Ideal and Abstention models. Bimodal Distribution, 3-Candidate Elections.}
\label{fig:avg_dist_bimodal_3cands_abstention}
\end{figure}

\begin{figure}[!h]
\includegraphics[width=\linewidth]{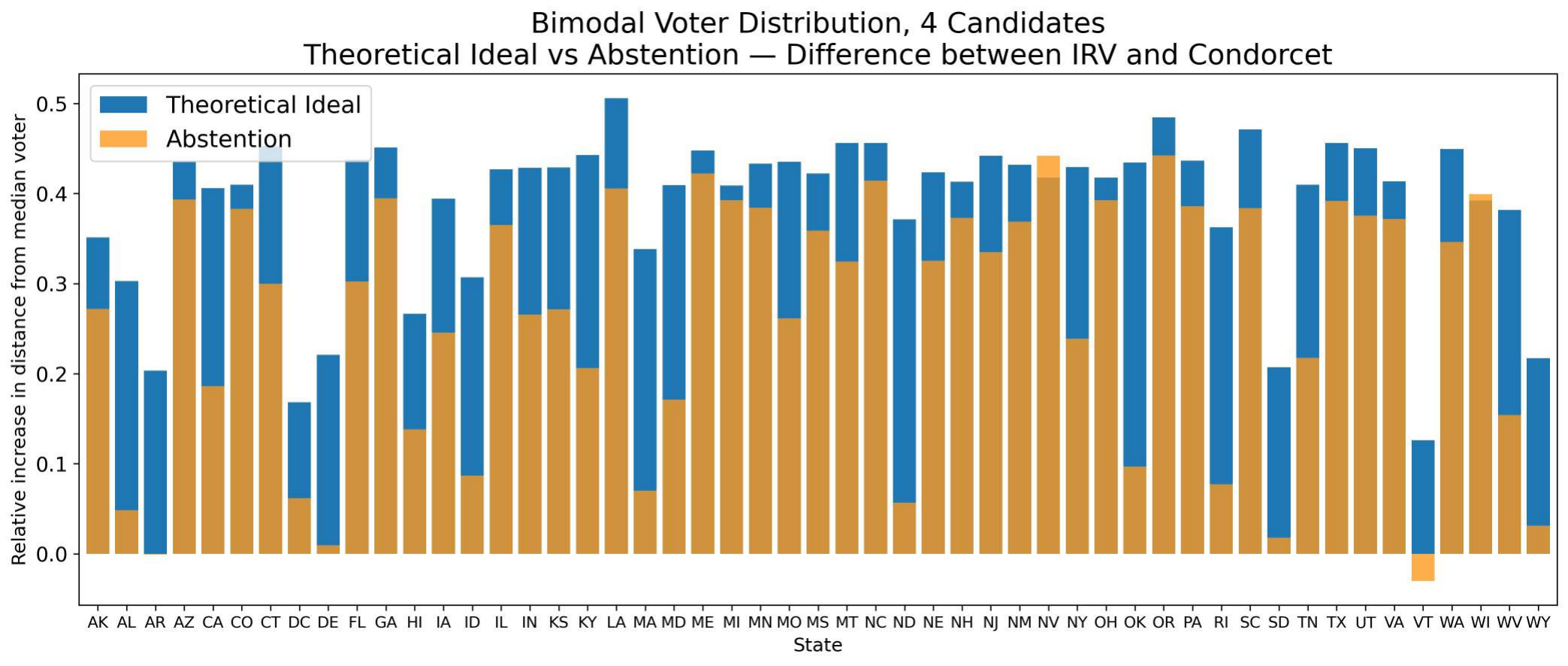}
\caption{The relative difference of the average distance of the IRV winner and the average distance of the Condorcet winner to the median voter, under the Theoretical Ideal and Abstention models. Bimodal Distribution, 4-Candidate Elections.}
\label{fig:avg_dist_bimodal_4cands_abstention}
\end{figure}

\begin{figure}[!h]
\includegraphics[width=\linewidth]{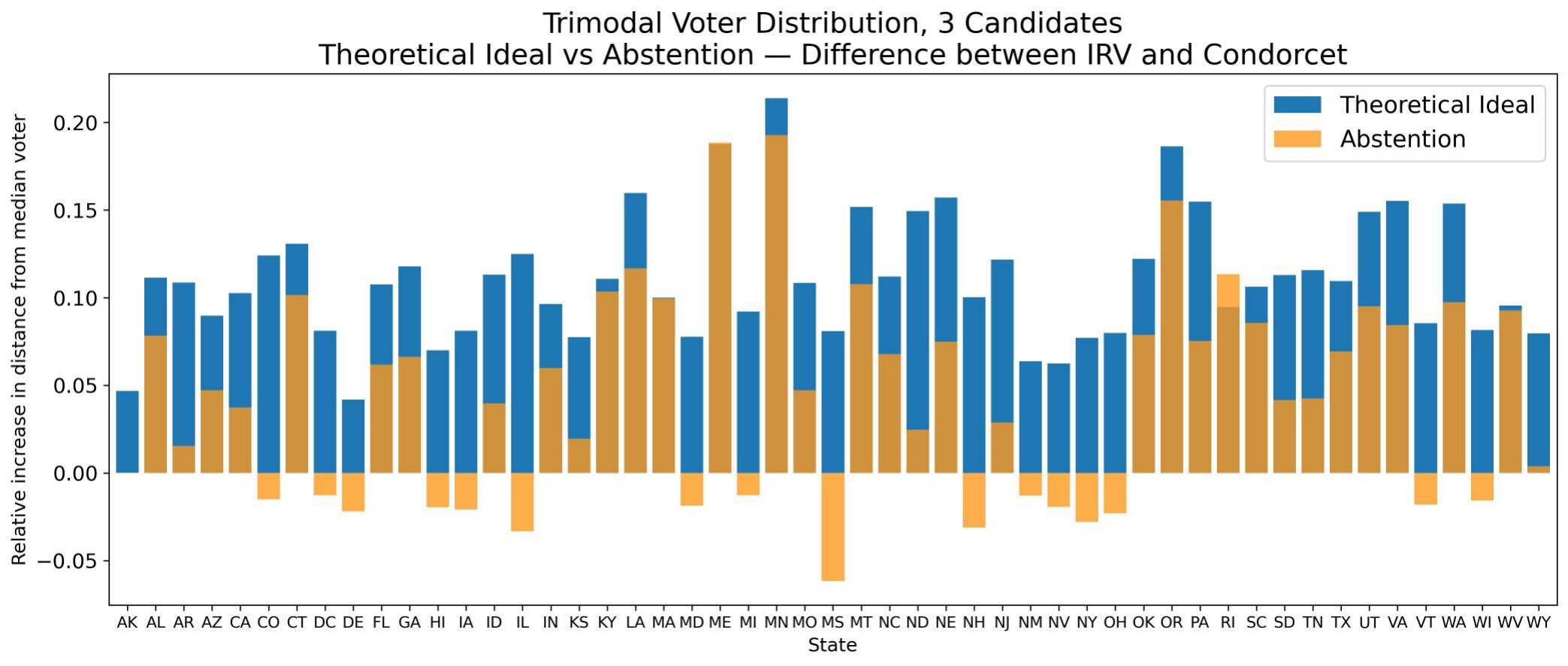}
\caption{The relative difference of the average distance of the IRV winner and the average distance of the Condorcet winner to the median voter, under the Theoretical Ideal and Abstention models. Trimodal Distribution, 3-Candidate Elections.}
\label{fig:avg_dist_trimodal_3cands_abstention}
\end{figure}

\begin{figure}[!h]
\includegraphics[width=\linewidth]{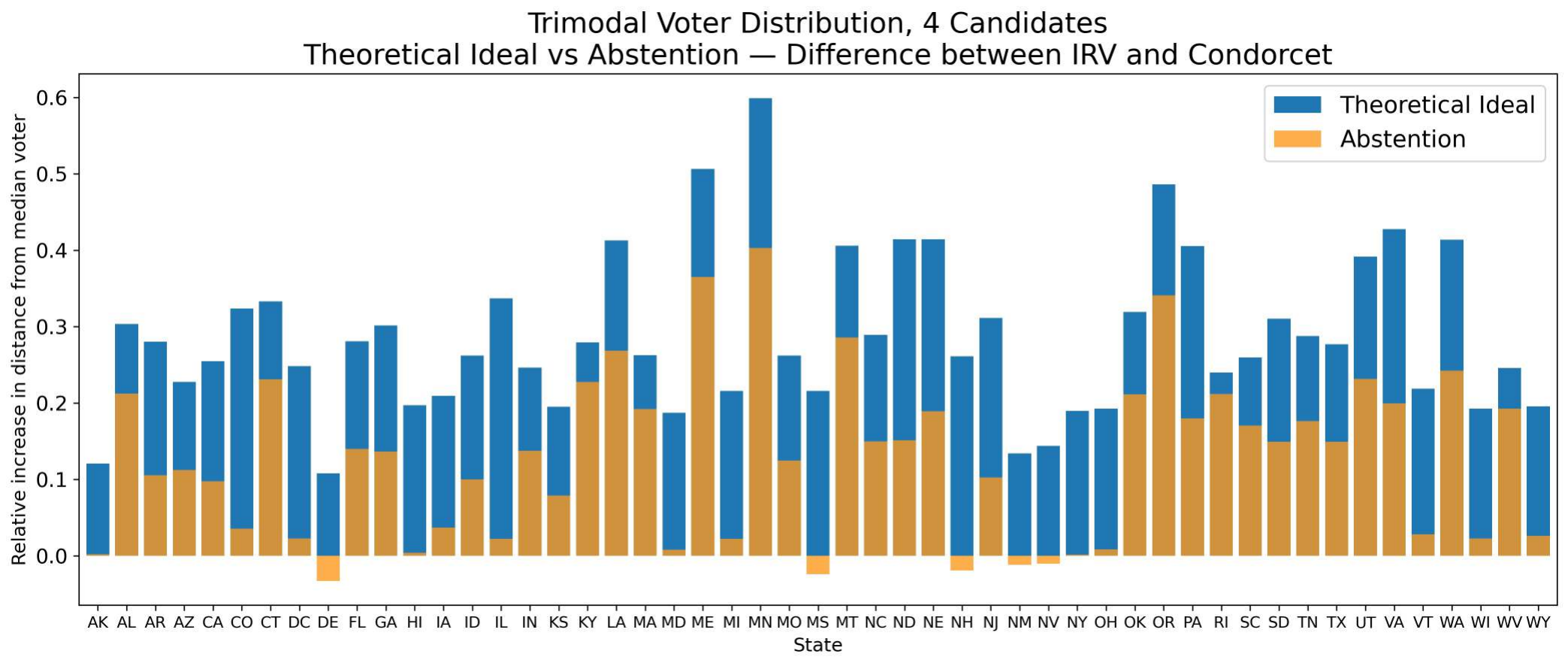}
\caption{The relative difference of the average distance of the IRV winner and the average distance of the Condorcet winner to the median voter, under the Theoretical Ideal and Abstention models. Trimodal Distribution, 4-Candidate Elections.}
\label{fig:avg_dist_trimodal_4cands_abstention}
\end{figure}

\begin{figure}[!h]
\includegraphics[width=\linewidth]{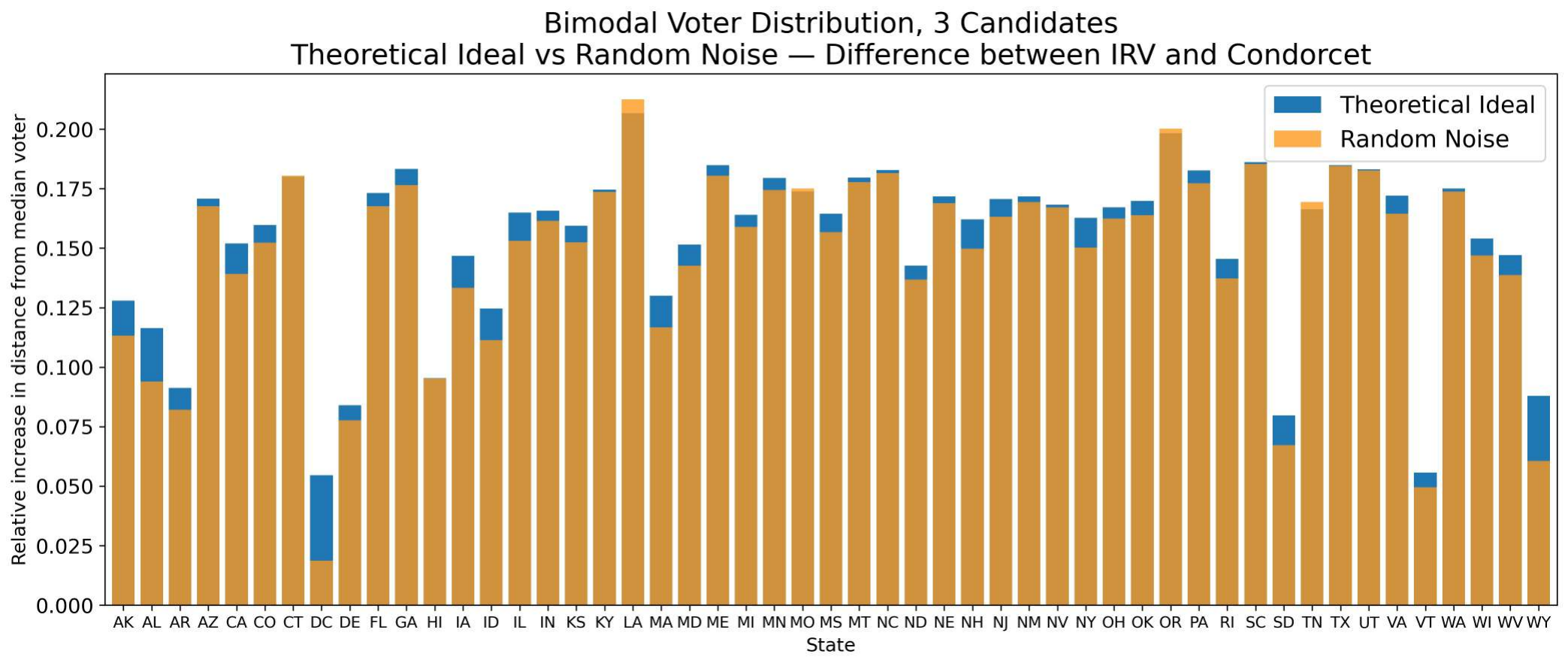}
\caption{The relative difference of the average distance of the IRV winner and the average distance of the Condorcet winner to the median voter, under the Theoretical Ideal and Noise models. Bimodal Distribution, 3-Candidate Elections.}
\label{fig:avg_dist_bimodal_3cands_noise}
\end{figure}

\begin{figure}[!h]
\includegraphics[width=\linewidth]{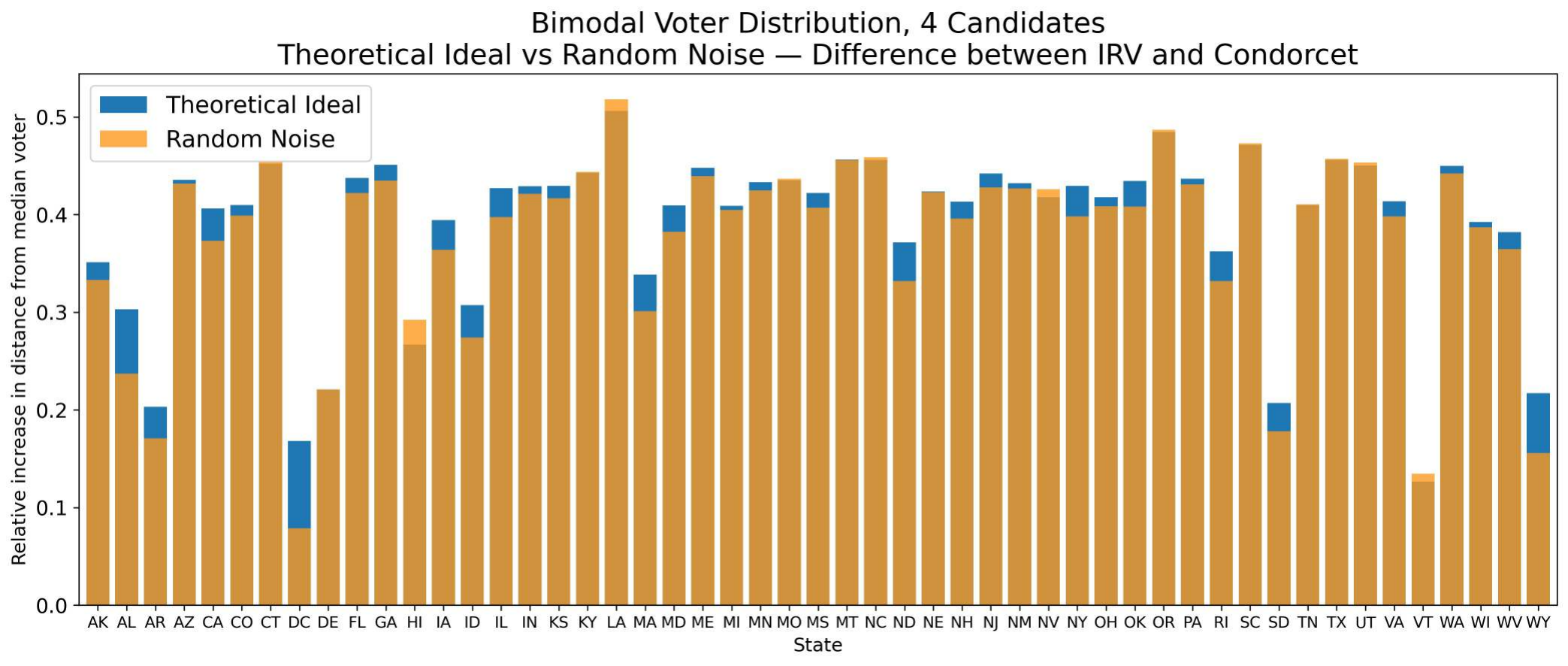}
\caption{The relative difference of the average distance of the IRV winner and the average distance of the Condorcet winner to the median voter, under the Theoretical Ideal and Noise models. Bimodal Distribution, 4-Candidate Elections.}
\label{fig:avg_dist_bimodal_4cands_noise}
\end{figure}

\begin{figure}[!h]
\includegraphics[width=\linewidth]{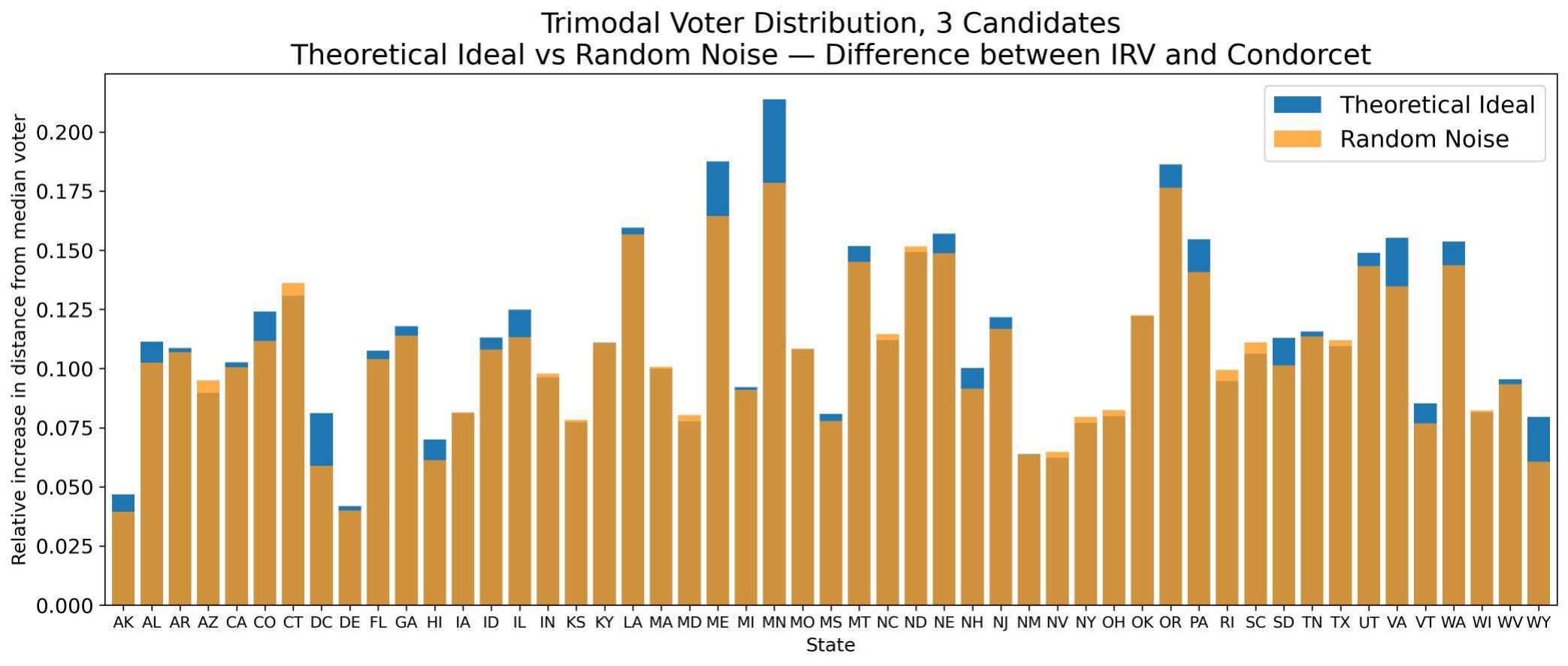}
\caption{The relative difference of the average distance of the IRV winner and the average distance of the Condorcet winner to the median voter, under the Theoretical Ideal and Noise models. Trimodal Distribution, 3-Candidate Elections.}
\label{fig:avg_dist_trimodal_3cands_noise}
\end{figure}

\begin{figure}[!h]
\includegraphics[width=\linewidth]{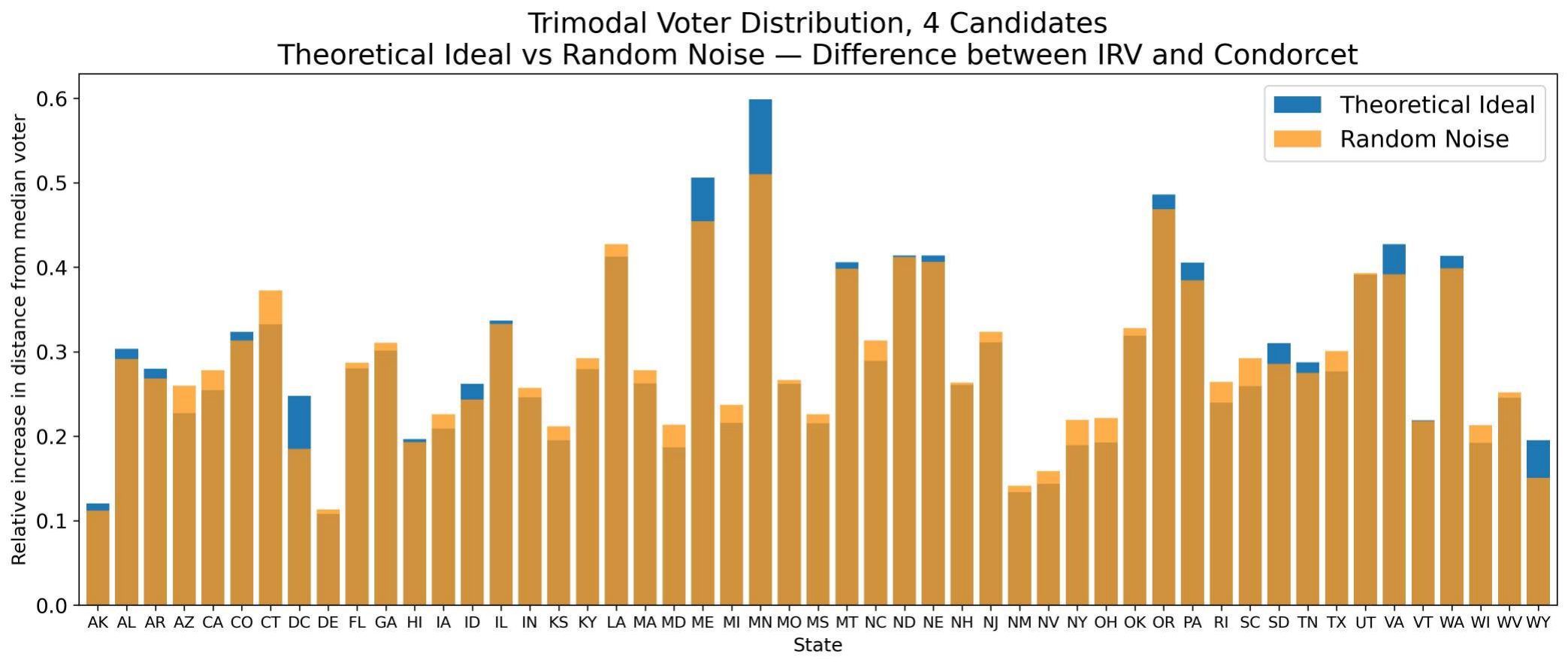}
\caption{The relative difference of the average distance of the IRV winner and the average distance of the Condorcet winner to the median voter, under the Theoretical Ideal and Noise models. Trimodal Distribution, 4-Candidate Elections.}
\label{fig:avg_dist_trimodal_4cands_noise}
\end{figure}

\begin{figure}[!h]
\includegraphics[width=\linewidth]{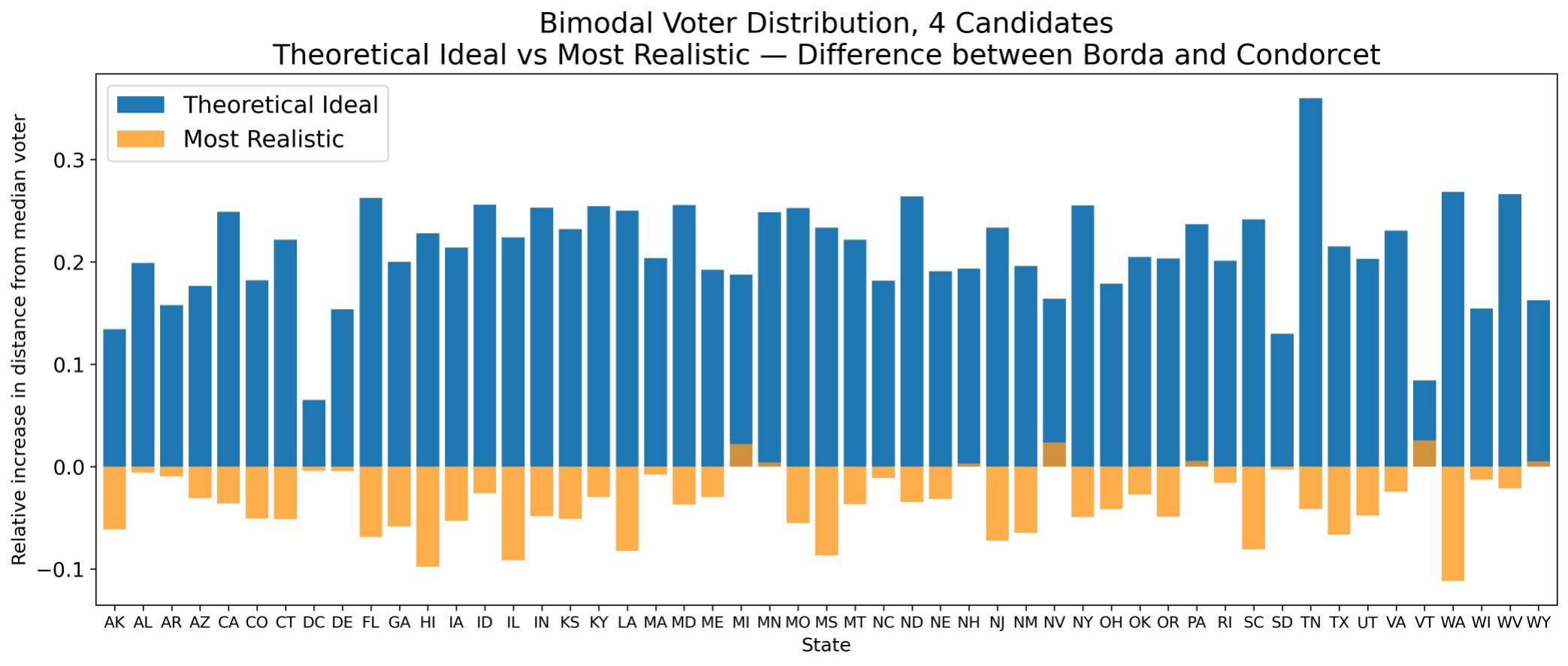}
\caption{The relative difference of the average distance of the Borda winner and the average distance of the Condorcet winner to the median voter, under the Theoretical Ideal and Most Realistic models. Bimodal Distribution, 4-Candidate Elections.}
\label{fig:avg_dist_Borda_bimodal_4cands}
\end{figure}

\begin{figure}[!h]
\includegraphics[width=\linewidth]{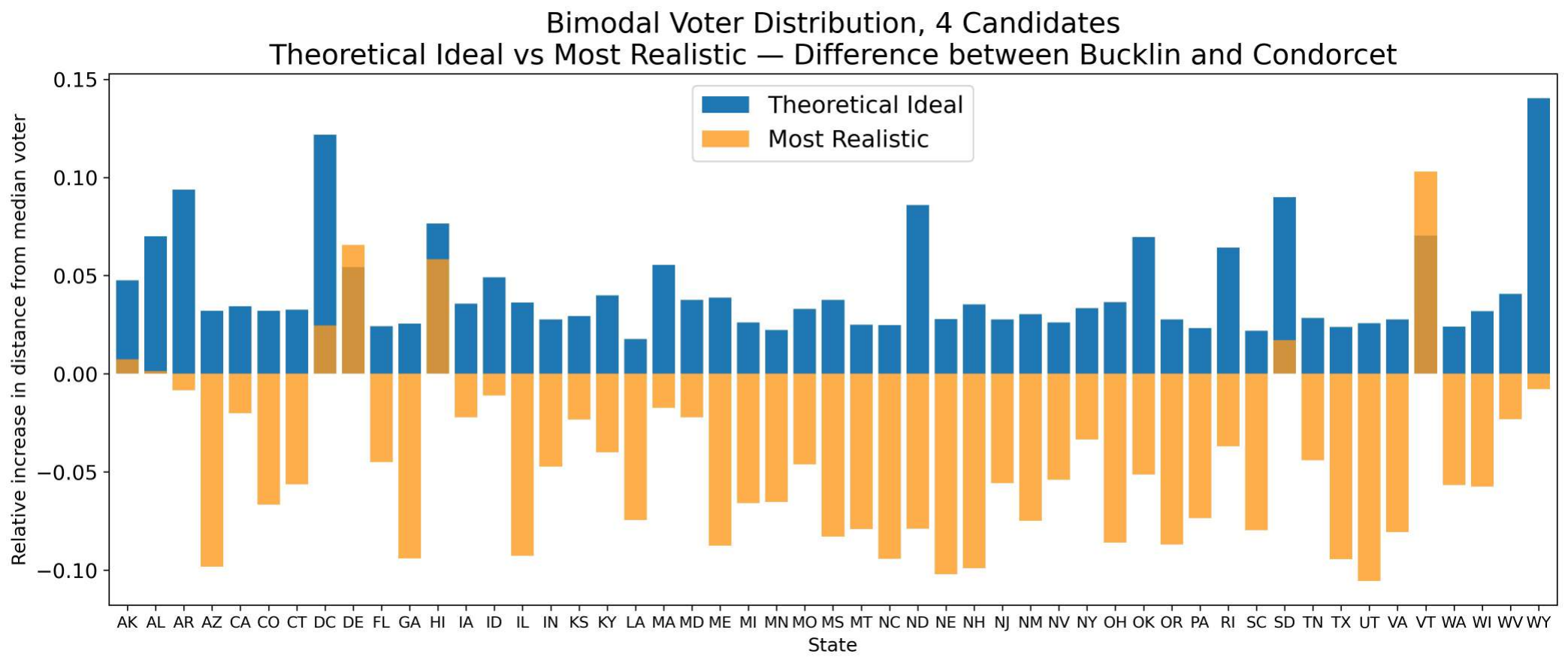}
\caption{The relative difference of the average distance of the Bucklin winner and the average distance of the Condorcet winner to the median voter, under the Theoretical Ideal and Most Realistic models. Bimodal Distribution, 4-Candidate Elections.}
\label{fig:avg_dist_Bucklin_bimodal_4cands}
\end{figure}
\clearpage

\section{Relative Differences and Changes for all models}\label{sec:all_mean_vars_plots}

\begin{figure}[!htbp]
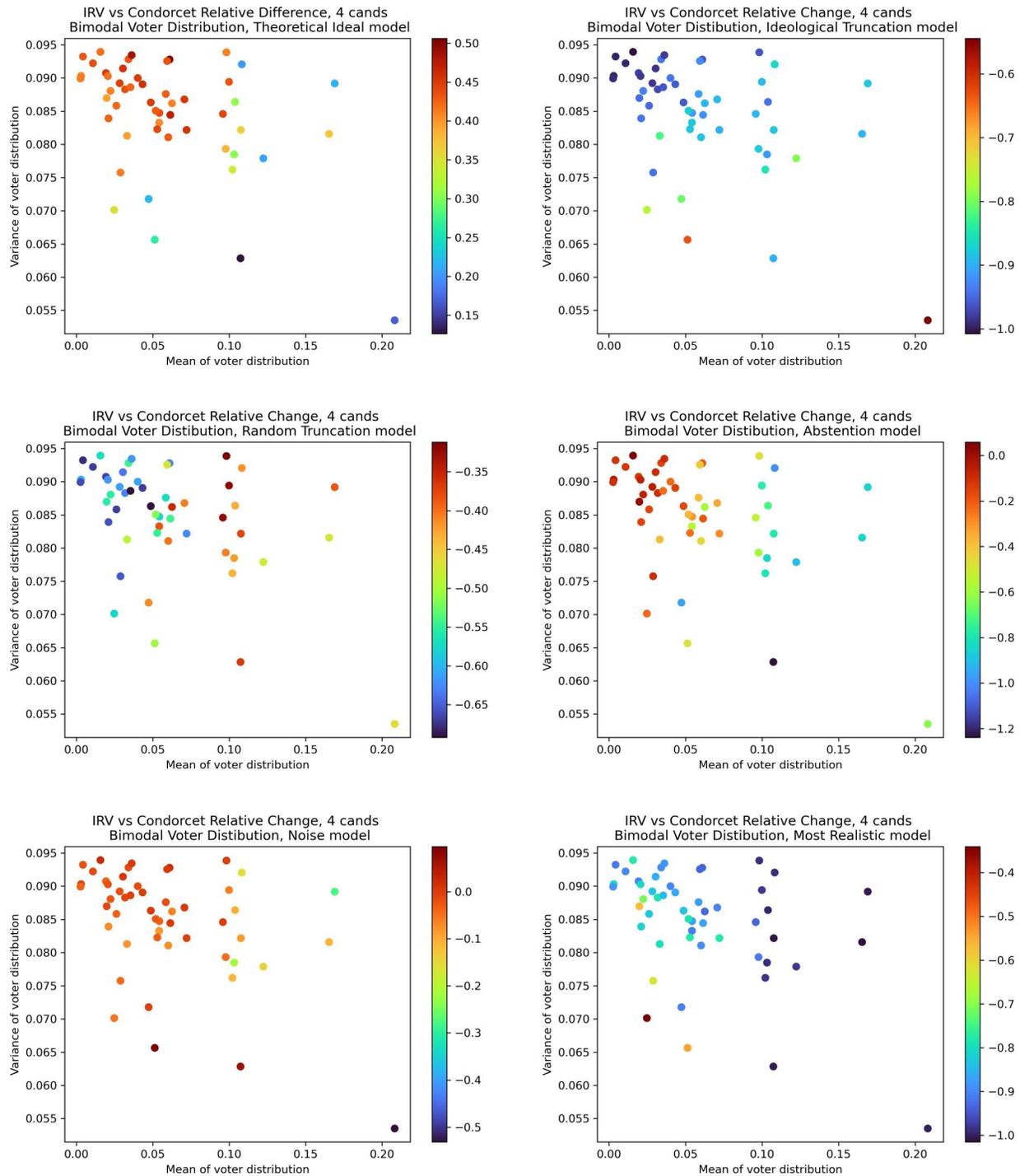

    \centering
    \begin{tabular}{@{}p{0.5\textwidth}p{0.5\textwidth}@{}}

    \includegraphics[width=\linewidth]{paper_1_images/bimodal_4cands_TI_base.pdf}
    &
    
    \includegraphics[width=\linewidth]{paper_1_images/bimodal_4cands_Ideological_Truncation.pdf}
    \\[6pt]

    \includegraphics[width=\linewidth]{paper_1_images/bimodal_4cands_Random_Truncation.pdf}
    &
    
    \includegraphics[width=\linewidth]{paper_1_images/Bimodal_4cands_Abstention.pdf}
    \\[6pt]

    \includegraphics[width=\linewidth]{paper_1_images/bimodal_4cands_Noise.pdf}
    &
    
    \includegraphics[width=\linewidth]{paper_1_images/Bimodal_4cands_Most_Realistic.pdf}
    \\
    
    \end{tabular}
    \caption{Plots showing the relative difference between IRV and Condorcet under the Theoretical Ideal model and the relative change under each variation of the model. Each point is a state (or D.C.) plotted according to the (symmetrized) mean and variance of the bimodal distribution, with electoral results coming from four candidate elections.}
\end{figure}

\begin{figure}[!htbp]
    \centering

    \begin{tabular}{@{}p{0.5\textwidth}p{0.5\textwidth}@{}}

    \includegraphics[width=\linewidth]{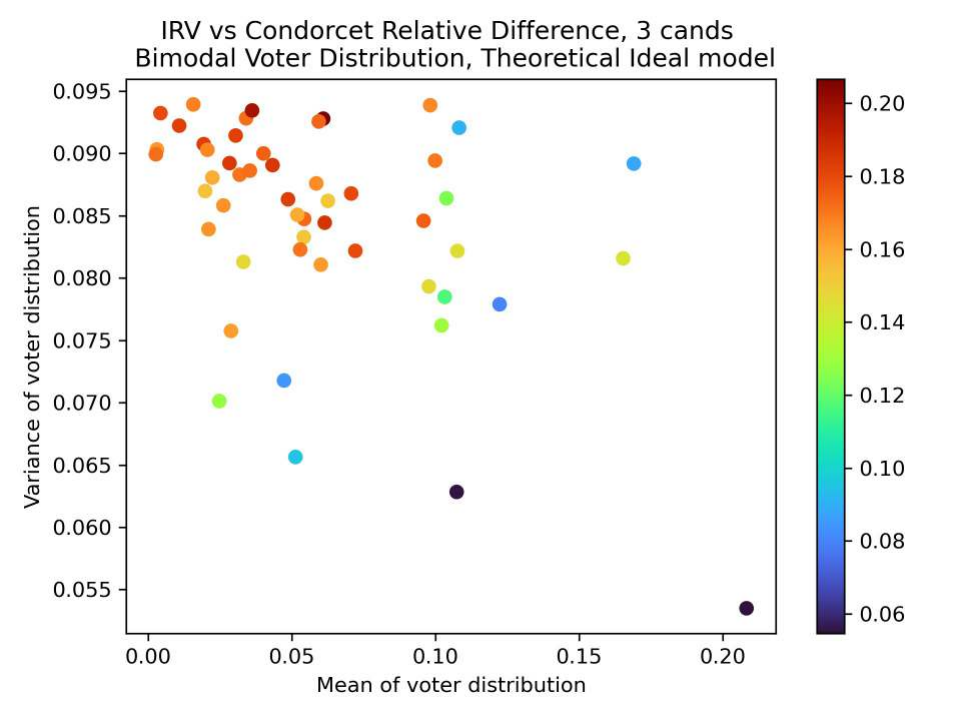}
    &
    
    \includegraphics[width=\linewidth]{paper_1_images/bimodal_3cands_Ideological_Truncation.pdf}
    \\[6pt]

    \includegraphics[width=\linewidth]{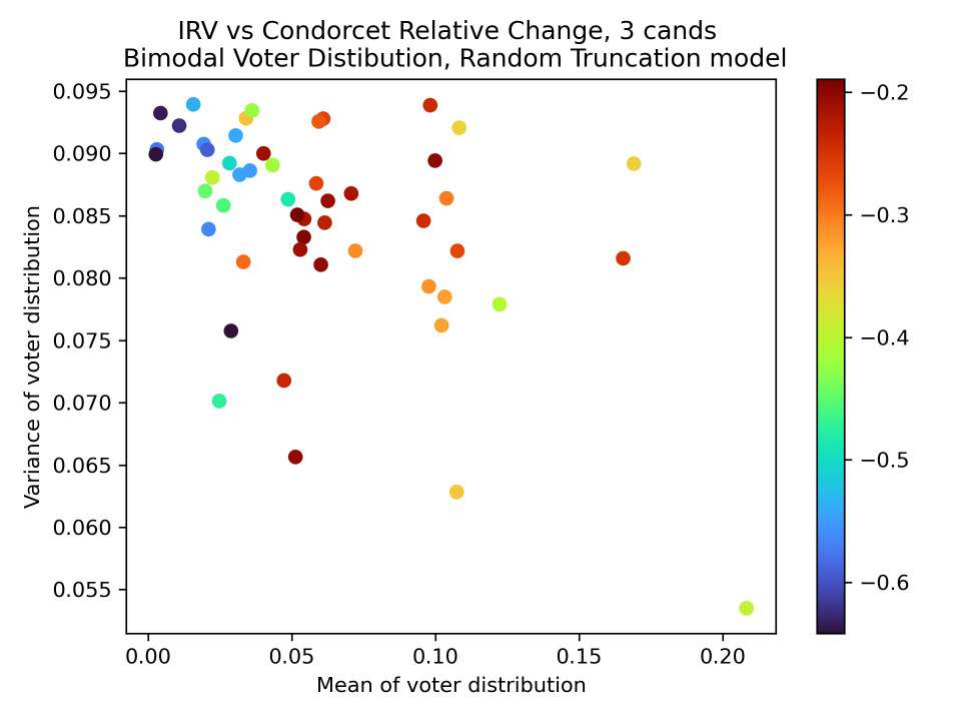}
    &
    
    \includegraphics[width=\linewidth]{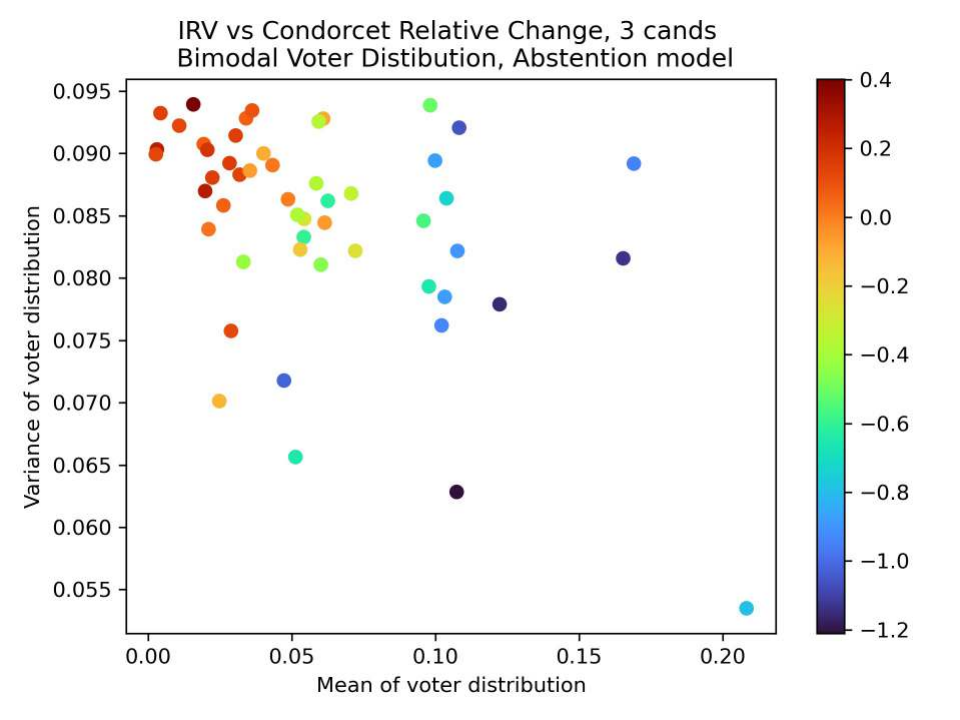}
    \\[6pt]

    \includegraphics[width=\linewidth]{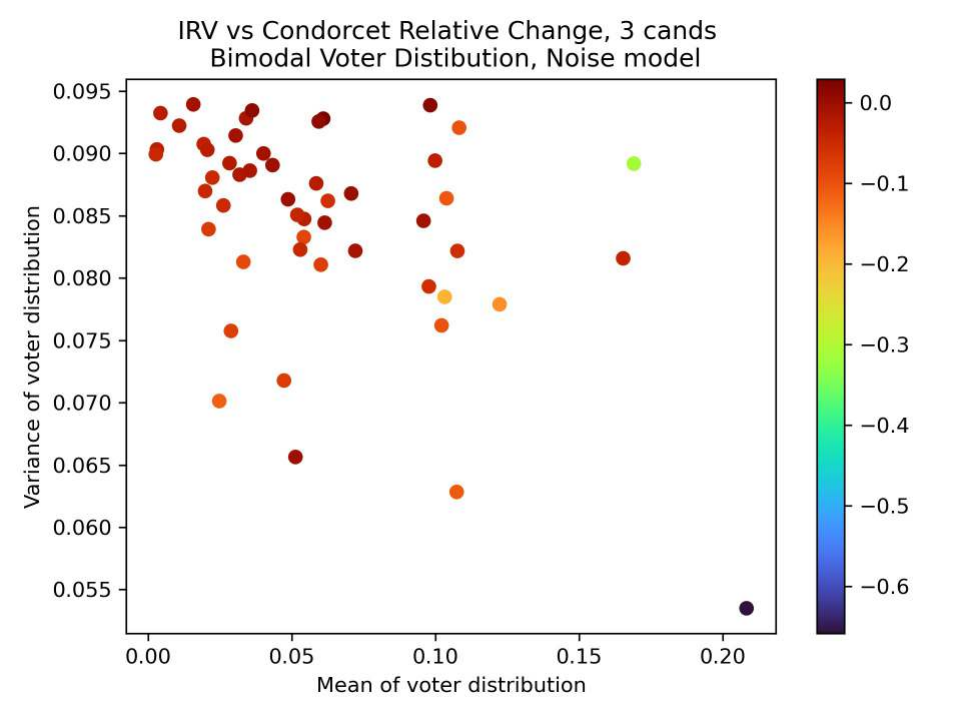}
    &
    
    \includegraphics[width=\linewidth]{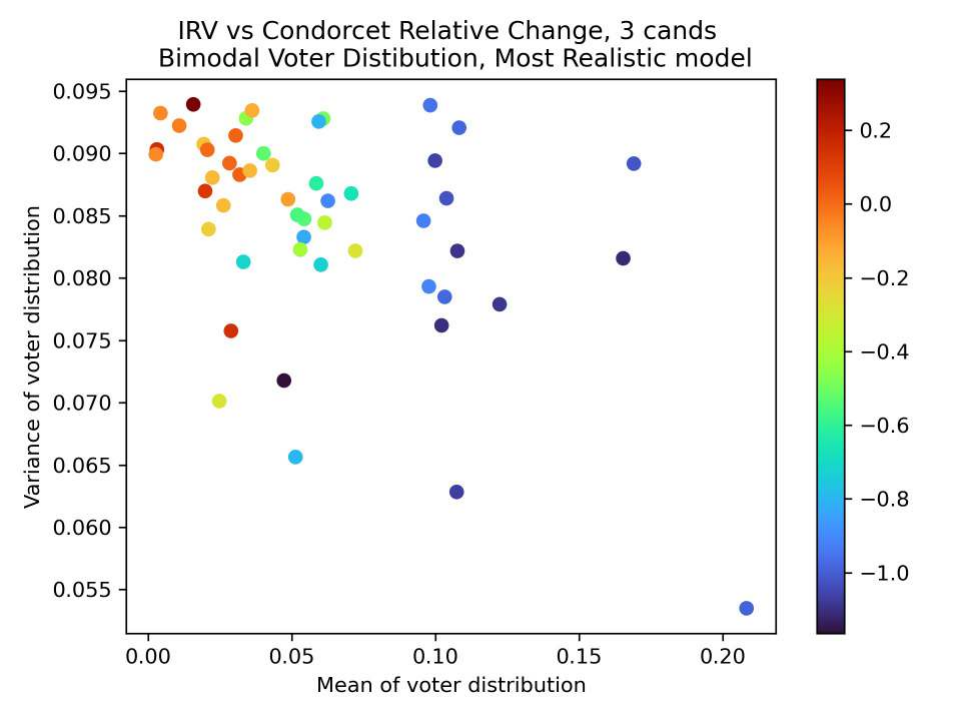}
    \\
    
    \end{tabular}

    \caption{Plots showing the relative difference between IRV and Condorcet under the Theoretical Ideal model and the relative change under each variation of the model. Each point is a state (or  D.C.) plotted according to the (symmetrized) mean and variance of the bimodal distribution, with electoral results coming from three candidate elections.}
\end{figure}

\begin{figure}[!htbp]
    \centering
    \begin{tabular}{@{}p{0.5\textwidth}p{0.5\textwidth}@{}}

    \includegraphics[width=\linewidth]{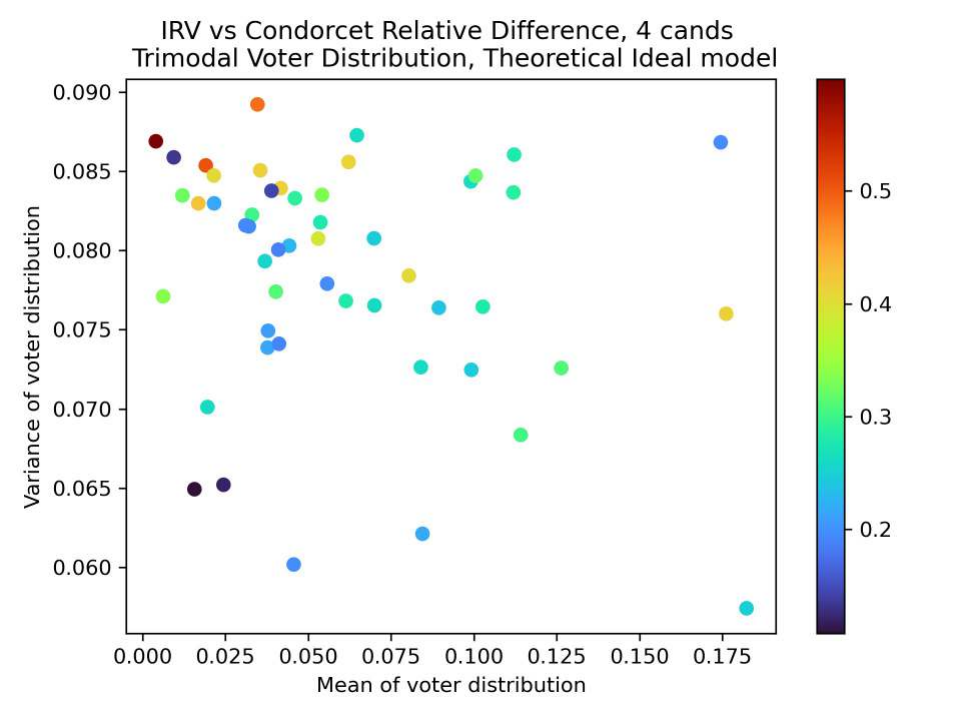}
    &
    
    \includegraphics[width=\linewidth]{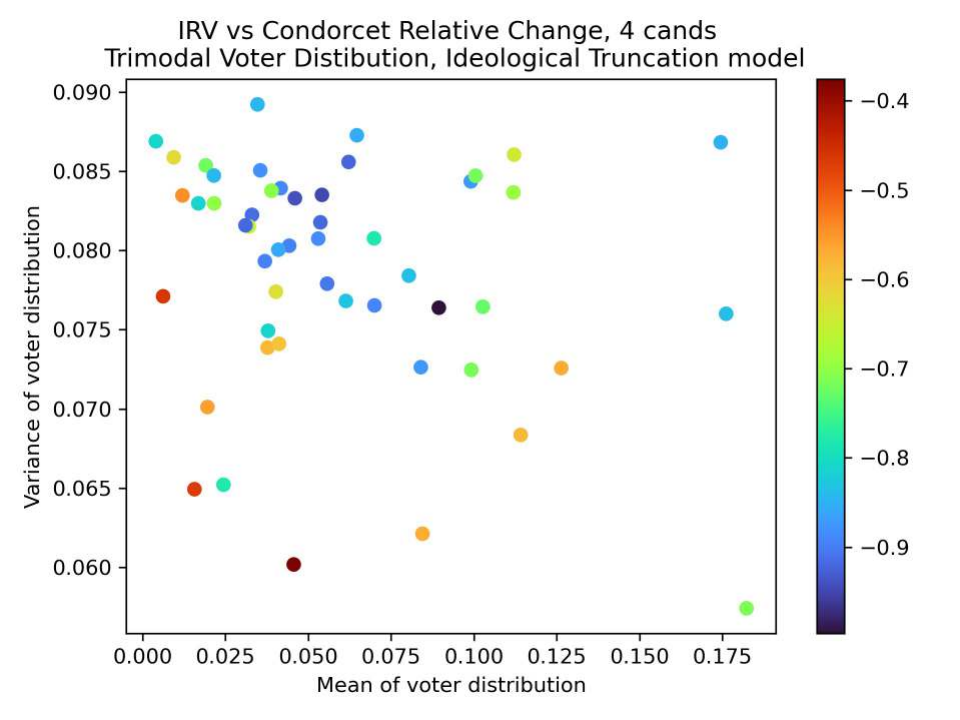}
    \\[6pt]

    \includegraphics[width=\linewidth]{paper_1_images/trimodal_4cands_Random_Truncation.pdf}
    &
    
    \includegraphics[width=\linewidth]{paper_1_images/trimodal_4cands_Abstention.pdf}
    \\[6pt]

    \includegraphics[width=\linewidth]{paper_1_images/trimodal_4cands_Noise.pdf}
    &
    
    \includegraphics[width=\linewidth]{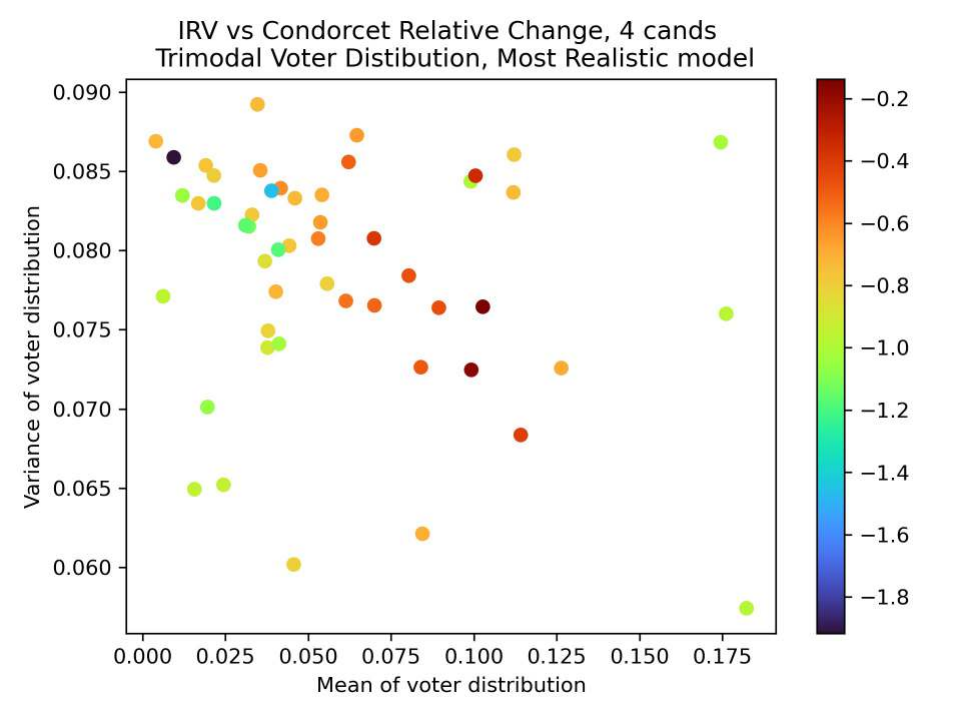}
    \\
    
    \end{tabular}

    \caption{Plots showing the relative difference between IRV and Condorcet under the Theoretical Ideal model and the relative change under each variation of the model. Each point is a state (or  D.C.) plotted according to the (symmetrized) mean and variance of the trimodal distribution, with electoral results coming from four candidate elections.}
\end{figure}

\begin{figure}[!htbp]
    \centering

    \begin{tabular}{@{}p{0.5\textwidth}p{0.5\textwidth}@{}}

    \includegraphics[width=\linewidth]{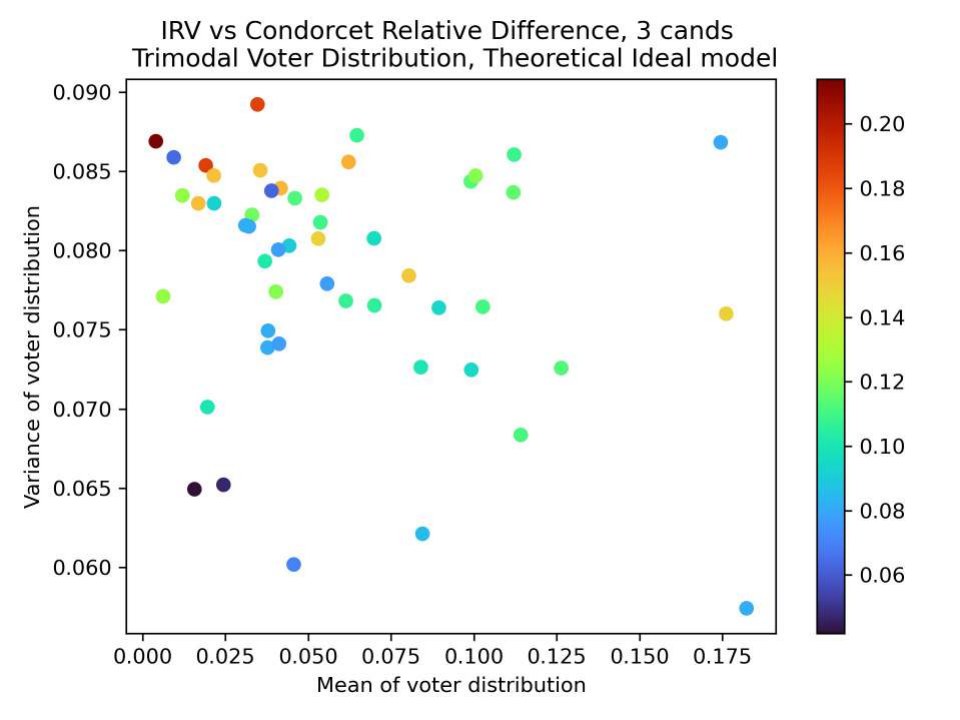}
    &
    
    \includegraphics[width=\linewidth]{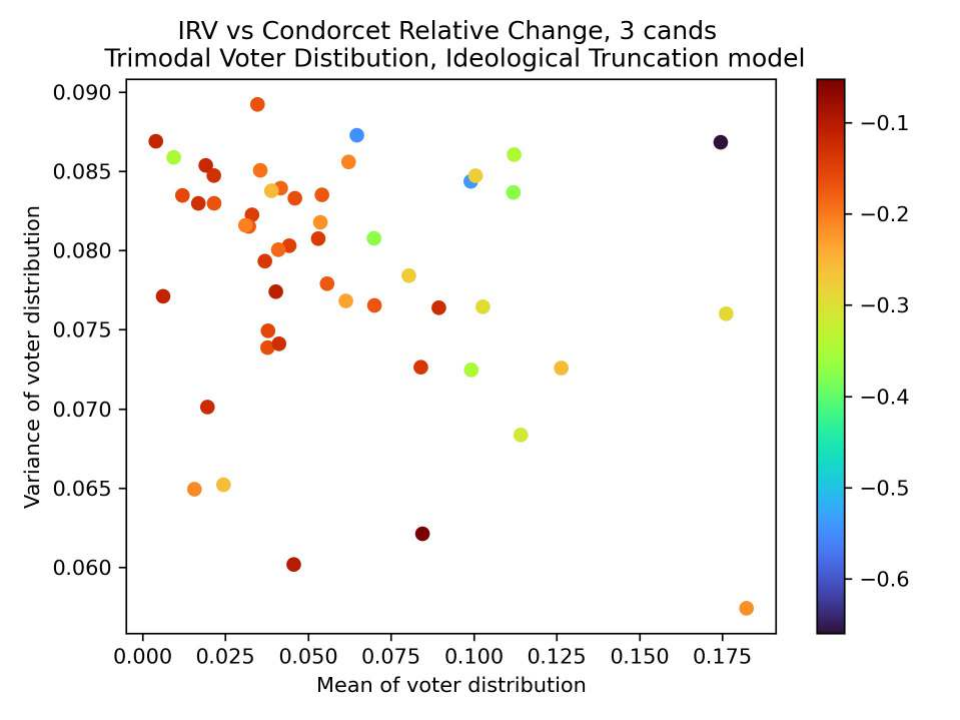}
    \\[6pt]

    \includegraphics[width=\linewidth]{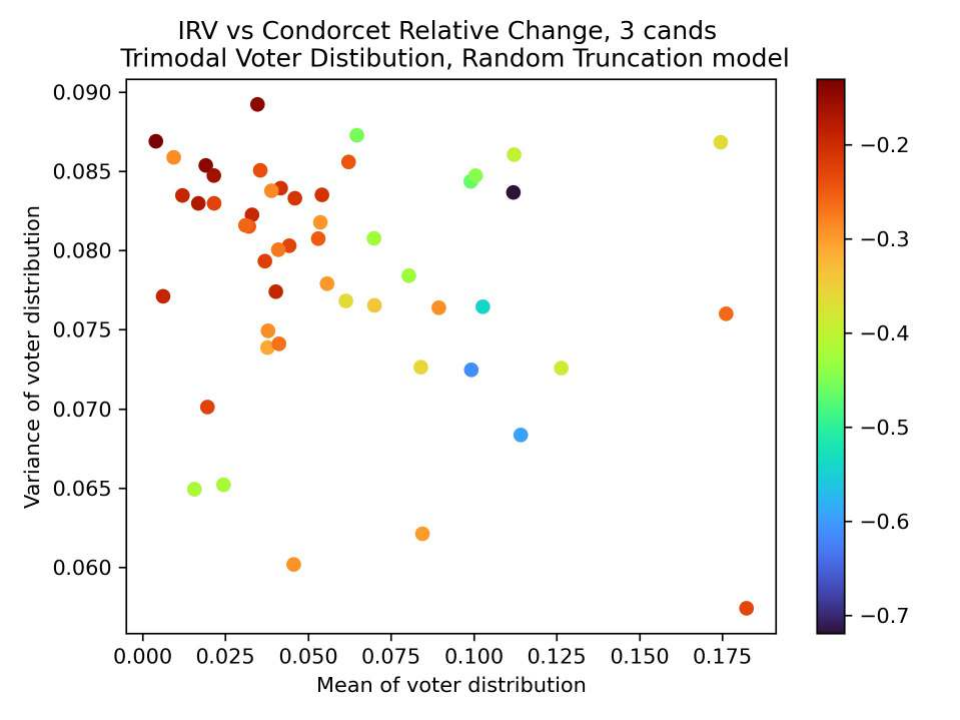}
    &
    
    \includegraphics[width=\linewidth]{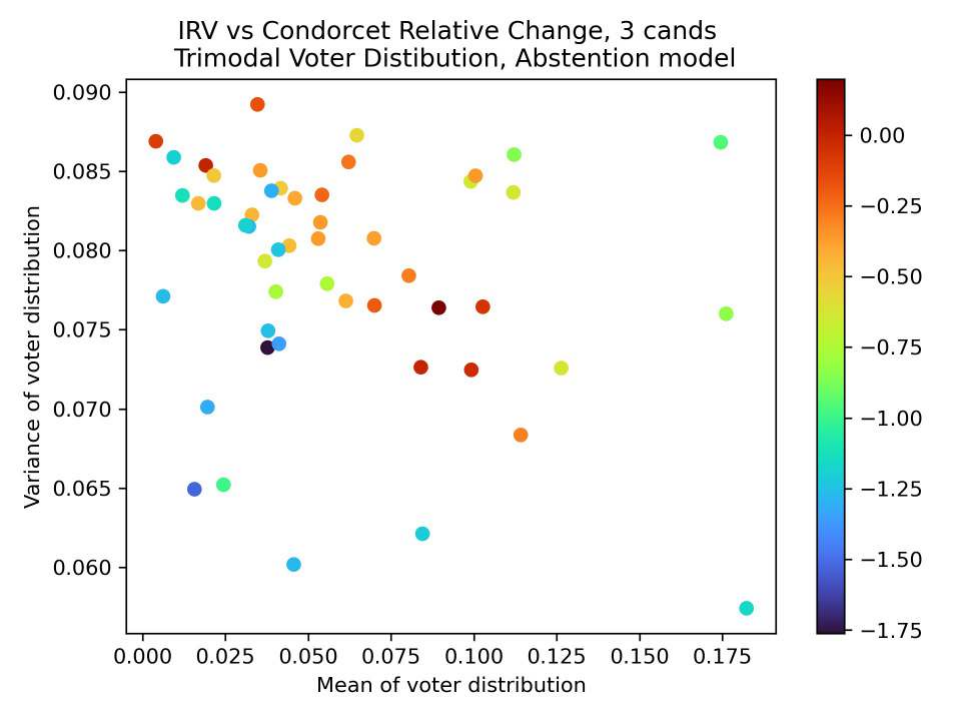}
    \\[6pt]

    \includegraphics[width=\linewidth]{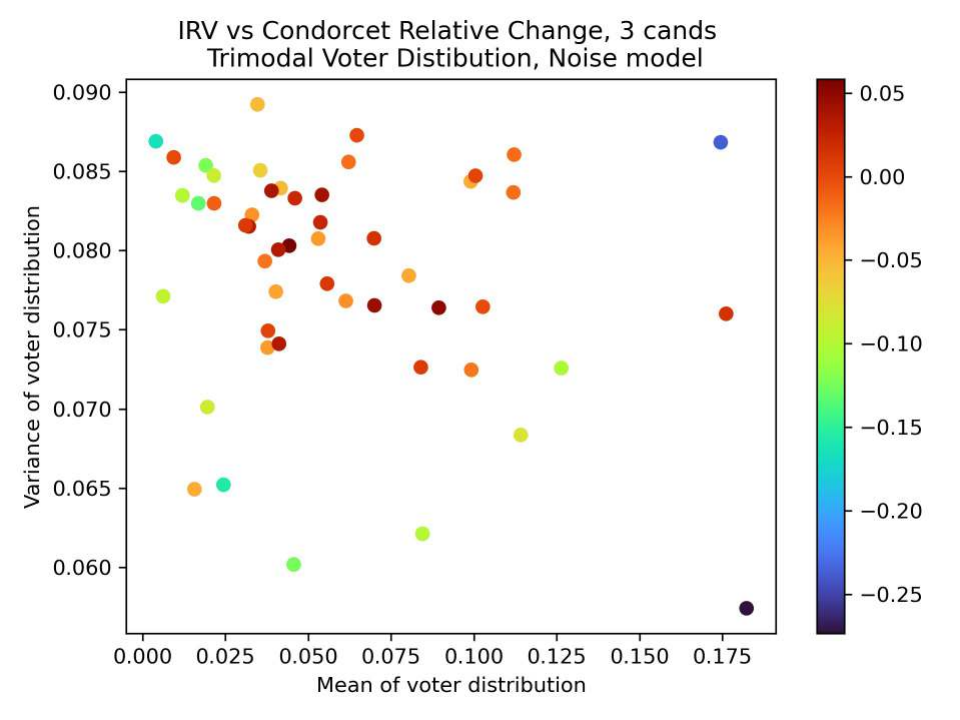}
    &
    
    \includegraphics[width=\linewidth]{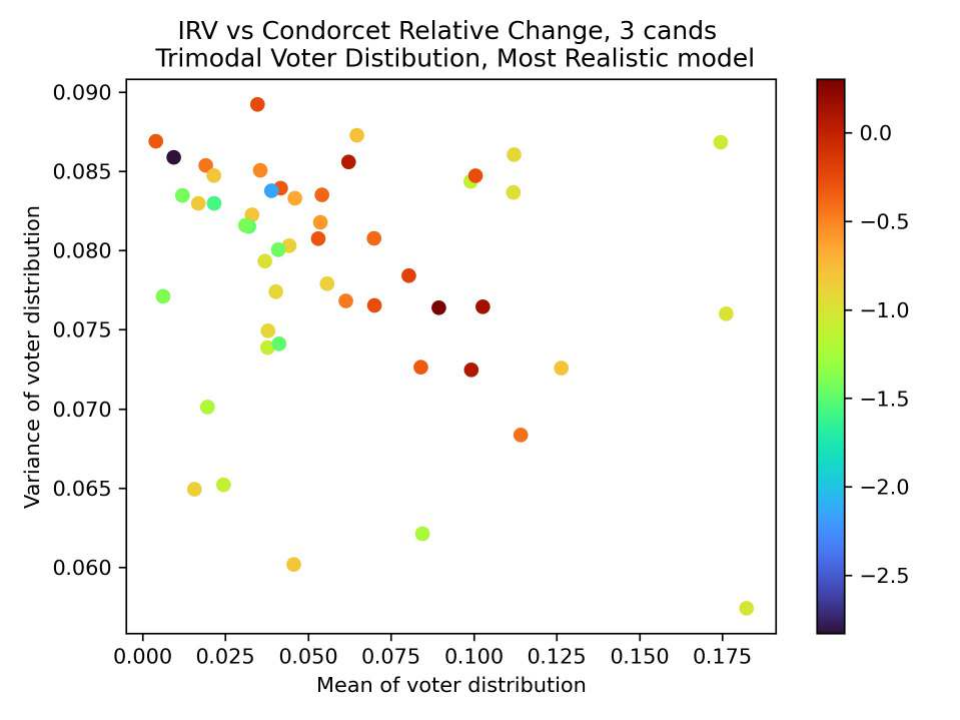}
    \\
    
    \end{tabular}
    
    \caption{Plots showing the relative difference between IRV and Condorcet under the Theoretical Ideal model and the relative change under each variation of the model. Each point is a state (or  D.C.) plotted according to the (symmetrized) mean and variance of the trimodal distribution, with electoral results coming from three candidate elections.}
\end{figure}

\newpage 

\section{State Distribution Summary Statistics}\label{sec:appendix_state_distribution_info}

\begin{table}[!htbp]
    \centering

    {\footnotesize
    \begin{tabular}{l|l|l|l|l|l|l}
State & Abstention & Bullet Rate & Bullet Rate & Distribution & Distribution & Median \\
& Rate & (Ideological) & (Random) & Mean & Variance & Voter \\
\hline

Alabama & (0.35, 0.35) & (0.33, 0.27) & (0.35, 0.35) & (0.1, 0.11) & (0.08, 0.07) & (0.18, 0.15) \\
Alaska & (0.38, 0.36) & (0.3, 0.27) & (0.35, 0.35) & (0.02, 0.02) & (0.07, 0.07) & (0.02, 0.0) \\
Arizona & (0.39, 0.4) & (0.39, 0.34) & (0.35, 0.35) & (0.03, 0.04) & (0.09, 0.08) & (0.08, 0.04) \\
Arkansas & (0.36, 0.37) & (0.35, 0.33) & (0.35, 0.35) & (0.11, 0.11) & (0.09, 0.09) & (0.2, 0.19) \\
California & (0.38, 0.4) & (0.36, 0.34) & (0.35, 0.35) & (-0.05, -0.04) & (0.08, 0.08) & (-0.13, -0.04) \\
Colorado & (0.4, 0.4) & (0.38, 0.36) & (0.35, 0.35) & (-0.02, -0.01) & (0.09, 0.08) & (-0.07, -0.01) \\
Connecticut & (0.38, 0.39) & (0.37, 0.36) & (0.35, 0.35) & (-0.07, -0.05) & (0.09, 0.08) & (-0.13, -0.05) \\
Delaware & (0.37, 0.37) & (0.3, 0.27) & (0.35, 0.35) & (-0.05, -0.02) & (0.07, 0.06) & (-0.11, -0.02) \\
District of Columbia & (0.23, 0.29) & (0.19, 0.22) & (0.35, 0.35) & (-0.21, -0.18) & (0.05, 0.06) & (-0.28, -0.26) \\
Florida & (0.37, 0.39) & (0.37, 0.32) & (0.35, 0.35) & (0.05, 0.06) & (0.08, 0.08) & (0.13, 0.08) \\
Georgia & (0.38, 0.4) & (0.4, 0.35) & (0.35, 0.35) & (0.02, 0.03) & (0.09, 0.08) & (0.07, 0.04) \\
Hawaii & (0.36, 0.35) & (0.27, 0.24) & (0.35, 0.35) & (-0.05, -0.05) & (0.07, 0.06) & (-0.1, -0.03) \\
Idaho & (0.35, 0.37) & (0.37, 0.36) & (0.35, 0.35) & (0.1, 0.1) & (0.09, 0.08) & (0.21, 0.19) \\
Illinois & (0.38, 0.39) & (0.38, 0.33) & (0.35, 0.35) & (-0.02, -0.01) & (0.08, 0.08) & (-0.09, -0.01) \\
Indiana & (0.38, 0.39) & (0.38, 0.34) & (0.35, 0.35) & (0.06, 0.07) & (0.09, 0.08) & (0.13, 0.09) \\
Iowa & (0.39, 0.39) & (0.35, 0.31) & (0.35, 0.35) & (0.03, 0.04) & (0.08, 0.07) & (0.1, 0.05) \\
Kansas & (0.39, 0.39) & (0.36, 0.33) & (0.35, 0.35) & (0.05, 0.06) & (0.09, 0.08) & (0.12, 0.06) \\
Kentucky & (0.37, 0.37) & (0.36, 0.33) & (0.35, 0.35) & (0.1, 0.1) & (0.08, 0.08) & (0.17, 0.12) \\
Louisiana & (0.37, 0.39) & (0.4, 0.37) & (0.35, 0.35) & (0.06, 0.06) & (0.09, 0.09) & (0.13, 0.06) \\
Maine & (0.39, 0.4) & (0.39, 0.37) & (0.35, 0.35) & (-0.03, -0.02) & (0.09, 0.09) & (-0.08, -0.01) \\
Maryland & (0.38, 0.4) & (0.37, 0.34) & (0.35, 0.35) & (-0.06, -0.04) & (0.09, 0.08) & (-0.14, -0.04) \\
Massachusetts & (0.36, 0.38) & (0.32, 0.31) & (0.35, 0.35) & (-0.1, -0.08) & (0.08, 0.07) & (-0.17, -0.08) \\
Michigan & (0.4, 0.4) & (0.39, 0.36) & (0.35, 0.35) & (0.0, 0.02) & (0.09, 0.08) & (-0.01, 0.03) \\
Minnesota & (0.38, 0.4) & (0.41, 0.37) & (0.35, 0.35) & (-0.0, 0.0) & (0.09, 0.09) & (-0.02, 0.01) \\
Mississippi & (0.38, 0.39) & (0.38, 0.31) & (0.35, 0.35) & (0.03, 0.04) & (0.09, 0.07) & (0.1, 0.04) \\
Missouri & (0.39, 0.39) & (0.39, 0.37) & (0.35, 0.35) & (0.06, 0.06) & (0.09, 0.09) & (0.14, 0.08) \\
Montana & (0.37, 0.38) & (0.36, 0.34) & (0.35, 0.35) & (0.07, 0.08) & (0.08, 0.08) & (0.13, 0.1) \\
Nebraska & (0.39, 0.4) & (0.39, 0.36) & (0.35, 0.35) & (0.04, 0.04) & (0.09, 0.08) & (0.07, 0.04) \\
Nevada & (0.4, 0.4) & (0.4, 0.36) & (0.35, 0.35) & (0.02, 0.04) & (0.09, 0.08) & (0.02, 0.03) \\
New Hampshire & (0.37, 0.38) & (0.34, 0.3) & (0.35, 0.35) & (-0.03, -0.02) & (0.08, 0.07) & (-0.05, -0.0) \\
New Jersey & (0.38, 0.39) & (0.36, 0.33) & (0.35, 0.35) & (-0.05, -0.04) & (0.08, 0.08) & (-0.11, -0.03) \\
New Mexico & (0.4, 0.4) & (0.39, 0.38) & (0.35, 0.35) & (-0.03, -0.01) & (0.09, 0.09) & (-0.09, -0.01) \\
New York & (0.37, 0.39) & (0.35, 0.31) & (0.35, 0.35) & (-0.06, -0.04) & (0.08, 0.07) & (-0.13, -0.04) \\
North Carolina & (0.39, 0.4) & (0.4, 0.36) & (0.35, 0.35) & (0.03, 0.05) & (0.09, 0.08) & (0.06, 0.05) \\
North Dakota & (0.34, 0.33) & (0.33, 0.31) & (0.35, 0.35) & (0.17, 0.18) & (0.08, 0.08) & (0.27, 0.27) \\
Ohio & (0.39, 0.4) & (0.39, 0.35) & (0.35, 0.35) & (0.02, 0.03) & (0.09, 0.08) & (0.07, 0.03) \\
Oklahoma & (0.38, 0.39) & (0.36, 0.35) & (0.35, 0.35) & (0.1, 0.1) & (0.09, 0.08) & (0.17, 0.14) \\
Oregon & (0.39, 0.4) & (0.4, 0.38) & (0.35, 0.35) & (-0.04, -0.03) & (0.09, 0.09) & (-0.09, -0.03) \\
Pennsylvania & (0.39, 0.4) & (0.4, 0.36) & (0.35, 0.35) & (0.01, 0.02) & (0.09, 0.08) & (0.02, 0.02) \\
Rhode Island & (0.37, 0.37) & (0.34, 0.34) & (0.35, 0.35) & (-0.11, -0.09) & (0.08, 0.08) & (-0.18, -0.06) \\
South Carolina & (0.37, 0.38) & (0.38, 0.33) & (0.35, 0.35) & (0.06, 0.07) & (0.08, 0.08) & (0.12, 0.07) \\
South Dakota & (0.35, 0.35) & (0.3, 0.28) & (0.35, 0.35) & (0.12, 0.13) & (0.08, 0.07) & (0.2, 0.19) \\
Tennessee & (0.35, 0.37) & (0.4, 0.35) & (0.35, 0.35) & (0.1, 0.11) & (0.09, 0.08) & (0.21, 0.18) \\
Texas & (0.38, 0.4) & (0.39, 0.35) & (0.35, 0.35) & (0.04, 0.05) & (0.09, 0.08) & (0.1, 0.06) \\
Utah & (0.38, 0.39) & (0.38, 0.35) & (0.35, 0.35) & (0.05, 0.05) & (0.09, 0.08) & (0.1, 0.05) \\
Vermont & (0.34, 0.35) & (0.26, 0.26) & (0.35, 0.35) & (-0.11, -0.08) & (0.06, 0.06) & (-0.15, -0.07) \\
Virginia & (0.38, 0.4) & (0.4, 0.36) & (0.35, 0.35) & (-0.0, 0.02) & (0.09, 0.08) & (-0.05, 0.02) \\
Washington & (0.38, 0.4) & (0.39, 0.36) & (0.35, 0.35) & (-0.04, -0.04) & (0.09, 0.09) & (-0.12, -0.05) \\
West Virginia & (0.36, 0.37) & (0.34, 0.3) & (0.35, 0.35) & (0.1, 0.1) & (0.08, 0.07) & (0.18, 0.1) \\
Wisconsin & (0.4, 0.4) & (0.37, 0.35) & (0.35, 0.35) & (0.02, 0.03) & (0.09, 0.08) & (0.05, 0.04) \\
Wyoming & (0.3, 0.31) & (0.35, 0.33) & (0.35, 0.35) & (0.17, 0.17) & (0.09, 0.09) & (0.29, 0.29) \\
\end{tabular}
}
    \caption{A table giving summary statistics relevant to the model parameters for three candidate elections. In each pair, the first number is for the bimodal distribution and the second is for the trimodal distribution. The first three columns give the rates for abstention and bullet voting. The last three columns give statistics for the distributions themselves.}
    \label{tab:state_stats_three}
\end{table}

\begin{table}[!htbp]
    \centering

    {\small 
\begin{tabular}{lllllll}
State & Abstention & Bullet Rate & Bullet Rate & Distribution & Distribution & Median \\
& Rate & (Ideological) & (Random) & Mean & Variance & Voter \\
\hline
Alabama & (0.26, 0.26) & (0.33, 0.32) & (0.34, 0.34) & (0.1, 0.11) & (0.08, 0.07) & (0.18, 0.15) \\
Alaska & (0.28, 0.27) & (0.34, 0.34) & (0.34, 0.34) & (0.02, 0.02) & (0.07, 0.07) & (0.02, 0.0) \\
Arizona & (0.29, 0.29) & (0.39, 0.39) & (0.34, 0.34) & (0.03, 0.04) & (0.09, 0.08) & (0.08, 0.04) \\
Arkansas & (0.27, 0.27) & (0.35, 0.36) & (0.34, 0.34) & (0.11, 0.11) & (0.09, 0.09) & (0.2, 0.19) \\
California & (0.28, 0.29) & (0.37, 0.39) & (0.34, 0.34) & (-0.05, -0.04) & (0.08, 0.08) & (-0.13, -0.04) \\
Colorado & (0.3, 0.3) & (0.39, 0.4) & (0.34, 0.34) & (-0.02, -0.01) & (0.09, 0.08) & (-0.07, -0.01) \\
Connecticut & (0.28, 0.29) & (0.38, 0.4) & (0.34, 0.34) & (-0.07, -0.05) & (0.09, 0.08) & (-0.13, -0.05) \\
Delaware & (0.27, 0.27) & (0.34, 0.33) & (0.34, 0.34) & (-0.05, -0.02) & (0.07, 0.06) & (-0.11, -0.02) \\
District of Columbia & (0.17, 0.21) & (0.19, 0.25) & (0.34, 0.34) & (-0.21, -0.18) & (0.05, 0.06) & (-0.28, -0.26) \\
Florida & (0.28, 0.29) & (0.37, 0.37) & (0.34, 0.34) & (0.05, 0.06) & (0.08, 0.08) & (0.13, 0.08) \\
Georgia & (0.28, 0.29) & (0.39, 0.4) & (0.34, 0.34) & (0.02, 0.03) & (0.09, 0.08) & (0.07, 0.04) \\
Hawaii & (0.27, 0.26) & (0.33, 0.3) & (0.34, 0.34) & (-0.05, -0.05) & (0.07, 0.06) & (-0.1, -0.03) \\
Idaho & (0.25, 0.27) & (0.34, 0.36) & (0.34, 0.34) & (0.1, 0.1) & (0.09, 0.08) & (0.21, 0.19) \\
Illinois & (0.28, 0.29) & (0.37, 0.38) & (0.34, 0.34) & (-0.02, -0.01) & (0.08, 0.08) & (-0.09, -0.01) \\
Indiana & (0.28, 0.29) & (0.38, 0.38) & (0.34, 0.34) & (0.06, 0.07) & (0.09, 0.08) & (0.13, 0.09) \\
Iowa & (0.29, 0.29) & (0.37, 0.37) & (0.34, 0.34) & (0.03, 0.04) & (0.08, 0.07) & (0.1, 0.05) \\
Kansas & (0.29, 0.29) & (0.38, 0.38) & (0.34, 0.34) & (0.05, 0.06) & (0.09, 0.08) & (0.12, 0.06) \\
Kentucky & (0.27, 0.28) & (0.36, 0.37) & (0.34, 0.34) & (0.1, 0.1) & (0.08, 0.08) & (0.17, 0.12) \\
Louisiana & (0.28, 0.29) & (0.39, 0.4) & (0.34, 0.34) & (0.06, 0.06) & (0.09, 0.09) & (0.13, 0.06) \\
Maine & (0.29, 0.3) & (0.39, 0.41) & (0.34, 0.34) & (-0.03, -0.02) & (0.09, 0.09) & (-0.08, -0.01) \\
Maryland & (0.28, 0.29) & (0.37, 0.39) & (0.34, 0.34) & (-0.06, -0.04) & (0.09, 0.08) & (-0.14, -0.04) \\
Massachusetts & (0.27, 0.28) & (0.34, 0.36) & (0.34, 0.34) & (-0.1, -0.08) & (0.08, 0.07) & (-0.17, -0.08) \\
Michigan & (0.29, 0.3) & (0.4, 0.4) & (0.34, 0.34) & (0.0, 0.02) & (0.09, 0.08) & (-0.01, 0.03) \\
Minnesota & (0.28, 0.3) & (0.39, 0.41) & (0.34, 0.34) & (-0.0, 0.0) & (0.09, 0.09) & (-0.02, 0.01) \\
Mississippi & (0.28, 0.29) & (0.38, 0.36) & (0.34, 0.34) & (0.03, 0.04) & (0.09, 0.07) & (0.1, 0.04) \\
Missouri & (0.29, 0.29) & (0.39, 0.4) & (0.34, 0.34) & (0.06, 0.06) & (0.09, 0.09) & (0.14, 0.08) \\
Montana & (0.28, 0.28) & (0.37, 0.37) & (0.34, 0.34) & (0.07, 0.08) & (0.08, 0.08) & (0.13, 0.1) \\
Nebraska & (0.29, 0.29) & (0.39, 0.4) & (0.34, 0.34) & (0.04, 0.04) & (0.09, 0.08) & (0.07, 0.04) \\
Nevada & (0.3, 0.3) & (0.41, 0.42) & (0.34, 0.34) & (0.02, 0.04) & (0.09, 0.08) & (0.02, 0.03) \\
New Hampshire & (0.27, 0.28) & (0.35, 0.35) & (0.34, 0.34) & (-0.03, -0.02) & (0.08, 0.07) & (-0.05, -0.0) \\
New Jersey & (0.28, 0.29) & (0.37, 0.38) & (0.34, 0.34) & (-0.05, -0.04) & (0.08, 0.08) & (-0.11, -0.03) \\
New Mexico & (0.3, 0.3) & (0.4, 0.43) & (0.34, 0.34) & (-0.03, -0.01) & (0.09, 0.09) & (-0.09, -0.01) \\
New York & (0.27, 0.29) & (0.36, 0.37) & (0.34, 0.34) & (-0.06, -0.04) & (0.08, 0.07) & (-0.13, -0.04) \\
North Carolina & (0.29, 0.3) & (0.39, 0.4) & (0.34, 0.34) & (0.03, 0.05) & (0.09, 0.08) & (0.06, 0.05) \\
North Dakota & (0.26, 0.24) & (0.34, 0.33) & (0.34, 0.34) & (0.17, 0.18) & (0.08, 0.08) & (0.27, 0.27) \\
Ohio & (0.29, 0.3) & (0.4, 0.4) & (0.34, 0.34) & (0.02, 0.03) & (0.09, 0.08) & (0.07, 0.03) \\
Oklahoma & (0.28, 0.29) & (0.38, 0.39) & (0.34, 0.34) & (0.1, 0.1) & (0.09, 0.08) & (0.17, 0.14) \\
Oregon & (0.29, 0.3) & (0.4, 0.42) & (0.34, 0.34) & (-0.04, -0.03) & (0.09, 0.09) & (-0.09, -0.03) \\
Pennsylvania & (0.29, 0.3) & (0.39, 0.4) & (0.34, 0.34) & (0.01, 0.02) & (0.09, 0.08) & (0.02, 0.02) \\
Rhode Island & (0.27, 0.28) & (0.35, 0.38) & (0.34, 0.34) & (-0.11, -0.09) & (0.08, 0.08) & (-0.18, -0.06) \\
South Carolina & (0.27, 0.28) & (0.37, 0.37) & (0.34, 0.34) & (0.06, 0.07) & (0.08, 0.08) & (0.12, 0.07) \\
South Dakota & (0.27, 0.26) & (0.32, 0.32) & (0.34, 0.34) & (0.12, 0.13) & (0.08, 0.07) & (0.2, 0.19) \\
Tennessee & (0.26, 0.28) & (0.36, 0.38) & (0.34, 0.34) & (0.1, 0.11) & (0.09, 0.08) & (0.21, 0.18) \\
Texas & (0.28, 0.29) & (0.38, 0.39) & (0.34, 0.34) & (0.04, 0.05) & (0.09, 0.08) & (0.1, 0.06) \\
Utah & (0.28, 0.29) & (0.38, 0.38) & (0.34, 0.34) & (0.05, 0.05) & (0.09, 0.08) & (0.1, 0.05) \\
Vermont & (0.24, 0.25) & (0.31, 0.32) & (0.34, 0.34) & (-0.11, -0.08) & (0.06, 0.06) & (-0.15, -0.07) \\
Virginia & (0.28, 0.3) & (0.39, 0.4) & (0.34, 0.34) & (-0.0, 0.02) & (0.09, 0.08) & (-0.05, 0.02) \\
Washington & (0.28, 0.3) & (0.38, 0.4) & (0.34, 0.34) & (-0.04, -0.04) & (0.09, 0.09) & (-0.12, -0.05) \\
West Virginia & (0.27, 0.27) & (0.34, 0.35) & (0.34, 0.34) & (0.1, 0.1) & (0.08, 0.07) & (0.18, 0.1) \\
Wisconsin & (0.3, 0.3) & (0.39, 0.4) & (0.34, 0.34) & (0.02, 0.03) & (0.09, 0.08) & (0.05, 0.04) \\
Wyoming & (0.23, 0.23) & (0.31, 0.3) & (0.34, 0.34) & (0.17, 0.17) & (0.09, 0.09) & (0.29, 0.29) \\
\end{tabular}
}
    \caption{A table giving summary statistics relevant to the model parameters for four candidate elections. In each pair, the first number is for the bimodal distribution and the second is for the trimodal distribution. The first three columns give the rates for abstention and bullet voting. The last three columns give statistics for the distributions themselves.}
    \label{tab:state_stats_four}
\end{table}

\section{Data from noise tuning simulations}\label{sec:appendix_state_distribution_info}

\begin{figure}[!htbp]
    \centering

    \begin{tabular}{c}      \includegraphics[width=0.8\linewidth]{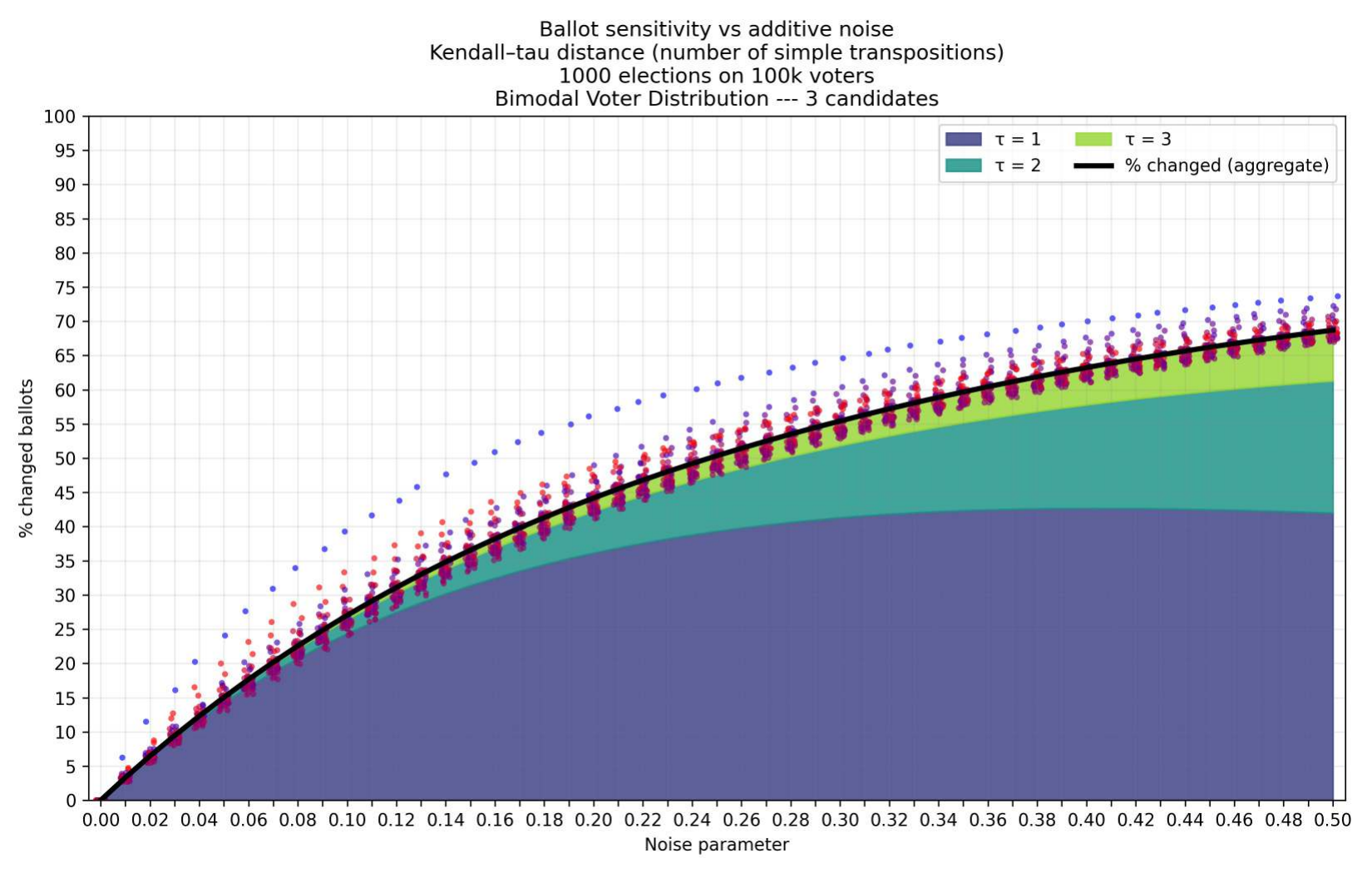} \\        \includegraphics[width=0.8\linewidth]{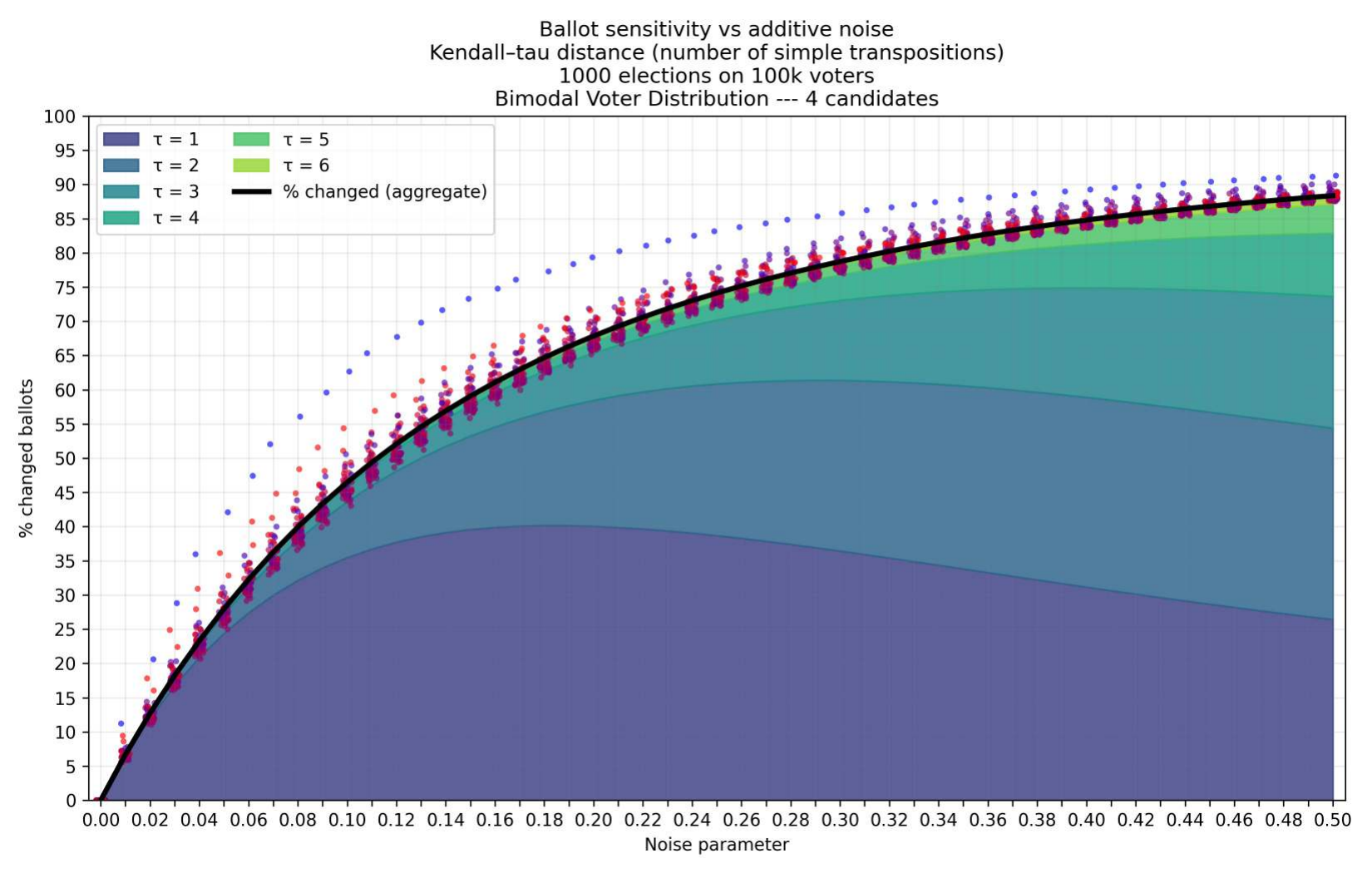} \\
    \end{tabular}

    \caption{For noise parameter values between 0.00 and 0.50 (in increments of 0.01), we simulated 1000 elections of 100k voters in each of the 50 states (plus DC). We then counted the number of ballots that changed after the inclusion of noise. The graph shows the average values for the Kendall-tau distance (i.e., the minimum number of simple transpositions to change the original ballot to the new ballot). The results for trimodal voter distributions are very similar.}\label{fig:noise_tuning}
\end{figure}

\newpage
\section*{Author Information}

\begin{flushleft}
David McCune\\
Division of Analytical Sciences, William Jewell College\\
\texttt{mccuned@william.jewell.edu}

\medskip

Matthew I. Jones\\
Department of Mathematics, Colby College

\medskip

Andrew Schultz\\
Department of Mathematics \& Statistics, Wellesley College

\medskip

Adam Graham-Squire\\
Department of Mathematics, High Point University

\medskip

Ismar Voli\'{c}\\
Department of Mathematics \& Statistics, Wellesley College

\medskip

Belle See\\
Wellesley College

\medskip

Karen Xiao\\
Wellesley College

\medskip

Malavika Mukundan\\
Department of Mathematics \& Statistics, Boston University
\end{flushleft}


\begin{thebibliography}{99}

\bibitem{AFG} N. Atkinson, N. Foley, and S. Ganz. (2024). Beyond the Spoiler Effect: Can Ranked-Choice Voting Solve the Problem of Political Polarization?. \emph{University of Illinois Law Review} no. 5: 1655-1698.

\bibitem{BFLR} D. Baumeister, P. Faliszewski, J. Lang, J. Rothe. Campaigns for lazy voters: truncated ballots. \emph{AAMAS '12: Proceedings of the 11th International Conference on Autonomous Agents and Multiagent Systems} \textbf{2}: 577-584.

\bibitem{B48} D. Black. (1948). On the Rationale of Group Decision-making. \emph{Journal of Political Economy} \textbf{56}(1): 23-34.

\bibitem{B13} B. Beuchel. (2013). Condorcet winners on median spaces. \emph{Social Choice and Welfare} \textbf{42}: 735-750.

\bibitem{C26} J. Clelland. (2026). Ranked Choice Voting And Condorcet Failure in the Alaska 2022 Special Election: How Might Other Voting Systems Compare?. \emph{The Mathematics Enthusiast} \textbf{23}(1): 185-196.



\bibitem{DDGGHMW25} C. Donnay, M. Duchin, J. Gibson, Z. Glaser, A. Hong, M. Mukundan, \& J. Wang. VoteKit: A Python package for computational social choice research. (2025). \emph{Journal of Open Source Software}, \textbf{10}(109), 7477

\bibitem{D57} A. Downs. (1957). \emph{An Economic Theory of Democracy}. Harper \& Row Publishers.

\bibitem{EH84} J. Enelow \& M. Hinich. (1984). \emph{The Spatial Theory of Voting: An Introduction.} Cambridge University Press.

\bibitem{EH90} J. Enelow \& M. Hinich. (1990). \emph{Advances in the Spatial Theory of Voting.} Cambridge University Press.

\bibitem{F25} E.B. Foley. (2025). Ballot Structures. In: L. Diamond, E. Foley, \& R. Pildes (Eds). (2025). \emph{Electoral Reform in the United States: Proposals for Combating Polarization and Extremism}. Lynne Rienner Publishers, Boulder, CO.

\bibitem{FM25} E. B. Foley \& E. Maskin. (2025). Condorcet Voting. Ohio State Legal Studies Research Paper No. 951, Available at SSRN: \url{http://dx.doi.org/10.2139/ssrn.5541259}.

\bibitem{GSM24} A. Graham-Squire and D. McCune. (2024). Ranked Choice Wackiness in Alaska. \emph{Math Horizons} \textbf{31}(1): 24-27.

\bibitem{HKRW} C. Hoffman, J.A. Kauba, J.C. Reidy, \& T. Weighill. (2024). Statistical models of ballot truncation in ranked choice elections. \emph{Communications in Statistics - Simulation and Computation}. \url{https://doi.org/10.1080/03610918.2024.2397032}

\bibitem{HP25} W. Holliday \& E. Pacuit. (2025). pref\_voting: The Preferential Voting Tools package for Python. \emph{Journal of Open Source Software}, \textbf{10}(105), 7020.

\bibitem{HP24} W. Holliday and E. Pacuit. (2024).  The Social Utility of Voting Revisited. Preprint: \url{https://ssrn.com/abstract=5073085}

\bibitem{KMT23} E. Kamwa, V. Merlin, \& F.M. Top. (2023). Scoring Run-off Rules, Single-peaked Preferences and Paradoxes of Variable Electorate.  Preprint: \url{hal-03143741v2}.

\bibitem{KGF20} D.M. Kilgour, J.C. Gr\`{e}goire, \& A.M. Foley. (2020). The prevalence and consequences of ballot truncation in ranked-choice elections. \emph{Public Choice} \textbf{184}: 197-218.

\bibitem{KGF25} D.M. Kilgour, J.C. Gr\`{e}goire, \& A.M. Foley. (2025).Condorcet efficiency: Weighted Bucklin vs. weighted scoring and Borda. \emph{Mathematical Social Sciences} \textbf{135}. \url{https://doi.org/10.1016/j.mathsocsci.2025.102420}.


\bibitem{Schwab_report} D. McCune, A. Schultz, I. Voli\'{c}, A. Graham-Squire, M.I. Jones, M. Mukundan, B. See, \& K. Xiao. (2025). Empirical Investigation of Ranked Voting Methods. Technical report, available at \url{https://mathematics-democracy-institute.org/empirical-analysis-of-ranked-choice-voting-methods/}.

\bibitem{MW} D. McCune and J. Wilson. (2025). Instant Runoff Voting and the Reinforcement Paradox. Preprint: \url{arXiv:2502.05185}.

\bibitem{M84} S. Merrill III. (1984). A Comparison of Efficiency of Multicandidate Electoral Systems. \emph{American Journal of Political Science} \textbf{28}(1): 23-48.

\bibitem{M80} H. Moulin. (1980). On strategy-proofness and single peakedness. \emph{Public Choice} \textbf{35}(4): 437-455.

\bibitem{M84} H. Moulin. (1984). Generalized Condorcet winners for single-peaked
and single-plateau preferences. \emph{Social Choice and Welfare} \textbf{1}: 127-147.

\bibitem{OS96} M. J. Osborne \& A. Slivinsky. (1996). A Model of Political
Competition with Citizen-Candidates. \emph{The Quarterly Journal of Economics} \textbf{111}(1): 65-96.

\bibitem{PH25} G.M. Parsons \& R. Hutchinson. (2025). Reform for Realists: The False Promise of Condorcet Voting. Forthcoming in \emph{Marquette Law Review}. Available at SSRN: \url{http://dx.doi.org/10.2139/ssrn.5101402}

\bibitem{PT14} F. Plassmann \& T.N. Tideman. (2014). How frequently do different voting rules encounter voting paradoxes in three-candidate elections?. \emph{Social Choice and Welfare} \textbf{42}: 31-75.

\bibitem{RT24} R. Robinette \& N. Tideman. (2024). The Importance of Condorcet Consistency in Preserving Democracy. In: Babst, G.A., Souris, R.N., McGregor, J. (eds) \emph{Liberal Constitutionalism and its Contemporary Challenges}. AMINTAPHIL: The Philosophical Foundations of Law and Justice, vol 12. Springer, Cham.


\bibitem{SAS} B. Schaffner, S. Ansolabehere, \& M. Shih. (2023). ``Cooperative Election Study Common Content, 2022", \url{https://doi.org/10.7910/DVN/PR4L8P}, Harvard Dataverse, V4.

\bibitem{T1} K. Tomlinson, J. Ugander, J. Kleinberg. (2023). The Moderating Effect of Instant Runoff Voting. Preprint: \url{arxiv.org/abs/2303.09734}

\bibitem{T2} K. Tomlinson, J. Ugander, J. Kleinberg. (2025). Exclusion Zones of Instant Runoff Voting. Preprint: \url{arxiv.org/abs/2502.16719}

\end{thebibliography}
\end{document}